\documentclass[12pt]{article}

\newlength{\abstractwidth}
\renewcommand{\thefootnote}{\fnsymbol{footnote}} 
\renewcommand{\thanks}[1]{\footnote{#1}} 

\newcommand{\ts}{\thinspace}

\makeatletter
\newcounter{fig}
\renewcommand\thefig{\arabic{fig}}

\def\fps@fig{tbp}
\def\ftype@fig{1}
\def\ext@fig{lof}
\def\fnum@fig{\figurename~\thefig}
\newenvironment{fig}
               {\@float{fig}}
               {\end@float}
\newenvironment{fig*}
               {\@dblfloat{fig}}
               {\end@dblfloat}

\makeatother

\usepackage{epsfig}

\newcommand{\be}{\begin{equation}}
\newcommand{\bea}{\begin{eqnarray}}
\newcommand{\eea}{\end{eqnarray}}
\newcommand{\ee}{\end{equation}}
\newcommand{\<}{\langle}
\renewcommand{\>}{\rangle}
\def\ba{\begin{eqnarray}}
\def\ea{\end{eqnarray}}

\def\ap{{\alpha^{\prime}}}

\def\eg{{\it e.g.~}}
\def\ie{{\it i.e.~}}

\def\Tr{{\rm Tr}}
\def\Re{{\rm Re}}
\def\Im{{\rm Im}}
\def\tr{{\rm tr}}
\def\str{{\rm str}}
\def\det{{\rm det}}
\def\mod{{\rm mod}}
\def\half{ {1\over 2}}
\def\p{\partial}
\def\AdS{${\rm AdS}_5 \times {\rm S}^5$}

\def\unit{1 \hskip-.3em \raise2pt\hbox{$ \scriptstyle |$ } }

\def\a{\alpha}
\def\b{\beta}
\def\c{\gamma}
\def\d{\delta}
\def\e{\epsilon}           
\def\f{\phi}               
  
\def\g{\gamma}

\def\k{\kappa}                    
\def\l{\lambda}
\def\m{\mu}
\def\n{\nu}
  
\def\q{\theta}                    
\def\r{\rho}                                     
\def\s{\sigma}                                   

\def\x{\xi}

\def\D{\Delta}
\def\F{\Phi}
\def\G{\Gamma}

\def\L{\Lambda}
\def\O{\Omega}  
\def\P{\Pi}

\def\L{{\cal L}}
\def\N{{\cal N}}
\def\O{{\cal O}}
\def\G{{\cal G}}

\def\cj{{\cal J}}

\def\bop#1{\setbox0=\hbox{$#1M$}\mkern1.5mu
        \vbox{\hrule height0pt depth.04\ht0
        \hbox{\vrule width.04\ht0 height.9\ht0 \kern.9\ht0
        \vrule width.04\ht0}\hrule height.04\ht0}\mkern1.5mu}
\def\Box{{\mathpalette\bop{}}}                        
\def\rarrow{\rightarrow}

\def\Bar#1{\overline{#1}}                      

\def\leftrighthookfill#1{$\mathsurround=0pt \mathord\hook#1
       \hrulefill\mathord\hook#1$}
\def\underhook#1{\vtop{\ialign{##\crcr                 
       $\hfil\displaystyle{#1}\hfil$\crcr
       \noalign{\kern-1pt\nointerlineskip\vskip2pt}
       \leftrighthookfill5\crcr}}}

\def\to{\rightarrow}

\makeatletter\newtheorem{ex}{Ex.}

\newcommand{\lf}[2]{\hbox{\large${\frac{ #1 }{ #2 }}$}}

\def\underset#1#2{\mathrel{\mathop{#2}\limits_{#1}}}

\makeatletter
\def\ronums{%
  \ifnum \@enumdepth >\thr@@\@toodeep\else
    \advance\@enumdepth\@ne
    \edef\@enumctr{enum\romannumeral\the\@enumdepth}%
      \list
        {\csname label\@enumctr\endcsname}%
        {\usecounter\@enumctr
          \addtolength{\leftmargin}{-\leftmargin}
          \settowidth{\labelwidth}{(99)}
          \itemindent = \labelwidth
          \addtolength{\itemindent}{\labelsep}
\renewcommand{\theenumi}{\roman{enumi}}
          }%
  \fi}

\makeatother

\newenvironment{alphal}{\begin{enumerate}
\renewcommand{\theenumi}{\alph{enumi}}
}{\end{enumerate}}

\newcommand{\lvert}{\mbox{\LARGE{$\vert$}}}

\begin{document}
\baselineskip=16pt

\begin{flushright}
UCLA/02/TEP/3 \\
MIT-CTP-3242
\end{flushright}

\begin{center}

{\Large \bf Supersymmetric Gauge Theories and

\medskip

\Large \bf the AdS/CFT Correspondence\footnote{Research
supported in part by the National Science Foundation under grants
PHY-98-19686 and PHY-00-96515.}}

\bigskip

{\Large\sl TASI 2001 Lecture Notes }

\bigskip \bigskip

{\bf Eric D'Hoker$^{a}$ and Daniel Z. Freedman$^b$}

\medskip

  ${}^a$ \textsl{Department of Physics and Astronomy} \\
  \textsl{University of California, Los Angeles, CA 90095, USA}

  \medskip

  ${}^b$ \textsl{Department of Mathematics and Center for Theoretical Physics} 
\\
  \textsl{Massachusetts Institute of Technology, Cambridge, MA
         02139, USA}
\end{center}

\vskip .8in

\centerline{\bf Abstract}

\medskip

In these lecture notes we first assemble the basic ingredients
of supersymmetric gauge theories (particularly N=4 super-Yang-Mills
theory), supergravity, and superstring theory. Brane solutions are
surveyed. The geometry and symmetries of anti-de Sitter space
are discussed. The AdS/CFT correspondence of Maldacena and its application
to correlation functions in the the conformal phase of N=4 SYM are
explained in considerable detail. A pedagogical treatment of
holographic RG flows is given including a review of the conformal
anomaly in four-dimensional quantum field theory and its calculation from
five-dimensional gravity. Problem sets and exercises await the reader.

\vfill\eject

\setcounter{footnote}{0}
\renewcommand{\thefootnote}{\roman{footnote}}

\tableofcontents

\vfill\eject

\baselineskip=14pt
\setcounter{equation}{0}
\setcounter{footnote}{0}

\section{Introduction}
\setcounter{equation}{0}

These lecture notes describe one of the most exciting developments
in theoretical physics of the past decade, namely Maldacena's bold conjecture
concerning the equivalence between superstring theory on certain 
ten-dimensional 
backgrounds involving Anti-de Sitter space-time and four-dimensional 
supersymmetric Yang-Mills theories. This AdS/CFT conjecture was unexpected 
because it relates a theory of gravity, such as string theory, to
a theory with no gravity at all. Additionally, the conjecture
related highly non-perturbative problems in Yang-Mills theory 
to questions in classical superstring theory or supergravity.
The promising advantage of the correspondence is that problems 
that appear to be intractable on one side may stand a chance of 
solution on the other side. We describe the initial conjecture,
the development of evidence that it is correct, and some further 
applications.

\medskip

The conjecture has given rise to a tremendous number of exciting
directions of pursuit and to a wealth of promising results.
In these lecture notes, we shall present a quick introduction to
supersymmetric Yang-Mills theory (in particular of $\N=4$ theory).
Next, we give a concise description of just enough supergravity and 
superstring theory to allow for an accurate description of the 
conjecture and for discussions of correlation functions
and holographic flows, namely the two topics that constitute
the core subject of the lectures. 

\medskip

The notes are based on the loosely coordinated lectures of both authors
at the 2001 TASI Summer School. It was decided to combine written
versions in order to have a more complete treatment. The bridge between
the two sets of lectures is Section 8 which presents a
self-contained introduction to the subject and a more detailed treatment 
of some material from earlier sections. 

\medskip

The AdS/CFT correspondence is a broad principle and the present notes
concern one of several pathways through the subject. An effort has
been made to cite a reasonably complete set of references on the subjects
we discuss in detail, but with less coverage of other aspects
and of background material.

\medskip

Serious readers will take the problem sets and exercises seriously!

\subsection{Statement of the Maldacena conjecture}

The Anti-de Sitter/Conformal Field Theory (AdS/CFT) correspondence, as
originally conjectured by Maldacena, advances a remarkable equivalence
between two seemingly unrelated theories. On one side (the {\sl AdS-side})
of the correspondence, we have 10-dimensional Type IIB string theory on the
product space \AdS, where the Type IIB 5-form flux through $S^5$ is an
integer $N$ and the equal radii $L$ of AdS$_5$ and $S^5$ are given by
$L^4= 4\pi g_s N \alpha '^2$, where $g_s$ is the string coupling. On the
other side (the {\sl SYM-side}) of the correspondence, we have 4-dimensional
super-Yang-Mills (SYM) theory with maximal $\N=4$ supersymmetry, gauge
group $SU(N)$, Yang-Mills coupling
$g_{YM}^2 =g_s$ in the conformal phase. The AdS/CFT conjecture states that
these two theories, including operator observables, states, correlation
functions and full dynamics, are equivalent to one another \cite{Malda,
Gubs, witten}. Indications of the equivalence had appeared in earlier
work, \cite{igor1,igor2,igor3}. For a general review of the
subject, see \cite{magoo}.

\medskip

In the strongest form of the conjecture, the correspondence is to hold for
all values of $N$ and all regimes of coupling $g_s = g_{YM}^2$. Certain
limits of the conjecture are, however, also highly non-trivial. The `t~Hooft 
limit on the SYM-side \cite{largeN}, in which $\lambda \equiv
g_{YM}^2 N$ is fixed as $N\to \infty$ corresponds to {\sl classical
string theory on
\AdS} (no string loops) on the AdS-side. In this sense, classical string
theory on
\AdS\ provides with a classical Lagrangian formulation of the large $N$
dynamics of $\N=4$ SYM theory, often referred to as the {\sl masterfield
equations}. A further limit $\lambda \to \infty$ reduces classical string
theory to classical Type IIB supergravity on \AdS. Thus, strong coupling
dynamics in SYM theory (at least in the large $N$ limit) is mapped onto
classical low energy dynamics in supergravity and string theory, a problem
that offers a reasonable chance for solution.

\medskip

The conjecture is remarkable because its correspondence is between a
10-dimensional theory of gravity and a 4-dimensional theory without
gravity at all, in fact, with spin $\leq 1$ particles only. The fact that
all the 10-dimensional dynamical degrees of freedom can somehow be encoded
in a 4-dimensional theory living at the boundary of AdS$_5$ suggests that
the gravity bulk dynamics results from a {\sl holographic image} generated
by the dynamics of the boundary theory \cite{holo}, see also
\cite{Bigatti:1999dp}. Therefore, the correspondence is also often
referred to as {\sl holographic}.

\subsection{Applications of the conjecture}

The original correspondence is between a $\N=4$ SYM theory in its conformal
phase and string theory on \AdS. The power of the correspondence is further
evidenced by the fact that the conjecture may be adapted to situations
without conformal invariance and with less or no supersymmetry on the SYM
side. The \AdS\ space-time is then replaced by other manifold or orbifold
solutions to Type IIB theory, whose study is usually more involved than was
the case for \AdS\ but still reveals useful information on SYM theory.

\vfill\eject

\section{Supersymmetry and Gauge Theories}
\setcounter{equation}{0}

We begin by reviewing the particle and field contents and invariant
Lagrangians in 4 dimensions, in preparation for a fuller discussion of
$\N=4$ super-Yang-Mills (SYM) theory in the next section. Standard
references include \cite{wessbagger, weinberg, west}; our conventions
are those of \cite{wessbagger}.

\subsection{Supersymmetry algebra in 3+1 dimensions}

Poincar{\'e} symmetry of flat space-time ${\bf R}^4$ with metric $\eta
_{\mu \nu} = {\rm diag} (- + + +)$, $\mu, \nu = 0,1,2 ,3$, is
generated by the translations ${\bf R}^4$ and  Lorentz transformations
$SO(1,3)$, with generators $P_\mu$ and $L_{\mu \nu}$ respectively. The
complexified Lorentz algebra is isomorphic to the complexified algebra of
$SU(2) \times SU(2)$, and its  finite-dimensional representations are
usually labeled by two positive (or zero) half integers $(s_+, s_-)$,
$s_\pm \in {\bf Z}/2$. Scalar, 4-vector, and rank 2 symmetric tensors
transform under
$(0,0)$,
$(\half, \half)$ and
$(1,1)$ respectively, while left and right chirality fermions and
self-dual and anti--self-dual rank 2 tensors transform under $(\half,0)$ and
$(0,\half)$ and $(1,0)$ and $(0,1)$ respectively.

Supersymmetry (susy) enlarges the Poincar{\'e} algebra by including spinor
{\sl supercharges},
\bea
a=1,\cdots , \N \qquad
\cases{Q_\alpha ^a  \quad  \ \ \alpha =1,2 & {\rm left \ Weyl \ spinor}
\cr & \cr
\bar Q_{\dot \alpha a}
=(Q_\alpha ^a)^\dagger & {\rm right \ Weyl \ spinor}
\cr}
\eea
Here, $\N$ is the number of independent supersymmetries of the algebra.
Two-component spinor notation, $\alpha =1,2$, is related to
4-component Dirac spinor notation by
\bea
\gamma ^\mu = \left ( \matrix{0 & \sigma ^\mu \cr \bar \sigma ^\mu & 0 \cr}
\right )
\qquad \qquad
Q^a = \left ( \matrix{Q^a _\alpha \cr \bar Q_a^{\dot \alpha}} \right )
\eea
The supercharges transform as Weyl spinors of $SO(1,3)$ and commute with
translations. The remaining susy structure relations are
\bea \label{susy}
\{ Q_\alpha ^a, \bar Q _{\dot \beta b}\}
= 2 \sigma ^\mu _{\alpha \dot \beta} P_\mu \delta ^a _b
\qquad \qquad
\{ Q_\alpha ^a,  Q _\beta  ^b \}
= 2 \epsilon _{\alpha  \beta} Z^{ab}
\eea
By construction, the generators $Z^{ab}$ are anti-symmetric in the indices
$I$ and $J$, and commute with all generators of the supersymmetry algebra.
For the last reason, the $Z^{ab}$ are usually referred to as {\sl central
charges}, and we have
\be
Z^{ab} = - Z^{ba}
\qquad \qquad
[Z^{ab}, {\rm anything} ] =0
\ee
Note that for $\N=1$, the anti-symmetry of $Z$ implies that $Z=0$.

The supersymmetry algebra is invariant under a global phase rotation
of  all supercharges $Q_\alpha ^a$, forming a group $U(1)_R$. In
addition, when $\N >1$, the different supercharges may be rotated into one
another under the unitary group $SU(\N)_R$.
These (automorphism) symmetries of the supersymmetry algebra are called
{\sl R-symmetries}. In quantum  field theories, part or all of these
R-symmetries may be broken by anomaly effects.

\subsection{Massless Particle Representations}

To study massless representations, we choose a Lorentz frame in which the
momentum takes the form $P^\mu = (E,0,0,E)$, $E>0$. The susy algebra
relation (\ref{susy}) then reduces to
\be \label{massless}
\{ Q_\alpha ^a, (Q _\beta ^b)^\dagger \}
= 2 (\sigma ^\mu P_\mu)_{\alpha \dot \beta} \delta _b ^a
= \left ( \matrix{4E & 0 \cr 0 & 0 \cr } \right ) _{\alpha \dot \beta}
\delta _b ^a
\ee
We consider only unitary particle representations, in which the operators
$Q_\alpha ^a$ act in a positive definite Hilbert space. The relation for
$\alpha = \dot \beta =2$ and $a=b$ implies
\be \label{vanishing}
\{ Q_2 ^a , (Q_2 ^a)^\dagger \} = 0
\qquad \Longrightarrow \qquad
Q_2 ^a =0, \quad Z^{ab}=0
\ee
The relation $Q_2 ^a=0$ follows because the left hand side of
(\ref{vanishing}) implies that the norm of $Q_2 ^a |\psi\rangle$ vanishes
for any state $|\psi\rangle$ in the Hilbert space. The relation $Z^{ab}=0$
then follows  from (\ref{susy}) for $\alpha =2$ and $\dot \beta =1$. The
remaining supercharge operators are

\begin{itemize}

\item $Q_1 ^a$ \ which lowers helicity by $1/2$;

\item $\bar Q_{\dot 1} ^a = (Q_1 ^a )^\dagger$ \ which raises
helicity by $1/2$.

\end{itemize}

\noindent
Together, $Q_1 ^a$ and $(Q_1 ^a)^\dagger$, with $a=1,\cdots , \N$ form a
representation of dimension $2^\N$ of the Clifford algebra associated with
the Lie algebra $SO(2\N)$. All the states in the representation may be
obtained by starting from the highest helicity state $|h \rangle$ and
applying products of $Q_1 ^a$ operators for all possible values of $a$.

We shall only be interested in CPT invariant theories, such as
quantum field theories and string theories, for which the particle
spectrum must be symmetric under a sign change in helicity. If the
particle spectrum obtained as a Clifford representation in the above
fashion is not already CPT self-conjugate, then we shall take instead the
direct sum with its CPT conjugate. For helicity $\leq 1$, the spectra are
listed in table 1. The $\N=3$ and $\N=4$ particle spectra then coincide,
and their quantum field theories are identical.

\begin{table}[b]
\begin{center}
\begin{tabular}{|c|c| c|c|c|c|c|} \hline
Helicity  & $\N=1$ & $\N=1$ & $\N=2$ & $\N=2$ & $\N=3$ & $\N=4$ \\
$\leq 1$  & gauge  & chiral & gauge  & hyper  & gauge  & gauge  \\
\hline \hline
1         & 1      & 0      & 1      & 0      & 1      & 1      \\ \hline
$1/2$     & 1      & 1      & 2      & 2      & 3+1    & 4      \\ \hline
0         & 0      & 1+1    & 1+1    & 4      & 3+3    & 6      \\ \hline
$-1/2$    & 1      & 1      & 2      & 2      & 1+3    & 4      \\ \hline
$-1$      & 1      & 0      & 1      & 0      & 1      & 1      \\ \hline
\hline Total $\#$ & $2 \times 2$ & $2 \times 2$ & $ 2 \times 4$ & 8 & $ 2
\times 8$ & 16  \\ \hline
\end{tabular}
\end{center}
\caption{Numbers of Massless States as a function of $\N$ and helicity}
\label{table:1}
\end{table}

\subsection{Massive Particle Representations}

For massive particle representations, we choose the rest frame with  $P^
\mu = (M,0,0,0)$, so that the first set of susy algebra structure
relations takes the form
\be
\{ Q_\alpha ^a, (Q_\beta ^b)^\dagger \} = 2M \delta _\alpha ^\beta
\delta _b ^a
\ee
To deal with the full susy algebra, it is convenient to make use of the
$SU(\N)_R$ symmetry to diagonalize in blocks of $2\times 2$
the anti-symmetric matrix $Z^{ab} = - Z^{ba}$. To do so, we split the
label $a$ into two labels : $a=(\hat a,\bar a)$ where $\hat a=1,2$ and
$\bar a =1,\cdots ,r$, where $\N=2r$ for $\N$ even  (and we append a further
single label when
$\N$ is odd). We then have
\be
Z = {\rm diag} (\epsilon Z_1, \cdots , \epsilon Z_r, \# )
\qquad
\epsilon ^{12} = - \epsilon ^{21} =1
\ee
where $\# $ equals 0 for $\N=2r+1$ and $\#$ is absent for $\N=2r$. The
$Z_{\bar a}$, ${\bar a} =1,\cdots ,r$ are real {\sl central charges}. In
terms of linear combinations $ {\cal Q} ^{\bar a}  _{\alpha \pm} \equiv
\half \bigl ( Q_\alpha ^{1\bar a} \pm \sigma ^0 _{\alpha \dot \beta} (Q_\beta
^{2\bar a})^\dagger \bigr ), $ the only non-vanishing susy structure
relation left is (the $\pm$ signs  below  are correlated)
\be
\label{BPS}
\{ {\cal Q}^{\bar a} _{\alpha \pm} , ( {\cal Q}^{\bar b} _{\beta \pm}
)^\dagger
\} =
\delta _{\bar b} ^{\bar a} \delta _\alpha ^\beta (M \pm Z_{\bar a})
\ee
In any {\sl unitary particle representation},  the operator on the left
hand side of (\ref{BPS}) must be positive, and thus we obtain the famous
{\sl BPS bound} (for Bogomolnyi-Prasad-Sommerfield, \cite{bps}) giving a
lower bound on the mass in terms of the central charges,
\be
M \geq |Z_{\bar a}| \qquad {\bar a} =1,\cdots , r = \big [ \N / 2 \big ]
\ee
Whenever one of the values $|Z_{\bar a}|$ equals $M$, the BPS bound is
(partially) saturated and either the supercharge ${\cal Q}^{\bar a} _{\alpha
+}$ or ${\cal Q}^{\bar a} _{\alpha -}$ must vanish. The supersymmetry
representation then suffers {\sl multiplet shortening}, and is usually
referred to as BPS. More precisely, if we have $M = |Z_{\bar a}|$ for
${\bar a} =1,\cdots, r_o$, and $M>|Z_{\bar a}|$ for all other values of
$\bar a$, the susy  algebra is effectively a Clifford algebra associated
with $SO(4\N -4r_o)$, the corresponding representation is said to be
$1/2^{r_o}$ BPS, and has dimension  $2^{2\N-2r_o}$.

\begin{table}[b]
\begin{center}
\begin{tabular}{|c|c| c|c|c|c|c|} \hline
Spin      & $\N=1$ & $\N=1$ & $\N=2$ & $\N=2$     & $\N=2$     & $\N=4$    \\
$\leq 1$  & gauge  & chiral & gauge  & BPS gauge  & BPS hyper  & BPS gauge \\
\hline \hline
1         & 1      & 0      & 1      & 1          & 0          & 1    \\
\hline
$1/2$     & 2      & 1      & 4      & 2          & 2          & 4    \\
\hline
0         & 1      & 2      & 5      & 1          & 4          & 5    \\
\hline
\hline
Total $\#$& 8      & 4      & 16     & 8          & 8          & 16   \\
\hline
\end{tabular}
\end{center}
\caption{Numbers of Massive States as a function of $\N$ and spin}
\label{table:2}
\end{table}

\subsection{Field Contents and Lagrangians}

The analysis of the preceding two subsections has revealed that the
supersymmetry particle representations for $1 \leq \N \leq 4$, with spin
less or equal to 1, simply consist of customary spin 1 vector particles,
spin 1/2  fermions and spin 0 scalars. Correspondingly, {\sl the fields in
supersymmetric theories with spin less or equal to 1 are customary spin 1
gauge fields, spin 1/2 Weyl fermion fields and spin 0 scalar fields, but
these fields are restricted to enter in multiplets of the relevant
supersymmetry algebras.}

\medskip

Let $\G$ denote the gauge algebra, associated with a compact Lie group $G$.
For any $1\leq \N \leq 4$, we have a gauge multiplet, which transforms
under the adjoint representation of $\G$. For $\N=4$, this is the only
possible multiplet. For $\N=1$ and $\N=2$, we also have {\sl matter
multiplets} : for $\N=1$, this is the {\sl chiral multiplet}, and for
$\N=2$ this is the {\sl hypermultiplet}, both of which may transform under
an arbitrary unitary, representation ${\cal R}$ of $\G$. Component fields
consist of the customary gauge field $A_\mu$, left Weyl fermions
$\psi _\alpha$ and $\lambda _\alpha$ and scalar fields $\phi$, $H$ and
$X$.

\begin{itemize}

\item $\N=1$ {\sl Gauge Multiplet} \ $(A_\mu \ \lambda_\alpha)$, where
$\lambda _\alpha$ is the gaugino Majorana fermion;

\item $\N=1$ {\sl Chiral Multiplet} \ $(\psi _\alpha \ \phi)$, where $\psi
_\alpha$ is a left Weyl fermion and $\phi$ a complex scalar, in the
representation ${\cal R}$ of $\G$.

\item $\N=2$ {\sl Gauge Multiplet} \ $(A_\mu \ \lambda _{\alpha \pm} \ \phi)
$,  where $\lambda _{\alpha \pm}$ are left Weyl fermions, and $\phi$ is the
complex {\it gauge scalar}. Under $SU(2)_R$ symmetry, $A_\mu$ and $\phi$ are
singlets, while $\lambda _+$ and $\lambda _-$ transform as a doublet.

\item $\N=2$ {\sl Hypermultiplet} \
$(\psi_{\alpha \pm}\ H_\pm )$,
where $\psi _{\alpha \pm}$ are left Weyl fermions and $H_\pm$ are two
complex  scalars, transforming under the representation ${\cal R}$ of $\G$.
Under $SU(2)_R$ symmetry, $\psi _\pm$ are singlets, while $H_+$ and $H_-$
transform as a doublet.

\item $\N=4$ {\sl Gauge Multiplet} \ $(A_\mu \ \lambda _\alpha ^a \ X^i)
$, where $\lambda _\alpha ^a$, $a=1,\cdots ,4$ are left Weyl fermions
and $X^i$, $i=1,\cdots , 6$ are real scalars. Under $SU(4)_R$
symmetry, $A_\mu$ is a singlet, $\lambda _\alpha ^a$ is a {\bf 4} and the
scalars $X^i$ are a rank 2 anti-symmetric {\bf 6}.

\end{itemize}

Lagrangians invariant under supersymmetry are customary Lagrangians of
gauge, spin 1/2 fermion and scalar fields, (arranged in multiplets of the
supersymmetry algebra) with  certain special relations amongst the coupling
constants and masses. We shall restrict attention to local Lagrangians in
which each term has a total of no more than two derivatives on all boson
fields and no more than one derivative on all fermion fields. All
renormalizable Lagrangians are of this form, but all low energy effective
Lagrangians are also of this type.

\medskip

The case of the $\N=1$ gauge multiplet $(A_\mu \ \lambda _\alpha)$ by
itself is particularly simple. We proceed by writing down all possible
gauge invariant polynomial terms of dimension 4 using minimal gauge
coupling,
\be
\label{lag}
\L =
- {1 \over 2 g^2} \tr F_{\mu \nu} F^{\mu \nu}
+{\theta _I \over 8 \pi ^2} \tr F_{\mu \nu} \tilde F^{\mu \nu}
-{i \over 2} \tr \bar \lambda \bar \sigma ^\mu D_\mu \lambda
\ee
where $g$ is the gauge coupling, $\theta _I$ is the instanton angle, the
field strength is $F_{\mu \nu}
= \partial _\mu A_\nu - \partial _\nu A_\mu + i [A_\mu , A_\nu]$,
$\tilde F_{\mu\nu}=\half \epsilon_{\mu\nu\rho\sigma} F^{\rho\sigma}$
is the Poincar{\'e} dual of $F$, and $D_\mu = \partial _\mu
\lambda + i [A_\mu , \lambda]$. Remarkably, $\L$ is automatically invariant
under the $\N=1$ supersymmetry transformations
\bea \label{xi}
\delta _\xi A_\mu
& = &
i \bar \xi \bar \sigma _\mu \lambda - i \bar \lambda \bar \sigma _\mu \xi
\nonumber \\
\delta _\xi \lambda
& = &
\sigma ^{\mu \nu } F_{\mu \nu } \xi
\eea
where $\xi$ is a spin 1/2 valued infinitesimal supersymmetry parameter. Note
that the addition in (\ref{lag}) of a Majorana mass term $m\lambda
\lambda$ would spoil supersymmetry.

\medskip

As soon as scalar fields are to be included, such as is the  case for any
other multiplet, it is no longer so easy to guess supersymmetry
invariant Lagrangians and one is led to the use of superfields. {\sl
Superfields} assemble all component fields of a given supermultiplet
(together with auxiliary fields) into a supersymmetry multiplet field on
which supersymmetry transformations act linearly. Superfield methods
provide a powerful tool for producing supersymmetric field equations for
any degree of supersymmetry. For $\N=1$ there is a standard off-shell
superfield formulation as well (see \cite{wessbagger, weinberg, west} for
standard treatments). Considerably more involved off-shell superfield
formulations are also available for $\N=2$ in terms of harmonic and
analytic superspace \cite{gikos}, see also the review of \cite{DP99}. For
$\N=4$ supersymmetry, no off-shell formulation is known at present; one
is thus forced to work either in components or in the $\N=1$ or $\N=2$
superfield formulations. A survey of theories with extended supersymmetry may
be found in \cite{Seiberg98}.

\subsection{The $\N$=1 Superfield Formulation}

The construction of field multiplets containing all fields that transform
linearly into one another under supersymmetry requires the introduction of
anti-commuting spin 1/2 coordinates. For $\N=1$ supersymmetry, we introduce
a (constant) left Weyl spinor coordinate $\theta _\alpha$ and its complex
conjugate $\bar \theta ^{\dot \alpha} = (\theta _\alpha)^\dagger$,
satisfying $ [x^\mu, \theta _\alpha] = \{ \theta _\alpha , \theta _\beta \}
= \{ \theta _\alpha, \bar \theta  ^{\dot \beta}\} = \{ \bar \theta ^{\dot
\alpha} , \bar \theta ^{\dot \beta} \}=0$.
Superderivatives are defined by
\be
D_\alpha \equiv {\partial \over \partial \theta ^\alpha} + i \sigma ^\mu
_{\alpha \dot \alpha} \bar \theta ^{\dot \alpha} \partial _\mu
\qquad \qquad
\bar D_{\dot \alpha} \equiv -{\p \over \p \bar \theta ^{\dot \alpha}}
- i \theta ^{ \alpha} \sigma ^\mu _{\alpha \dot \alpha}  \partial _\mu
\ee
where differentiation and integration of $\theta$ coordinates are defined by
\be
{\partial \over \partial \theta ^\alpha}
(1,\theta ^\beta, \bar \theta ^{\dot \beta})
\equiv
\int \! d\theta ^\alpha (1,\theta ^\beta, \bar \theta ^{\dot \beta})
\equiv
(0,\delta _\alpha {} ^\beta, 0)
\ee
For general notations and conventions for spinors and their contractions,
see \cite{wessbagger}.

\medskip

A {\sl general superfield} is defined as a general function of the
superspace coordinates $x^\mu, \theta _\alpha, \bar \theta ^{\dot \alpha}$.
Since the square of each $\theta ^\alpha$ or of each $\bar \theta ^{\dot
\alpha}$ vanishes, superfields admit finite Taylor expansions in powers of
$\theta$  and $\bar \theta$. A general superfield $S
(x,\theta, \bar \theta)$ yields the following {\sl component expansion}
\bea
\label{sfield}
S (x,\theta, \bar \theta)
& = &
\phi (x) + \theta \psi(x) + \bar \theta \bar \chi (x) + \bar \theta \bar
\sigma ^\mu \theta A_\mu (x) + \theta \theta f(x) + \bar \theta \bar \theta
 g^* (x) \nonumber \\
&& + i \theta \theta \bar \theta \bar \lambda (x) - i \bar \theta \bar
\theta  \theta \rho (x) + \half \theta \theta \bar \theta \bar \theta D(x)
\eea
A bosonic superfield obeys $[S, \theta ^\alpha]=[S, \bar \theta _{\dot
\alpha}]=0$, while a fermionic superfield obeys $\{S, \theta ^\alpha\}=\{S,
\bar \theta _{\dot \alpha}\}=0$.
If $S$ is bosonic (resp. fermionic), the component fields $\phi$, $A_\mu$,
$f$, $g$ and $D$ are bosonic (resp. fermionic) as well, while the fields
$\psi$, $\chi$, $\lambda$ and  $\rho$ are fermionic (resp. bosonic).
The superfields belong to a ${\bf Z}_2$ graded algebra of functions on
superspace, with the even grading for bosonic odd grading for fermionic
fields. We shall denote the grading by $g(S)$, or sometimes just $S$.
Superderivatives on superfields satisfy the following graded
differentiation rule
\bea
D_\alpha (S _1 S _2)
= (D_\alpha S _1) S _2
+ (-)^{g(S _1)  g(S _2)} S _1 (D_\alpha S _2)
\eea
where $g(S _i)$ is the grading of the field $S _i$.

\medskip

On superfields, supersymmetry transformations are realized in a linear way
via super-differential operators. The infinitesimal supersymmetry
parameter is still a constant left Weyl spinor $\xi$, as in (\ref{xi}) and
we have
\be \label{tfon}
\delta _\xi S = (\xi Q + \bar \xi \bar Q) S
\ee
with the supercharges defined by
\be
Q_\alpha = {\partial \over \partial \theta ^\alpha} - i \sigma ^\mu
_{\alpha \dot \alpha} \bar \theta ^{\dot \alpha} \partial _\mu
\qquad \qquad
\bar Q_{\dot \alpha} = -{\p \over \p \bar \theta ^{\dot \alpha}}
+ i \theta ^{ \alpha} \sigma ^\mu
_{\alpha \dot \alpha}  \partial _\mu
\ee
The super-differential operators $D_\alpha$ and $Q_\alpha$ differ only by a
sign change. They generate left and right actions of supersymmetry
respectively. Their relevant structure relations are
\be
\{ Q_\alpha, \bar Q_{\dot \beta}\}= 2 \sigma ^\mu _{\alpha \dot \beta} P_\mu
\qquad \qquad
\{ D_\alpha, \bar D_{\dot \beta}\}= - 2 \sigma ^\mu _{\alpha \dot \beta}
P_\mu
\ee
where $P_\mu = i \p _\mu$.  Since left and right actions mutually commute,
all components of $D$ anti-commute with all components of $Q$ : $\{
Q_\alpha , D_\beta\}= \{Q_\alpha , \bar D^{\dot \beta}\}=0$.
Furthermore, the product of any three $D$'s or any  three $Q$'s vanishes,
$ D_\alpha D_\beta D_\gamma = Q_\alpha Q_\beta Q_\gamma =0$.
The general superfield is reducible; the irreducible components are as
follows.

\medskip

\noindent
(a) The {\sl Chiral Superfield} $\Phi$ is obtained by imposing the
condition
\be
\label{dphi}
\bar D_{\dot \alpha} \Phi =0
\ee
This condition is invariant under the supersymmetry transformations  of
(\ref{tfon}) since $D$ and $Q$ anti-commute.
Equation (\ref{dphi}) may be solved in terms of the composite coordinates
$ x^\mu _\pm = x^\mu \pm i \theta \sigma ^\mu \bar
\theta $ which satisfy $ \bar D_{\dot \alpha} x^\mu _+=0$ and $
D_\alpha x_-^\mu =0$. We have
\bea
\label{chiral}
\Phi (x,\theta, \bar \theta)  =
\phi (x_+) + \sqrt 2 \theta \psi (x_+) + \theta \theta F(x_+)
\eea
The component fields $\phi$ and $\psi$ are the scalar and left Weyl spinor
fields of the chiral multiplet respectively, as discussed previously. The
field equation for $F$ is a non-dynamical or {\sl auxiliary field} of the
chiral multiplet.

\medskip

\noindent
(b) The {\sl Vector Superfield} is obtained by imposing on a general
superfield of the type (\ref{sfield}) the condition
\be
V=V^\dagger
\ee
thereby setting $\chi = \psi$, $g=f$ and $\rho=\lambda$  and requiring
$\phi$, $A_\mu$ and $D$ to be real.

\medskip

\noindent
(c) The {\sl Gauge Superfield} is a special case of a vector superfield,
where $V$ takes values in the gauge algebra $\G$. The reality condition
$V=V^\dagger$ is preserved by the gauge transformation
\be
\label{nonabgauge}
e^ V \longrightarrow e^ {V'} = e^{-i\Lambda ^\dagger} e^V e^{i \Lambda} \,.
\ee
where $\Lambda$ is a chiral superfield taking values also in $\G$.
Under the above gauge transformation law, the component fields $\phi$,
$\psi=\chi$,  and $f=g$ of a general real superfield may be gauged  away
in an algebraic way. The gauge in which this is achieved is the {\sl
Wess-Zumino gauge}, where the gauge superfield  is given by
\be
\label{vector}
V(x,\theta, \bar \theta)
=
\bar \theta \bar \sigma ^\mu \theta A_\mu (x)
+ i \theta \theta \bar \theta \bar \lambda (x) - i \bar \theta \bar \theta
\theta \lambda (x) + \half \theta \theta \bar \theta \bar \theta D(x)
\ee
The component fields $A_\mu$ and $\lambda$ are the gauge and gaugino fields
of the gauge multiplet respectively, as discussed previously. The field $D$
has not appeared previously and is an {\sl auxiliary field}, just as $F$
was an auxiliary field for the chiral multiplet.

\subsection{General $\N$=1 Susy Lagrangians via Superfields}

Working out the supersymmetry transformation (\ref{tfon}) on  chiral and
vector superfields in terms of components, we see that the only
contribution to the auxiliary fields is from the $\theta \partial$ term of
$Q$ and thus takes the form of a total derivative. However, because the form
(\ref{vector}) was restricted to Wess-Zumino gauge, $F$ and $D$
transform by a total derivative only if $F$ and $D$ are themselves gauge
singlets, in which case we have
\bea
\delta _\xi F &=& i \sqrt 2 \partial _\mu (\bar \xi \bar \sigma ^\mu \psi)
\nonumber \\
\delta _\xi D &=& \partial _\mu (i\bar \xi \bar \sigma ^\mu \lambda
-i \bar \lambda \bar \sigma ^\mu \xi)
\eea
These transformation properties guarantee that the $F$ and
$D$ auxiliary fields yield supersymmetric invariant Lagrangian terms,
\bea
{\rm F-terms} & \qquad & \L_F = F = \int d^2 \theta \ \Phi
\nonumber \\
{\rm D-terms} & \qquad & \L_D = \half D = \int d^4 \theta \  V
\eea
The $F$ and $D$ terms used to construct invariants need not be elementary
fields, and may be gauge invariant composites of elementary fields instead.
Allowing for this possibility, we may now derive the most general possible
$\N=1$ invariant Lagrangian in terms of superfields. To do so, we need the
following ingredients $\L _U$, $\L _G$ and $\L_K$. Putting together
contributions from these terms, we have the most general
$\N=1$ supersymmetric Lagrangian with the restrictions of above.

\medskip

(1) \ Any complex analytic function $U$ depending only on left chiral
superfields $\Phi ^i$ (but not on their complex conjugates) is itself a
left chiral superfield, since $\bar D _{\dot \alpha} \Phi ^i =0$ implies
that
$ \bar D_{\dot \alpha} U(\Phi ^i)=0$.
Thus, for any complex analytic function $U$, called the  {\sl
superpotential}, we may construct an invariant contribution to the
Lagrangian by forming an $F$-term
\be
\L _U = \int d^2 \theta \ U(\Phi ^i) + \int d^2 \bar \theta
\overline{U(\Phi ^i)} = \sum _i F^i {\partial U \over \partial \phi ^i}
-\half \sum _{i,j} \psi ^i \psi ^j {\partial ^2 U \over \partial \phi ^i
\partial
\phi ^j} + {\rm cc}
\ee

(2) \ The gauge field strength is a fermionic left chiral spinor
superfield $W_\alpha$, which is constructed out of the gauge superfield
$V$  by
\be
4 W_\alpha = - \bar D \bar D \big (e^{-V} D_\alpha e^{+V} \big )
\ee
The gauge field strength may be used as a chiral superfield along with
elementary (scalar) chiral superfields to build up $\N=1$ supersymmetric
Lagrangians via $F$-terms. In view of our restriction to Lagrangians
with no more than two derivatives on Bose fields, $W$ can enter at most
quadratically. Thus, the most general gauge
kinetic and self-interaction term is from the $F$-term of the gauge field
strength $W_\alpha$ and the elementary (scalar) chiral superfields  $\Phi
^i$ as  follows,
\be
\L _G = \int d^2 \theta \ \tau _{cc'}(\Phi ^i) W^c W^{c'} + {\rm complex \
conjugate}
\ee
Here, $c$ and $c'$ stand for the gauge index running over the adjoint
representation of $\G$, and are contracted in a gauge invariant way. The
functions $\tau _{cc'}(\Phi ^i)$ are  complex analytic.

\medskip

(3) \ The left and right chiral superfields  $\Phi ^i $ and $(\Phi ^i)
^\dagger$, as well as the gauge superfield $V$, may be combined into a
gauge invariant vector superfield $K(e^V \Phi ^i , (\Phi ^i) ^\dagger)$,
provided the gauge algebra is realized linearly on the fields $\Phi ^i$.
The function $K$ is called the {\sl K{\"a}hler potential}. Assuming that
the gauge  transformations $\Lambda$ act on $V$ by (\ref{nonabgauge}),
the chiral superfields $\Phi$ transform as $
\Phi \longrightarrow \Phi ' = e^{-i \Lambda } \Phi $,
so that $e^V \Phi$ transforms as $\Phi$. An invariant Lagrangian may
be constructed via a $D$-term,
\be
\L _K = \int d^4 \theta  K(e^V \Phi ^i, (\Phi ^i) ^\dagger)
\ee
Upon expanding $\L_K$ in components, one sees immediately that the leading
terms already generates an action with two derivatives on boson fields.
As a result, $K$ must be a function only of the superfields
$\Phi ^i$ and $(\Phi ^ i) ^\dagger$ and $V$, but not of their derivatives.

\subsection{$\N$\ts=\ts1 non-renormalization theorems}

Non-renormalization theorems provide very strong results on the structure
of the effective action at low energy as a function of the bare action.
Until recently, their validity was restricted to perturbation theory and
the proof of the theorems was based on supergraph methods
\cite{pertnon}. Now, however, non-renormalization theorems have been
extended to the non-perturbative regime, including the effects of
instantons \cite{seiberg88}. The assumptions underlying  the theorems are
that (1) a supersymmetric renormalization is carried out, (2) the
effective action is constructed by Wilsonian renormalization
(see \cite{kwilson} for a review). The last assumption allows one to
circumvent
any possible singularities  resulting from the integration over massless
states.

\medskip

The non-renormalization theorems state that the superpotential $\L _U$ is
{\sl unrenormalized}, or more precisely that it receives no quantum
corrections, infinite or finite. Furthermore, the gauge field term $\L
_G$ is renormalized only through the gauge coupling $\tau _{cc'}$, such
that its complex analytic dependence on the chiral superfields is
preserved. In perturbation theory, $\tau _{cc'}$ receives quantum
contributions only through 1-loop graphs, essentially because the
$U(1)_R$ axial anomaly is a 1-loop effect in view of the Adler-Bardeen
theorem. Non-perturbatively, instanton corrections also enter, but in a
very restricted way. The special renormalization properties of these two
$F$-terms are closely related to their holomorphicity \cite{seiberg88}.
The K\" ahler potential term $\L _K$ on the other hand does receive
renormalizations both at the perturbative and non-perturbative levels.

\vfill\eject

\section{\protect\boldmath$\N$\ts=\ts4 Super Yang-Mills}
\setcounter{equation}{0}

The Lagrangian for the $\N=4$ super-Yang Mills theory is unique and given
by \cite{Grimm}
\bea
\L & = & \tr \biggl \{
-{1 \over 2 g^2} F_{\mu \nu} F^{\mu \nu}
+ {\theta _I \over 8 \pi ^2} F_{\mu \nu} \tilde F^{\mu \nu}
- \sum _a i \bar \lambda ^a \bar \sigma ^\mu D_\mu \lambda _a
- \sum _i D_\mu X^i D^\mu X^i
\nonumber \\ && \qquad
+  \sum _{a,b,i} g C^{ab} _i \lambda _a [X^i, \lambda _b] +
\sum _{a,b,i} g \bar C_{iab}  \bar \lambda ^a [X^i, \bar \lambda ^b]
+ {g^2 \over 2} \sum _{i,j} [X^i , X^j]^2 \biggr \}
\eea
The constants $C^{ab}_i$ and $C_{iab}$ are related to the Clifford
Dirac matrices for $SO(6) _R \sim SU(4)_R$. This is evident when
considering $\N=4$ SYM in $D=4$ as the dimensional reduction on $T^6$ of
$D=10$ super-Yang-Mills theory (see problem set 4.1). By construction,
the Lagrangian is invariant under $\N=4$ Poincar\' e supersymmetry, whose
transformation laws are given by
\bea
\delta X ^i & = &
[Q ^a _\alpha , X^i] =
C^{iab} \lambda _{\alpha b}
\nonumber \\
\delta \lambda _b & = &
\{ Q^a _\alpha , \lambda _{\beta b} \} =
F^+ _{\mu \nu} (\sigma ^{\mu \nu} )^\alpha {} _\beta \delta _b ^a +
[X^i,X^j] \epsilon _{\alpha \beta} (C_{ij}) ^a {} _b
\nonumber \\
\delta \bar \lambda ^b _{\dot \beta} & = &
\{ Q^a _\alpha  , \bar \lambda ^b _{\dot \beta} \} = C^{ab} _i \bar
\sigma _{\alpha \dot \beta} ^\mu D_\mu X^i
\nonumber \\
\delta A_\mu & = &
[Q^a _\alpha , A_\mu] = (\sigma _\mu) _\alpha {} ^{\dot \beta} \bar \lambda
^a _{\dot \beta}
\eea
The constants $(C_{ij}) ^a {} _b$ are related to bilinears in Clifford
Dirac matrices of $SO(6)_R$.

\medskip

Classically, $\L$ is {\sl scale invariant}. This may be
seen by assigning the standard mass-dimensions to the fields and couplings
\be
[A_\mu ]=[X^i] =1 \qquad
[\lambda _a] = {3 \over 2} \qquad
[g]=[\theta _I]=0
\ee
All terms in the Lagrangian are of dimension 4, from which scale invariance
follows. Actually, in a relativistic field theory, scale invariance and
Poincar\' e invariance combine into a larger {\sl conformal symmetry},
forming the group $SO(2,4)\sim SU(2,2)$. Furthermore, the combination of
$\N=4$ Poincar\' e supersymmetry and conformal invariance produces an even
larger {\sl superconformal symmetry} given by the supergroup $SU(2,2|4)$.

\medskip

Remarkably, upon perturbative quantization, $\N=4$ SYM theory exhibits no
ultraviolet divergences in the correlation functions of its canonical
fields. Instanton corrections also lead to finite contributions and
is believed that the theory is UV finite. As a result, the renormalization
group $\beta$-function of the theory vanishes identically (since no
dependence on any renormalization scale is introduced during the
renormalization process). The theory is exactly scale invariant at the
quantum level, and the superconformal group $SU(2,2|4)$ is a fully quantum
mechanical symmetry.

\medskip

The {\sl Montonen-Olive or S-duality conjecture} in addition posits a
discrete global symmetry of the theory \cite{montolive}. To state this
invariance, it is standard to combine the real coupling $g$ and the real
instanton angle
$\theta _I$ into a single complex coupling
\be
\tau \equiv {\theta _I \over 2 \pi} + { 4 \pi i \over g^2}
\ee
The quantum theory is invariant under $\theta _I \to \theta _I + 2 \pi$,
or $\tau \to \tau+1$. The {\sl Montonen-Olive conjecture} states that the
quantum theory is also invariant under the $\tau \to - 1/\tau$. The
combination of both symmetries yields the S-duality  group
SL(2,{\bf Z}), generated by
\be
\tau \to {a \tau + b \over c \tau + d}
\qquad\qquad ad-bc=1, \quad a,b,c,d \in {\bf Z}
\ee
Note that when $\theta _I =0$, the S-duality transformation amounts to
$g\to 1/g$, thereby exchanging strong and weak coupling.

\subsection{Dynamical Phases}

To analyze the dynamical behavior of $\N=4$ theory, we look at the
potential energy term,
$$
 - g^2 \sum _{i,j} \int \tr [X^i, X^j]^2
$$
In view of the positive definite behavior of the Cartan - Killing form on
the compact gauge algebra $\G$, each term in the sum is positive or zero.
When the full potential is zero, a minimum is thus automatically attained
corresponding to a $\N=4$ supersymmetric ground state. In turn, any
$\N=4$ supersymmetric ground state is of this form,
\be
[X^i, X^j]=0, \qquad i,j=1,\cdots ,6
\ee
There are two classes of solutions to this equation,

\begin{itemize}

\item The {\sl superconformal phase}, for which $\< X^i \> =0$ for all
$i=1,\cdots ,6$. The gauge algebra $\G$ is unbroken. The superconformal
symmetry $SU(2,2|4)$ is also unbroken. The physical states and operators
are gauge invariant (i.e. $\G$-singlets) and transform under unitary
representations of $SU(2,2|4)$.

\item The {\sl spontaneously broken or Coulomb phase}, where $\< X^i\>
\not=0$ for at least one $i$. The detailed dynamics will depend upon the
degree of residual symmetry. Generically, $ \G \to  U(1) ^r$ where $ r =
{\rm rank}\  \G$, in which case the low energy behavior is that of $r$
copies of $\N=4$ $U(1)$ theory. Superconformal symmetry is spontaneously
broken  since the non-zero vacuum expectation value $\<X^i\>$ sets a scale.

\end{itemize}

\subsection{Isometries and Conformal Transformations}

In the first part of these lectures, we shall consider the SYM theory in
the conformal phase and therefore make heavy use of superconformal
symmetry. In the present subsection, we begin by reviewing conformal
symmetry first. Let $M$ be a Riemannian (or pseudo-Riemannian) manifold of
dimension $D$ with metric $G_{\mu \nu}$, $\mu, \nu =0,1,\cdots ,D-1$. We
shall now review the notions of diffeomorphisms, isometries and conformal
transformations.

\medskip

$\bullet$ \ A {\sl diffeomorphism} of $M$ is a differentiable map of local
coordinates $x^\mu$, $\mu = 1,\cdots ,D$, of $M$ given either globally by
$x^\mu \to x'^\mu  (x)$ or infinitesimally by a vector field $v^\mu(x)$
so that $\delta _v x^\mu = - v^\mu (x) $. Under a general diffeomorphism,
the metric on $M$ transforms as
\bea
G'_{\mu \nu} (x') dx'^\mu dx'^\nu & = & G_{\mu \nu} dx^\mu dx^\nu
 \\
\delta _v G_{\mu \nu} & = & \nabla _\mu v_\nu + \nabla _\nu v_\mu
\qquad \qquad
\nabla _\mu v_\nu \equiv \p_\mu v_\nu - \Gamma _{\mu \nu} ^\rho v_\rho
\nonumber
\eea

\medskip

$\bullet$ \ An {\sl isometry} is a diffeomorphism under which the metric is
invariant,
\bea
G'_{\mu \nu} (x) = G_{\mu \nu} (x)
\qquad {\rm or} \qquad
\delta _v G_{\mu \nu } = \nabla _\mu v_\nu + \nabla _\nu v_\mu =0
\eea

\medskip

$\bullet$ A \ {\sl conformal transformation} is a diffeomorphism that
preserves the metric up to an overall (in general $x$-dependent) scale
factor, thereby preserving all angles,
\bea
G'_{\mu \nu} (x) = w(x) G_{\mu \nu} (x)
\qquad {\rm or} \qquad
\delta _v G_{\mu \nu } = \nabla _\mu v_\nu + \nabla _\nu v_\mu =
 {2 \over D} G_{\mu \nu} \nabla _\rho v^\rho
\eea

\medskip

The case of $M={\bf R}^D$, $D\geq 3$, flat Minkowski space-time with flat
metric  $ \eta _{\mu \nu } = {\rm diag} (- + \cdots +)$ is
an illuminating example. (When $D=2$, the conformal algebra is isomorphic
to the infinite-dimensional Virasoro algebra.) Since now $\nabla _\mu =
\p _\mu$, the equations for isometries reduce to $\p _\mu v_\nu + \p _\nu
v_\mu=0$, while those for conformal transformations become $\p _\mu v_\nu +
\p _\nu v_\mu - 2/D \eta _{\mu \nu } \p _\rho v^\rho=0$. The solutions are
\bea
{\rm isometries} && (1) \quad v^\mu \ {\rm constant} \ : \ {\rm
translations}
\nonumber \\
&& (2) \quad v_\mu = \omega _{\mu \nu} x^\nu \ : \ {\rm Lorentz}
\nonumber \\
{\rm conformal} && (3) \quad v^\mu = \lambda x^\mu \ : \ {\rm dilations}
\nonumber \\
&& (4) \quad v_\mu = 2 c_\rho x^\rho x_\mu - x_\rho x^\rho c_\mu
\ : \ {\rm special \ conformal}
\eea
In a local field theory, continuous symmetries produce conserved currents,
according to Noether's Theorem. Currents associated with isometries and
conformal transformations may be expressed in terms of the stress tensor
$T_{\mu \nu}$. This is because the stress tensor for any local field theory
encodes the response of the theory to a change in metric; as a result, it
is automatically symmetric $T^{\mu \nu} = T^{\nu \mu}$. We have
\be
j^\mu _v \equiv T^{\mu \nu} v_\nu
\ee
Covariant conservation of this current requires that $ \nabla _\mu j^\mu_v
= (\nabla _\mu T^{\mu \nu}) v_\nu + T^{\mu \nu} \nabla _\mu v_\nu =0$.
For an {\sl isometry}, conservation thus requires that $\nabla _\mu T^{\mu
\nu}=0$. For a {\sl conformal
transformation}, conservation in addition requires that $T_\mu {}^\mu=0$.
Starting out with Poincar\' e and scale invariance, all
of the above conditions will have to be met so that special conformal
invariance will be automatic.

\subsection{(Super) Conformal $\N$=4 Super Yang-Mills}

In this subsection, we show that the global continuous symmetry group of
$\N=4$ SYM is given by the supergroup $SU(2,2|4)$, see \cite{minwalla}. The
ingredients are as follows.

\begin{itemize}

\item {\sl Conformal Symmetry}, forming the group $SO(2,4) \sim SU(2,2)$
is generated by translations $P^\mu$, Lorentz transformations $L_{\mu
\nu}$, dilations $D$ and special conformal transformations $K^\mu$;

\item {\sl R-symmetry}, forming the group $SO(6)_R \sim SU(4)_R$, generated
by $T^A$, $A=1,\cdots ,15$;

\item {\sl Poincar\' e supersymmetries} generated  by the supercharges $Q^a
_\alpha$ and their complex conjugates $\bar Q _{\dot \alpha a}$,
$a=1,\cdots,4$. The presence of these charges results immediately from
$\N=4$ Poincar\' e supersymmetry;

\item {\sl Conformal supersymmetries} generated by the supercharges
$S_{\alpha a}$ and their complex conjugates $\bar S ^a _{\dot \alpha}$. The
presence of these symmetries results from the fact that the Poincar\' e
supersymmetries and the special conformal transformations $K_\mu$ do not
commute. Since both are symmetries, their commutator must also be a
symmetry, and these are the $S$ generators.

\end{itemize}

The two bosonic subalgebras $SO(2,4)$ and $SU(4)_R$ commute. The
supercharges $Q_\alpha ^a$ and $\bar S ^a _{\dot \alpha}$ transform under
the {\bf 4} of $SU(4)_R$, while $\bar Q_{\dot \alpha a}$ and $S_{\alpha a}$
transform under the {\bf 4}$^*$. From these data, it is not hard to
see how the various generators fit into a super algebra,
\bea
\left ( \matrix{
P_\mu \ K_\mu \ L_{\mu \nu} \ D && Q^a _\alpha \ \bar S ^a _{\dot \alpha}
\cr & \cr
\bar Q _{\dot \alpha a} \ S_{\alpha a}  && T^A \cr} \right )
\eea
All structure relations are rather straightforward, except the relations
between the supercharges, which we now spell out. To organize the structure
relations, it is helpful to make use of a natural grading of the algebra
given by the dimension of the generators,
\bea
[D]=[L_{\mu \nu}]= [T^A] =0 \qquad && [P^\mu ]=+1 \qquad [K_\mu ]=-1
\nonumber \\
&&
[Q] = + 1/2 \qquad [S]=-1/2
\eea
Thus, we have
\bea
\{ Q _\alpha ^a , Q_\beta ^b \} & = & \{ S _{\alpha a }, S_{\beta b} \}= \{
Q_{\alpha} ^a, \bar S_{\dot \beta }^b\} = 0
\nonumber \\
\{ Q_\alpha ^a , \bar Q _{\dot \beta b} \} & = & 2 \sigma _{\alpha \dot
\beta }^\mu P_\mu \delta _b {}^a
\nonumber \\
\{ S_{\alpha a}, \bar S ^b _{\dot \beta} \} & = & 2 \sigma _{\alpha \dot
\beta} ^\mu K_\mu \delta _a {} ^b
\nonumber \\
\{ Q_\alpha ^a , S_{\beta b} \} & = & \epsilon _{\alpha \beta} \bigl (
\delta _b {}^a D + T ^a {}_b \bigr ) + \half \delta _b {}^a L_{\mu \nu}
\sigma ^{\mu \nu} _{\alpha \beta}
\eea

\subsection{Superconformal Multiplets of Local Operators}

We shall be interested in constructing and classifying all local, gauge
invariant operators in the theory that are polynomial in the canonical
fields. The restriction to polynomial operators stems from the fact that it
is those operators that will have definite dimension.

\medskip

The canonical fields $X^i$, $\lambda _a$ and $A_\mu$ have unrenormalized
dimensions, given by 1, 3/2 and 1 respectively.
Gauge invariant operators will be constructed rather from the gauge
covariant objects $X^i$, $\lambda _a$, $F^\pm _{\mu \nu}$ and the covariant
derivative $D_\mu$, whose dimensions are
\be
[X^i ] = [D_\mu ] =1
\qquad
[F^\pm _{\mu \nu} ] =2
\qquad
[\lambda _a ] = {3 \over 2}
\ee
Here, $F^\pm _{\mu \nu}$ stands for the (anti) self-dual gauge field
strength. Thus, if we temporarily ignore the renormalization effects of
composite operators, we see that all operator dimensions will be positive
and that the number of operators whose dimension is less than a given
number is finite. The only operator with dimension 0 will be the unit
operator.

\medskip

Next, we introduce the notion of {\sl superconformal primary operator}.
Since the conformal supercharges $S$ have dimension $-1/2$,
successive application of $S$ to any operator of definite dimension must at
some point yield 0 since otherwise we would start generating operators of
negative dimension, which is impossible in a unitary representation.
Therefore one defines a {\sl superconformal primary operator} $\O$ to be a
non-vanishing operator such that
\be
[S, \O]_\pm =0 \qquad \O \not=0
\ee
An equivalent way of defining a superconformal primary operator is as the
lowest dimension operator in a given superconformal multiplet or
representation. It is important to distinguish a superconformal primary
operator from a {\sl conformal primary operator}, which is instead
annihilated by the special conformal generators $K^\mu$, and is thus
defined by a weaker condition. Therefore, every superconformal primary is
also a conformal primary operator, but the converse is not, in general,
true.

\medskip

Finally, an operator $\O$ is called a {\sl superconformal descendant
operator} of an operator $\O'$ when it is of the form,
\be
\O = [Q, \O']_\pm
\ee
for some well-defined local polynomial gauge invariant operator $\O'$.
If $\O$ is a descendant of $\O'$, then these two operators belong to the
same superconformal multiplet. Since the dimensions are related by $\Delta
_\O = \Delta _{\O'} + 1/2$, the operator $\O$ can never be a conformal
primary operator, because there is in the same multiplet at least one
operator $\O'$ of dimension lower than $\O$. As a result, in a given
irreducible superconformal multiplet, there is a single superconformal
primary operator (the one of lowest dimension) and all others are
superconformal descendants of this primary.

\medskip

It is instructive to have explicit forms for the superconformal primary
operators in $\N=4$ SYM. The construction is most easily carried out by
using the fact that a superconformal primary operator is NOT the
$Q$-commutator of another operator. Thus, a key ingredient in the
construction is the $Q$ transforms of the canonical fields. We shall
need these here only schematically,
\bea
\{ Q, \lambda \} = F^+ + [X,X]
&& \qquad
[Q, X] = \lambda
\nonumber \\
\{ Q, \bar \lambda \} = DX \hskip .55in
&& \qquad
[Q, F] = D \lambda
\eea
A local polynomial operator containing any of the elements on the rhs of
the above structure relations cannot be primary. As a result, chiral primary
operators can involve neither the gauginos $\lambda$ nor the gauge field
strengths $F^\pm$. Being thus only functions of the scalars $X$, they can
involve neither derivatives nor commutators of $X$. As a result,
superconformal primary operators are gauge invariant scalars involving only
$X$ in a symmetrized way.

\medskip

The simplest are the {\sl single trace operators}, which are of the form
\bea
\str \biggl ( X^{i_1} X^{i_2} \cdots X^{i_n} \biggr )
\eea
where $i_j$, $j=1,\cdots ,n$ stand for the $SO(6)_R$ fundamental
representation indices. Here, ``str" denotes the symmetrized trace over the
gauge algebra and as a result of this operation, the above operator is
totally symmetric in the $SO(6)_R$-indices $i_j$. In general, the above
operators transform under a reducible representation (namely the
symmetrized product of $n$ fundamentals) and irreducible operators may be
obtained by isolating the traces over $SO(6)_R$ indices. Since $\tr X^i=0$,
the simplest operators are
\bea
\sum _i \tr X^i X^i & \sim & {\rm Konishi \ multiplet}
\nonumber \\
\tr X^{\{ i} X^{j\}} & \sim & {\rm supergravity \ multiplet}
\eea
where $\{ij\}$ stands for the traceless part only. The reasons for these
nomenclatures will become clear once we deal with the AdS/CFT
correspondence.

\medskip

More complicated are the {\sl multiple trace operators}, which are obtained
as products of single trace operators. Upon taking the tensor product of
the individual totally symmetric representations, we may now also encounter
(partially) anti-symmetrized representations of $SO(6)_R$. There is a
one-to-one correspondence between chiral primary operators and unitary
superconformal multiplets, and so all state and operator multiplets may be
labeled in terms of the superconformal chiral primary operators.

\subsection{$\N=4$ Chiral or BPS Multiplets of Operators}

The unitary representations of the superconformal algebra $SU(2,2|4)$ may
be labeled by the quantum numbers of the bosonic subgroup, listed below,
\bea
&&
SO(1,3) \times SO(1,1) \times SU(4)_R
\nonumber \\
&&
\ (s_+,s_-) \quad \qquad \Delta \quad \qquad [r_1,r_2,r_3]
\eea
where $s_\pm$ are positive or zero half integers, $\Delta $ is the positive
or zero dimension and $[r_1,r_2,r_3]$ are the Dynkin labels of the
representations of $SU(4)_R$. It is sometimes preferable to refer to
$SU(4)_R$ representations by their dimensions, given in terms of $\bar r_i
\equiv r_i +1$ by
\be
\dim [r_1,r_2,r_3] = {1 \over 12} \bar r_1 \bar r_2 \bar r_3 (\bar r_1 +
\bar r_2)  (\bar r_2 + \bar r_3) (\bar r_1+ \bar r_2+ \bar r_3)
\ee
instead of their Dynkin labels.  The complex conjugation
relation is $[r_1,r_2,r_3]^* = [r_3,r_2,r_1]$.

\medskip

In unitary representations, the dimensions $\Delta$ of the operators are
bounded from below by the spin and $SU(4)_R$ quantum numbers. To see this,
it suffices to restrict to primary operators since they have the lowest
dimension in a given irreducible multiplet. As shown previously, such
operators are scalars, so that the spin quantum numbers vanish,
and the dimension is bounded from below by the $SU(4)_R$ quantum
numbers. A systematic analysis of \cite{Dobrev85}, (see also
\cite{And99, Fer00}) for this case reveals the existence of 4 distinct
series,
\begin{enumerate}

\item $\Delta = r_1 + r_2 + r_3$;

\item $ \Delta = {3 \over 2} r_1 + r_2 + \half r_3 \geq 2 + \half r_1 +
r_2 +{3\over 2} r_3$ \qquad this requires $ r_1 \geq r_3 +2$;

\item $ \Delta = \half r_1 + r_2 + {3\over 2} r_3 \geq 2 + {3\over 2} r_1
+ r_2 + \half  r_3$ \qquad this requires $r_3 \geq r_1+2$;

\item $ \Delta \geq {\rm Max} \biggl [
2 + {3\over 2} r_1 + r_2 +\half r_3 ; 2 + \half r_1 + r_2
+{3\over 2} r_3 \biggr ]$

\end{enumerate}
Clearly, cases 2. and 3. are complex conjugates of one another.

\medskip

Cases 1. 2. and 3. correspond to {\sl discrete series of representations},
for which at least one supercharge $Q$ commutes with the primary operator.
Such representations are shortened and usually referred to as {\sl chiral
multiplets} or {\sl BPS multiplets}. The term BPS multiplet arises from the
analogy with the BPS multiplets of Poincar\' e supersymmetry discussed in
subsections \S 2.3. Since these representations are shortened, their
dimension is {\sl unrenormalized} or protected from receiving quantum
corrections.

\medskip

Case 4. corresponds to {\sl continuous series of representations}, for
which no supercharges $Q$ commute with the primary operator. Such
representations are referred to as {\sl non-BPS}. Notice that the dimensions
of the operators in the continuous series is separated from the dimensions
in the discrete series by a gap of at least 2 units of dimension.

\medskip

The BPS multiplets play a special role in the AdS/CFT correspondence. In
Table 3 below, we give a summary of properties of various BPS and non-BPS
multiplets. In the column labeled by $\# Q$ is listed the number of
Poincar\' e supercharges that leave the primary invariant.

\begin{table}[t]
\begin{center}
\begin{tabular}{|c|c| c|c|c|c|c|} \hline
 Operator type & $\# Q$  & spin range & $SU(4)_R$ primary & dimension
$\Delta$ \\ \hline \hline
identity & 16  & 0  & $[0,0,0]$             & 0        \\ \hline
1/2 BPS  & 8   & 2  & $[0,k,0]$, \ $k\geq 2$  & $k $     \\ \hline
1/4 BPS  & 4   & 3  & $[\ell,k,\ell]$, \ $\ell \geq 1$  & $k+2\ell$
\\ \hline
1/8 BPS  & 2   & 7/2& $[\ell,k,\ell+2m]$ & $k+2\ell+3m$, \ $m\geq 1$
\\ \hline
non-BPS  & 0       & 4      & any          & unprotected \\ \hline
\end{tabular}
\end{center}
\caption{Characteristics of BPS and Non-BPS multiplets}
\label{table:3}
\end{table}

\medskip

\noindent {\bf Half-BPS operators}

It is possible to give an explicit description of all 1/2 BPS operators.
The simplest series is given by single-trace operators of the form
\bea
\O _k (x) \equiv {1 \over n_k} \str \biggl ( X^{\{ i_1} (x) \cdots X^{i_k\}}
(x) \biggr )
\eea
where ``str" stands for the symmetrized trace introduced previously,
$\{ i_1 \cdots i_k\}$ stands for the $SO(6)_R$ traceless part of the
tensor, and $n_k$ stands for an overall normalization of the operator which
will be fixed by normalizing its 2-point function. The dimension of these
operators is unrenormalized, and thus equal to $k$.

\medskip

However, it is also possible to have multiple trace 1/2 BPS operators. They
are built as follows. The tensor product of $n$ representations
$[0,k_1,0] \otimes \cdots \otimes [0,k_n,0]$, always contains the
representation $[0,k,0]$, $k=k_1 + \cdots + k_n$, with multiplicity 1. (The
highest weight of the representation $[0,k,0]$ is then the sum of the
highest weights of the component representations.) The most general 1/2 BPS
gauge invariant operators are  given by the projection onto the
representation $[0,k,0]$ of the corresponding product of operators,
\bea
\O _{(k_1,\cdots ,k_n)} (x) \equiv
\biggl [\O _{k_1} (x) \cdots \O _{k_n} (x) \biggr ]  _{[0,k,0]}
\qquad
k=k_1 + \cdots + k_n
\eea
Here the brackets $[\ ]$ stand for the operators product of the operators
inside. This product is in general singular and thus ambiguous, but the
projection onto the representation $[0,k,0]$ is singularity free and thus
unique.

\medskip

\noindent {\bf 1/4 and 1/8 BPS Operators}

There are no single-trace 1/4 BPS operators. The simplest construction is
in terms of double trace operators. It is easiest to list all possibilities
in a single expression, using the notations familiar already from the 1/2
BPS case. The operators are of the form
\bea
\biggl [\O _{k_1} (x) \cdots \O _{k_n} (x) \biggr ]
_{[\ell ,k,\ell ]}
\qquad
k+ 2 \ell = k_1 + \cdots + k_n
\eea
In the free theory, the above operators will be genuinely 1/4 BPS, but in
the interacting theory, the operators will also contain an admixture of
descendants of non-BPS operators \cite{ryzhov}.
The series of 1/8 BPS operators starts with triple trace operators, and
are generally of the form
\bea
\biggl [\O _{k_1} (x) \cdots \O _{k_n} (x) \biggr ]
_{[\ell ,k,\ell +2m ]}
\qquad
k+ 2 \ell +3m = k_1 + \cdots + k_n
\eea
In the interacting theory, admixtures with descendants again have to
be included.

\subsection{Problem Sets}

(3.1) Show that the 1-loop renormalization group $\beta$-function for
$\N=4$ SYM vanishes.

\bigskip

\noindent
(3.2) Express the $\N=4$ SYM Lagrangian in terms of $\N=1$ superfields.

\bigskip

\noindent
(3.3) Work out the full conformal $SO(2,4)\sim SU(2,2)$ and superconformal
$SU(2,2|4)$ structure relations (commutators and anti-commutators of the
generators).

\bigskip

\noindent
(3.4) Derive the Noether currents associated with the Poincar\' e $Q ^a
_\alpha$ and conformal $\bar S _{\dot \alpha a} $ supercharges (and
complex conjugates) in terms of the canonical fields of $\N=4$ SYM.

\bigskip

\noindent
(3.5) In the Abelian Coulomb phase of $\N=4$ SYM, where the gauge algebra
$\G$ is spontaneously broken to $U(1)^r$, $r={\rm rank} \ \G$, the
global superconformal algebra $SU(2,2|4)$ is also spontaneously broken. To
simplify matters, you may take $\G = SU(2)$. (a) Identify the generators
of $SU(2,2|4)$ which are preserved and (b) those which are spontaneously
broken, thus producing Goldstone bosons and fermions. (c) Express the
Goldstone boson and fermion fields in terms of the canonical fields of
$\N=4$ SYM.

\vfill\eject

\section{Supergravity and Superstring Theory}
\setcounter{equation}{0}

In this section, we shall review the necessary supergravity and superstring
theory to develop the theory of D-branes and D3-branes in particular.

\subsection{Spinors in general dimensions}

Consider $D$-dimensional Minkowski space-time $M_D$ with flat metric $\eta
_{\mu \nu} = {\rm diag} (- + \cdots +)$, $\mu, \nu =0,1,\cdots ,D-1$. The
Lorentz group is $SO(1,D-1)$ and the generators of the Lorentz algebra
$J_{\mu \nu}$ obey the standard structure relations
\bea
[J_{\mu \nu} , J_{\rho \sigma} ] =
- i \eta _{\mu \rho} J_{\nu \sigma}
+ i \eta _{\nu \rho} J_{\mu \sigma}
- i \eta _{\nu \sigma} J_{\mu \rho}
+ i \eta _{\mu \sigma} J_{\nu \rho}
\eea
The {\sl Dirac spinor} representation, denoted $S_D$, is defined in
terms of the standard Clifford-Dirac matrices $\Gamma _\mu$,
\bea
J_{\mu \nu} = {i \over 4} [\Gamma _\mu , \Gamma _\nu]
\qquad \qquad
\{ \Gamma _\mu , \Gamma _\nu \} = 2 \eta _{\mu \nu}
\eea
Its (complex) dimension is given by ${\rm dim}_{\bf C} S_D = 2 ^{[D/2]}$.

\medskip

For $D$ even, the Dirac spinor representation is always reducible because
in that case there exists a chirality matrix $\bar \Gamma$, with square
$\bar \Gamma ^2 = I$, which anti-commutes with all $\Gamma _\mu$ and
therefore commutes with $J_{\mu \nu}$,
\bea
\bar \Gamma \equiv i^{\half D(D-1)+1} \Gamma _0 \Gamma _1 \cdots \Gamma
_{d-1}
\qquad \qquad
\{ \bar \Gamma , \Gamma _\mu\} =0  \quad \Rightarrow \quad  [\bar
\Gamma , J_{\mu
\nu}]=0
\eea
As a result, the Dirac spinor is the direct sum of two {\sl Weyl spinors}
$S_D = S_+ \oplus S_-$. The reality properties of the Weyl spinors depends
on $D \ (\mod \ 8)$, and is given as follows,
\bea
D \equiv 0,4 \ (\mod \ 8) && S_- = S_+ ^* \qquad {\rm both \ complex}
\nonumber \\
D \equiv 2,6 \ (\mod \ 8) && S_+ \quad \ S_- \qquad {\rm self-conjugate}
\eea
For both even and odd $D$, the {\sl charge conjugate} $\psi ^c$ of a Dirac
spinor
$\psi$ is defined by
\bea
\label{chargeconj}
\psi ^c \equiv C \Gamma _0 \psi ^*
\qquad \qquad
C \Gamma _\mu C^{-1} = - (\Gamma _\mu)^T
\eea
Requiring that a spinor be real is a basis dependent condition and thus
not properly Lorentz covariant. The proper Lorentz invariant condition for
reality is that a spinor be its own charge conjugate $\psi ^c =\psi$; such
a spinor is called a {\sl Majorana spinor}. The Majorana condition requires
that $(\psi ^c)^c = \psi$, or $C \Gamma _0 (C\Gamma _0 )^* = I$, which is
possible only in dimensions $D\equiv 0,1,2,3,4 \ (\mod \ 8)$. In
dimensions $D \equiv 0,4 \ (\mod \ 8)$, a Majorana spinor is equivalent to
a Weyl spinor, while in dimension $D \equiv 2 \ (\mod \ 8)$ it is
possible to impose the Majorana and Weyl conditions at the same time,
resulting in {\sl Majorana-Weyl spinors}. In dimensions $D\equiv 5,6,7$
(mod 8), one may group spinors into doublets $\Psi _\pm$ and it is
possible to impose a {\sl symplectic Majorana condition} given by $\Psi
_\pm ^c = \mp \Psi _\mp$. Useful reviews are in \cite{weinberg,
polchinski}.

\medskip

\subsection{Supersymmetry in general dimensions}

The basic Poincar\' e supersymmetry algebra in $M_D$ is obtained by
supplementing the Poincar\' e algebra with $\N$ supercharges $Q_\alpha ^I$,
$I=1,\cdots , \N$. Here $Q$ transforms in the spinor representation $S$,
which could be a Dirac spinor, a Weyl spinor, a Majorana spinor or a
Majorana-Weyl spinor, depending on $D$. Thus, $\alpha$ runs over the spinor
indices $\alpha = 1,\cdots , {\rm dim} S$. Whatever the spinor is, we shall
always  write it as a Dirac spinor. The fundamental supersymmetry algebra
could include central charges just as was the case for $D=4$. However, we
shall here be interested mostly in a restricted class of supersymmetry
representations in which we have a massless graviton, such as we have in
supergravity and in superstring theory. Therefore, we may ignore the
central charges.

\medskip

A general result, valid in dimension $D\geq 4$, states that {\sl
interacting massless fields of spin $>2$ cannot be causal, and are
excluded on physical grounds}. Considering theories with a
massless graviton, and assuming that supersymmetry is realized linearly, the
massless graviton must be part of a massless supermultiplet of states and
fields. By the above general result, this multiplet cannot contain fields
and states of spin $>2$. This fact puts severe restrictions on which
supersymmetry algebras can be realized in various dimensions.

\medskip

The existence of massless unitary representations of the supersymmetry
algebra requires vanishing central charges, just as was the case in $d=4$.
Thus, we shall consider the Poincar\' e supersymmetry algebras of the form
(useful reviews are in \cite{weinberg, polchinski}, see also
\cite{Gliozzi:1976qd} and \cite{Nahm:1977tg}),
\bea
\{ Q^I _\alpha , (Q^J _\beta)^\dagger \} = 2 \delta ^I{} _J (\Gamma _\mu)
_\alpha ^\beta P^\mu
\qquad \qquad
\{ Q^I _\alpha , Q^J _\beta \} = 0
\eea
To analyze
massless representations, choose $P^\mu = (E,0,\cdots,0,E), \ E>0$, so that
the supersymmetry algebra in this representation simplifies and becomes
\bea
\{ Q^I _\alpha , (Q^J _\beta)^\dagger \} = 2 \delta ^I{} _J \left (
\matrix{4E & 0 \cr 0 & 0 } \right )_\alpha  ^ \beta
\eea
On this unitary massless representation, half of the supercharges
effectively vanish $Q^I_\alpha =0$, $\alpha = \half \dim S +1, \cdots ,
\dim S$. Half of the remaining supercharges may be viewed as lowering
operators  for the Clifford algebra, while the other half may be viewed as
raising operators. Thus, the total number of raising operators is $
1/4 \cdot \N \cdot \dim_{\bf R} S $. Each operator raising helicity by 1/2,
and total helicity ranging at most from $-2$ to $+2$, we should have at
most 8 raising operators and this produces an important bound,
\bea
\N \dim_{\bf R} S \leq 32
\eea
In other words, the maximum number of Poincar\' e supercharges is always
32.

\medskip

The largest dimension $D$ for which the bound may be satisfied is $D=11$
and $\N=1$, for which there are precisely 32 Majorana supercharges. In
$D=10$, the bound is saturated for $\N=2$ and 16-dimensional Majorana-Weyl
spinors. There is indeed a unique $D=11$ supergravity theory discovered by
Cremmer, Julia and Scherk \cite{cjs}. Many of the lower dimensional
theories may be constructed by Kaluza-Klein compactification on a circle
or on a torus of the $D=11$ theory and we shall therefore treat this
method first \cite{trieste}.

\medskip

\subsection{Kaluza-Klein compactification on a circle}

We wish to compactify one space dimension on a circle $S^1 _R$ of
radius $R$. Accordingly, we decompose the coordinates $x^\mu$ of ${\bf
R}^D$ into a coordinate $y$ on the circle and the remaining coordinates
$x^{\bar \mu}$. The wave operator with flat metric in $D$ dimensions $\Box
_D$ then becomes
\bea
\Box _D = \Box _{D-1} + {\p ^2 \over \p y^2}
\eea
We shall be interested in finding out how various fields behave, in
particular in the limit $R\to 0$, referred to as {\sl dimensional
reduction}.

\medskip

We begin with a scalar field $\phi (x^\mu)$ obeying periodic boundary
conditions on $S^1 _R$, which has the following Fourier decomposition,
\bea
\phi (x^{\bar \mu}, y) = \sum _{n\in {\bf Z}} \phi _n (x^{\bar \mu}) e ^{2
\pi i n y /R}
\eea
The $d$-dimensional kinetic term of a scalar field with mass $m$ then
decomposes as follows,
\bea
\int \! d^dx \phi (-\Box _d + m^2 ) \phi
=
\sum _{n \in {\bf Z}} 2 \pi R \int \! d^{d-1}x \phi _n \biggl (-\Box _{d-1}
+ m^2 + {4 \pi ^2 n^2 \over R^2} \biggr ) \phi _n
\eea
As $R\to 0$, all modes except $n=0$ acquire an infinitely heavy mass and
decouple. The zero mode $n=0$ is the unique mode invariant under
translations on $S^1 _R$. Thus, {\sl the dimensional reduction on a circle
of a scalar field with periodic boundary conditions is again a scalar
field.} Under dimensional reduction with any other boundary condition,
there will be no zero mode left and thus the scalar field will completely
decouple.

\medskip

Next, consider a bosonic field with periodic boundary conditions
transforming under an arbitrary tensor representation of the Lorentz group
$SO(1,D-1)$ on $M_D$. Let us begin with a vector field $A_\mu(x^\nu)$ in the
fundamental of $SO(1,D-1)$. The index $\mu$ must now also be split into a
component along the direction $y$ and the remaining $D-1$ directions $\bar
\mu$. The first results in a scalar $A_y (x^{\bar \nu})$, while the
second results in a vector $A_{\bar \mu} (x^{\bar \nu})$ of the $D-1$
dimensional Lorentz group $SO(1,D-2)$. We notice that this decomposition is
nothing but the branching rule for the fundamental representation of
$SO(1,D-1)$ decomposing under the subgroup $SO(1,D-2)$.
For a field $A$ obeying period boundary conditions and transforming under a
general tensor representation $T$ of $SO(1,D-1)$, dimensional reduction on a
circle will produce a direct sum of representations $T_i$ of $SO(1,D-2)$,
which is the restriction of $T$ to the subgroup $SO(1,D-2)$.

\medskip

For a spinor field obeying periodic boundary conditions and
transforming under a general spinor representation $S$ of $SO(1,D-1)$,
dimensional reduction will produce a direct sum of representations $S_i$ of
$SO(1,D-2)$ which is the restriction of $S$ to the subgroup $SO(1,D-2)$.
Finally, assembling bosons and fermions with periodic boundary conditions
in a supersymmetry multiplet, we see that dimensional reduction will
preserve all Poincar\' e supersymmetries, and that the supercharges will
behave as the spinor fields described above under this reduction.

\medskip

An important example is the rank 2 symmetric tensor, i.e. the metric
$G_{\mu \nu}$,
\bea
G_{\mu \nu} & \to &
\left \{ \matrix{G_{y y} & {\rm scalar \ mixing \ with \ dilaton} \cr
                 G_{\bar \mu  y} & {\rm graviphoton} \cr
                 G_{\bar \mu \bar \nu} & {\rm metric} }
\right .
\eea
Again, fields obeying boundary conditions other than periodic will
completely decouple.

\medskip

\subsection{D=11 and D=10 Supergravity Particle and Field Contents}

In this subsection, we begin by listing the field contents and the number
of physical degrees of freedom of the $\N=1$, $D=11$ supergravity theory.
By dimensional reduction on a circle, we find the $\N=2$, $D=10$ Type IIA
theory, which is parity conserving and has two Majorana-Weyl gravitini of
opposite chiralities. Finally, we list the field and particle contents for
the $\N=2$, $D=10$ Type IIB theory, which is chiral and has two
Majorana-Weyl gravitini of the same chirality.

\medskip

The $\N=1$, $D=11$ supergravity theory has the following field and particle
contents,
\bea
D=11 \left \{ \matrix{
G_{\mu \nu}      & SO(9) &  44_B & {\rm metric \ - \ graviton} \cr
A_{\mu \nu \rho} &       &  84_B & {\rm antisymmetric \ rank \ 3} \cr
\psi  _{\mu \alpha} &    & 128_F & {\rm Majorana \ gravitino} }
\right .
\eea
Here and below, the numbers following the little group (for the massless
representations) $SO(9)$ represent the number of physical degrees of
freedom in the multiplet. For example, the graviton in $D=11$ is given by
the rank 2 symmetric traceless representation of $SO(9)$, of dimension $9
\times 10 /2 -1 =44$. The Majorana spinor $\psi _{\mu \alpha}$ as a vector
has 9 physical components, but it also satisfies the $\Gamma$-tracelessness
condition $(\Gamma ^\mu)^{\beta \alpha} \psi _{\mu \alpha}=0$, which cuts
the number down to 8. The 32 component spinor satisfies a Dirac equation,
which cuts its number of physical components down to 16, yielding a total
of $8 \times 16 = 128$. The subscripts $B$ and $F$ refer to the bosonic or
fermionic nature of the state.

\medskip

The $\N=2$, $D=10$ Type IIA theory is obtained by dimensional
reduction on a circle,
\bea
{\rm Type \ IIA} \left \{ \matrix{
G_{\mu \nu}      & SO(8) &  35_B & {\rm metric \ - \ graviton} \cr
\Phi             &       &   1_B & {\rm dilaton} \cr
B_{\mu \nu}      &       &  28_B & {\rm NS-NS \ rank \ 2 \ antisymmetric}
\cr
A_{3 \mu \nu \rho} &     &  56_B & {\rm antisymmetric \ rank \ 3} \cr
A_{1 \mu }       &       &   8_B & {\rm graviphoton} \cr
\psi _{\mu \alpha} ^\pm && 112_F & {\rm Majorana-Weyl \ gravitinos} \cr
\lambda _\alpha ^\pm    &&  16_F & {\rm Majorana-Weyl \ dilatinos} }
\right .
\eea
Here, the gravitinos are again $\Gamma$-traceless. The two gravitinos $\psi
^\pm _{\mu \alpha}$ as well as the two dilatinos $\lambda ^\pm _\alpha$ have
opposite chiralities and the theory is parity conserving.

\medskip

The $\N=2$, $D=10$ Type IIB theory has the following field and particle
contents,
\bea
{\rm Type \ IIB} \left \{ \matrix{
G_{\mu \nu}      & SO(8) &  35_B & {\rm metric \ - \ graviton} \cr
C+ i\Phi         &       &   2_B & {\rm axion-dilaton} \cr
B_{\mu \nu} + i A_{2 \mu \nu} &&  56_B & {\rm  rank \ 2 \ antisymmetric}\cr
A_{4 \mu \nu \rho \sigma} ^+ &  &  35_B & {\rm antisymmetric \ rank \ 4} \cr
\psi _{\mu \alpha} ^I \ \scriptstyle{I=1,2} && 112_F & {\rm Majorana-Weyl \
gravitinos} \cr
\lambda _\alpha ^I  \ \scriptstyle{I=1,2}  &&  16_F & {\rm Majorana-Weyl \
dilatinos} }
\right .
\eea
The rank 4 antisymmetric tensor $A^+ _{\mu \nu \rho \sigma}$ has self-dual
field strength, a fact that is indicated with the + superscript. The
gravitinos are again $\Gamma$-traceless. The two gravitinos
$\psi ^I _{\mu \alpha}$ have the same chirality,  while the two dilatinos
$\lambda ^I _\alpha$ also have the same chirality but opposite to that of
the gravitinos. The theory is chiral or parity violating.

\subsection{D=11 and D=10 Supergravity Actions}

Remarkably, the $D=11$ supergravity theory has a relatively simple action.
It is convenient to use exterior differential notation for all
anti-symmetric tensor fields, such as the rank 3 tensor $A_3 \equiv 1/3!
A_{3 \mu \nu \rho} dx^\mu dx^\nu dx^\rho$, with field strength $F_4\equiv
dA_3$,
\bea
S_{11} = {1 \over 2 \kappa _{11} ^2} \int \biggl [ \sqrt G (R_G -\half
|F_4|^2) - {1 \over 6} A_3 \wedge F_4 \wedge F_4  \biggr ] + {\rm fermions}
\eea
where $\kappa _{11}^2$ is the 11-dimensional Newton constant.
The action for the Type IIA theory may be deduced from this action by
dimensional reduction, but we shall not need it here. There are also $D=10$
supergravities with only $\N=1$ supersymmetry, which in particular may
couple to $D=10$ super-Yang-Mills theory.

\medskip

There exists no completely satisfactory action for the Type IIB theory,
since it involves an antisymmetric field $A^+_4$ with self-dual field
strength. However, one may write an action involving both dualities of
$A_4$ and then impose the self-duality as a supplementary field equation.
Doing so, one obtains\footnote{We use the notation $G\equiv -\det G_{\mu
\nu}$ and $\int \sqrt G |F_p|^2 \equiv {1 \over p!} \int \sqrt G
G^{\mu _1 \nu _1} \cdots G^{\mu _p \nu _p} \bar F_{\mu _1 \cdots \mu_p}
F_{\nu _1 \cdots \nu_p}$ where $\bar F$ denotes the complex conjugate of
$F$. For real fields, this definition coincides with that of
\cite{polchinski}.} (see for example \cite{IIB, polchinski})
\bea
\label{IIBaction}
S_{IIB} &=& +{ 1 \over 4 \kappa_B ^2} \int \sqrt G e^{-2\Phi} (2R_G + 8
\p _\mu \Phi \p ^\mu \Phi -  |H_3|^2 )
\\
&& - {1 \over 4 \kappa_B ^2} \int \biggl [ \sqrt G (|F_1|^2 +
|\tilde F_3|^2 + \half  |\tilde F_5|^2) + A_4 ^+ \wedge H_3 \wedge F_3
\biggr ] + {\rm fermions}
\nonumber
\eea
where the field strengths are defined by
\bea
\cases{
F_1  = dC \cr H_3 = dB \cr F_3  = dA_2 \cr F_5  = dA_4 ^+}
\qquad \qquad
\cases{\tilde F_3  = F_3 - C H_3 \cr
\tilde F_5  = F_5 -\half A_2 \wedge
H_3 + \half B \wedge F_3}
\eea
and we have the supplementary self-duality condition $* \tilde F_5 = \tilde
F_5$.

\medskip

The above form of the action naturally arises from the string low energy
approximation. The first line in (\ref{IIBaction}) originates from the
NS-NS sector while the second line (except for the fermions) originates
from the RR sector, as we shall see shortly. Type IIB supergravity is
invariant under the non-compact symmetry group $SU(1,1) \sim SL(2,{\bf
R})$, but this symmetry is not manifest in (\ref{IIBaction}). To render the
symmetry manifest, we redefine fields from the {\sl string metric}
$G_{\mu \nu}$ used in (\ref{IIBaction}) to the {\sl Einstein metric}
$G_{E\mu \nu}$, along with expressing the tensor fields in terms
of complex fields,\footnote{The detailed relation with the $SU(1,1)$
formulation of Type IIB supergravity is given as follow : the $SU(1,1)$
frame $V_\pm ^\alpha$, $\alpha =1,2$ is given by $V_+ ^1 = \tau /\sqrt {\Im
\tau}$, $V_- ^1 = \bar \tau /\sqrt {\Im \tau}$, and $V_\pm ^2 = 1 /\sqrt
{\Im \tau}$. The frame transforms as a $SU(1,1)$ doublet and satisfied $V_-
^1 V_+ ^2 - V_- ^2 V_+ ^1 =1$. The complex 3-form is defined by $G_3 = V_+
^2 F_3 - V_+ ^1 H_3$ and is a $SU(1,1)$ singlet. The complex variable $\tau$
parametrizes the coset $SU(1,1)/U(1)$; under this local $U(1)$ group,
$V_\pm$ have charge $\pm 1$ while $G_3$ has charge $+1$. }
\bea
G_{E\mu \nu}  \equiv  e^{-\Phi /2} G_{\mu \nu}
\qquad && \ \,
\tau  \equiv  C + i e ^{-\Phi}
\nonumber \\ &&
G_3  \equiv  (F_3 - \tau H_3) /\sqrt {\Im \tau}
\eea
The action may then be written simply as,
\bea
S_{\rm IIB} & = & {1 \over 4 \kappa _B ^2} \int \sqrt{G_E} \biggl (
2 R_{G_E} - {\p_\mu \bar \tau \p^\mu \tau \over  (\Im \tau )^2}
- \half | F_1 |^2  -  | G_3|^2
- \half | \tilde F_5 |^2 \biggr )
\nonumber \\ &&
- {1 \over 4i \kappa _B ^2} \int A_4 \wedge \bar G_3 \wedge G_3
\eea
Under the $SU(1,1) \sim SL(2,{\bf R})$ symmetry of Type IIB supergravity,
the metric and $A_4 ^+$ fields are left invariant. The dilaton-axion field
$\tau$ changes under a M\" obius transformation,
\bea
\tau \to \tau ' = {a \tau + b \over c \tau + d}
\qquad \qquad
ad-bc=1, \ a,b,c,d \in {\bf R}
\eea
Finally, the $B_{\mu \nu}$ and $A_{2 \mu \nu}$ fields rotate into one
another under the linear transformation associated with the above M\" obius
transformation, and this may most easily be re-expressed in terms of the
complex 3-form field $G_3$,
\bea
G_3 \to G_3 ' =  {c \bar \tau + d \over |c  \tau +d|} \ G_3
\eea

\medskip

The susy transformation laws of Type IIB supergravity \cite{IIB, gsw}
on the fermion fields -- the dilatino $\lambda$ and the gravitino $\psi
_M$ -- are of the form, (we shall not need the transformation laws on
bosons),
\bea
\delta \lambda &=& {i \over \kappa _B} \Gamma ^\mu \eta ^* {\p _\mu \tau
\over \Im \tau}   - {i \over 24}
\Gamma ^{\mu \nu \rho} \eta G_{3 \mu \nu \rho} + ({\rm Fermi})^2
 \\
\delta \psi _\mu &=&
{1 \over \kappa _B} D_\mu \eta + {i \over 480 } \Gamma ^{\mu_1 \cdots \mu_5}
\Gamma _\mu \eta F_{5\mu _1 \cdots \mu_5}
+{1 \over 96} ( \Gamma _\mu {}^{\rho \sigma \tau} G_{3 \rho \sigma \tau} - 9
\Gamma ^{\nu \rho} G_{3 \mu \nu \rho} )
\eta ^*
+  ({\rm Fermi})^2
\nonumber
\eea
Note that in the $SU(1,1)$ formulation, the supersymmetry
transformation parameter $\eta$ has $U(1)$ charge 1/2, so that $\lambda$
has charge 3/2 and $\psi _\mu$ has charge 1/2.

\subsection{Superstrings in D\ts =\ts 10}

The geometrical data of superstring theory in the Ramond-Neveu-Schwarz
(RNS) formulation are the bosonic worldsheet field $x^\mu$ and the
fermionic worldsheet fields $\psi ^\mu _\pm$, which may both be viewed as
functions of local worldsheet coordinates $\xi ^1$, $\xi ^2$. The
subscript
$\pm$ indicates the two worldsheet chiralities. Both $x ^\mu$ and
$\psi_\pm ^\mu$ transform under the vector representation of the
space-time Lorentz group. The theory has two sectors, the Neveu-Schwarz
(NS) and Ramond (R) sectors. The NS ground state is a space-time boson,
while the R ground state is a space-time fermion. The full space-time
bosonic (resp. fermionic) spectrum of the theory is obtained by applying
$x^\mu$ and $\psi ^\mu$ fields to the NS (resp. R) ground states.
Space-time supersymmetry is achieved by imposing a suitable
Gliozzi-Scherk-Olive (GSO) projection \cite{Gliozzi:1976qd}. For
simplicity, we shall only consider theories with orientable strings; the
Type II and heterotic string theories fit in this category. Interactions
arise from the joining and splitting of the worldsheets, so that the
number of handles (which equals the genus for orientable worldsheets)
corresponds to the number of loops in a field theory reinterpretation of
the string diagram. (Standard references on superstring theory include
\cite{gsw, polchinski}, lecture notes \cite{D97} and a review on
perturbation theory \cite{dp88}.)

\begin{fig}[htp]
\centering
\epsfxsize=6in
\epsfysize=1.6in
\epsffile{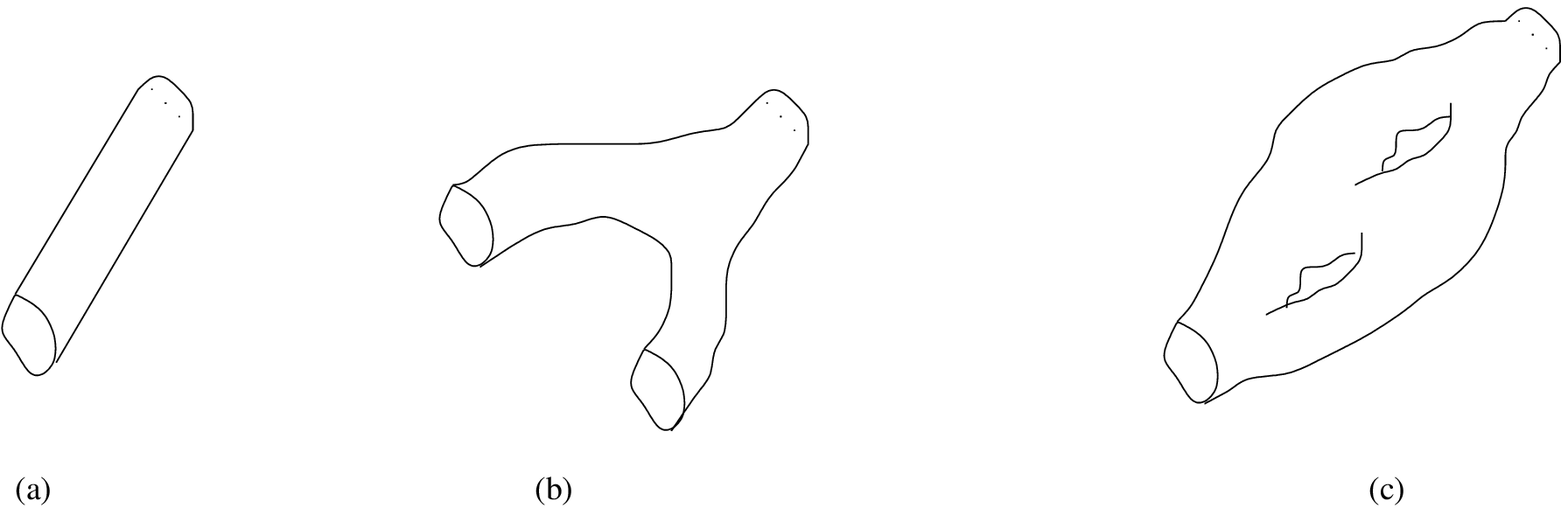}
\caption{Propagating closed strings (a) free, (b) interaction, (c) two-loop}
\label{fig:1}
\end{fig}

One aspect of string theory that we shall make use of in these lectures is
the fact that (1) the low energy limit of string theory is supergravity
and that (2) string theory produces definite and calculable higher
derivative corrections to the supergravity action and field equations. To
explain these facts, it is easiest to concentrate on the space-time
bosonic fields, since space-time fermionic fields require the use of the
more complicated fermion vertex operator. For Type II theories, the
space-time bosons arise from two sectors in turn; the NS-NS sector and
the R-R sector. Fields in the R-R sector again couple to the string
worldsheet through the use of the fermion vertex operator, and for
simplicity we shall ignore also these fields here (even though they will
of course be very important for the AdS/CFT conjecture). The remaining
fields are now the same for all four closed orientable string theories,
Type IIA, Type IIB and the two heterotic strings, namely the metric
$G_{\mu \nu}$, the NS-NS antisymmetric rank 2 tensor $B_{\mu \nu}$ and
the dilaton $\Phi$. The full worldsheet action for the coupling of these
fields is still very complicated on a  worldsheet with general worldsheet
metric and worldsheet gravitino fields $\chi _m$. The contribution from
the worldsheet bosonic field $x^\mu$ gives rise to a generalized
non-linear sigma model,
\bea
S_x
= { 1 \over 4 \pi \alpha '} \int _\Sigma \! \sqrt \gamma  \biggl [
\bigl \{ \gamma ^{mn} G_{\mu \nu} (x)
+ \epsilon ^{mn} B_{\mu \nu} (x) \bigr \} \p_m x^\mu \p_n x^\nu
+ \alpha ' R^{(2)} _\gamma \Phi (x) \biggr ]
\eea
where $\alpha '$ is the square of the Planck length, $\gamma_{mn}$ is the
worldsheet metric, $\gamma ^{mn}$ its inverse and $R^{(2)}_\gamma$ its
associated Gaussian curvature. The contribution from the worldsheet
fermionic field $\psi _\pm ^\mu$ gives rise to a worldsheet supersymmetric
completion of the above non-linear sigma model. Here, we quote its form
only for a flat worldsheet metric and vanishing worldsheet gravitino
field,
\bea
S_\psi
= { 1 \over 4 \pi \alpha '} \int _\Sigma d^2 \xi \biggl (
G_{\mu \nu} (x) (\psi _+ ^\mu D_{\bar z} \psi ^\nu _+ + \psi _- ^\mu
D_z \psi ^\nu _-) + \half R_{\mu \nu \rho \sigma} \psi _+ ^\mu \psi _+ ^\nu
\psi _- ^\rho \psi _- ^\sigma \biggr )
\eea
where $R_{\mu \nu \rho \sigma}$ is the Riemann tensor for the metric
$G_{\mu \nu}$ and the covariant derivatives are given by
\bea
D_{\bar z} \psi _+ ^\mu & = & \p _{\bar z} \psi _+ ^\mu + \biggl ( \Gamma
^\mu  _{\rho \sigma} (x) + \half H_3 {}^\mu {}_{\rho \sigma} (x) \biggr )
\p _{\bar z} x^\rho \psi _+ ^\sigma
\nonumber \\
D_z \psi _- ^\mu & = & \p _z \psi _- ^\mu + \biggl ( \Gamma ^\mu
_{\rho \sigma} (x) - \half H_3{}^\mu {}_{\rho \sigma} (x) \biggr ) \p _z
x^\rho \psi _- ^\sigma
\eea
where $H_{3 \mu \rho \sigma} $ is the field strength of $B_{\mu
\nu}$ and $\Gamma ^\mu _{\rho \sigma}$ is the Levi-Civita
connections for $G$.

\medskip

The non-chiral scattering amplitudes are given by the functional integral
over all $x^\mu$ and $\psi _\pm$ as well as over all worldsheet metrics
$\gamma _{mn}$ and all worldsheet gravitini fields $\chi _m$ by
\bea
{\rm amplitude} = \sum _{{\rm topologies}} \int D\gamma_{mn} D\chi_m
\int Dx^\mu D\psi \ e^{-S_x + S_\psi}
\eea
The full amplitudes must then be obtained by first chirally splitting
\cite{dp88,dpXX} the non-chiral amplitudes in terms of the conformal
blocks of the corresponding conformal field theories of the left and
right movers and imposing the GSO projection.

\medskip

The quantization prescription given by the above formula for the amplitude
is in the first quantized formulation of string theory. There, a given
string configuration (a given worldsheet topology) is quantized in the
presence of external background fields, such as the metric $G_{\mu \nu}$,
the rank 2 anti-symmetric tensor field $B_{\mu \nu}$ and the dilaton
$\Phi$. The quantization of the string produces excitations of these very
fields as well as of all the other string modes. In comparison with the
first quantized formulation of particles is field theory, the background
fields may be interpreted as vacuum expectation values of the
corresponding field operators.

\medskip

If the vacuum expectation value of the dilaton field is $\phi = \< \Phi
\>$, then the contribution of the vacuum expectation value to the string
amplitude is governed by the Euler number $\chi (\Sigma)$ of the worldsheet
$\Sigma$,
\bea
{1 \over 2 \pi} \int _\Sigma \sqrt \gamma R^{(2)} _\gamma = \chi (\Sigma)
= 2 - 2h -b
\eea
where $h$ is the genus or number of handles and $b$ is the number of
boundaries or punctures. Therefore, a genus $h$ worldsheet (without
boundary) will receive a multiplicative contribution of $ e^{-(2-2h) \phi}
= g_s ^{2h-2}$ which gives reason to identify $g_s = e^{\phi}$ with the
(closed) string coupling constant. For open string theories, the expansion
is rather in integer powers of the open string coupling constant $g_o =
e^{\phi /2}$.

\subsection{Conformal Invariance and Supergravity Field Equations}

As a two-dimensional quantum field theory, the generalized non-linear sigma
model makes sense for any background field assignment. However, when the
non-linear sigma model is to define a consistent string theory, further
physical conditions need to be satisfied. The most crucial one is that the
single string spectrum be free of negative norm states. Such states always
appear because Poincar\' e invariance of the theory forces the string map
$x^\mu$ to obey the following canonical relations $[x^\mu, \dot x^\nu]\sim
G^{\mu \nu}$, so that $x^0$ creates negative norm states.

\medskip

The decoupling of negative norm states out of the Fock space construction
occurs via {\sl worldsheet conformal invariance} of the non-linear sigma
model. In particular, conformal invariance requires worldsheet scale
invariance of the full quantum mechanical non-linear sigma model.
Transformations of the worldsheet scale $\Lambda$ are broken by quantum
mechanical anomalies whose form is encoded by the $\beta$-functions of the
renormalization group (RG). As will be explained in the next paragraph, each
background field has a $\beta$-function, and worldsheet scale and conformal
invariance thus require the vanishing of these $\beta$-functions.

\medskip

The background fields $ G_{\mu \nu}(x)$, $B_{\mu \nu}(x)$ and $\Phi(x)$ may
be viewed as generating functions for an infinite series of coupling
constants. For example, for the metric we have,
\bea
G_{\mu \nu} (x) = \sum _{n=0} ^\infty {1 \over n!}
(x-x_0)^{\mu _1} \cdots (x-x_0)^{\mu _n} \p_{\mu _1} \cdots \p_{\mu _n}
G_{\mu \nu} (x_0)
\eea
where each of the Taylor expansion coefficients $\p_{\mu _1} \cdots  \p_{mu
_n} G_{\mu \nu} (x_0)$ may be viewed as an independent set of couplings.
Under renormalization, and thus under RG flow, this infinite number of
couplings flows into itself, and the corresponding flows may again be
described by generating functions $\beta ^G_{\mu \nu}(x)$, $\beta ^B _{\mu
\nu}(x)$ and $\beta ^\Phi (x)$ defined, for example, for the metric by
\bea
\beta ^G_{\mu \nu}(x) = {\p G_{\mu \nu} \over \p \ln \Lambda}
 \equiv
\sum _{n=0} ^\infty {1 \over n!}
(x-x_0)^{\mu _1} \cdots (x-x_0)^{\mu _n} {\p \p_{\mu _1} \cdots \p_{\mu _n}
G_{\mu \nu} (x_0) \over \p \ln \Lambda}
\eea
 Customarily, when an infinite
number of couplings occur in a quantum field theory, it is termed {\sl
non-renormalizable}, because the prediction of any physical observable
would require an infinite number of input data to be specified at the
renormalization point. In string theory, however, this infinite number of
couplings is exactly what is required to describe the dynamics of a string
in a consistent background. We now explain how this comes about.

\medskip

First, we assume that the whole renormalization process of the non-linear
sigma model will preserve space-time diffeomorphism invariance. The number
of terms that can appear in the RG flow is then finite, order by order in
the $\alpha '$ expansion \cite{friedan85}. Second, the presence of an
infinite number of couplings makes it possible to have the string
propagate in an infinite family of space-times. The leading order
$\beta$-functions are given by \cite{cfmp}
\bea
\beta ^G _{\mu\nu} & = &
\half R_{\mu \nu} - {1 \over 8} H_{\mu \rho \sigma } H_\nu {}^{\rho \sigma}
+\p _\mu \Phi \p _\nu \Phi + \O (\alpha ')
 \\
\beta ^B _{\mu\nu} & = &  - \half D_\rho H^\rho{}_{\mu \nu} + \p_\rho H^\rho
{}_{\mu \nu} + \O (\alpha ')
\nonumber \\
\beta ^\Phi  & = &
{1 \over 6} (D-10) + \alpha ' \biggl (
2 \p _\mu \Phi \p ^\mu \Phi - 2 \nabla ^\mu \p _\mu \Phi + \half
R_G - {1 \over 24} H_{\mu \nu \rho} H^{\mu \nu \rho} \biggr ) + \O (\alpha
')^2
\nonumber
\eea
To leading order in $\alpha'$, the requirement of scale invariance reduces
precisely to the supergravity field equations for the Type II theory where
all RR $A$-fields have been (consistently) set to 0. String theory provides
higher $\alpha'$ corrections to the supergravity field equations, which by
dimensional analysis must be also terms with higher derivatives in $x^\mu$.

\subsection{Branes in Supergravity}

A rank $p+1$ antisymmetric tensor field $A_{\mu _1 \cdots \mu _{p+1}}$ may
be identified with a $(p+1)$-form,
\bea
A_{p+1} \equiv {1 \over (p+1)!} A_{\mu_1 \cdots \mu_{p+1}} dx^{\mu _1}
\wedge \cdots \wedge dx^{\mu _{p+1}}
\eea
A $(p+1)$-form naturally couples to geometrical objects $\Sigma _{p+1}$ of
space-time dimension $p+1$, because a diffeomorphism invariant action may be
constructed as follows
\bea
S_{p+1} = T_{p+1} \int _{\Sigma _{p+1}} A_{p+1}
\eea
The action is invariant under Abelian gauge transformations
$\rho _p (x)$ of rank $p$
\bea
A_{p+1}  \to  A_{p+1} + d\rho _p
\eea
because $S_{p+1}$ transforms with a total derivative. The field $A_{p+1}$
has a gauge invariant field strength $F_{p+2}$, which is a $p+2$ form
whose flux is conserved. {\sl Solutions to supergravity with non-trivial
$A_{p+1}$ charge are referred to as $p$-branes, after the space-dimension
of their geometry.}

\medskip

Each $A_{p+1}$ gauge field has a {\sl magnetic dual} $A^{\rm
magn}_{D-3-p}$ which is a differential form field of rank $D-3-p$, whose
field strength is related to that of $A_{p+1}$ by Poincar\' e duality
\bea
d A^{\rm magn} _{D-3-p} \equiv * d A_{p+1}
\eea
Accordingly, each $p$-brane also has a magnetic dual, which is a
$(D-4-p)$ brane and which now couples to the field $A^{\rm magn} _{D-3-p}$.

\medskip

The possible branes in $D=11$ supergravity are very restricted because the
only antisymmetric tensor field in the theory is $A_{\mu \nu \rho}$ of
rank 3, so that we have a 2-brane, denoted $M2$ and its magnetic dual $M5$.
The branes in Type IIA/B theory are further distinguished as follows. When
the antisymmetric field whose charge they carry is in the R-R sector, the
brane is referred to as a {\sl D-brane}. D-branes were introduced first in
string theory in  \cite{dbranes}. On the other hand, the
$1$-brane that couples to the NS-NS field $B_{\mu \nu}$ is nothing but the
fundamental string, denoted F1, whose magnetic dual is NS5 \cite{NS5}.
Below we present a Table of the branes occurring for various $p$ in the
$D=11$ supergravity and in the Type IIA/B supergravities in $D=10$.

\begin{table}[b]
\begin{center}
\begin{tabular}{|c|c|c|c|c|} \hline
 name & $D=11$ & Type IIA & Type IIB & Magnetic Dual \\ \hline \hline
 D(-1) instanton & --- & --- & $A_0=C+ie^{- \Phi}$  & D7 \\ \hline
 D0 particle & --- & $A_{1 \mu}$ & ---             & D6 \\ \hline
 F1 string   & --- & $B_{\mu \nu}$ & $B_{\mu \nu}$ & NS5 \\
 D1 string   & --- & --- & $A_{2 \mu \nu}$         & D5 \\ \hline
 M2 membrane & $A_{\mu \nu \rho}$ & --- & ---      & M5 \\
 D2 brane    &  --- & $A_{3 \mu \nu \rho}$ & ---   & D4 \\ \hline
 D3 brane    &  --- & --- & $A^+ _{4 \mu \nu \rho \sigma}$ & D3 \\
\hline \hline
\end{tabular}
\end{center}
\caption{Branes in various theories}
\label{table:3A}
\end{table}

\subsection{Brane Solutions in Supergravity}

Each brane is realized as a 1/2 BPS solution in supergravity. The geometry
of these solutions will be important, and we describe it now. A
$p$-brane has a $(p+1)$-dimensional flat hypersurface, with Poincar\' e
invariance group ${\bf R}^{p+1} \times SO(1,p)$. The transverse space is
then of dimension $D-p-1$ and solutions may always be found with maximal
rotational symmetry $SO(D-p-1)$ in this transverse space. Thus,
$p$-branes in supergravity may be thought of as solutions with symmetry
groups
\bea
\cases{
$D=11$ & ${\bf R}^{p+1}  \times SO(1,p) \times SO(10-p)$ \cr
$D=10$ & ${\bf R}^{p+1}  \times SO(1,p) \times SO(9-p)$ \cr}
\eea
For example the $M2$ brane has symmetry group ${\bf R}^3 \times SO(1,2)
\times SO(8)$ while the $D3$ brane has instead ${\bf R}^4 \times SO(1,3)
\times SO(6)$. We shall denote the coordinates as follows
\bea
{\rm Coordinates \ /\! / \ to \ brane} & x^\mu & \mu = 0,1,\cdots ,p
\nonumber \\
{\rm Coordinates \ \perp \ to \ brane} & y^u =
x^{p + u} & u = 1,2,\cdots , D-p-1
\nonumber
\eea
Poincar\' e invariance in $p+1$ dimensions forces the metric in those
directions to be a rescaling of the Minkowski flat metric, while rotation
invariance in the transverse directions forces the metric in those
directions to be a rescaling of the Euclidean metric in those dimensions.
Furthermore, the metric rescaling functions should be independent of
$x^\mu$, $\mu =0,1,\cdots ,p$. Substituting an Ansatz with the above
restrictions into the field equations, one finds that the solution may be
expressed in terms of a single function $H$ as follows, \cite{Horowitz:cd}
\bea
{\rm Dp} & \qquad & ds^2 = H(\vec{y})^{-1/2} dx^\mu dx_\mu + H(\vec{y})^{1/2}
d\vec{y}^2
\qquad \qquad
e^\Phi = H(\vec{y})^{(3-p)/4}
\nonumber \\
{\rm NS5} & \qquad & ds^2 = dx^\mu dx_\mu + H(\vec{y}) d\vec{y}^2
\qquad \hskip 1in
e^{2 \Phi} = H(\vec{y})
\nonumber \\
{\rm M2} & \qquad & ds^2 = H(\vec{y})^{-2/3} dx^\mu dx_\mu + H(\vec{y})^{1/3}
d\vec{y}^2
\nonumber \\
{\rm M5} & \qquad & ds^2 = H(\vec{y})^{-1/3} dx^\mu dx_\mu + H(\vec{y})^{2/3}
d\vec{y}^2
\eea
Here, the $Dp$ metric is expressed in the string frame. The single function
$H$ must be harmonic with respect to $\vec{y}$.

\medskip

Assuming maximal rotational symmetry by $SO(D-p-1)$ in the transversal
dimensions, and using the fact that the metric should tend to flat
space-time as $y\to \infty$, the most general solution is parametrized by a
single scale factor $L$ and is given by
\bea
H(y) = 1 + { L ^{D-p-3} \over y^{D-p-3} } \qquad \qquad
\eea
Since $\alpha'$ is the only dimensionful parameter of the theory, $L$
must be a numerical constant (possibly dependent on the dimensionless
string couplings) times the above $\alpha'$ dependence. Of particular
interest will be the solution of $N$ coincident branes, for which we have
$L ^{D-p-3}=  N \rho _p $. For $Dp$ branes, we have $\rho _p
= g_s \ (4 \pi)^{(5-p)/2} \Gamma ((7-p)/2) (\alpha ')^{(D-p-3)/2} $.

\medskip

It is easy to see that one still has a solution when $H$ is harmonic without
insisting on rotation invariance in the transverse space, so that the
general solution is of the form,
\bea
\label{DbraneH}
H(\vec{y}) = 1 + \sum _{I=1} ^N {C_I \over |\vec{y} - \vec{y}_I|^{D-3-p}}
\qquad \qquad
C_I = N_I \rho _p\, ,\ N_I \in {\bf N}
\eea
for any array of $N$ points $\vec{y}$.

\medskip

It is very important in the theory of branes in Type IIA/B string theory to
understand the dependence of the string coupling $g_s$ of the various
brane solutions, in particular of $c_p$. To do so, we return to the
supergravity field equations, (omitting derivative terms in the dilaton and
axion fields for simplicity),
\bea
{\rm IIA} &&
R_{\mu \nu} = {1 \over 4} H_{\mu \rho \sigma} H_\nu {}^{\rho \sigma}
+ e^{2\Phi} \biggl ( F_{2\mu \rho } F_{2\nu} {}^\rho
+ {1 \over 6} F_{4\mu \sigma \rho \tau} F_{4\nu} {}^{\rho \sigma \tau}
\biggr )
\\
{\rm IIB} &&
R_{\mu \nu} =
{1 \over 4} H_{\mu \rho \sigma} H_\nu {}^{\rho \sigma}
+ e^{2\Phi} \biggl ( F_{1\mu} F_{1\nu}
+ {1 \over 4} \tilde F_{3\mu \sigma \rho} \tilde F_{3\nu} {}^{\rho \sigma}
+ {1 \over 24} \tilde F_{5\mu \rho \sigma \tau \upsilon} ^+ \tilde F_{5\nu}
^{+ \rho \sigma \tau \upsilon} \biggr )
\nonumber
\eea
Recall that the string coupling is given by $g_s=e^\phi$ where $\phi =
\< \Phi \>$.
In both Type IIA and Type IIB, the fundamental string F1 and the NS5 brane
have non-vanishing $H_{\mu \rho\sigma}$ fields, but {\sl vanishing RR
fields} $F_i$. Therefore, these brane solutions do not involve the string
coupling constant $g_s$ and $\rho _p$ is independent of $g_s$. D-brane
solutions on the other hand will have $H_{\mu \rho \sigma}=0$, but have at
least one of the R-R antisymmetric fields $F_i \not=0$. Such solutions will
involve the string coupling explicitly and therefore $\rho _p \sim g_s$.
This leads for example to the expression given for $\rho _p$ above. Each
brane solution breaks precisely half of the supersymmetries of the
corresponding theory, as is shown in Problem Set (4.1).

\subsection{Branes in Superstring Theory}

While originally found as solutions to supergravity field equations, the
$p$-branes of Type IIA/B supergravity are expected to extend to solutions
of the full Type IIA/B string equations. These solutions will then break
precisely half of the supersymmetries of the string theory. As compared to
the supergravity solutions, the full string solutions may, of course, be
subject to $\alpha'$ corrections of their metric and other fields. Often,
it is useful to compare these semi-classical solutions of string theory
with solitons in quantum field theory, such as the familiar `t~Hooft--Polyakov
magnetic monopole. The fundamental string F1 and the NS5 brane indeed very
much behave as large size semi-classical solitons, whose energy depends on the
string coupling via $1 / g_s^2$, as is familiar from solitons in quantum field
theory.

\medskip

Besides its supergravity low energy limit, the only other well-understood
limit of string theory is that of weak coupling where $g_s \to 0$. It is in
this approximation that string theory may be defined in terms of a genus
expansion in string worldsheets. Remarkably, D-branes (but not the F1 string
or NS5 branes) admit a special limit as well. As may be seen from
(\ref{DbraneH}), in the limit where $g_s \to 0$, the metric becomes flat
everywhere, except on the $(p+1)$-dimensional hyperplane characterized by
$\vec{y}=0$, where the metric appears to be singular. Thus, in the
weak-coupling limit, the D-brane solution of supergravity reduces to a
localized defect in flat space-time. Strings propagating in this background
are moving in flat space-time, except when the string reaches the
D-brane. The interaction of the string with the D-brane is summarized by a
boundary condition on the string dynamics. The correct conditions turn out
to be Dirichlet boundary conditions in the directions perpendicular to the
brane and Neumann conditions parallel to the brane. The $Dp$-brane may
alternatively be described in string perturbation theory as a
$(p+1)$-dimensional hypersurface in flat 10-dimensional space-time on which
open strings end with the above boundary conditions. The open string end
points are thus tied to be on the brane, but can move freely along the
brane. This was indeed the original formulation \cite{dbranes}; see
also~\cite{pcj}.

\subsection{The Special Case of D3 branes}

The D3-brane solution is of special interest for a variety of reasons~:
(1) its worldbrane has 4-dimensional Poincar\' e invariance; (2) it has
constant axion and dilaton fields; (3) it is regular at $y=0$; (4) it is
self-dual. Given its special importance, we shall present here a more
complete description of the D3-brane. The solution is characterized by
\bea
\cases{
g_s = e^{\phi}, \  C \  {\rm constant}
\cr B_{\mu \nu } = A_{2\mu \nu}=0 \cr
ds^2 = H(y) ^{-1/2} dx^\mu dx_\mu + H(y)^{1/2} (dy^2 + y^2 d\Omega _5^2)
\cr F^+ _{5 \mu \nu \rho \sigma \tau} = \epsilon _{\mu \nu \rho \sigma
\tau
\upsilon} \p ^\upsilon H \cr}
\eea
Here, $\epsilon _{\mu \nu \rho \sigma \tau \upsilon}$ is the volume element
transverse to the 4-dimensional Minkowski D3-brane in $D=10$. The $N$-brane
solution with general locations of $N_I$ parallel D3-branes located at
transverse position $\vec{y}_i$ is given by
\bea
H(\vec{y}) =
1 + \sum _{I=1} ^N { 4 \pi g_s N_I (\alpha ')^2 \over |\vec{y} - \vec{y}_I|
^4}
\eea
where the total number of D3-branes is $N= \sum _I N_I$. The fact that the
geometry is regular as $\vec{y} \to \vec{y}_I$ despite the apparent
singularity in the metric will be shown in the next section.

\medskip

It is useful to compare the scales involved in the D3 brane solution and
their relations with the coupling constant.\footnote{The discussion given
here may be extended to $Dp$ branes to some extent. However, when $p\not=
3$, the dilaton is not constant and the strength of the coupling will
depend upon the distance to the brane.} The radius $L$ of the D3 brane
solution to string theory is a scale that is not necessarily of the same
order of magnitude as the Planck length $\ell_P$, which is defined by
$\ell_P ^2 = \alpha '$. Their ratio is given instead by $L^4 = 4 \pi g_s N
\ell_P^4$. For $g_s N \ll 1$, the radius $L$ is much smaller than the
string length $\ell_P$, and thus the supergravity approximation is not
expected to be a reliable approximation to the full string solution. In
this regime we have $g_s \ll 1$, so that string perturbation theory is
expected to be reliable and the D3 brane may be treated using conformal
field theory techniques. For $g_s N \gg 1$,  the radius $L$ is much larger
than the string length $\ell_P$, and thus the supergravity approximation is
expected to be a good approximation to the full string solution. It is
possible to have at the same time $g_s \ll 1$ provided $N$ is very large,
so string perturbation theory may be {\sl simultaneously} a good
approximation.

\medskip

The D3 brane solution is more properly a two-parameter family of solutions,
labeled by the string coupling $g_s$ and the instanton angle $\theta _I =
2\pi C$, or the single complex parameter $\tau = C + i e^{-\phi}$. The
$SU(1,1)\sim SL(2,{\bf R})$ symmetry of Type IIB supergravity acts
transitively on $\tau$, so all solutions lie in a single orbit of this
group.
In superstring theory, however, the range of $\theta_I$ is quantized so
that the identification $\theta _I\sim \theta _I +2 \pi$ may be made, and
as a result also $\tau \sim \tau +1$. Therefore, the allowed M\" obius
transformations must be elements of the $SL(2,{\bf Z})$ subgroup of
$SL(2,{\bf R})$, for which $a,b,c,d \in {\bf Z}$. These transformations
map between equivalent solutions in string theory. Thus, the string
theories defined on D3 backgrounds which are related by an $SL(2,{\bf
Z})$ duality will be equivalent to one another. This property will be of
crucial importance in the AdS/CFT correspondence where it will emerge as
the reflection of Montonen-Olive duality in $\N=4$ SYM theory.

\subsection{Problem Sets}

(4.1) The Lagrangian for $D=10$ super-Yang-Mills theory (which is
constructed to be invariant under $\N=1$ supersymmetry) is given by
\bea
\L = - {1 \over 2 g^2} \tr  (F_{\mu \nu} F^{\mu \nu}
-2i \bar \lambda \Gamma ^\mu D_\mu \lambda  )
\eea
The supersymmetry transformations are given by ($\Gamma ^{\mu \nu} \equiv
\half [\Gamma ^\mu, \Gamma ^\nu]$)
\bea
\delta A_\mu = - i \bar \zeta \Gamma _\mu  \lambda
\qquad \qquad
\delta \lambda = \half F_{\mu \nu} \Gamma ^{\mu \nu} \zeta
\eea
for a Majorana-Weyl spinor gaugino $\lambda$. Show that under dimensional
reduction on a flat 6-dimensional torus, (with periodic boundary conditions
on all fields), the theory reduces to $D=4$, $\N=4$ super-Yang-Mills. Use
this reduction to relate the matrices $C_i$ in the Lagrangian for the $D=4$
theory to the Clifford Dirac matrices of $SO(6)$, and to derive the
supersymmetry transformations of the theory.

\bigskip

(4.2) Assume the following Ansatz for a D3 brane solution to the Type IIB
sugra field equations~: constant dilaton $\phi$, vanishing axion $C=0$,
vanishing two-forms $A_{2\mu \nu} = B_{\mu \nu}=0$, $F_{5 \mu \nu \rho
\sigma \tau} \sim \epsilon _{\mu \nu \rho \sigma \tau \upsilon}
\p ^\upsilon H$ and metric of the form
$$
ds^2 = H^{-\half} (\vec{y}) dx^\mu d x_\mu + H^\half (\vec{y}) d\vec{y}^2
$$
Here, $x^\mu$, $\mu =0,\cdots, 3$ are the coordinates along the brane,
while $\vec{y} \in {\bf R}^6$ are the coordinates perpendicular to the
brane.  Show that the sugra equations
hold provided $H$ is harmonic in the transverse directions (i.e.
satisfies $\Box _y H= 0 $, except at the position of the brane,
where a pole will occur).

\bigskip

\noindent
(4.3) Continuing with the set-up of (4.2), show that regularity of the
solution requires the poles of $H$ to have integer strength.

\bigskip

\noindent
(4.4)
Show that the D3 brane solution preserves 16 supersymmetries (i.e. half of
the total number).

\vfill\eject

\section{The Maldacena  AdS/CFT Correspondence}
\setcounter{equation}{0}

In the preceding sections, we have provided descriptions of $D=4$, $\N=4$
super-Yang-Mills theory on the one hand and of D3 branes in supergravity
and superstring theory on the other hand. We are now ready to exhibit the
{\sl Maldacena or near-horizon limit} close to the D3 branes and formulate
precisely the Maldacena or AdS/CFT correspondence which conjectures the
identity or duality between $\N=4$ SYM and Type IIB superstring theory on
\AdS. We shall also present the three different forms of the conjecture,
the first being a correspondence with the full quantum string theory, the
second being with classical string theory and finally the weakest form
being with classical supergravity on \AdS. In this section, the precise
mapping between both sides of the conjecture will be made for the global
symmetries as well as for the fields and operators. The mapping between the
correlation functions will be presented in the next section. For a
general review see \cite{magoo}; see also \cite{duff} and \cite{peet}.

\subsection{Non-Abelian Gauge Symmetry on D3 branes}

Open strings whose both end points are attached to a single brane can have
arbitrarily short length and must therefore be massless. This excitation
mode induces a massless $U(1)$ gauge theory on the worldbrane which is
effectively 4-dimensional flat space-time \cite{verlinde}. Since the brane
breaks half of the total number of supersymmetries (it is 1/2 BPS), the
$U(1)$ gauge theory must have $\N=4$ Poincar\' e supersymmetry. In the
low energy approximation (which has at most two derivatives on bosons and
one derivative on fermions in this case), the $\N=4$ supersymmetric $U(1)$
gauge theory is free.

\begin{fig}[htp]
\centering
\epsfxsize=6in
\epsfysize=3in
\epsffile{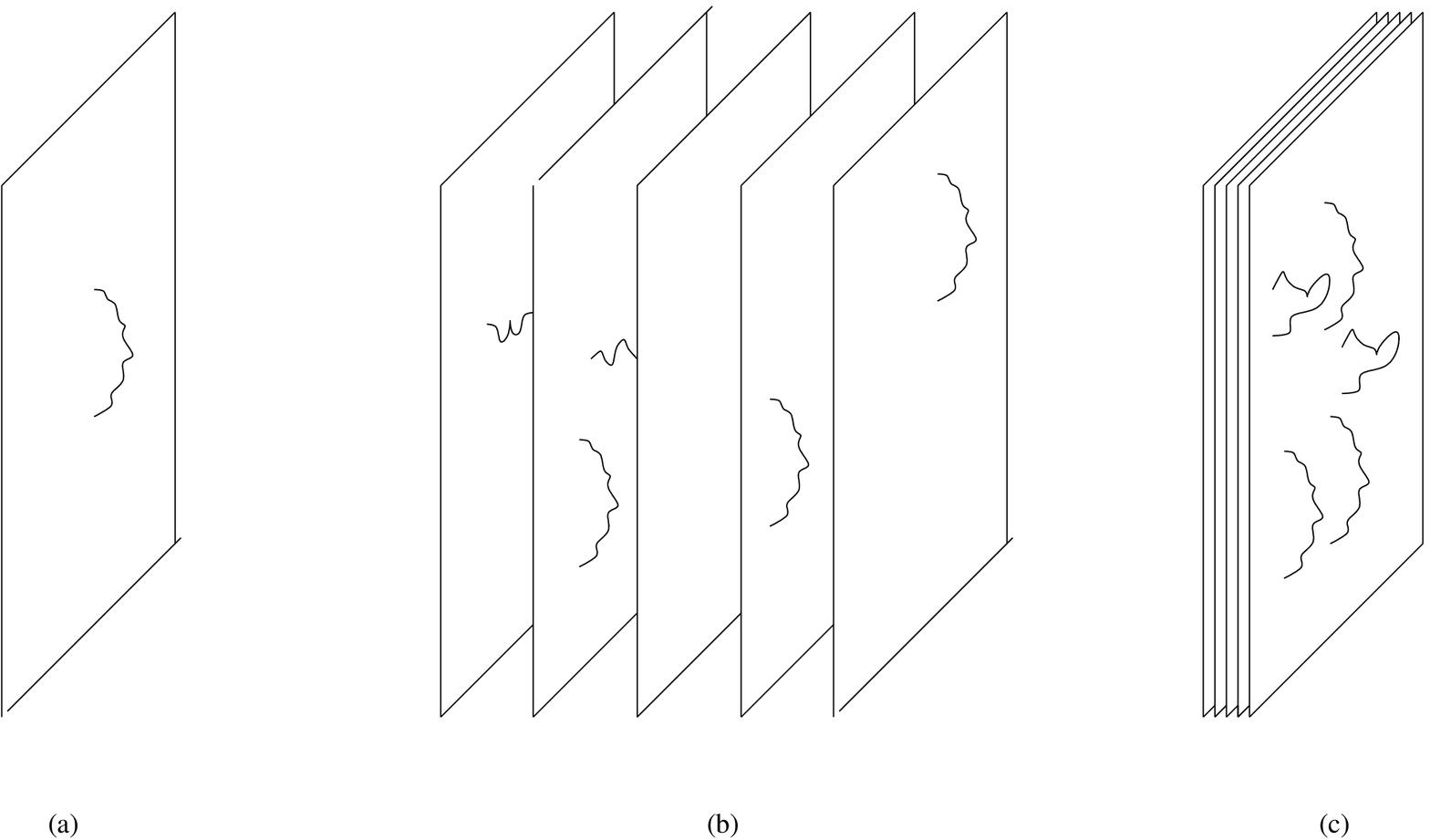}
\caption{D-branes~: (a) single, (b) well-separated, (c) (almost) coincident }
\label{fig:2}
\end{fig}

With a number $N >1$ of parallel separated D3-branes, the end points of an
open string may be attached to the same brane. For each brane, these
strings can have arbitrarily small length and must therefore be massless.
These excitation modes induce a massless $U(1)^N$ gauge theory with $\N=4$
supersymmetry in the low energy limit. An open string can also, however,
have one of its ends attached to one brane while the other end is attached
to a different brane. The mass of such a string cannot get arbitrarily
small since the length of the string is bounded from below by the
separation distance between the branes (see however problem set (5.4)).
There are $N^2 -N$ such possible strings. In the limit where the $N$ branes
all tend to be coincident, all string states would be massless and the
$U(1)^N$ gauge symmetry is enhanced to a full $U(N)$ gauge symmetry.
Separating the branes should then be interpreted as Higgsing the gauge
theory to the Coulomb branch where the gauge symmetry is spontaneously
broken (generically to $U(1)^N$). The overall $U(1)=U(N)/SU(N)$ factor
actually corresponds to the overall position of the branes and may be
ignored when considering dynamics on the branes, thereby leaving only a
$SU(N)$ gauge symmetry \cite{witten97}. These various configurations are
depicted in Fig. 2.

\medskip

In the low energy limit, $N$ coincident branes support an $\N=4$
super-Yang-Mills theory in 4-dimensions with gauge group $SU(N)$.

\subsection{The Maldacena limit}

The space-time metric of $N$ coincident D3-branes may be recast in the
following form,\footnote{In this section, we shall denote 10-dimensional
indices by $M,N, \cdots$, 5-dimensional indices by $\mu, \nu ,\cdots$ and
4-dimensional Minkowski indices by $i,j,\cdots$, and the
Minkowski metric by $\eta _{ij} = {\rm diag} (-+++)$.}
\bea
\label{D3metric}
ds^2 = \biggl ( 1 + {L^4 \over y^4} \biggr ) ^{-\half} \eta _{ij} dx^i
dx^j + \biggl ( 1 + {L^4 \over y^4} \biggr ) ^\half (dy^2 +y^2 d\Omega _5
^2)
\eea
where the {\sl radius $L$ of the D3-brane} is given by
\bea
L^4 = 4 \pi g_s N (\alpha ')^2
\eea
To study this geometry more closely, we consider its limit in two regimes.

\begin{fig}[htp]
\centering
\epsfxsize=6in
\epsfysize=3in
\epsffile{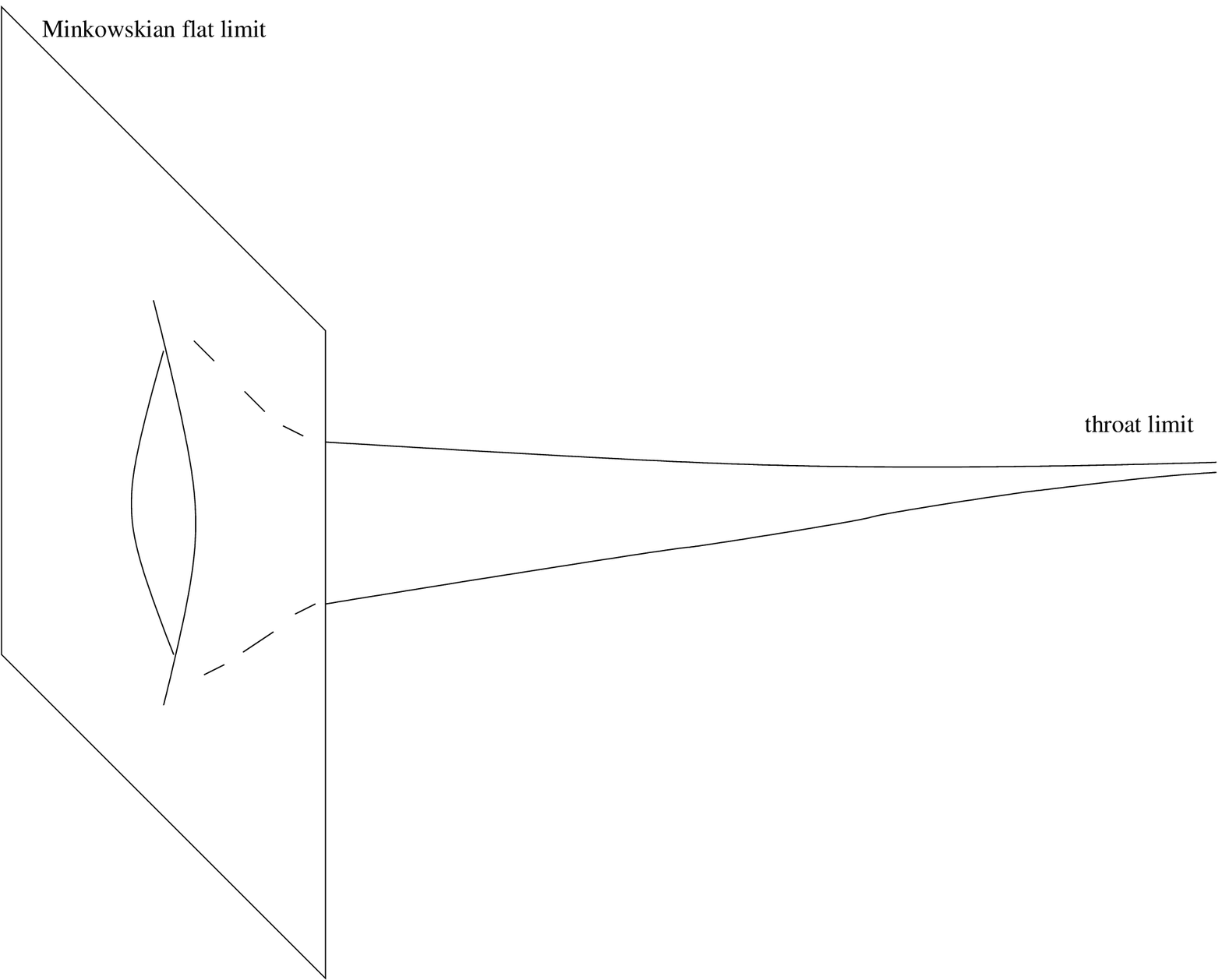}
\caption{Minkowski region of AdS (a), and throat region of AdS (b)}
\label{fig:3}
\end{fig}

As $y \gg L$, we recover flat space-time ${\bf R}^{10}$. When $y<
L$, the geometry is often referred to as the {\sl throat} and would at first
appear to be singular as $y\ll L$. A redefinition of the coordinate
\bea
u \equiv  L^2 /y
\eea
and the large $u$ limit, however, transform the metric into the following
asymptotic form
\bea
ds^2 = L^2 \biggl [ {1 \over u^2} \eta _{ij} dx^i dx^j  + {du^2 \over
u^2}  + d\Omega _5 ^2  \biggr ]
\eea
which corresponds to a product geometry.  One component is the
five-sphere $S^5$ with metric $L^2 d\Omega _5 ^2$. The  remaining component
is the hyperbolic space AdS$_5$ with constant negative curvature metric
$ L^2 u^{-2} ( du^2 + \eta _{ij} dx^i dx^j )$. In conclusion, the
geometry close to the brane ($y\sim 0$ or $u\sim \infty$) is regular and
highly symmetrical, and may be summarized as \AdS\ where both components
have identical radius
$L$.

\medskip

The Maldacena limit \cite{Malda} corresponds to keeping fixed $g_s$ and
$N$ as well as all physical length scales, while letting $\alpha ' \to 0$.
Remarkably, this limit of string theory exists and is (very !)
interesting. In the Maldacena  limit, only the \AdS\ region of the
D3-brane geometry survives the limit and contributes to the string
dynamics of physical processes, while the  dynamics in the asymptotically
flat region decouples from the theory.

\medskip

To see this decoupling in an elementary way, consider a physical quantity,
such as the effective action $\L$ and carry out its $\alpha '$ expansion in
an arbitrary background with Riemann tensor, symbolically denoted by $R$.
The expansion takes on the schematic form
\bea
\label{alphaprime}
\L = a_1 \alpha ' R + a_2 (\alpha ')^2 R ^2 + a_3 (\alpha ')^3 R^3 +
\cdots
\eea
Now physical objects and length scales in the asymptotically flat region are
characterized by a scale $y \gg L$, so that by simple scaling
arguments we have $R \sim 1/y^2$. Substitution this behavior into
(\ref{alphaprime}) yields the following expansion of the effective action,
\bea
\L = a_1 \alpha ' {1 \over y^2} + a_2 (\alpha ')^2 {1 \over y^4} + a_3
(\alpha ')^3 {1 \over y^6} + \cdots
\eea
Keeping the physical size $y$ fixed, the entire contribution to the
effective action from the limit $\alpha ' \to 0$ is then seen to vanish.

\medskip

A more precise way of establishing this decoupling is by taking the
Maldacena limit directly on the string theory non-linear sigma model in the
D3 brane background. We shall concentrate here on the metric part, thereby
ignoring the contributions from the tensor field $F^+_5$. We denote the
$D=10$ coordinates by $x^M$, $M=0,1,\cdots ,9$, and the metric by $G_{MN}
(x)$. The first 4 coordinates coincide with $x^\mu$ of the Poincar\' e
invariant D3 worldvolume, while the coordinates on the 5-sphere are
$x^M$ for $M=5,\cdots ,9$ and $x^4 = u$. The full D3 brane metric of
(\ref{D3metric}) takes the form $ds^2  =  G_{MN} dx^M dx^N = L^2 \bar G
_{MN} (x;L) dx^M dx^N$, where the rescaled metric $\bar G_{MN}$ is given by
\bea
\bar G _{MN} (x;L) dx^M dx^N = \biggl ( 1 + {L^4 \over u^4} \biggr )
^\half ({ du^2 \over u^2}  + d\Omega _5 ^2) + \biggl ( 1 + {L^4 \over u^4}
\biggr ) ^{-\half}  {1 \over u^2} \eta _{ij} dx^i dx^j
\eea
Inserting this metric into the non-linear sigma model, we obtain
\bea
S_G= {1 \over 4 \pi \alpha '} \int _\Sigma \sqrt \gamma \gamma ^{mn} G_{MN}
(x) \p _m x^M \p_n x^N = {L^2 \over 4\pi \alpha '} \int _\Sigma \sqrt
\gamma \gamma ^{mn} \bar G_{MN} (x;L) \p _m x^M \p_n x^N
\eea
The overall coupling constant for the sigma model dynamics is given by
\bea
{L^2 \over 4\pi \alpha '}  = \sqrt{ {\lambda \over 4 \pi}}
\qquad \qquad
\lambda \equiv g_s N
\eea
Keeping $g_s$ and $N$ fixed but letting $\alpha ' \to 0$ implies that
$L\to 0$. Under this limit the sigma model action admits a smooth
limit, given by
\bea
\label{adssigma}
S_G=  \sqrt{ {\lambda \over 4 \pi}} \int _\Sigma \sqrt
\gamma \gamma ^{mn} \bar G_{MN} (x;0) \p _m x^M \p_n x^N
\eea
where the metric $\bar G_{MN} (x;0)$ is the metric on \AdS,
\bea
\bar G _{MN} (x;L) dx^M dx^N =  {1 \over u^2} \eta _{ij} dx^i dx^j + {
du^2 \over u^2}  + d\Omega _5 ^2
\eea
rescaled to unit radius. Manifestly, the coupling $1/\sqrt \lambda$ has
taken over the role of $\alpha '$ as the non-linear sigma model coupling
constant and the radius $L$ has cancelled out.

\subsection{Geometry of Minkowskian and Euclidean AdS}

Before moving on to the actual Maldacena conjecture, we clarify the
geometry of AdS space-time, both with Minkowskian and Euclidean signatures.
Minkowskian AdS$_{d+1}$ (of unit radius) may be defined in ${\bf R}^{d+1}$
with coordinates $(Y_{-1}, Y_0, Y_1, \cdots , Y_d)$ as the $d+1$ dimensional
connected hyperboloid with isometry $SO(2,d)$ given by the equation
\bea
- Y_{-1} ^2 -Y_0^2 + Y_1 ^2 +\cdots + Y_d ^2 = -1
\eea
with induced metric $ds^2 = - dY_{-1} ^2 - dY_0^2 + dY_1 ^2 +\cdots + dY_d
^2$. The topology of the manifold is that of the cylinder $S^1 \times {\bf
R}$ times the sphere $S^{d-1}$, and is therefore not simply connected. The
topology of the boundary is consequently given by $\p {\rm AdS}_{d+1} = S^1
\times S^{d-1}$. The manifold may be represented by the coset $SO(2,d) /
SO(1,d)$. A schematic rendition of the manifold is given in Fig. 4 (a),
with $r^2 = Y_1 ^2 + \cdots + Y_d^2$.

\begin{fig}[htp]
\centering
\epsfxsize=6in
\epsfysize=3in
\epsffile{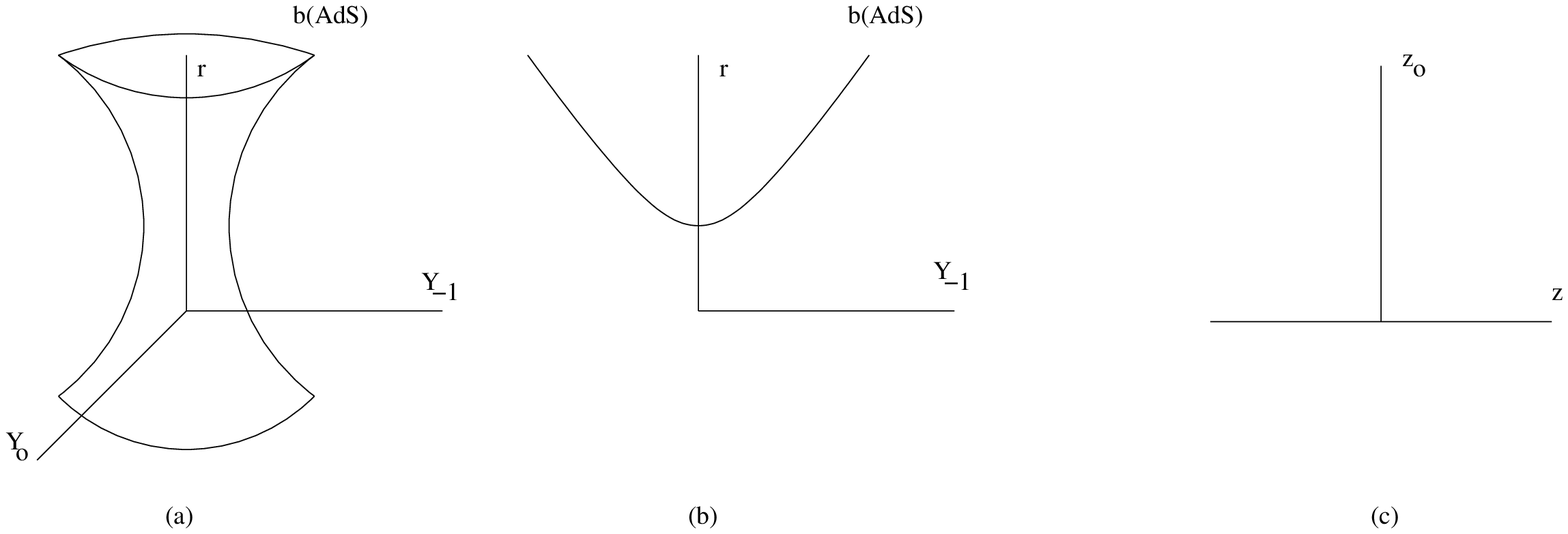}
\caption{Anti-de Sitter Space (a) Euclidean, (b) Minkowskian, (c)
upper half space}
\label{fig:4}
\end{fig}

Euclidean AdS$_{d+1}$ (of unit radius) may be defined in Minkowski flat
space ${\bf R}^{d+1}$ with coordinates $(Y_{-1}, Y_0, Y_1, \cdots , Y_d)$
as the $d+1$ dimensional disconnected hyperboloid with isometry $SO(1,d)$
given by the equation
\bea
- Y_{-1} ^2 + Y_0^2 + Y_1 ^2 +\cdots + Y_d ^2 = -1
\eea
with induced metric $ds^2 = - dY_{-1} ^2 + dY_0^2 + dY_1 ^2 +\cdots + dY_d
^2$. The topology of the manifold is that of ${\bf R}^{d+1}$. The topology
of the boundary is that of the $d$-sphere, $ \p {\rm AdS}_{d+1} = S^d$.
The manifold may be represented by the coset $SO(1,d+1) / SO(d+1)$.
A schematic rendition of the manifold is given in Fig. 4 (b),
with $r^2 = Y_0^2 + Y_1 ^2 + \cdots + Y_d^2$.
Introducing the coordinates $Y_{-1} + Y_0 = {1 \over z_0}$ and $ z_i = z_0
Y_i$ for $ i=1,\cdots ,d$, we may map Euclidean AdS$_{d+1}$ onto the upper
half space $H_{d+1}$ with Poincar\' e metric $ds^2$, defined by
\bea
H_{d+1} = \{ (z_0, \vec{z}), \ z_0 \in {\bf R}^+, \vec{z} \in {\bf R}^d \}
\qquad \qquad
ds^2 = {1 \over z_0^2} (dz_0^2 + d\vec{z}^2)
\eea
A schematic rendition is given in Fig. 4 (c).
A standard stereographic transformation may be used to map $H_{d+1}$ onto
the unit ball.

\subsection{The AdS/CFT Conjecture}

The AdS/CFT or Maldacena conjecture states the {\sl equivalence} (also
referred  to as {\sl duality}) between the following theories \cite{Malda}
\begin{itemize}
\item Type IIB superstring theory on \AdS\ where both AdS$_5$ and $S^5$
have the same radius $L$, where the 5-form $F^+ _5$ has integer flux
$N=\int _{S^5} F_5^+$ and where the string coupling is $g_s$;
\item $\N=4$ super-Yang-Mills theory in 4-dimensions, with gauge group
$SU(N)$ and Yang-Mills coupling $g_{YM}$ in its (super)conformal phase;
\end{itemize}
with the following identifications between the parameters of both theories,
\bea
g_s = g_{YM}^2 \qquad \qquad L^4 = 4 \pi g_s N (\alpha ') ^2
\eea
and the axion expectation value equals the SYM instanton angle $\< C \>
=\theta _I$. Precisely what is meant by {\sl equivalence} or {\sl duality}
will be the subject of the remainder of this section, as well as of the
next one. In brief, {\sl equivalence} includes a precise map between the
states (and fields) on the superstring side and the local gauge invariant
operators on the $\N=4$ SYM side, as well as a correspondence between the
correlators in both theories.

\medskip

The above statement of the conjecture is referred to as the {\sl strong
form}, as it is to hold for all values of $N$ and of $g_s = g_{YM}^2$.
String theory quantization on a general curved manifold (including
\AdS), however, appears to be very difficult and is at present out of
reach. Therefore, it is natural to seek limits in which the Maldacena
conjecture becomes more tractable but still remains non-trivial.

\subsubsection{The `t~Hooft Limit}

The `t~Hooft limit consists in keeping the `t~Hooft coupling $\lambda
\equiv g_{YM}^2 N = g_s N$ fixed and letting $N\to \infty$. In Yang-Mills
theory, this limit is well-defined, at least in perturbation theory, and
corresponds to a topological expansion of the field theory's Feynman
diagrams. On the AdS side, one may interpret the `t~Hooft limit as follows.
The string coupling may be re-expressed in terms of the `t~Hooft coupling
as $g_s = \lambda /N$. Since $\lambda $ is being kept fixed, the `t~Hooft
limit corresponds to {\sl weak coupling string perturbation theory}.

\medskip

This form of the conjecture, though weaker than the original version is
still a very powerful correspondence between classical string theory and
the large $N$ limit of gauge theories. The problem of finding an action
built out of classical fields to which the large $N$ limit of gauge
theories are classical solutions is a challenge that had been outstanding
since `t~Hooft's original paper \cite{largeN}. The above correspondence
gives a concrete, though still ill-understood, realization of this
``large $N$ master-equation".

\begin{table}[t]
\begin{center}
\begin{tabular}{|c|c|c|} \hline \hline
$\bullet$ $\N=4$ conformal SYM & & $\bullet$ Full Quantum  Type IIB string
\\
all $N$, $g_{YM}$ & $\Leftrightarrow $ & theory on \AdS \\
$\bullet$ $g_s =g_{YM}^2$ &   & $\bullet$ $L^4 = 4 \pi g_s N \alpha '^2$ \\
 \hline \hline
$\bullet$ `t~Hooft limit of $\N=4$ SYM & & $\bullet$ Classical Type IIB
string theory \\
\qquad $\lambda= g_{YM}^2 N$ fixed, $N \to \infty$ & $\Leftrightarrow $ &
\quad  on \AdS \\
$\bullet$ $1/N$ expansion && $\bullet$ $g_s$ string loop expansion
\\ \hline \hline
$\bullet$ Large $\lambda$ limit of $\N=4$ SYM &  & $\bullet$ Classical
Type IIB supergravity  \\
(for $N\to \infty$) & $\Leftrightarrow $ & on \AdS \\
$\bullet$ $\lambda ^{-1/2} $ expansion && $\bullet$ $\alpha '$ expansion
\\ \hline \hline
\end{tabular}
\end{center}
\caption{The three forms of the AdS/CFT conjecture in order of
decreasing strength}
\end{table}

\subsubsection{The Large $\lambda$ Limit}

In taking the `t~Hooft limit, $\lambda = g_s N$ is kept fixed while $N\to
\infty$. Once this limit has been taken, the only parameter left is
$\lambda$. Quantum field theory perturbation theory corresponds to
$\lambda \ll 1$. On the AdS side of the correspondence, it is actually
natural to take $\lambda \gg 1$ instead. It is very instructive to
establish the meaning of an expansion around $\lambda$ large. To  do, we
expand in powers of $\alpha '$ a physical quantity such as the effective
action, as we already did in (\ref{alphaprime}),
\bea
\L = a_1 \alpha ' R + a_2 (\alpha ')^2 R^2 + a_3
(\alpha ')^3 R^3 + \cdots
\eea
The distance scales in which we are now interested are those typical of
the throat, whose scale is set by the AdS radius $L$. Thus, the scale of
the Riemann tensor is set by
\bea
R \sim 1 / L^2 =  (g_s N)^{-\half} /\alpha ' = \lambda
^{-\half}/ \alpha '
\eea
and therefore, the expansion of the effective action in powers of $\alpha
'$ effectively becomes an expansion in powers of $\lambda ^{-\half}$,
\bea
\L = a_1 \lambda ^{-\half} + a_2 \lambda ^{-1} + a_3
(\alpha ')^3 \lambda ^{-{3 \over 2}} + \cdots
\eea
The interchange of the roles of $\alpha '$ and $\lambda ^{-1/2}$ may also
be seen directly from the worldsheet non-linear sigma model action of
(\ref{adssigma}).
Clearly, any $\alpha '$ dependence has disappeared from the string theory
problem and the role of $\alpha'$ as a scale has been replaced by the
parameter $\lambda ^{-1/2}$.

\subsection{Mapping Global Symmetries}

A key necessary ingredient for the AdS/CFT correspondence to hold is that
the global unbroken symmetries of the two theories be identical.
The continuous global symmetry of $\N=4$ super-Yang-Mills theory in its
conformal phase was previously shown to be the superconformal group
$SU(2,2|4)$, whose maximal bosonic subgroup is $SU(2,2) \times SU(4)_R
\sim SO(2,4) \times SO(6)_R$. Recall that the bosonic subgroup arises as the
product of the conformal group $SO(2,4)$ in 4-dimensions by the $SU(4)_R$
automorphism group of the $\N=4$ Poincar\' e supersymmetry algebra. This
bosonic group is immediately recognized on the AdS side as the isometry
group of the \AdS\ background. The completion into the full supergroup
$SU(2,2|4)$ was discussed for the SYM theory in subsection \S 3.3, and
arises on the AdS side because 16 of the 32 Poincar\' e supersymmetries are
preserved by the array of $N$ parallel D3-branes, and in the AdS limit, are
supplemented by another 16 conformal supersymmetries (which are broken in
the full D3-brane geometry). Thus, the global symmetry $SU(2,2|4)$ matches
on both sides of the AdS/CFT correspondence.

\medskip

$\N=4$ super-Yang-Mills theory also has Montonen-Olive or S-duality
symmetry, realized on the complex coupling constant $\tau$ by M\" obius
transformations in $SL(2,{\bf Z})$. On the AdS side, this symmetry is a
global discrete symmetry of Type IIB string theory, which is unbroken by
the D3-brane solution, in the sense that it maps non-trivially only the
dilaton and axion expectation values, as was shown earlier. Thus, S-duality
is also a symmetry of the AdS side of the AdS/CFT correspondence. It must
be noted, however, that S-duality is a useful symmetry only in the
strongest form of the AdS/CFT conjecture. As soon as one takes the `t~Hooft 
limit $N\to \infty$ while keeping $\lambda = g_{YM}^2 N$
fixed, S-duality no longer has a consistent action. This may be seen for
$\theta _I=0$, where it maps $g_{YM} \to 1/g_{YM}$ and thus $\lambda \to
N^2 /\lambda$.

\subsection{Mapping Type IIB Fields and CFT Operators}

Given that we have established that the global symmetry groups on both
sides of the AdS/CFT correspondence coincide, it remains to show that the
actual representations of the supergroup $SU(2,2|4)$ also coincide on both
sides. The spectrum of operators on the SYM side was explained already in
subsection \S 3.5. Suffice it to recall here the special significance of the
short multiplet representations, namely 1/2 BPS representations with a span
of spin 2, 1/4 BPS representations with a span of spin 3 and 1/8 BPS
representations with a span of spin 7/2. Non-BPS representations in general
have a span of spin 4.

\medskip

A special role is played by the {\sl single color trace operators} because
out of them, all higher trace operators may be constructed using the OPE.
Thus one should expect single trace operators on the SYM side to correspond
to single particle states (or canonical fields) on the AdS side \cite{Malda};
see also \cite{Andrianopoli:1998jh}. Multiple
trace states should then be interpreted as bound states of these one
particle states. Multiple trace BPS operators have the property that their
dimension on the AdS side is simply the sum of the dimensions of the
BPS constituents. Such bound states occur in the spectrum at the lower
edge of the continuum threshold and are therefore called {\sl threshold
bound states}. A good example to keep in mind when thinking of threshold
bound states in ordinary quantum field theory is another case of BPS
objects~: magnetic monopoles \cite{thp} in the Prasad-Sommerfield limit
\cite{bps} (or exactly in the Coulomb phase of $\N=4$ SYM). A collection
of $N$ magnetic monopoles with like charges forms a static solution of the
BPS equations and therefore form a threshold bound state. Very recently, a
direct coupling of double-trace operators to AdS supergravity has been studied
in \cite{Aharony:2001pa}.

\medskip

To identify the contents of irreducible representations of $SU(2,2|4)$ on
the AdS side, we describe all Type IIB massless supergravity and massive
string degrees of freedom by fields $\varphi$ living on \AdS.
We introduce coordinates $z^\mu$, $\mu =0,1, \cdots ,4$ for AdS$_5$ and
$y^u$, $u =1,\cdots ,5$ for $S^5$, and decompose the metric as
\bea
ds^2 = g_{\mu \nu} ^{AdS} dz^\mu dz^\nu + g_{uv } ^S dy^u
dy^v
\eea
The fields then become functions $\varphi (z,y)$ associated with the
various $D=10$ degrees of freedom. It is convenient to decompose $\varphi
(z,y)$ in a series on $S^5$,
\bea
\varphi (z,y) = \sum _{\Delta =0} ^\infty \varphi _\Delta (z) Y_\Delta (y)
\eea
where $Y_\Delta$ stands for a basis of spherical harmonics on $S^5$.
For scalars for example, $Y_\Delta$ are labelled by the rank $\Delta$ of
the totally symmetric traceless representations of $SO(6)$. Just as fields
on a circle received a mass contribution from the momentum mode on the
circle, so also do fields compactified on $S^5$ receive a contribution to
the mass. From the eigenvalues of the Laplacian on $S^5$, for various
spins, we find the following relations between mass and scaling
dimensions,
\bea
{\rm scalars } & \qquad & m^2 = \Delta (\Delta - 4) \nonumber \\
{\rm spin \ 1/2, \ 3/2} & \qquad & |m| = \Delta -2           \nonumber \\
p-{\rm form} & \qquad & m^2 = (\Delta -p) (\Delta +p - 4) \nonumber \\
{\rm spin 2 } & \qquad &  m^2 = \Delta (\Delta - 4)
\eea
The complete correspondence between the representations of $SU(2,2|4)$ on
both sides of the correspondence is given in Table 6.
The mapping of the descendant states is also very interesting. For the
$D=10$ supergravity multiplet, this was worked out in \cite{krvn}, and is
given in Table 7. Generalizations to AdS$_4 \times S^7$ were discussed in
\cite{Biran, Gun86, deWit} while those to AdS$_7 \times S^4$ were
discussed in \cite{Pilch84, GNW}, with recent work on AdS/CFT for
M-theory on these spaces in \cite{Aharony98, Corrado, DP00, arsok}. General
reviews may be found in \cite{Polyakov:2001af}, \cite{deWit:1999ui}. Recently,
conjectures involving also de Sitter space-times have been put forward in
\cite{Spradlin:2001pw} and references therein. Finally, we point out that the
existence of singleton and doubleton representations of the conformal group
SO(2,4) is closely related with the AdS/CFT correspondence; for recent
accounts, see \cite{Flato:1998iy}, \cite{Gunaydin:1998sw}, \cite{Flato:1999yp}
and \cite{Kogan:1999bn} and references therein. Additional references on the
(super)symmetries of AdS are in \cite{Bilal:1998ck}, \cite{Imamura:1998gk} and
\cite{Shuster:1999zf}.

\begin{table}[t]
\begin{center}
\begin{tabular}{|c|c|} \hline \hline
Type IIB string theory &   $\N=4$ conformal super-Yang-Mills
\\ \hline \hline
Supergravity Excitations   & Chiral primary + descendants
\\
1/2 BPS, spin $\leq 2$ & $\O _2 = \tr X^{\{ i} X^{j\}}$ + desc.
\\ \hline
Supergravity Kaluza-Klein & Chiral primary + Descendants
\\
1/2 BPS, spin $\leq 2$ & $\O _\Delta = \tr X^{\{ i_1} \cdots X^{i_\Delta
\}}$ + desc.
\\ \hline
Type IIB massive string modes & Non-Chiral operators, dimensions $\sim
\lambda ^{1/4}$ \\
non-chiral, long multiplets & e.g. Konishi $\tr X^i X^i$
\\ \hline
Multiparticle states & products of operators at distinct points
\\
 & $\O _{\Delta _1} (x_1) \cdots \O _{\Delta _n} (x_n)$
\\ \hline
Bound states & product of operators at same point
\\
 & $\O _{\Delta _1} (x) \cdots \O _{\Delta _n} (x)$
\\ \hline \hline
\end{tabular}
\end{center}
\caption{Mapping of String and Sugra states onto SYM Operators}
\label{table:6}
\end{table}

\subsection{Problem Sets}

\noindent
(5.1)
The Poincar\' e upper half space is defined by $H_{d+1} = \{ (z_0,\vec{z})
\in {\bf R}^{d+1}, z_0 >0 \}$ with metric $ds^2 = (dz_0^2 + d\vec{z}^2 )/
z_0 ^2$. (a) Show -- by solving the geodesic equations -- that the geodesics
of $H_{d+1}$ are the half-circles of arbitrary radius $R$, centered at an
arbitrary point $(0,\vec{c})$ on the boundary of $H_{d+1}$. Compute the
geodesic distance between any two arbitrary points.

\medskip

\noindent
(5.2) We now represent Euclidean AdS$ _{d+1}$ as the manifold in ${\bf
R}^{d+2}$ given by the equation $ - Y_{-1} ^2 + Y_0^2 + \vec{Y}^2 =-1 $,
with induced metric $ds^2 = - dY_{-1} ^2 + dY_0^2 + d\vec{Y}^2 $. Show that
the geodesics found in problem (5.1) above are simply the sections by planes
through the origin, given by the equation
$$
Y_{-1} - Y_0 = (R^2 - \vec{c}^2) (Y_{-1} + Y_0) + 2 \vec{c} \cdot \vec{Y}
$$
(You may wish to explore the analogy with the geometry and geodesics of
the sphere $S^{d+1}$.)

\medskip

\noindent
(5.3) The geodesic distance between two separate D3 branes is actually
infinite, as may be seen by integrating the infinitesimal distance $ds$ of
the D3 metric. Using the worldsheet action of a string suspended between
the two D3 branes, explain why this string still has a finite mass spectrum.

\medskip

\noindent
(5.4) Consider a classical bosonic string in AdS$_{d+1}$ space-time,
with its dynamics governed by the Polyakov action, namely in the presence
of the AdS$_{d+1}$ metric $G_{\mu \nu } (x)$. (We ignore the
anti-symmetric tensor fields for simplicity.)
$$
S[x] = \int _\Sigma \! d^2 \xi \sqrt{\gamma} \gamma^{mn} \p _m x^\mu \p _n
x^\nu G_{\mu \nu} (x)
$$
Solve the string equations assuming a special Ansatz that the solution be
spherically symmetric, i.e. invariant under the $SO(d)$ subgroup of
$SO(2,d)$.

\vfill\eject

\begin{table}[htp]
\begin{center}
\begin{tabular}{|l|c|c|c|c|c|c|c|} \hline
SYM Operator   & desc & SUGRA & dim & spin  & $Y$ & $SU(4)_R$ &
 lowest reps
                \\ \hline \hline
$\O _k         \sim   \tr X^k$, $k\geq 2$
                  & --
                  & $h_{\alpha}^{\alpha}\ \ a_{\alpha \beta \gamma \delta}$
                  & $k$
                  & $(0,0)$
                  & 0
                  & $(0,k,0)$
                  & {\bf 20'},{\bf 50},{\bf 105}
                \\ \hline
$\O _k ^{(1)}    \sim \tr \lambda X^k$, $k\geq 1$
                  & $Q$
                  & $\psi _{(\alpha)}$
                  & $k+{3\over 2}$
                  & $(\half, 0)$
                  & $\half$
                  & $(1,k,0)$
                  & {\bf 20},{\bf 60},{\bf 140'}
                \\ \hline
$\O _k ^{(2)}    \sim \tr \lambda \lambda X^k$
                  & $Q^2$
                  & $A_{\alpha \beta} $
                  & $k+3$
                  & $(0,0)$
                  & $1$
                  & $(2,k,0)$
                  & {\bf 10$_c$},{\bf 45$_c$},{\bf 126$_c$}
                \\ \hline
$\O _k ^{(3)}   \sim  \tr \lambda \bar \lambda X^k$
                  & $Q \bar Q$
                  & $ h_{\mu \alpha} \ \ a_{\mu \alpha \beta \gamma}$
                  & $k+3$
                  & $(\half,\half)$
                  & $0$
                  & $(1,k,1)$
                  & {\bf 15},{\bf 64},{\bf 175}
                \\ \hline
$\O _k ^{(4)} \!  \sim \! \tr F_+ X^k$, $k\geq 1$
                  & $Q^2$
                  & $A_{\mu \nu}$
                  & $k+2$
                  & $(1,0)$
                  & $1$
                  & $(0,k,0)$
                  & {\bf 6$_c$},{\bf 20$_c$},{\bf 50$_c$}
                \\ \hline
$\O _k ^{(5)}   \sim  \tr F_+ \bar \lambda X^k$
                  & $Q^2\bar Q$
                  & $\psi _\mu$
                  & $k+{7 \over 2}$
                  & $(1, \half)$
                  & $\half$
                  & $(0,k,1)$
                  & ${\bf 4}^*,{\bf 20}^*,{\bf 60}^*$
                \\ \hline
$\O _k ^{(6)}    \sim \tr F_+ \lambda X^k$
                  & $Q^3$
                  & ``$\lambda$"
                  & $k+{7 \over 2}$
                  & $(\half ,0)$
                  & ${3 \over 2}  $
                  & $(1, k,0)$
                  & {\bf 4},{\bf 20},{\bf 60}
                \\ \hline
$\O _k ^{(7)}   \sim  \tr \lambda \lambda \bar \lambda X^k$
                  & $Q^2\bar Q$
                  & $\psi _{(\alpha)}$
                  & $k+{9 \over 2}$
                  & $(0,\half)$
                  & $\half$
                  & $(2,k,1)$
                  & {\bf 36},{\bf 140},{\bf 360}
                \\ \hline
$\O _k ^{(8)}    \sim \tr F_+^2 X^k$
                  & $Q^4$
                  & $B$
                  & $k+4$
                  & $(0,0)$
                  & $2$
                  & $(0,k,0)$
                  & {\bf 1$_c$},{\bf 6$_c$},{\bf 20'$_c$}
                \\ \hline
$\O _k ^{(9)}   \sim \tr F_+ F_- X^k$
                  & $Q^2\bar Q^2$
                  & $h_{\mu \nu}'$
                  & $k+4$
                  & $(1,1)$
                  & $0$
                  & $(0,k,0)$
                  & {\bf 1},{\bf 6},{\bf 20'}
                \\ \hline
$\O _k ^{(10)}   \sim \tr F_+ \lambda \bar \lambda X^k$
                  & $Q^3\bar Q$
                  & $A_{\mu \alpha}$
                  & $k+5$
                  & $(\half,\half)$
                  & $1$
                  & $(1,k,1)$
                  & {\bf 15},{\bf 64},{\bf 175}
                \\ \hline
$\O _k ^{(11)}   \sim \tr F_+ \bar \lambda \bar \lambda X^k$
                  & $Q^2\bar Q ^2$
                  & $a_{\mu \nu \alpha \beta}$
                  & $k+5$
                  & $(1, 0)$
                  & $0$
                  & $(0,k,2)$
                  & {\bf 10$_c$},{\bf 45$_c$},{\bf 126$_c$}
                \\ \hline
$\O _k ^{(12)}   \sim \tr \lambda \lambda \bar \lambda \bar \lambda X^k$
                  & $Q^2\bar Q ^2$
                  & $h_{(\alpha \beta)}$
                  & $k+6$
                  & $(0, 0)$
                  & $0$
                  & $(2,k,2)$
                  & {\bf 84},{\bf 300},{\bf 2187}
                \\ \hline
$\O _k ^{(13)}   \sim \tr F_+^2 \bar \lambda X^k$
                  & $Q^4\bar Q$
                  & ``$\lambda$"
                  & $k+{11\over 2}$
                  & $(0, \half)$
                  & ${3\over 2}$
                  & $(0,k,1)$
                  & ${\bf 4}^*,{\bf 20}^*,{\bf 60}^*$
                \\ \hline
$\O _k ^{(14)}   \sim \tr F_+ \lambda \bar \lambda \bar \lambda X^k$
                  & $Q^3\bar Q^2$
                  & $\psi _{(\alpha)}$
                  & $k+{13\over 2}$
                  & $(\half, 0)$
                  & $\half$
                  & $(1,k,2)$
                  & {\bf 36}$^*$,{\bf 140}$^*$,{\bf 360}$^*$
                \\ \hline
$\O _k ^{(15)}   \sim \tr F_+ F_- \lambda X^k$
                  & $Q^3\bar Q^2$
                  & $\psi _\mu$
                  & $k+{11\over 2}$
                  & $(\half, 1)$
                  & $\half$
                  & $(1,k,0)$
                  & {\bf 4},{\bf 20},{\bf 60}
                \\ \hline
$\O _k ^{(16)}   \sim \tr F_+ F_- ^2 X^k$
                  & $Q^4\bar Q^2$
                  & $A_{\mu \nu}$
                  & $k+6$
                  & $(1, 0)$
                  & $1$
                  & $(0,k,0)$
                  & {\bf 1$_c$},{\bf 6$_c$},{\bf 20'$_c$}
                \\ \hline
$\O _k ^{(17)} \!  \sim  \!\tr \!  F_+ \! F_- \lambda \bar \lambda X^k$
                  & $Q^3\bar Q^3$
                  & $h_{\mu \alpha} \ \ a_{\mu \alpha \beta \gamma}$
                  & $k+7$
                  & $(\half, \half)$
                  & $0$
                  & $(1,k,1)$
                  & {\bf 15},{\bf 64},{\bf 175}
                \\ \hline
$\O _k ^{(18)}   \! \sim \! \tr F_+ ^2 \bar \lambda \bar \lambda X^k$
                  & $Q^4\bar Q^2$
                  & $A_{\alpha \beta}$
                  & $k+7$
                  & $(0, 0)$
                  & $1$
                  & $(0,k,2)$
                  & {\bf 10$_c$},{\bf 45$_c$},{\bf 126$_c$}
                \\ \hline
$\O _k ^{(19)}  \! \sim \! \tr F_+^2 F_- \bar \lambda    X^k$
                  & $Q^4\bar Q^3$
                  & $\psi _{(\alpha)}$
                  & $k+{15\over 2}$
                  & $(0, \half)$
                  & $\half$
                  & $(0,k,1)$
                  & ${\bf 4}^*,{\bf 20}^*,{\bf 60}^*$
                \\ \hline
$\O _k ^{(20)}  \! \sim \! \tr F_+ ^2 F_- ^2 X^k$
                  & $Q^4\bar Q^4$
                  & $h_\alpha ^\alpha \ \ a_{\alpha \beta \gamma \delta}$
                  & $k+8$
                  & $(0, 0)$
                  & $0$
                  & $(0,k,0)$
                  & {\bf 1},{\bf 6},{\bf 20'}
                \\ \hline
\end{tabular}
\end{center}
\caption{Super-Yang-Mills Operators, Supergravity Fields and
$SO(2,4)\times U(1)_Y \times SU(4)_R$ Quantum Numbers. The range of $k$ is
$k\geq 0$, unless otherwise specified. }
\label{table:7}
\end{table}

\vfill\eject

\section{AdS/CFT Correlation Functions}
\setcounter{equation}{0}

In the preceding section, evidence was presented for the Maldacena
correspondence between $\N=4$ super-conformal Yang-Mills theory with
$SU(N)$ gauge group and Type IIB superstring theory on \AdS. The evidence
was based on the precise matching of the global symmetry group
$SU(2,2|4)$, as well as of the specific representations of this group. In
particular, the single trace 1/2 BPS operators in the SYM theory matched in
a one-to-one way with the canonical fields of supergravity, compactified
on \AdS. In the present section, we present a more detailed version of the
AdS/CFT correspondence by mapping the correlators on both sides of the
correspondence.

\subsection{Mapping Super Yang-Mills and AdS Correlators}

We work with Euclidean AdS$_5$, or $H = \{ (z_0,\vec{z}), z_0 >0,
\vec{z}\in {\bf R}^4\}$ with Poincar\' e metric $ds^2 = z_0 ^{-2} (dz_0^2 +
d\vec{z}^2)$, and boundary $\p H= {\bf R}^4$. (Often, this space will be
graphically represented as a disc, whose boundary is a circle; see Fig.
5.) The metric diverges at the boundary $z_0=0$, because the overall scale
factor blows up there. This scale factor may be removed by a Weyl
rescaling of the metric, but such rescaling is not unique. {\sl A unique
well-defined limit to the boundary of AdS$_5$ can only exist if the
boundary theory is scale invariant} \cite{witten}. For finite values of
$z_0>0$, the geometry will still have 4-dimensional Poincar\' e
invariance but need not be scale invariant.

\medskip

Superconformal $\N=4$ Yang-Mills theory is scale invariant and may thus
consistently live at the boundary $\p H$. The dynamical observables
of $\N=4$ SYM are the local gauge invariant polynomial operators described
in section 3; they naturally live on the boundary $\p H$, and are
characterized by their dimension, Lorentz group $SO(1,3)$ and $SU(4)_R$
quantum numbers \cite{witten}.

\medskip

On the AdS side, we shall decompose all 10-dimensional fields onto
Kaluza-Klein towers on $S^5$, so that effectively all fields $\varphi
_\Delta (z)$ are on AdS$_5$, and labeled by their dimension $\Delta$
(other quantum number are implicit). Away from the bulk interaction region,
it is assumed that the bulk fields are free asymptotically (just as
this  is assumed in the derivation of the LSZ formalism in flat space-time
quantum field theory). The free field then satisfies $(\Box + m
_\Delta ^2)\varphi ^0 _\Delta =0$  with $m_\Delta ^2 = \Delta (\Delta -4)$
for scalars. The two independent solutions are characterized by the
following asymptotics as
$z_0 \to 0$,
\bea
\varphi ^0 _\Delta (z_0,\vec{z}) = \left \{  \matrix{
z_0 ^\Delta & {\rm normalizable} \cr
&\cr
z_0 ^{4-\Delta} & \mbox{\rm non-normalizable} } \right .
\eea
Returning to the interacting fields in the fully interacting theory,
solutions will have the same asymptotic behaviors as in the free case. It
was argued in \cite{balakraus} that the normalizable modes determine the
vacuum expectation values of operators of associated dimensions and
quantum numbers. The non-normalizable solutions on
the other hand do not correspond to bulk excitations because they are not
properly square normalizable. Instead, they represent the coupling of
external sources to the supergravity or string theory. The precise
correspondence is as follows \cite{witten}. The non-normalizable solutions
$\varphi _\Delta$ define {\sl associated boundary fields} $\bar \varphi
_\Delta$ by the following relation
\bea
\label{barphilimit}
\bar \varphi _\Delta (\vec{z}) \equiv \lim _{z_0 \to 0}
\varphi _\Delta (z_0,\vec{z}) z_0 ^{4 -\Delta}
\eea
Given a set of boundary fields $\bar \varphi _\Delta (\vec{z})$, it is
assumed that a complete and unique bulk solution to string theory exists.
We denote the fields of the associated solution $\varphi _\Delta$.

\medskip

The mapping between the correlators in the SYM theory and the dynamics of
string theory is given as follows \cite{witten, Gubs}. First, we introduce
a generating functional $\Gamma [\bar \varphi _\Delta]$ for all the
correlators of  single trace operators $\O _\Delta$ on the SYM side in
terms of the source fields $\bar \varphi _\Delta$,
\bea
\exp \{ - \Gamma [\bar \varphi _\Delta]\}
\equiv
\< \exp \biggl \{\int _{\p H} \bar \varphi _\Delta \O _\Delta \biggr \}
\>
\eea
This expression is understood to hold order by order in a perturbative
expansion in the number of fields $\bar \varphi _\Delta$. On the AdS side,
we assume that we have an action $S[\varphi _\Delta]$ that summarizes the
dynamics of Type IIB string theory on \AdS. In the supergravity
approximation, $S[\varphi _\Delta]$ is just the Type IIB supergravity
action on \AdS. Beyond the supergravity approximation, $S[\varphi _\Delta]$
will also include $\alpha '$ corrections due to massive string effects.
The mapping between the correlators is given by
\bea
\Gamma [\bar \varphi _\Delta] = {\rm extr} \ S[\varphi _\Delta]
\eea
where the extremum on the rhs is taken over all fields $\varphi _\Delta$
that satisfy the asymptotic behavior (\ref{barphilimit}) for the boundary
fields $\bar \varphi _\Delta$ that are the sources to the SYM operators
$\O _\Delta$ on the lhs. Additional references on the field-state-operator
mapping may be found in \cite{Arutyunov:1998ve}, \cite{Muck:1999kk},
\cite{Muck:1999gi}, \cite{Arut00A}, \cite{Bena:1999jv} and
\cite{Boschi-Filho:2000vd}.

\subsection{Quantum Expansion in 1/N -- Witten Diagrams}

The actions of interest to us will have an overall coupling constant
factor. For example, the part of the Type IIB supergravity action for the
dilaton $\Phi$ and the axion $C$ in the presence of a metric $G_{\mu \nu}$
in the Einstein frame, is given by
\bea
S[G,\Phi, C] = {1 \over 2 \kappa _5 ^2} \int _H \sqrt G
\biggl [ - R_G + \Lambda + \half \p_\mu \Phi \p ^\mu \Phi +
\half e^{2 \Phi} \p_\mu C \p ^\mu C \biggr ]
\eea
and the 5-dimensional Newton constant $\kappa _5^2$ is given by $\kappa _5
^2 = 4 \pi ^2 /N^2$, a relation that will be explained and justified in
(\ref{G5}). For large $N$, $\kappa_5$ will be small and one may perform a
small $\kappa_5$, i.e. a semi-classical expansion of the correlators
generated by this action. The result is a set of rules, analogous to
Feynman rules, which may be summarized by {\sl Witten diagrams}. The
Witten diagram is represented by a disc, whose interior corresponds to
the interior of AdS while the boundary circle corresponds to the boundary
of AdS \cite{witten}. The graphical rules are as follows,
\begin{itemize}
\item
Each external source to $\bar \varphi _\Delta (\vec{x}_I)$ is located at
the boundary circle of the Witten diagram at a point $\vec{x}_I$.
\item
>From the external source at $\vec{x}_I$ departs a propagator to either
another boundary point, or to an interior interaction point via a {\sl
boundary-to-bulk propagator}.
\item
The structure of the interior interaction points is governed by the
interaction vertices of the action $S$, just as in Feynman diagrams.
\item
Two interior interaction points may be connected by {\sl bulk-to-bulk
propagators}, again following the rules of ordinary Feynman diagrams.
\end{itemize}

\begin{fig}[htp]
\centering
\epsfxsize=6.3in
\epsfysize=1.4in
\epsffile{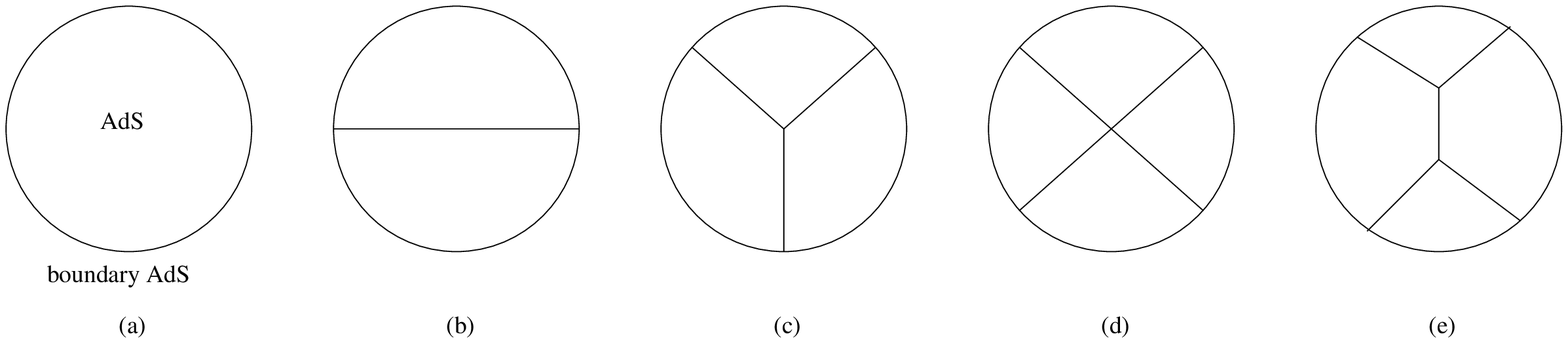}
\caption{Witten diagrams (a) empty, (b) 2-pt, (c) 3-pt, (d) 4-pt contact,
(e)  exchange}
\label{fig:5}
\end{fig}

Tree-level 2-, 3- and 4-point function contributions are given as an example in
figure \ref{fig:5}. The approach that will be taken here is based on the
component formulation of sugra. It is possible however to make progress
directly in superspace \cite{superspace}, but we shall not discuss this here.

\subsection{AdS Propagators}

We shall define and list the solution for the propagators of general
scalar fields, of massless gauge fields and massless gravitons. The
propagators are considered in Euclidean AdS$_{d+1}$, a space that
we shall denote by $H$. Recall that the Poincar\' e metric is given by
\bea
ds ^2 =g_{\mu \nu} dz^\mu dz^\nu = z_0^{-2} (dz_0^2 + d\vec{z}^2)
\eea
Here, we have set the AdS$_{d+1}$ radius to unity.
By $SO(1,d+1)$ isometry of $H$, the Green functions essentially depend
upon the $SO(1,d+1)$-invariant distance between two points in $H$. The
{\sl geodesic distance} is given by (see problem 5.1)
\bea
d(z,w) = \int _w ^z \! ds =  \ln \biggl ({1 + \sqrt {1-\xi ^2} \over \xi }
\biggr )
\hskip 0.7in
\xi \equiv {2 z_0 w_0 \over z_0^2 + w_0^2 +(\vec{z}-\vec{w})^2}
\eea
Given its algebraic dependence on the coordinates, it is more convenient to
work with the object $\xi$ than with the geodesic distance.  The {\sl
chordal distance} is given by $u=\xi ^{-1} -1$. The distance relation may
be inverted to give $ u= \xi ^{-1} -1 = \cosh d -1 $.

\bigskip

\noindent {\bf The massive scalar bulk-to-bulk propagator}

\medskip

Let $\varphi _\Delta(z)$ be a scalar field of conformal weight $\Delta$ and
mass$^2$ $m^2=\Delta (\Delta -d)$ whose linearized dynamics is given by a
coupling to a scalar source $J$ via the action
\bea
S_{\varphi _\Delta}
= \int _H \! d^{d+1} z \sqrt g \biggl [ \half g^{\mu \nu}
\p _\mu \varphi _\Delta \p_\nu \varphi _\Delta + \half m^2 \varphi
_\Delta ^2 -
\varphi _\Delta J\biggr ]
\eea
The field is then given in response to the source by
\bea
\varphi _\Delta (z) = \int _H \! d^{d+1} z' \sqrt g G_\Delta (z,z') J(z')
\eea
where the scalar Green function satisfies the differential equation
\bea
(\Box _g   + m^2) G_\Delta (z,z') = \delta (z,z')
\qquad
\delta (z,z') \equiv {1\over \sqrt g} \delta (z-z')
\eea
The (positive) scalar Laplacian is given by
\bea
\Box _g = - {1 \over \sqrt g} \p_\mu \sqrt g g^{\mu \nu} \p_\nu
= -z_0^2 \p_0^2 + (d-1) z_0 \p_0 - z_0^2 \sum _{i=1}^d \p_i^2
\eea
The scalar Green function is the solution to a hypergeometric equation,
given by \cite{fronsdal},\footnote{The study of quantum Liouville theory
with a SO(2,1) invariant vacuum \cite{dfj} is closely related to the
study of AdS$_2$, as was shown in \cite{strom}. Propagators and
amplitudes were studied there long ago \cite{dfj} and the $\N=1$
supersymmetric generalization is also known \cite{d83}.}
\bea
G_\Delta (z,w) = G_\Delta (\xi)
=   {2^{-\Delta}  C _\Delta \over 2\Delta -d} \xi ^\Delta
F \biggl ({\Delta \over 2}, {\Delta \over 2} + \half; \Delta -{d \over 2}
+1;\xi ^2 \biggr )
\eea
where the overall normalization constant is defined by
\bea
 C_\Delta = {\Gamma (\Delta)
\over \pi^{d/2} \Gamma ( \Delta -{d\over 2})}
\eea
Since $0\leq \xi \leq 1$, the hypergeometric function is
defined by its convergent Taylor series for all $\xi$ except at the
coincident point $\xi=1$ where $z=w$.

\bigskip

\noindent {\bf The massive scalar boundary-to-bulk propagator}

\medskip

An important limiting case of the scalar bulk-to-bulk propagator is when
the source is on the boundary of $H$. The action to linearized order is
given by
\bea
S_{\varphi _\Delta}
= \int _H \! d^{d+1} \! z \ \sqrt g \biggl [
\half g^{\mu \nu} \p _\mu \varphi _\Delta \p_\nu \varphi _\Delta
+ \half m^2 \varphi _\Delta ^2 \biggr ]
- \int _{\p H} \! d^d \! \vec{z} \ \bar \varphi _\Delta (\vec{z})  \bar J
(\vec{z})
\eea
where the bulk field $\varphi _\Delta$ is related to the boundary field
$\bar \varphi _\Delta$ by the limiting relation,
\bea
\bar \varphi _\Delta (\vec{z}) = \lim _{z_0 \to \infty} z_0 ^{\Delta -d}
\varphi _\Delta (z_0,\vec{z})
\eea
The corresponding {\sl boundary-to-bulk propagator} is the Poisson
kernel, \cite{witten},
\bea
K_\Delta (z,\vec{x}) =  C_\Delta \biggl ( {z_0  \over z_0^2 +
(\vec{z} - \vec{x})^2}
\biggr ) ^\Delta
\eea
The bulk field generated in response to the boundary
source $\bar J$ is given by
\bea
\varphi _\Delta (z) = \int _{\p H} \! d^d \! \vec{z} \ K_\Delta (z,\vec{x})
\bar J(\vec{x})
\eea
This propagator will be especially important in the AdS/CFT correspondence.

\bigskip

\noindent {\bf The gauge propagator}

\medskip

Let $A_\mu (z)$ be a massless or massive gauge field, whose linearized
dynamics is given by a coupling to a covariantly conserved bulk current
$j^\mu$ via the action
\bea
S_A = \int _H \! d^{d+1} \! z \ \sqrt g \biggl [ {1 \over 4} F_{\mu \nu}
F^{\mu \nu} + \half m^2 A_\mu A^\mu - A_\mu j^\mu \biggr ]
\eea
It would be customary to introduce a gauge fixing term, such as Feynman
gauge, to render the second order differential operator acting on $A_\mu$
invertible when $m=0$. A more convenient way to proceed is to remark that
the differential operator needs to be inverted only on the subspace of all
$j^\mu$ that are covariantly conserved. The gauge propagator is a bivector
$G_{\mu \nu'} (z,z')$ which satisfies
\bea
- {1 \over \sqrt g} \p _\sigma \biggl (\sqrt g g^{\sigma \rho}
\p _{[\rho} G_{\mu ] \nu'} (z,z')\biggr ) + m^2 G_{\mu \nu'} (z,z')
= g_{\mu \nu} \delta (z,z') + \p_\mu \p_{\nu'} \Lambda (u)
\eea
The term in $\Lambda$ is immaterial when integrated against a
covariantly conserved current. For the massless case, the gauge propagator
is given by, \cite{DF98, DFMMR99}, see also \cite{aj},
\bea
G_{\mu \nu'} (z,z') = - (\p_\mu \p_{\nu'} u) F(u)  + \p_\mu \p_{\nu'}
S(u)
\eea
where $S$ is a gauge transformation function, while the physical part of
the propagator takes the form,
\bea
 F(u) = {\Gamma ((d-1)/ 2) \over 4 \pi ^{(d+1)/2}} { 1 \over
[u(u+2)]^{(d-1)/2}}
\eea

\bigskip

\noindent {\bf The massless graviton propagator}

\medskip

The action for matter coupled to gravity in an AdS background is given by
\bea
S_g = \half \int _H \! d^{d+1} \! z \ \sqrt g (-R_g + \Lambda) + S_m
\eea
where $R_g$ is the Ricci scalar for the metric $g$ and $\Lambda$ is the
``cosmological constant".  $S_m$ is the matter action, whose variation with
respect to the metric is, by definition, the stress tensor $T_{\mu \nu}$.
The stress tensor is covariantly conserved $\nabla _\mu T^{\mu \nu}=0$.
Einstein's equations read
\bea
R_{\mu \nu} - \half g_{\mu \nu} (R_g - \Lambda) = T_{\mu \nu}
\qquad \qquad
T_{\mu \nu } = {\delta S_m \over \sqrt g \delta g^{\mu \nu}}
\eea
We take $\Lambda = - d(d-1)$, so that in the absence of matter sources, we
obtain Euclidean AdS$=H$ with $R_g=-d(d+1)$ as the maximally symmetric
solution. To obtain the equation for the graviton propagator $G_{\mu
\nu;\mu'\nu'}(z,w)$, it suffices to linearize Einstein's equations around
the AdS metric in terms of small deviations $h_{\mu \nu} = \delta g_{\mu
\nu}$ of the metric. One find
\bea
h_{\mu \nu} (z) = \int _H \! d^{d+1} \! w \ \sqrt g G_{\mu \nu; \mu' \nu'}
(z,w) T^{\mu' \nu'}(w)
\eea
where the graviton propagator satisfies
\bea
\label{defW}
W_{\mu \nu} {}^{\kappa \lambda} G_{\kappa \lambda \mu'\nu'} =
\biggl ( g_{\mu \mu'} g_{\nu \nu'} + g_{\mu \nu'} g_{\nu \mu'}
-{ 2 g_{\mu \nu} g_{\mu'\nu'} \over d-1} \biggr ) \delta (z,w)
+ \nabla _{\mu'} \Lambda _{\mu \nu ; \nu'} + \nabla _{\nu'} \Lambda _{\mu
\nu ; \mu'}
\nonumber
\eea
and the differential operator $W$ is defined by
\bea
W_{\mu \nu} {}^{\kappa \lambda} G_{\kappa \lambda \mu'\nu'}
& \equiv &
- \nabla ^\sigma \nabla _\sigma G_{\mu \nu;\mu' \nu'}
- \nabla _\mu \nabla _\nu G^\sigma {}_{ \sigma;\mu' \nu'}
+ \nabla _\mu \nabla ^\sigma G_{\sigma \nu ;\mu' \nu'}
\nonumber \\
&&
+ \nabla _\nu \nabla ^\sigma G_{\mu \sigma ;\mu' \nu'}
-2 G_{\mu \nu;\mu' \nu'} + 2 g_{\mu \nu} G^\sigma {}_{ \sigma;\mu' \nu'}
\eea
The solution to this equation is obtained by decomposing $G$ onto a basis
of 5 irreducible $SO(1,d)$-tensors, which may all be expressed in terms of
the metric $g_{\mu \nu}$ and the derivatives of the chordal distance
$\p_\mu u$, $\p_\mu \p_{\nu'}u$ etc. One finds that three linear
combinations of these 5 tensors correspond to diffeomorphisms, so that we
have
\bea
G_{\mu \nu ; \mu' \nu'} & = &
(\p_\mu \p_{\mu'}u \ \p_\nu \p_{\nu'} u + \p_\mu \p_{\nu'}u \ \p_\nu
\p_{\mu'} u ) G(u) + g_{\mu \nu} g_{\mu'\nu'} H(u)
\nonumber \\
&&
+ \nabla _{(\mu} S_{\nu);\mu'\nu'} + \nabla _{(\mu'} S_{\mu \nu);\nu')}
\eea
The functions $G$ precisely obeys the equation for a massless scalar
propagator $G_\Delta (u)$  with $\Delta =d$, so that $G(u)=G_d(u)$. The
function $H(u)$ is then given by
\bea
-(d-1) H(u) = 2(1+u)^2 G(u) + 2 (d-2) (1+u) \int ^u _\infty  dv G(v)
\eea
which may also be expressed in terms of hypergeometric functions. The
graviton propagator was derived using the above methods, or alternatively
in De~Donder gauge in \cite{DFMMR99}. Propagators for other fields, such
as massive tensor and form fields were constructed in \cite{props1} and
\cite{props2}; see also \cite{Caldarelli:1998wk} and \cite{Chang:2001xh}.

\subsection{Conformal Structure of 1- 2- and 3- Point Functions}

Conformal invariance is remarkably restrictive on correlation functions
with 1, 2, and 3 conformal operators \cite{mp}. We illustrate this point
for correlation functions of superconformal primary operators, which are
all scalars.

\medskip

The {\sl 1-point function} is given by
\bea
\< \O _\Delta (x) \> = \delta _{\Delta , 0}
\eea
Indeed, by translation invariance, this object must be independent of $x$,
while by scaling invariance, an $x$-independent quantity can have dimension
$\Delta$ only when $\Delta =0$, in which case when have the identity
operator.

\medskip

The {\sl 2-point function} is given by
\bea
\< \O _{\Delta _1} (x_1) \O _{\Delta _2} (x_2) \>
= { \delta _{\Delta _1, \Delta _2} \over |x_1 - x_2|^{2 \Delta _1}}
\eea
Indeed, by Poincar\' e symmetry, this object only depends upon $(x_1 -
x_2)^2$; by inversion symmetry, it must vanish unless $\Delta _1 = \Delta
_2$; by scaling symmetry one fixes the exponent; and by properly
normalizing the operators, the 2-point function may be put in diagonal
form with unit coefficients.

\medskip

The {\sl 3-point function} is given by
\bea
\label{threept}
\< \O _{\Delta _1} (x_1) \O _{\Delta _2} (x_2) \O _{\Delta _3}
(x_3)\> = { c_{\Delta _1 \Delta _2 \Delta _3} (g_s, N) \over
|x_1 - x_2|^{\Delta - 2 \Delta _3} |x_2 - x_3|^{\Delta -2 \Delta _1}  |x_3 -
x_1|^{\Delta - 2 \Delta _2}  }
\eea
where $\Delta  = \Delta _1 + \Delta _2 + \Delta _3$.
The coefficient $c_{\Delta _1 \Delta _2 \Delta _3}$ is independent of the
$x_i$ and will in general depend upon the coupling $g_{YM}^2$ of the
theory and on the Yang-Mills gauge group through $N$.

\subsection{SYM Calculation of 2- and 3- Point Functions}

All that is needed to compute the SYM correlation functions of the
composite operators
\bea
\O _\Delta (x) \equiv {1 \over n_\Delta} \str X^{i_1}(x) \cdots
X^{i_\Delta }(x)
\eea
to Born level (order $g_{YM}^0$) is the propagator of the scalar field
\bea
\< X ^{ic} (x_1) X^{jc'} (x_2) \> = {\delta ^{ij} \delta ^{cc'} \over 4
\pi^2 (x_1 - x_2)^2}
\eea
where $c$ is a color index running over the adjoint representation of
$SU(N)$ while $i=1,\cdots ,6$ labels the fundamental representation of
$SO(6)$. Clearly, the 2- and 3- point functions have the space-time
behavior  expected from the preceding discussion of conformal invariance.
Normalizing the 2-point function as below, we have $n_k^2 = \str (T^{c_1}
\cdots T^{c_k}) \str (T^{c_1} \cdots T^{c_k})$.
\bea
\< \O _{\Delta _1} (x_1) \O _{\Delta _2} (x_2) \>
& = &
 {\delta _{\Delta _1, \Delta _2} \over (x_1 - x_2)^{2 \Delta _1}}
\nonumber \\
\< \O _{\Delta _1} (x_1) \O _{\Delta _2} (x_2) \O _{\Delta _3}
(x_3) \> & \sim &
{1 \over
(x_1 - x_2)^{\Delta _{12}} (x_2 - x_3)^{\Delta _{23}}  (x_3 - x_1)^{\Delta
_{31}}  }
\eea
Using the fact that the number $\Delta _i$ of propagators emerging from
operator $\O _{\Delta _i}$ equals the sum $\Delta_{ij}+\Delta_{ik}$, we find
$2 \Delta _{ij} = \Delta_i + \Delta _j - \Delta _k$, in agreement with
(\ref{threept}). The precise numerical coefficients may be worked out with
the help of the contractions of color traces.

\subsection{AdS Calculation of 2- and 3- Point Functions}

On the AdS side, the 2-point function to lowest order is obtained by taking
the boundary-to-bulk propagator $K_\Delta(z,\vec{x})$ for a field with
dimension $\Delta$ and extracting the $z_0 ^\Delta$ behavior as $z_0 \to
0$, which gives
\bea
\lim _{z_0 \to 0} z_0 ^{-\Delta} K_\Delta (z,\vec{x})
\sim {1 \over (\vec{z} - \vec{x})^{2 \Delta}}
\eea
in agreement with the behavior predicted from conformal invariance
\cite{fmmr}.

\medskip

The 3-point function involves an integral over the intermediate
supergravity interaction point, and is given by
\bea
\G (\Delta_1,\Delta_2,\Delta_3) \int _{S^5} Y_{\Delta _1} Y_{\Delta _2}
Y_{\Delta _3}
\int _H \! {d^5 \! z \over z_0^5} \prod _{i=1} ^3 C_{\Delta _i} \biggl (
{z_0 \over z_0^2 + (\vec{z} -\vec{x_i})^2 } \biggr )^{\Delta _i}
\eea
where $\G (\Delta_1,\Delta_2,\Delta_3)$ stands for the supergravity 3-point
coupling and the second factor is the integrals over the spherical
harmonics of $S^5$. To carry out the integral over $H$, one proceeds in
three steps. First, use a translation to set $\vec{x}_3=0$. Second, use an
inversion about 0, given by $z^\mu \to z^\mu /z^2$ to set $\vec{x}_3 '
=\infty$. Third, having one point at $\infty$, one may now use translation
invariance again, to obtain for the $H$-integral
\bea
\sim (x_{13}')^{2 \Delta _1} (x_{23} ')^{2 \Delta _2}
\int _H \! {d^5 \! z \over z_0^5} {z_0 ^{\Delta _1 + \Delta _2 + \Delta _3}
\over z^{2 \Delta _1} [ z_0^2 + (\vec{z} - \vec{x}_{13}' -
\vec{x}_{23}')^2 ] ^{\Delta _2} }
\eea
Carrying out the $\vec{z}$ integral using a Feynman parametrization of the
integral and then carrying out the $z_0$ integral, one recovers again the
general space-time dependence of the 3-point function \cite{fmmr}. A more
detailed account of the AdS calculations of the 2- and 3-point functions
will be given in \S 8.4.

\subsection{Non-Renormalization of 2- and 3- Point Functions}

Upon proper normalization of the operators $\O _\Delta$, so that their
2-point function is canonically normalized, the three point couplings
$c_{\Delta _1, \Delta _2,\Delta_3} (g_{YM}^2,N)$ may be computed in a unique
manner. On the SYM side, small coupling $g_{YM}$ perturbation theory
yields results for $g_{YM}\ll 1$, but all $N$. On the AdS side, the only
calculation available in practice so far is at the level of
classical supergravity, which means the large $N$ limit (where quantum
loops are being neglected), as well as large `t~Hooft coupling $\lambda =
g_{YM}^2 N$ (where $\alpha'$ string corrections to supergravity are being
neglected). Therefore, a direct comparison between the two calculations
cannot be made because the calculations hold in mutually exclusive regimes
of validity.

\medskip

Nonetheless, one may compare the results of the calculations in both
regimes. This involves obtaining a complete normalization of the
supergravity three-point couplings $\G (\Delta_1,\Delta_2,\Delta_3)$, which
was worked out in \cite{lmrs}. It was found that
\bea
\lim _{N, \lambda =g_sN \to \infty} c_{\Delta _1, \Delta _2,\Delta_3}
(g_s,N) \bigg |_{AdS}
= \lim _{N\to \infty} c_{\Delta _1, \Delta _2,\Delta_3} (0,N)
\bigg |_{SYM}
\eea
Given that this result holds irrespectively of the dimensions $\Delta _i$,
it was conjectured in \cite{lmrs} that this result should be viewed as
emerging from a {\sl non-renormalization effect} for 2- and 3-point
functions of 1/2 BPS operators. Consequently, it was conjectured that
the equality should hold for all couplings, at large $N$,
\bea
\lim _{N \to \infty} c_{\Delta _1, \Delta _2,\Delta_3} (g_s,N)
\bigg |_{AdS}
= \lim _{N\to \infty} c_{\Delta _1, \Delta _2,\Delta_3} (g_{YM}^2,N)
\bigg |_{SYM}
\eea
and more precisely that $c_{\Delta _1, \Delta _2,\Delta_3} (g_s,N)$ be
independent of $g_s$ in the $N\to \infty $ limit.

\medskip

Independence on $g_{YM}$ of the three point coupling $c_{\Delta _1,
\Delta _2,\Delta_3} (g_{YM}^2,N)$ is now a problem purely in $\N=4$ SYM
theory, and may be studied there in its own right. This issue has been
pursued since by performing calculations of the same correlators to order
$g_{YM}^2$. It was found that to this order, neither the 2- nor the 3-point
functions receive any corrections \cite{DFS}. Consequently, a stronger
conjecture was proposed to hold for all $N$,
\bea
 c_{\Delta _1, \Delta _2,\Delta_3} (g_s,N)
\bigg |_{AdS}
= c_{\Delta _1, \Delta _2,\Delta_3} (g_{YM}^2,N)
\bigg |_{SYM}
\eea
Further evidence that this relation holds has been obtained
using $\N=1$ superfields \cite{Sk99, pen99} and $\N=2$ off-shell
analytic/harmonic superfield methods \cite{Eden00, Eden00A}. The problem has 
also been
investigated using $\N=4$ on-shell superspace methods \cite{int99,
Howe:1998zi}, via the study of nilpotent superconformal invariants, which
had been introduced for OSp(1,N) in \cite{d84}. Similar
non-renormalization effects may be derived for 1/4 BPS operators and
their correlators as well \cite{quarter}. Two and three point correlators
have also been investigated for superconformal descendant fields; for the
R-symmetry current in \cite{fmmr} and later in \cite{Chalmers:1998xr}; see also
\cite{Muck:1998rr} and  \cite{Muck:1998iz}. Additional references include
\cite{Bastianelli:1999vm} and \cite{Bastianelli:2000mk}. A further test of the
Maldacena conjecture involving the Weyl anomaly is in \cite{Mansfield:2000zw}.

\subsection{Extremal 3-Point Functions}

We now wish to investigate the dependence of the 3-point function of 1/2
BPS single trace operators
\bea
\< \O _{\Delta _1} (x_1) \O _{\Delta _2} (x_2) \O _{\Delta _3}
(x_3) \>
\eea
on the dimensions $\Delta _i$ a little more closely. Recall that these
operators transform under the irreducible representations of $SU(4)_R$ with
Dynkin labels $[0,\Delta _i,0]$. As a result, the correlators must vanish
whenever $\Delta _i > \Delta _j + \Delta _k$ for any one of the labels
$i\not= j,k$, since in this case no $SU(4)_R$ singlet exists. Whenever
$\Delta _i \leq \Delta _j + \Delta _k$, for all $i,j,k$, the correlator is
allowed by $SU(4)_R$ symmetry.

\medskip

These facts may also be seen at Born level in SYM perturbation theory by
matching the number of $X$ propagators connecting different operators. If
$\Delta _i > \Delta _j + \Delta _k$, it will be impossible to match up the
$X$ propagator lines and the diagram will have to vanish.

\medskip

The case where $\Delta _i = \Delta _j + \Delta _k$ for one of the labels
$i$ is of special interest and is referred to as an {\sl extremal
correlator} \cite{DFMMR99B}. Although allowed by $SU(4)_R$ group theory,
its Born graph effectively factorizes into two 2-point functions, because
no $X$ propagators directly connect the vertices  operators $j$ and $k$.
Thus, the extremal 3-point function is non-zero. However, the
supergravity coupling $\G (\Delta_1,\Delta_2,\Delta_3)\sim \Delta _1 -
\Delta _2 -\Delta _3$ vanishes in the extremal case as was shown in
\cite{lmrs}. The reason that all these statements can be consistent with
the AdS/CFT correspondence is because the AdS$_5$ integration actually
has a pole at the extremal dimensions, as may indeed be seen by taking a
closer look at the integrals,
\bea
\int _H \! {d^5 \! z \over z_0^5} \prod _{i=1}^3 {z_0 ^{\Delta _i}
\over (z_0^2 + (\vec{z} - \vec{x}_i)^2 )^{\Delta _i}}
\sim { 1 \over \Delta _1 - \Delta _2 - \Delta _3}
\eea
Thus, the AdS/CFT correspondence for extremal 3-point functions holds
because a zero in the supergravity coupling is compensated by a pole in the
AdS$_5$ integrals.

\medskip

Actually, the dimensions $\Delta _i$ are really integers (which is why
``pole" was put in quotation marks above) and direct analytic
continuation in them is not really justified. It was shown in
\cite{DFMMR99B} that when keeping the dimensions $\Delta _i$ integer, it
is possible to study the supergravity integrands more carefully and to
establish that while the bulk contribution vanishes, there remains a
boundary contribution (which was immaterial for non-extremal correlators).
A careful analysis of the boundary contribution allows one to recover
agreement with the SYM calculation directly.

\subsection{Non-Renormalization of General Extremal Correlators}

Extremal correlators may be defined not just for 3-point functions, but for
general $(n+1)$-point functions. Let $\O _\Delta$ and $\O _{\Delta
_i}$ with $i=1,\cdots ,n$ be 1/2 BPS chiral primary operators obeying the
relation $\Delta = \Delta _1 + \cdots + \Delta _n$, which generalizes the
extremality relation for the 3-point function. We have the
{\sl extremal correlation non-renormalization conjecture}, stating the
form of the following correlator \cite{DFMMR99B},
\bea
\label{exform}
\< \O _\Delta (x) \O _{\Delta _1} (x_1) \cdots \O _{\Delta
_n}(x_n) \>
= A(\Delta _i;N) \prod _{i=1} ^n {1 \over (\vec{x} - \vec{x}_i)^{2 \Delta
_i}}
\eea
The conjecture furthermore states that the overall function $A(\Delta
_i;N)$ is independent of the points $x_i$ and $x$ and is also independent
of the string coupling constant $g_s=g_{YM}^2$. {\sl The  conjecture also
states that the associated supergravity bulk couplings $\G (\Delta;
\Delta_1,\cdots ,\Delta_n)$ must vanish} \cite{DFMMR99B}.

\medskip

There is by now ample evidence for the conjecture and we shall briefly
review it here. First, there is evidence from the SYM side. To Born level
(order $\O(g_{YM}^0)$, the factorization of the space-time dependence in a
product of 2-point functions simply follows from the fact that no
$X$-propagator lines can connect different points $x_i$; instead all
$X$-propagator lines emanating from any vertex $x_i$ flow into the point
$x$. The absence of $\O(g_{YM}^2)$ perturbative corrections was
demonstrated in \cite{Bianchi99}. Off-shell $\N=2$ analytic/harmonic
superspace methods have been used to show that $g_{YM}$ corrections are
absent to all orders of perturbation theory \cite{Eden00}, \cite{Eden00A}.

\medskip

On the AdS side,  the simplest diagram that contributes to the extremal
correlator is the contact graph, which is proportional to
\bea
\G(\Delta; \Delta _1,\cdots,\Delta_n)
\int _H \! {d^5 \! z \over z_0^5} {z_0 ^\Delta \over (z-\vec{x})^{2\Delta}}
\prod _{i=1} ^n { z_0 ^{\Delta _i} \over (z-\vec{x}_i)^{2 \Delta _i}}
\eea
In view of the relation $\Delta = \Delta _1 + \cdots + \Delta _n$, the
integration is convergent everywhere in $H$, except when $\vec{z} \to
\vec{x}$ and $z_0 \to 0$, where a simple pole arises in $\Delta - \Delta _1
- \cdots - \Delta _n$. Finiteness of Type IIB superstring theory on \AdS
(which we take as an assumption here) guarantees that the full correlator
must be convergent. Therefore, the associated supergravity bulk coupling
must vanish,
\bea
\G(\Delta; \Delta _1,\cdots,\Delta_n) \sim
\Delta - \Delta _1 - \cdots - \Delta _n
\eea
as indeed stated in the conjecture. Assuming that it makes sense to
``analytically continue in the dimensions $\Delta$", one may proceed
as follows. The pole of the $z$-integration and the zero of the
supergravity coupling $\G$ compensate one another and the contribution of
the contact graph to the extremal correlator will be given by the residue
of the pole, which is precisely of the form (\ref{exform}). It is also
possible to carefully treat the boundary contributions generated by the
supergravity action in the extremal case, to recover the same result
\cite{DFMMR99B}.

\medskip

The analysis of all other AdS graph, which have at least one bulk-to-bulk
exchange in them, was carried out in detail in \cite{DFMMR99B}. For the
exchange of chiral primaries in the graph, the extremality condition
$\Delta = \Delta _1 + \cdots + \Delta _n$ implies that each of the exchange
bulk vertices must be extremal as well. A non-zero contribution can then
arise only if the associated integral is divergent, produces a pole in
the dimensions, and makes the interaction point collapse onto the boundary
$\p H$. Dealing with all intermediate external vertices in this way, one
recovers that all intermediate vertices have collapsed onto $\vec{x}$,
thereby reproducing the space-time behavior of (\ref{exform}). The exchange
of descendants may be dealt with in an analogous manner.

\medskip

Assuming non-renormalization of 2- and 3-point functions for all (single
and multiple trace) 1/2 BPS operators, and assuming the space-time form
(\ref{exform}) of the extremal correlators, it is possible to {\sl prove}
that the overall factor $A(\Delta _i;N)$ is independent of $g_s=g_{YM}^2$,
as was done in \cite{DFMMR99B} in a special case. We present only the
simplest non-trivial case of $n=3$ and $\Delta =6$;  the general case may
be proved by induction. Assuming the space-time form, we have
\bea
\label{corr1}
 \< \O _6 (x) \O _2 (x_1) \O_2 (x_2) \O _2 (x_3) \>
= A \prod _{i=1} ^3 {1 \over (\vec{x} - \vec{x}_i)^4 }
\eea
We begin with the OPE
\bea
\O _6 (x) \O _2 (x_1) \sim { c \O _4 (x) + c' [\O _2 \O
_2]_{\rm max} (x) \over (x-x_1)^4} + {\rm less \ singular}
\eea
Using non-renormalization of the 3-point functions $ \< \O _6
\O _2 \O_4 \>$ and $ \< \O _6 \O _2 [\O _2 \O
_2]_{\rm max}$, we find that $c$ and $c'$ are independent of the coupling
$g_{YM}$. Now substitute the above OPE into the correlator (\ref{corr1}),
and use the fact that the 3-point functions $ \< \O _4
\O _2 \O_2 \>$ and $ \< \O _2 \O _2 [\O _2 \O
_2]_{\rm max}$ are not renormalized. It immediately follows that $A$ in
(\ref{corr1}) is independent of the coupling.

\subsection{Next-to-Extremal Correlators}

The space-time dependence of extremal correlators was characterized by its
factorization into a product of $n$ 2-point functions. The space-time
dependence of {\sl Next-to-extremal correlators} $\< \O _\Delta (x)
\O _{\Delta _1} (x_1) \cdots \O _{\Delta _n}(x_n) \>$, with the
dimensions satisfying $\Delta = \Delta _1 + \cdots + \Delta_n -2$
is characterized by its factorization into a product of $n-2$ two-point
functions and one 3-point function.  Therefore, the conjectured space-time
dependence of {\sl next-to-extremal correlators} is given by
\cite{Eden00A}
\bea
\label{nextform}
\< \O _\Delta (x) \O _{\Delta _1} (x_1) \cdots \O _{\Delta
_n}(x_n) \>
= {B(\Delta _i;N) \over x_{12}^2 (x-x_1)^{2 \Delta _1 -2} (x-x_2)^{2
\Delta _2 -2}}
\prod _{i=3}^n { 1 \over (x-x_i) ^{2 \Delta _i}}
\eea
where the overall strength $B(\Delta _i;N)$ is independent of $g_{YM}$.
This form is readily checked at Born level and was verified at order
$\O(g_{YM}^2)$ by \cite{Erd99}.

\medskip

On the AdS side, the exchange diagrams, say with a single exchange, are
such that one vertex is extremal while the other vertex is not
extremal. A divergence arises when the extremal vertex is attached to the
operator of maximal dimension $\Delta$ and its collapse onto the point $x$
now produces a 3-point correlator times $n-2$ two-point correlators,
thereby reproducing the space-time dependence of (\ref{nextform}).
Other exchange diagrams may be handled analogously. However, there is also
a contact graph, whose AdS integration is now {\sl convergent}. Since the
space-time dependence of this contact term is qualitatively different from
the factorized form of (\ref{nextform}), the only manner in which
(\ref{nextform}) can hold true is if the supergravity bulk coupling
associated with next-to-extremal couplings vanishes,
\bea
\G (\Delta ; \Delta _1 ,\cdots ,\Delta _n) =0
\qquad {\rm whenever} \qquad
\Delta = \Delta _1 +\cdots +\Delta _n -2
\eea
which is to be included as part of the conjecture \cite{DEFP}. This type
of cancellation has been checked to low order in \cite{Arut99}.

\subsection{Consistent Decoupling and Near-Extremal Correlators}

The vanishing of the extremal and next-to-extremal supergravity couplings
has a direct interpretation, at least in part, in supergravity. Recall that
the operator $\O _2$ and its descendants are dual to the 5-dimensional
supergravity multiplet on AdS$_5$, while the operators $\O _\Delta$
with $\Delta \geq 3$ and its descendants are dual to the Kaluza-Klein
excitations on $S^5$ of the 10-dimensional supergravity multiplet. Now,
prior work on gauged supergravity \cite{Gun85, Pern85} has shown that the
5-dimensional gauged supergravity theory on AdS$_5$ all by itself exists and
is consistent. Thus, there must exist a {\sl consistent truncation of the
Kaluza-Klein modes of supergravity on \AdS\ to only the supergravity on
AdS$_5$}; see also \cite{Nastase:2000pp}. In a perturbation expansion, this
means that if only AdS$_5$
supergravity modes are excited, then the Euler-Lagrange equations of the
full \AdS\ supergravity must close on these excitations alone without
generating Kaluza-Klein excitation modes. This means that the one 1-point
function of any Kaluza-Klein excitation operator in the presence of
AdS$_5$ supergravity alone must vanish, or
\bea
\G (\Delta , \Delta _1 ,\cdots , \Delta _n)=0,
\qquad  \Delta _i=2, \ i=1,\cdots ,n
\qquad
{\rm for \ all}
\qquad
\Delta \geq 4
\eea
When $\Delta > 2n$, the cancellation takes place by $SU(4)_R$ group theory
only. For $\Delta =2n$ and $\Delta =2n-2$, we have special cases of
extremal and next-to-extremal correlators respectively, but for $4\leq
\Delta \leq 2n-4$, they belong to a larger class. We refer to these as
{\sl near-extremal correlators} \cite{DEFP},
\bea
\< \O _\Delta (x) \O _{\Delta _1} (x_1) \cdots \O _{\Delta
_n}(x_n) \>
\qquad
\Delta = \Delta _1 + \cdots + \Delta _n -2m
\qquad
0\leq m \leq n-2
\eea
The principal result on near-extremal correlators (but which are not of
the extremal or next-to-extremal type) is that they do receive coupling
dependent quantum corrections, but only through lower point functions
\cite{DEFP}. Associated supergravity couplings must vanish,
\bea
\G (\Delta , \Delta _1 ,\cdots, \Delta _n) =0
\qquad
\Delta = \Delta _1 + \cdots + \Delta _n -2m
\qquad
0\leq m \leq n-2
\eea
Arguments in favor of this conjecture may be given based on the divergence
structure of the AdS integrals and on perturbation calculations in SYM.

\medskip

\subsection{Problem Sets}

(6.1) Using infinitesimal special conformal symmetry (or global
inversion under which $x^\mu \to  x^\mu / x^2$) show that $\< \O
_\Delta (x) \O _{\Delta '} (x')\>=0$ unless $\Delta ' = \Delta$.

\bigskip

\noindent
(6.2) Gauge dependent correlators in gauge theories such as $\N=4$ SYM
theory will, in general, depend upon a renormalization scale $\mu$. (a)
Show that the general form of the scalar two point function to one loop
order is given by
$$
\< X^{ic} (x) X^{jc'} (y) \> = {\delta ^{cc'} \delta ^{ij} \over
(x-y)^2} \biggl (A + B \ln (x-y)^2 \mu ^2 \biggr )
$$
for some numerical constants $A$ and $B$. (b) Show that the 2-pt function
of the gauge invariant operator $\O _2 (x) \equiv \tr X^i(x) X^j(x) - {1
\over 6} \delta ^{ij} \sum _k \tr X^k(x) X^k(x)$ is $\mu$-independent. (c)
Show that the 2-pt  function of the gauge invariant operator $\O _K (x)
\equiv \tr X^i(x) X^i (x)$ (the Konishi operator) is $\mu$-dependent. (d)
Calculate the 1-loop anomalous dimensions of $\O_2$ and $\O_K$.

\bigskip

\noindent
(6.3) Consider the Laplace operator $\Delta$ acting on scalar
functions  on the sphere $S^d$ with round $SO(d+1)$-invariant metric and
radius $R$. Compute the eigenvalues of of $\Delta$. Suggestion :
$\Delta$ is related to the quadratic Casimir operator $ L^2 \equiv \sum
_{i,j=1} ^{d+1} L_{ij} ^2$ where $L_{ij}$ are the generators of
$d+1$-dimensional angular momentum, i.e. generators of $SO(d+1)$; thus
the problem may be solved by pure group theory methods, analogous to
those used for rotations  on $S^2$.

\bigskip

\noindent
(III.4) Continuing on the above problem, show that the eigenfunctions are
of the form $c_{i_1 \cdots i_p} x^{i_1} \cdots x^{i_p}$, where we have
now represented the sphere by the usual equation in ${\bf R}^{d+1}$~:
$\sum _i  (x^i)^2 = R^2$ and $c$ is totally symmetric and traceless.

\vfill\eject

\section{Structure of General Correlators}
\setcounter{equation}{0}

In the previous section, we have concentrated on matching between the SYM
side and the AdS side of the Maldacena correspondence correlation
functions that were not renormalized or were simply renormalized from their
free form. This led us to uncover a certain number of important {\sl
non-renormalization effects}, most of which are at the level of conjecture.

\medskip

However, $\N=4$ super-Yang-Mills theory is certainly not a free quantum
field theory, and generic correlators will receive quantum corrections from
their free field values, and therefore will acquire non-trivial coupling
$g_s = g_{YM}^2$ dependence. In this section, we analyze the behavior of
such correlators. We shall specifically deal with the 4-point function. The
relevant dynamical information available from correlators in conformal
quantum field theory is contained in the {\sl scaling dimensions of general
operators}, in the {\sl operator mixings} between general operators and in
the values of the {\sl operator product (OPE) coefficients}. As in the case
of the 3-point function, a direct quantitative comparison between the
results of weak coupling $g_{YM}$ perturbation theory in SYM and the large
$N$, large `t~Hooft coupling $\lambda = g_{YM}^2 N$ limit of supergravity
cannot be made, because the domains of validity of the expansions do not
overlap. Nonetheless, general properties lead to exciting and non-trivial
comparisons, which we shall make here.

\subsection{RG Equations for Correlators of General Operators}

It is a general result of quantum field theory that all renormalizations
of local operators are multiplicative. This is familiar for canonical
fields; for example the bare field $\phi_0 (x)$ and the renormalized field
$\phi(x)$ in a scalar field theory are related by the field renormalization
factor $Z_\phi$ via the relation $\phi _0 (x) = Z_\phi \phi (x)$. Composite
operators often requires {\sl additive renormalizations}; for example the
proper definition of the operator $\phi ^2 (x)$ requires the subtraction
of a constant $C$. If this constant is viewed as multiplying the identity
operator $I$ in the theory, then renormalization may alternatively be
viewed as multiplicative (by a matrix) on an array of two operators $I$ and
$\phi ^2 (x)$ as follows,
\bea
\label{matrixrenorm}
\left (\matrix{I \cr \phi ^2(x)} \right ) _0
=
 \left (\matrix{I & 0 \cr -C & Z_{\phi^2} } \right )
\left (\matrix{I \cr \phi ^2(x)} \right )
\eea
The general rule is that operators will renormalize with operators with the
same quantum numbers but of lesser or equal dimension.

\medskip

In more complicated theories such as $\N=4$ super-Yang-Mills theory,
renormalization will continue to proceed in a multiplicative way. If we
denote a basis of (local gauge invariant polynomial) operators by $\O
_I$, and their bare counterparts by $\O _{0I}$, then we have the
following multiplicative renormalization formula
\bea
\O _{0I} (x) = \sum _J Z_I {}^J  \O _J (x)
\eea
The field renormalization matrix $Z_I{}^J$ may be arranged in block lower
triangular form, in ascending value of the operator dimensions,
generalizing (\ref{matrixrenorm}). Consider now a general correlator of
such operators
\bea
G_{I_1, \cdots , I_n} (x_i;g,\mu)
\equiv
\< \O _{I_1} (x_1) \cdots \O _{I_n} (x_n) \>
\eea
and its bare counterpart $G_{0I_1, \cdots , I_n} (x_i ; g_0 ,\Lambda)$, in
a theory in which we schematically represent the dimensionless and
dimensionful couplings by $g$, and their bare counterparts by $g_0$. The
renormalization scale is $\mu$ and the UV cutoff is $\Lambda$.
Multiplicative renormalization implies the following relation between
the renormalized and bare correlators
\bea
G_{0I_1, \cdots , I_n} (x_i;g_0,\Lambda)
= \sum _{J_1,\cdots ,J_n} Z_{I_1} {}^{J_1} \cdots Z_{I_n} {}^{J_n}
(g,\mu,\Lambda) G_{J_1, \cdots , J_n} (x_i;g,\mu)
\eea
Keeping the bare parameters $g_0$ and $\Lambda$ fixed and varying the
renormalization scale
$\mu$, we see that the lhs is independent of $\mu$. Differentiating both
sides with respect to $\mu$ is the standard way of deriving the
renormalization group equations for the renormalized correlators, and we
find
\bea
\biggl ( {\p \over \p \ln \mu} + \beta {\p \over \p g}  \biggr )  G_{I_1,
\cdots , I_n} (x_i;g,\mu)
- \sum _{j=1} ^n \sum _J \gamma _{I_j} {}^J G_{I_1, \cdots,
I_{j-1},J, I_{j+1}, \cdots , I_n} (x_i;g,\mu) =0 \qquad
\eea
where the RG $\beta$-function and anomalous dimension matrix
$\gamma _I{}^J$  are defined by
\bea
\beta  (g) \equiv {\p g \over \p \ln \mu} \bigg |_{g_0, \Lambda}
\hskip 1in
\gamma _I{}^J (g) \equiv
- \sum _K (Z^{-1})_I {}^K {\p Z_K {}^J \over \p \ln \mu}
\bigg |_{g_0,\Lambda}
\eea
For each $I$, only a finite number of $J$'s are non-zero in the sum over
$J$. The diagonal entries $\gamma _I {}^I$ contribute to the anomalous
dimension of the operator $\O _I$, while the off-diagonal entries are
responsible for operator mixing. Operators that are eigenstates of the
dimension operator $D$ (at a given coupling $g$) correspond to the
eigenvectors of the matrix $\gamma$.

\subsection{RG Equations for Scale Invariant Theories}

Considerable simplifications occur in the RG equations for scale invariant
quantum field theories. Scale invariance requires in particular that
$\beta (g_*)=0$, so that the theory is at a fixed point $g_*$. In rare
cases, such as is in fact the case for $\N=4$ SYM, the theory is scale
invariant for all couplings. If no dimensionful couplings occur in the
Lagrangian, either from masses or from vacuum expectation vales of
dimensionful fields,
$\gamma _I {}^J$ is constant and the RG equation becomes a
simple scaling equation
\bea
 {\p \over \p \ln \mu}  G_{I_1, \cdots , I_n} (x_i;g_*,\mu)
- \sum _{j=1} ^n \sum _J \gamma _{I_j} {}^J (g_*) G_{I_1, \cdots,
I_{j-1},J, I_{j+1}, \cdots , I_n} (x_i;g_*,\mu) =0 \qquad
\eea
In conformal theories, the dilation generator may be viewed as a Hamiltonian
of the system conjugate to radial evolution \cite{fhj}. Therefore, in
unitary scale invariant theories, the dilation generator should be
self-adjoint, and hence the anomalous dimension matrix should be
Hermitian.\footnote{In non-unitary theories, the matrix $\gamma _I {}^J$
may be put in Jordan diagonal form, and this form will produce dependence
on $\mu$ through $\ln \mu$ terms. A fuller discussion is given in
\cite{DMMR99}.} As such, $\gamma _I{}^J$ must be diagonalizable with real
eigenvalues, $\gamma _i$. Standard scaling arguments then give the
behavior of the correlators
\bea
G_{I_1, \cdots , I_n} (\lambda x_i;g_*,\lambda ^{-1} \mu)
= \lambda ^{-\Delta _1} \cdots \lambda ^{ - \Delta _n}
G_{I_1, \cdots , I_n} (x_i;g_*, \mu)
\eea
where the full dimensions $\Delta _i$ are given in terms of the
canonical dimension $\delta _i$ by $\Delta _i = \gamma _i + \delta _i$.

\subsection{Structure of the OPE}

One of the most useful tools of local quantum field theory is the Operator
Product Expansion (OPE) which expresses the product of two local operators
in terms of a sum over all local operators in the theory,
\bea
\O _I (x) \O _J(y) = \sum _K C_{IJ}{}^K (x-y;g,\mu) \O _K (y)
\eea
The OPE should be understood as a relations that holds when evaluated
between states in the theory's Hilbert space or when inserted into
correlators with other operators,
\bea
\< \O _I (x) \O _J(y) \prod _L \O _L (z_L) \> = \sum _K
C_{IJ}{}^K (x-y;g,\mu)
\< \O _K (y) \prod _L \O _L (z_L) \>
\eea
From the latter, together with the RG equations for the correlators, one
deduces the RG equations for the OPE coefficient functions $C_{IJ}{}^K$,
\bea
\biggl ( {\p \over \p \ln \mu} + \beta {\p \over \p g} \biggr ) C_{IJ}{}^K
= \sum _L \biggl (
 \gamma _I {}^L C_{LJ} {}^K + \gamma _J {}^L C_{IL}{}^K - C_{IJ}{}^L
\gamma _L {}^K \biggr )
\eea
In a scale invariant theory, we have $\beta =0$ and $\gamma _I {}^J$
constant. Furthermore, if the theory is unitary, $\gamma _I {}^J$ may be
diagonalized in terms of operators $\O _{\Delta _I}$ of definite
dimension $\Delta _I$. The OPE then simplifies considerably and we have,
\cite{OPE},
\bea
\O _{\Delta _I} (x) \O _{\Delta _J} (y) =
\sum _K { c_{\Delta _I \Delta _J \Delta _K} \over (x-y)^{\Delta _I +
\Delta _J - \Delta _K} } \O _{\Delta _K}(y)
\eea
The operator product coefficients $c_{\Delta _I \Delta _J \Delta _K}$ are
now independent of $x$ and $y$, but they will depend upon the coupling
constants and parameters of the theory, such as $g_{YM}$ and $N$.

\subsection{Perturbative Expansion of OPE in Small Parameter}

Conformal field theories such as $\N=4$ SYM have coupling constants
$g_{YM}, \theta _I , N$ and the theory is (super)-conformal for any value of
these parameters. In particular, the scaling dimensions are fixed but may
depend upon these parameters in a non-trivial way,
\bea
\Delta _I = \Delta (g_{YM}, \theta _I, N)
\eea
The dependence of the composite operators on the canonical fields
will in general also involve these coupling dependences.

\medskip

It is interesting to analyze the effects of small variations in any of
these parameters on the structure of the OPE and correlation functions.
Especially important is the fact that infinitesimal changes in $\Delta _I$
produce {\sl logarithmic dependences} in the OPE. To see this, assume that
\bea
\Delta _I = \Delta _I ^0 + \delta _I
\qquad | \delta _I |\ll \Delta _I
\eea
and now observe that to first order in $\delta _I$, we have
\bea
\O _{\Delta _I} (x) \O _{\Delta _J} (y) =
\sum _K { c_{\Delta _I \Delta _J \Delta _K} \O _{\Delta _K}(y) \over
(x-y)^{\Delta ^0 _I + \Delta ^0 _J - \Delta ^0 _K} }
\biggl \{1 - (\delta _I + \delta _J - \delta _K) \ln |x-y|\mu \biggr \}
\eea
In the special case where the dimensions $\Delta _I ^0$ and $\Delta _J ^0$
are unchanged, because the operators are protected (e.g. BPS) then isolating
the logarithmic dependence allows one to compute $\delta _K$ and thus the
correction to the dimension of operators occurring in the OPE. A useful
reference on anomalous dimensions and the OPE, though not in conformal
field theory, is in \cite{witten76}.

\subsection{The 4-Point Function -- The Double OPE }

Recall that the AdS/CFT correspondence maps supergravity fields into
single-trace 1/2 BPS operators on the SYM side. Thus, the only correlators
that can be computed directly are the ones with one 1/2 BPS operator
insertions. To explore even the simplest renormalization effects of non-BPS
operators, such as their change in dimension, via the AdS/CFT
correspondence, we need to go beyond the 3-point function. The simplest
case is the 4-point function, which indeed can yield information on the
anomalous dimensions of single and double trace operators.

\medskip

Thanks to conformal symmetry, the 4-point function may be factorized into
a factor capturing its overall non-trivial conformal dependence times a
function $F(s,t)$ that depends only upon 2 conformal invariants $s,t$ of 4
points,
\bea
\< \O _{\Delta _1} (x_1) \O _{\Delta _2} (x_2) \O _{\Delta _3}
(x_3) \O _{\Delta _4} (x_4)\>
= {1 \over | x_{13}| ^{\Delta _1 + \Delta _3} |x_{24}|^{\Delta _2 + \Delta
_4}} F(s,t)
\eea
The conformal invariants of the 4-point function $s$ and $t$ may be chosen
as follows
\bea
s = \half { x_{13}^2 x_{24}^2 \over x_{12}^2 x_{34}^2 + x_{14}^2 x_{23}^2}
\hskip 1in
t = {x_{12}^2 x_{34}^2 - x_{14}^2 x_{23}^2 \over x_{12}^2 x_{34}^2 +
x_{14}^2 x_{23}^2}
\eea
The fact that there are only 2 conformal invariants may be seen as follows.
By a translation, take $x_4=0$; under an inversion, we then have
$x_4'=\infty$ and we may use translations again to choose $x_3'=0$. There
remain 3 Lorentz invariants, $x_1^2$, $x_2^2$ and $x_1 \cdot x_2$, and thus
2 independent scale-invariant ratios. Note that 2 is also the number of
Lorentz invariants of a massless 4-point function in momentum space.

\medskip

A specific representation for the function $F$ may be obtained by making
use of the OPE twice in the 4-point function, one on the pair  $\O
_{\Delta _1} (x_1) \O _{\Delta _3} (x_3)$ and once on the pair $ \O
_{\Delta _2} (x_2) \O _{\Delta _4} (x_4)$. One obtains the {\sl double
OPE}, first introduced in \cite{ferr72},
\bea
\< \O _{\Delta _1} (x_1) \O _{\Delta _2} (x_2) \O _{\Delta _3}
(x_3) \O _{\Delta _4} (x_4)\>
= \sum _{\Delta \Delta'}
{c_{\Delta _1 \Delta _3 \Delta} \over | x_{13}| ^{\Delta _1 \Delta _3
\Delta} }
{1 \over |x_{12}| ^{\Delta + \Delta '}}
{c_{\Delta _2 + \Delta _4 -\Delta '} \over  |x_{24}|^{\Delta _2 +
\Delta _4 - \Delta '}}
\eea
The OPE coefficients $c_{\Delta _1 \Delta _3 \Delta}$ appeared in
the simple OPE of the operators $\O _{\Delta _1}$ and $\O _{\Delta
_3}$. General properties of the OPE and double OPE have been studied recently
in \cite{Park:1999pd}, \cite{Hoffmann:2000dx}, \cite{Dolan:2000ut},
\cite{Ferrara:2001uj} and \cite{Eden:2001ec} and from a perturbative point of
view in \cite{Arutyunov:2000im}; see also \cite{Bianchi:2001cm}.

\subsection{4-pt Function of Dilaton/Axion System}

The possible intermediary fields and operators are restricted by the
$SU(4)_R$ tensor product formula for external
operators.\footnote{Actually, the possible intermediary fields and
operators are restricted by the full $SU(2,2|4)$ superconformal algebra
branching rules. Since no $\N=4$ off-shell superfield formulation is
available, however, it appears very difficult to make direct use of this
powerful fact.} Assuming external operators (such as the 1/2 BPS
primaries) in representations
$[0,\Delta,0]$ and $[0,\Delta ',0]$, their tensor product decomposes as
\bea
[0,\Delta,0] \otimes[0,\Delta ',0] =
\oplus _{\mu =0} ^{\Delta '} \oplus _{\nu =0} ^{\Delta '-\mu} [\nu, \Delta
+ \Delta ' - 2 \mu - 2 \nu, \nu]
\eea
For example, the product of two AdS$_5$ supergravity primaries in the
representation ${\bf 20'}=[0,2,0]$ is given by (the subscript $A$ denotes
antisymmetrization)
\bea
{\bf 20'} \otimes {\bf 20'} = {\bf 1 } \oplus {\bf 15}_A \oplus {\bf 20'}
\oplus {\bf 84} \oplus {\bf 105} \oplus {\bf 175}_A
\eea
Actually, the simplest group theoretical structure emerges when taking two
$SU(4)_R$ and Lorentz singlets which are $SU(2,2|4)$ descendants. We
consider the system of dimension $\Delta =4$ half-BPS operators dual to the
dilaton and axion fields in the bulk;
\bea
\O _\phi = \tr F_{\mu \nu } F^{\mu \nu} + \cdots
\hskip 1in
\O _C = \tr F_{\mu \nu} \tilde F^{\mu \nu} + \cdots
\eea
The further advantage of this system is that the classical supergravity
action is simple,
\bea
S[G,\Phi, C] = {1 \over 2 \kappa_5 ^2} \int _H \sqrt G \biggl [
- R_G + \Lambda + \half \p _\mu \Phi \p ^\mu \Phi + \half e^{2 \Phi}
\p _\mu C \p ^\mu C \biggr ]
\eea
In the AdS/CFT correspondence, $\kappa _5^2$ may be related to $N$ by
$\kappa _5^2 = 4\pi^2 /N^2$. This system was first examined in
\cite{Liu:1998ty} and \cite{Liu:1999kg}. An investigation directly of the
correlator of half-BPS chiral primaries may be found in 
\cite{Arutyunov:2000ku}.

\begin{fig}[htp]
\centering
\epsfxsize=4.5in
\epsfysize=1.5in
\epsffile{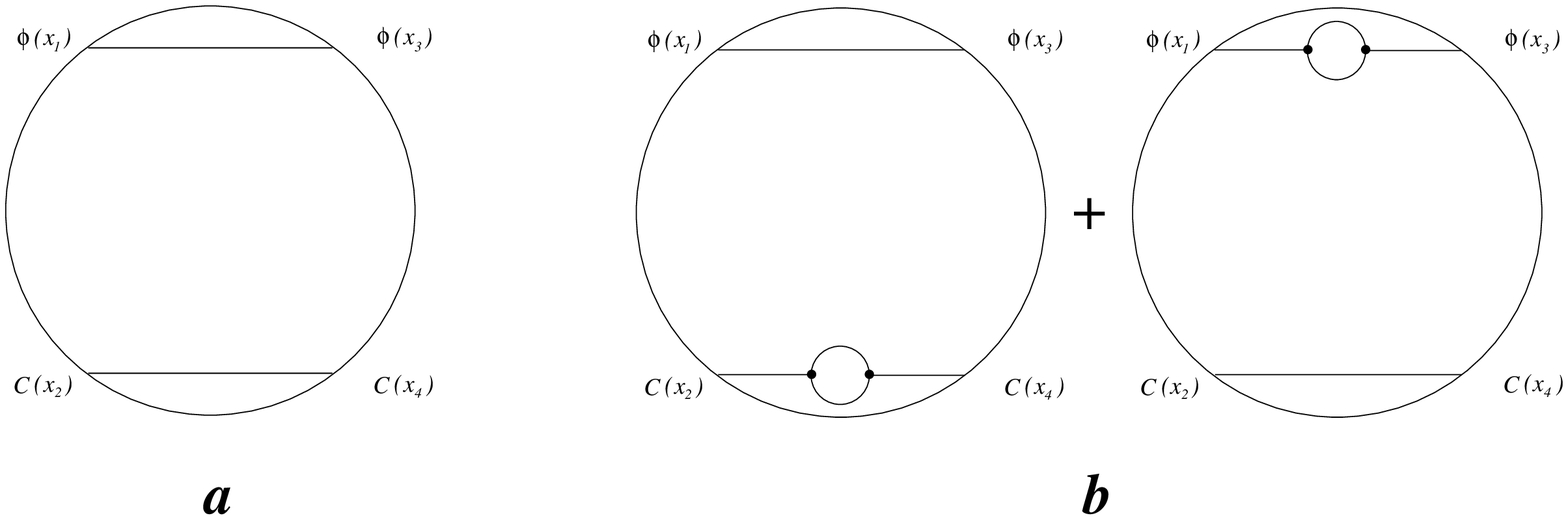}
\caption{Disconnected contributions to the correlator $\< \O_\Phi \O_C
\O_\Phi \O_C \>$ to order $1/N^2$}
\label{fig:6}
\end{fig}

\begin{fig}[htp]
\centering
\epsfxsize=4in
\epsfysize=2.5in
\epsffile{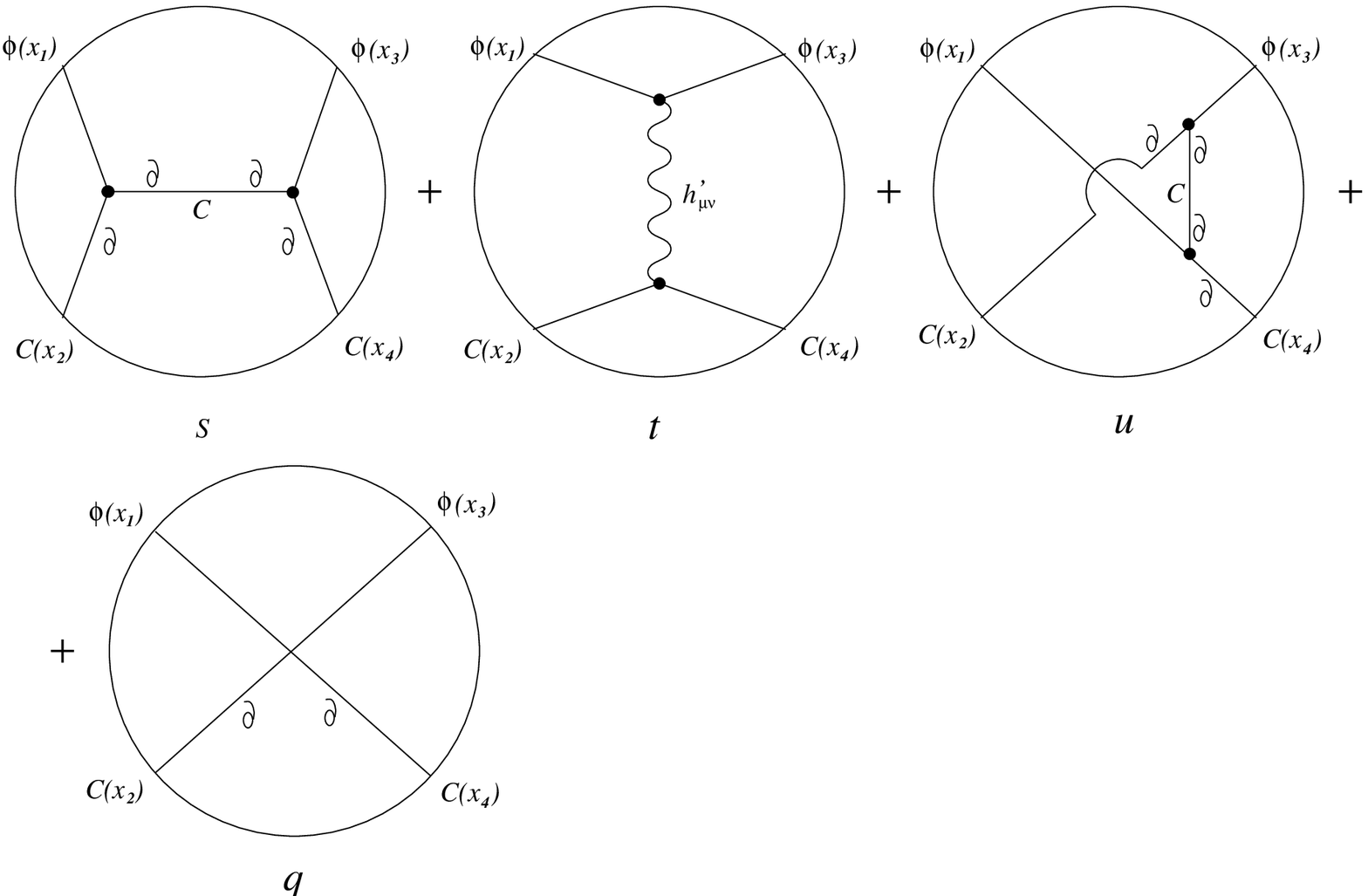}
\caption{Connected contributions to the correlator $\< \O_\Phi \O_C
\O_\Phi \O_C \>$ to order $1/N^2$}
\label{fig:7}
\end{fig}

\subsection{Calculation of 4-point Contact Graph}

The 4-point function receives contributions from the {\sl contact graph}
and from a number of {\sl exchange graphs}, which we now discuss in turn.
The most general 4-point contact term is given by the following integral,
\bea
D_{\Delta _1 \Delta _2 \Delta _3 \Delta _4} (x_i)
\equiv \int _H \! {d^{d+1} \! z \over z_0^{d+1}}
\prod _{i=1} ^4 \biggl ( { z_0
\over z_0^2 + (\vec{z} - \vec{x}_i)^2 } \biggr ) ^{\Delta _i}
\eea
This integral is closely related to the momentum space integration of the
box graph. In fact, we shall not need this object in all its generality,
but may restrict to the case $D_{\Delta \Delta \Delta'\Delta'}$. The
calculation in the general case is given in \cite{DF99} and
\cite{DFMMR99A}; see also \cite{Liu:1998th}.

\begin{fig}[htp]
\centering
\epsfxsize=2in
\epsfysize=1in
\epsffile{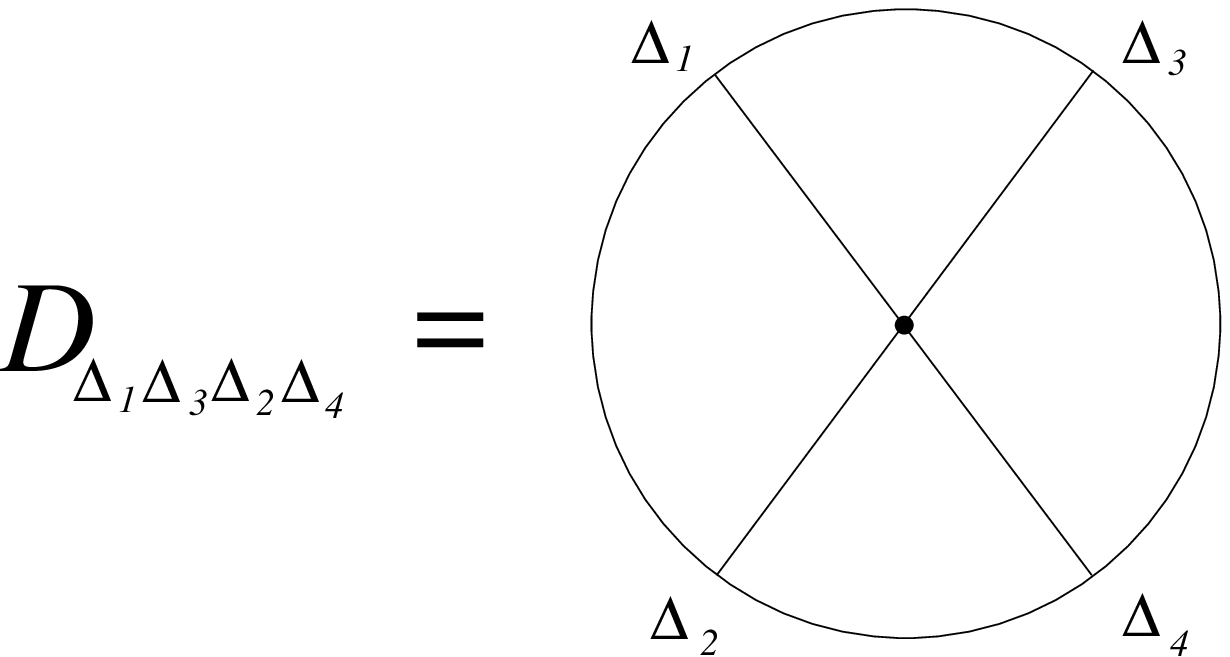}
\caption{Definition of the contact graph function $D$}
\label{fig:8}
\end{fig}

To compute this object explicitly, it is convenient to factor out the
overall non-trivial conformal dependence. This may be done by first
translating $x_1$ to 0,  then performing an inversion and then
translating also $x_3'$ to 0. The result may be expressed in terms of
\bea
x\equiv x_{13}' - x_{14}'
\qquad
y \equiv x_{13}' - x_{12}'
\eea and is found to be
\bea
D_{\Delta \Delta \Delta'\Delta'} (x_i)
=  x_{12} ^{2 \Delta '} x_{13}^{2 \Delta} x_{14}^{2 \Delta'}
\times \int _H \! {d^{d+1} \! z \over z_0^{d+1}}
{z_0 ^{2 \Delta + 2 \Delta '}
\over z^{2 \Delta} (z-x)^{2 \Delta '} (z-y)^{2\Delta '}}
\eea
Introducing two Feynman parameters, and carrying out the $z$-integration,
the integral may be re-expressed as
\bea
D_{\Delta \Delta \Delta'\Delta'} (x_i)
& = &
{x_{12} ^{2 \Delta '} x_{13}^{2 \Delta} x_{14}^{2 \Delta'}
\over (x^2 + y^2)^{\Delta '}}
{\pi ^{d/2} \Gamma (\Delta + \Delta ' -d/2)
\over 2 ^{\Delta'} \Gamma (\Delta ) \Gamma (\Delta ')}
 \\
&& \quad \times
\int _0 ^\infty \! d \rho \int _{-1} ^{+1} \! d\lambda \
{\rho ^{\Delta -1} (1-\lambda ^2)^{\Delta -1}
\over [1 + \rho (1-\lambda ^2)]^\Delta }
{1 \over (s+\rho +\rho \lambda t)^{\Delta'}}
\nonumber
\eea
Remarkably, for positive integers $\Delta$ and $\Delta '$, the integral for
any $\Delta$, $\Delta '$ and $d$ may be re-expressed in terms of successive
derivatives of a universal function $I(s,t)$,
\bea
I(s,t) \equiv \int _{-1} ^{+1} \! d\lambda {1 \over 1+\lambda t
-s(1-\lambda ^2)} \ln {1+\lambda t \over s(1-\lambda ^2)}
\eea
in the following way,
\bea
D_{\Delta \Delta \Delta'\Delta'} (x_i)
& = &
(-)^{\Delta + \Delta '} {x_{12} ^{2 \Delta '} x_{13}^{2 \Delta} x_{14}^{2
\Delta'}  \over (x^2 + y^2)^{\Delta '}}
{ \pi ^{d/2} \Gamma (\Delta + \Delta ' -d/2)
\over \Gamma (\Delta )^2 \Gamma (\Delta ')^2}
 \\
&& \quad \times
\biggl ({\p \over \p s} \biggr )^{\Delta '-1}
\biggl \{
s^{\Delta -1} \biggl ({\p \over \p s} \biggr )^{\Delta -1} I(s,t)
\biggr \}
\nonumber
\eea
While the function $I(s,t)$ is not elementary, its asymptotic behavior is
easily obtained.

\medskip

In the {\sl direct channel} or {\sl t-channel}, we have $|x_{13}| \ll
|x_{12}|$ and $|x_{24}|\ll |x_{12}|$, so that we have both $s,t\to 0$. Of
principal interest will be the contribution which contains logarithms of
$s$, and this part is given by (for the full asymptotics,
\cite{DFMMR99A}); see also \cite{Sanjay:1999uw},
\bea
I^{\rm log} (s,t) = -\ln s \sum _{k=0} ^\infty  a_k(t) s^k
\hskip 1in
a_k(t) = \int _{-1} ^{+1} \! d\lambda
{(1-\lambda ^2)^k \over (1+\lambda t)^{k+1} }
\eea

\medskip

In the two {\sl crossed channels}, we have $s\to 1/2$~: in the {\sl
s-channel}  $|x_{12}|,|x_{34}| \ll |x_{13}|$ for which $t \to -1$; in the
{\sl u-channel} $|x_{23}|, |x_{14}| \ll |x_{34}|$ for which $t \to +1$.
Of principal interest will be the contribution which contains logarithms
of $(1-t^2)$, and this part is given by (for the full asymptotics, see
\cite{DFMMR99A}),
\bea
I^{\rm log} (s,t) =
- \ln (1-t^2) \sum _{k=0} ^\infty (1-2s)^k \alpha _k (t)
\hskip .7in
\alpha _k(t) = \sum _{\ell=0} ^\infty {\Gamma (\ell +\half) (1-t^2)^\ell
\over \Gamma (\half)  (2\ell +k+1) \ell !} \quad
\eea

\subsection{Calculation of the 4-point Exchange Diagrams}

A direct approach to the calculation of the exchange graphs for scalar and
gravitons is to insert the scalar or graviton propagators computed
previously and then to perform the integrals over the 3-point interaction
vertices. This approach was followed in \cite{DF98, DF99, DFMMR99A}.
However, it is also possible to exploit the special space-time properties
of conformal symmetry to take a more convenient approach discussed in
\cite{DFR99}. This approach consists in first computing the 3-point
interaction integral with two boundary-to-bulk propagators (say to
vertices 1 and 3) with the bulk-to-bulk propagator between the same
interaction vertex and an arbitrary bulk point. Conformal invariance and
the assumption of integer dimension $\Delta \geq d/2$ makes this into a
very simple object. We shall follow the last method to evaluate the
exchange graphs.

\begin{fig}[htp]
\centering
\epsfxsize=1.7in
\epsfysize=1.6in
\epsffile{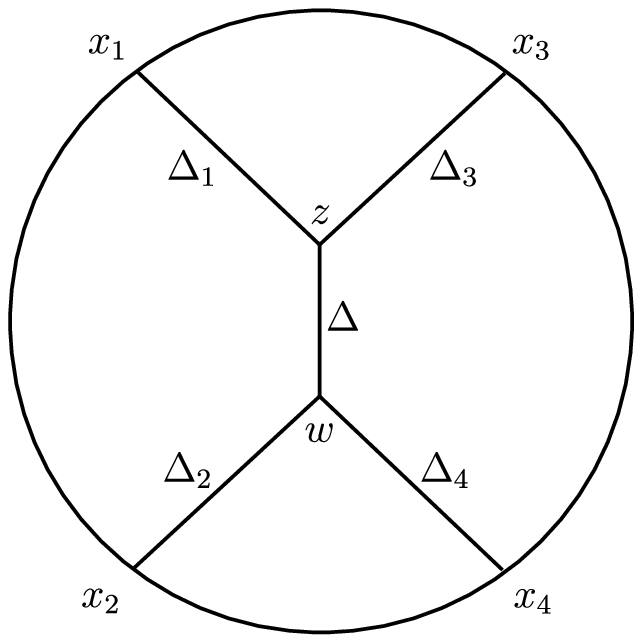}
\caption{The $t$-channel exchange graph}
\label{fig:9}
\end{fig}

For the {\sl scalar exchange diagram}, we need to compute the following
integral\footnote{In this subsection, we shall not write explicitly the
propagator normalization constants $C_\Delta$; however, they will be
properly restored in the next subsection.}
\bea
A(w,x_1,x_3) = \int _H \! {d^{d+1} \! z \over z_0^{d+1}} G_\Delta (w,z)
K_{\Delta _1} (z,x_1) K_{\Delta _3} (z,x_3)
\eea
As in the past, we simplify the integral by using translation invariance to
translate $x_1$ to 0, and then performing an inversion. As a result,
\bea
A(w,x_1,x_3) = |x_{13}|^{- 2 \Delta _3} I(w'-x_{13}')\, ,
\hskip .5in
I(w) = \int _H \! {d^5 \! z \over z_0^5} G_\Delta (w,z) {z_0 ^{\Delta _1 +
\Delta _3} \over z^{2 \Delta _3}}
\eea
We now use the fact that $G_\Delta$ is a Green function and satisfies
$(\Box_w + \Delta (\Delta -d) ) G_\Delta (w,z) = \delta (w,z)$, so that
\bea
(\Box_w + \Delta (\Delta -d) ) I(w) = {w_0 ^{\Delta _1 + \Delta _3} \over
w^{2\Delta _3}}
\eea
In terms of the scale invariant combination $\zeta = w_0^2 /w^2$, we have
$I(w) = w_0 ^{\Delta _{13}} f_S (\zeta)$, $\Delta _{13}=\Delta _1 - \Delta
_3$ and the function $f_S$ now satisfies the following differential
equation
\bea
\label{feq}
&&
4 \zeta ^2 (\zeta -1) f_S'' + 4 \zeta \big [(\Delta _{13}+1) \zeta -
\Delta_{13} + d/2 -1 \big ] f_S'
\\ && \hskip 1.4in
+ (\Delta - \Delta _{13}) (\Delta + \Delta _{13}-d) f_S
= \zeta ^{\Delta _3}
\nonumber
\eea
Making the change of variables $\sigma = 1/\zeta$, we find that the new
differential equation is manifestly of the hypergeometric type and is
solved by
\bea
f_S(\zeta ) = F\biggl ({\Delta - \Delta _{13}\over 2}, {d-\Delta -\Delta
_{13}\over 2}; {d \over 2}; 1-{1\over \zeta} \biggr )
\eea
The other linearly independent solution to the hypergeometric equation is
singular as $\zeta \to 1$, which is unacceptable since the original
integral was perfectly regular in this limit (which corresponds to
$\vec{w}\to 0$).

\medskip

It is easier, however, to find the solutions in terms of a power series, $
f_S(\zeta) = \sum _k  f_{Sk} \zeta ^k$.
Upon substitution into (\ref{feq}), we find solutions that truncate to a
finite number of terms in $\zeta$, provided $\Delta _1 + \Delta _3 -
\Delta$ is a positive integer. Notice that $k$ need not take integer
values, rather $k-\Delta _3$ must be integer. The series truncates from
above at $k_{\rm max} = \Delta _3 -1$, so that $f_{Sk}=0$ when $k \geq
\Delta _3$, and
\bea
f_{Sk} = {\Gamma (k) \Gamma (k +\Delta _{13})
\Gamma (\half\{\Delta _1 + \Delta _3 - \Delta\})
\Gamma (\half \{\Delta + \Delta _1 + \Delta _3 -d\})
\over
4 \Gamma (\Delta _1) \Gamma (\Delta _3)
\Gamma (k+1 +\half \{\Delta _{13} -\Delta\})
\Gamma (k+1 +\half \{\Delta _{13} +\Delta -d\})}
\eea
Still under the assumption that $\Delta _1 + \Delta _3 - \Delta$ is a
positive integer, the series also truncates from below at $k_{\rm min}=
\half (\Delta - \Delta _{13})$.

\medskip

It remains to complete the calculation and substitute the above partial
result into the full exchange graphs. The required integral is
\bea
S(x_i) = \int _H dw \sqrt g K_{\Delta _2}(w,x_2) K_{\Delta
_4}(w,x_4) A(w,x_1,x_3)
\eea
Remarkably, the expansion terms $w_0 ^{\Delta _{13}} \zeta ^k = w_0
^{\Delta _1 + \Delta _3 + 2k} /w^{2k}$ are precisely of the form of the
product of two boundary-to-bulk propagators, one with dimension $k$, the
other with dimension $\Delta _{13} +k$. Thus, the scalar exchange diagram
may be written as a sum over contact graphs in the following way,
\bea
S(x_i) =
\sum _{k=k_{\rm \min}} ^{k_{\rm max}} f_{Sk} |x_{13}|^{-2 \Delta _3 + 2
k} D_{k \, \Delta _{13}+k \, \Delta _2  \Delta _4}(x_i)
\eea
The evaluation of the contact graphs was carried out in the
preceding subsection for the special cases $ \Delta _1 = \Delta _3$ and
$\Delta _2 = \Delta _4$.

\medskip

For the {\sl massless graviton exchange diagram}, we need to compute the
integral,
\bea
\label{graveq}
A_{\mu \nu} (w,x_1,x_2) = \int _H {d^{d+1} \! z \over z_0^{d+1}}
G_{\mu \nu \mu' \nu'} (w,z) T^{\mu'\nu'}(z,x_1,x_3)
\eea
where the stress tensor is generated by two boundary-to-bulk scalar
propagators whish we assume both to be of dimension $\Delta _1$,
\bea
T^{\mu'\nu'}(z,x_1,x_3) & = & \nabla ^{\mu'} K_{\Delta _1} (z,x_1)
\nabla ^{\nu'} K_{\Delta _1} (z,x_3)
-\half g^{\mu'\nu'} \biggl [ \nabla _{\rho'} K_{\Delta _1} (z,x_1)
\nabla ^{\rho'} K_{\Delta _1} (z,x_3)
 \nonumber \\ && \hskip .5in
+ \Delta _1 (\Delta _1-d) K_{\Delta _1} (z,x_1)
\nabla ^{\rho'} K_{\Delta _1} (z,x_3) \biggr ]
\eea
Under translation of $x_1$ to 0 and inversion, then using the
symmetries of rank 2 symmetric tensors on AdS$_5$, and finally using the
operator $W$ on both sides of (ref{graveq}), we find
\bea
A_{\mu \nu}(w,x_1,x_3) & = & {1 \over w^4 |x_{13}|^{2\Delta _1}}
J_{\mu \kappa} (w) J_{\nu \lambda}(w) I_{\kappa \lambda}(w'-x_{13}')
\nonumber \\
I_{\kappa \lambda}(w) & = &
\biggl ( {\delta _{0\mu}\delta _{0\nu} \over w_0^2} -
{1 \over d-1} g_{\mu \nu} \biggr ) f_G (\zeta) + \nabla _{(\mu} v_{\nu)}
\eea
where the field $v_\mu$ represents an immaterial action of a diffeomorphism
while the function $f_G (\zeta)$ satisfies the first order differential
equation
\bea
2\zeta (1-\zeta) f_G'(\zeta) - (d-2) f_G (\zeta ) = \Delta _1 \zeta
^{\Delta _1}
\eea
It is again possible to solve this equation via a power series $f_G(\zeta)
= \sum _k f_{Gk} \zeta ^k$. The range of $k$ is found to be $d/2-1=k_{\rm
min}
\leq k \leq k_{\rm max}= \Delta _1 -1$, provided $\Delta _1 -d/2$ is a
non-negative integer and $d>2$. The coefficients are then given by
\bea
f_{Gk} = - {\Delta _1 \Gamma (k) \Gamma (\Delta _1 +1 -d/2) \over \Gamma
(\Delta _1) \Gamma (k+2 -d/2)}
\eea
The result is particularly simple for the case of interest here when $d=4$
and $\Delta _1 $ integer,
\bea
f_G(\zeta) = - {\Delta _1 \over 2\Delta _1 -2} (\zeta + \zeta^2 +\cdots +
\zeta^{\Delta _1-1} )
\eea
Again, this result may be substituted into the remaining integral in $w$
versus the boundary-to-bulk propagators from the interaction point $w$ to
$x_2$ and $x_4$, thereby yielding again contributions proportional to
contact terms.

\subsection{Structure of Amplitudes}

The full calculations of the graviton exchange amplitudes are quite
involved and will not be reproduced completely here \cite{DFMMR99A}.
Instead, we quote the contributions to the amplitudes from the correlator
$[\O_\phi (x_1)
\O_C (x_2) \O_\phi (x_3) \O_C (x_4) ]$, where the graviton is exchanged
in the
$t$-channel only. The sum of the axion exchange graph $I_s$ in the
$s$-channel, of the axion exchange $I_u$ in the $u$-channel and of the
quartic contact graph $I_q$ is listed separately from the graviton
contribution $I_g$,
\bea
I_s + I_u + I_q &=&
{6^4 \over \pi ^8} \biggl [ 64 x_{24}^2 D_{4455} - 32 D_{4444} \biggr ]
\\
I_{\rm grav} &=&
{6^4 \over \pi ^8} \biggl [8  ({1 \over s} -2) x_{24}^2 D_{4455}
+ {64 \over 9s} {x_{24}^2 \over x_{13}^2} D_{3355} + {16 \over 3s}
{x_{24}^2 \over x_{13}^4} D_{2255}
\nonumber \\
&& \hskip .3in
+18 D_{4444} - {46 \over 9 x_{13}^2}
D_{3344}  - {40 \over 9 x_{13}^4} D_{2244} -{8 \over 3 x_{13}^6} D_{1144}
\biggr ]
\nonumber
\eea
The most interesting information is contained in the power singularity part
of this amplitude as well as in the part containing logarithmic
singularities. Both are obtained from the singular parts of the universal
function $I(s,t)$ in terms of which the contact functions $D_{\Delta _1
\Delta _2 \Delta _4 \Delta _4}$ may be expressed.

\subsection{Power Singularities}

In the $s$-channel and $u$-channel, no power singularities occur in the
supergravity result. This is consistent with the fact that there are no
power singular terms in the OPE of $\O_\phi$ with $\O_C$, since the
resulting composite operator would have $U(1)_Y$ hypercharge 4, and the
lowest operator with those quantum numbers has dimension 8. (More details
on this kind of argument will be given in \S 7.12.)

\medskip

In the $t$-channel, where $|x_{13}|, |x_{24}| \ll |x_{12}|$, we have
$s,t\to 0$, with $s\sim t^2$. The power singularities in this channel come
entirely from the graviton exchange part, given by
\bea
I_{\rm grav} \bigg |_{\rm sing} =
{ 2^{10} \over 35\pi^6} {1 \over x_{13}^8 x_{24}^8} \biggl [
s(7t^2 + 6 t^4) + s^2 (-7 + 3 t^2) - 8 s^3 \biggr ]
\eea
To compare this behavior with the singularities expected from the OPE, we
derive first the behavior of the variables $s$ and $t$ in the $t$-channel
limit,
\bea
s \sim {x_{13}^2 x_{24}^2 \over 2 x_{12}^4}
\hskip 1in
t \sim - {x_{13} \cdot J(x_{12}) \cdot x_{24} \over x_{12}^2}
\eea
where $J_{ij} (x) \equiv \delta _{ij} - 2 x_i x_j /x^2$ is the conformal
inversion Jacobian tensor. Therefore, the leading singularity in the
graviton exchange contribution may be written as
\bea
I_{\rm grav} \bigg |_{\rm sing} =
{2^6 \over 5 \pi ^6} {1 \over x_{13}^6 x_{24}^6}
{4 (x_{13} \cdot J(x_{12}) \cdot x_{24} )^2 - x_{13}^2 x_{24}^2
\over x_{12}^8}
\eea
with further subleading terms suppressed by additional powers of $x_{13}^2
/x_{12}^2$ and $x_{24}^2 /x_{12}^2$. The leading contribution above
describes the exchanges of an operator of dimension 4, whose tensorial
structure is that of the stress tensor.

\medskip

Note that there is also a term corresponding to the exchange of the
identity operator, with behavior $ x_{13}^{-8} x_{24}^{-8}$,
which derives from the disconnected contribution to the correlator in Fig
5 (a). Note that there is no contribution in the singular terms that
corresponds to the exchange of an operator of dimension 2. One candidate
would be $\O _2$ which is a Lorentz scalar; however, it is a ${\bf 20'}$
under $SU(4)_R$, and therefore not allowed in the OPE of two singlets. The
other candidate is the Konishi operator, which is both a Lorentz and
$SU(4)_R$ singlet. The fact that it is not seen here is consistent with
the fact that its dimension becomes very large $\sim \lambda ^{1/4}$ in
the limit
$\lambda \to \infty$ and is dual to a massive string excitation.

\subsection{Logarithmic Singularities}

The logarithmic singularities in the $t$-channel are produced by both the
scalar exchange and contact graphs as well as by the graviton exchange
graph \cite{DFMMR99A}. They are given by
\bea
I_s + I_u + I_q \bigg |_{\rm log} &=&
{960 \over \pi ^6} {\ln s \over x_{13}^8 x_{24}^8}
\sum _{k=0} ^\infty s^{k+4} (k+1)^2 (k+2)^2 (k+3)^2 (3k+4)  a_{k+3}(t)
\\
I_{\rm grav} \bigg |_{\rm log} & = &
{24 \over \pi ^6}  {\ln s \over x_{13}^8 x_{24}^8} \sum _{k=0} ^\infty
s^{k+4} {\Gamma (k+4) \over \Gamma (k+1)} \biggl \{
(k+4)^2 (15k^2 +55k +42) a_{k+4}(t)
\nonumber \\ && \hskip 1.3in
-2(5k^2 + 20k+16) (3k^2 +15k+22) a_{k+3}(t)  \biggr \}
\nonumber
\eea
To leading order, these expressions simplify as follows,
\bea
I_s + I_u + I_q \bigg |_{\rm log} &=&
+ \ {2^7 \cdot 3^3 \over 7 \pi ^6 x_{12}^{16}}
\ln {x_{13}^2 x_{24}^2\over x_{12}^4}
\\
I_{\rm grav} \bigg |_{\rm log} & = &
- \ {2^7 \cdot 3 \over 7 \pi ^6 x_{12}^{16}}
\ln {x_{13}^2 x_{24}^2\over x_{12}^4}
\nonumber
\eea
Assembling all logarithmic contributions for the various correlators, we
get, \cite{DMMR99},
\bea
\< \O_\phi \O_\phi \O_\phi \O_\phi \>_{\rm log} & = & - { 208 \over 21 N^2}
{1 \over x_{12}^{16}} \ln {x_{13}^2 x_{24}^2 \over x_{12}^4}
\hskip 1in {\rm t-channel}
\nonumber \\
\< \O_C \O_C \O_C \O_C \>_{\rm log} & = & - { 208 \over 21 N^2}
{1 \over x_{12}^{16}} \ln {x_{13}^2 x_{24}^2 \over x_{12}^4}
\hskip 1in {\rm t-channel}
\nonumber \\
\< \O_\phi \O_C \O_\phi \O_C \>_{\rm log} & = & + { 128 \over 21 N^2}
{1 \over x_{12}^{16}} \ln {x_{13}^2 x_{24}^2 \over x_{12}^4}
\hskip 1in {\rm t-channel}
\nonumber \\
\< \O_\phi \O_C \O_\phi \O_C \>_{\rm log} & = & - { 8 \over N^2}
{1 \over x_{13}^{16}} \ln {x_{12}^2 x_{34}^2 \over x_{13}^4}
\hskip 1.2in {\rm s-channel}
\eea
Here, the overall coupling constant factor of $\kappa _5^2$ has been
converted to a factor of $1/N^2$ with the help of the relation $\kappa _5^2
= 4\pi^2 /N^2$, a relation that will be explained and justified in
(\ref{G5}). Further investigations of these log singularities may be found in
\cite{Bianchi:1999ge}.

\subsection{Anomalous Dimension Calculations}

We shall use the supergravity calculations of the 4-point functions for
the operators $\O_\phi$ and $\O_C$ to extract anomalous
dimensions of double-trace operators built out of linear combinations of
$[\O_\phi \O_\phi]$, $[\O_C \O_C ]$ and $[\O_\phi \O_C]$. This was
done in \cite{DMMR99} by taking the limits in various channels of the
three 4-point functions $\< \O_\phi (x_1) \O_\phi (x_2) \O_\phi (x_3)
\O_\phi (x_4)\> $
$\< \O_C (x_1) \O_C (x_2) \O_C (x_3) \O_C (x_4)\> $ and
$\< \O_\phi (x_1) \O_C (x_2) \O_\phi (x_3) \O_C (x_4)\> $. For example, we
extract the following simple behavior from the {\sl s-channel limit}
$x_{12}, x_{34} \to 0$ of the correlator $\< \O_\phi (x_1) \O_C (x_2)
\O_\phi (x_3) \O_C (x_4)\> $,
\bea
\O _\phi (x_1) \O_C (x_2) = A_{\phi c} (x_{12} \mu) [\O _\phi \O_C]_\mu
(x_2) +\cdots
\eea
where $\mu$ is an arbitrary renormalization scale for the composite
operators and $A_{\phi c}$ is the corresponding logarithmic coefficient
function, whose precise value in the large $N$, large $\lambda$ limit is
available from the logarithmic singularities of the correlator, and is
given by
\bea
\label{OPEA}
A_{\phi c} (x_{12}\mu) = 1 - {16 \over N^2} \ln (x_{12}\mu)
\eea
This leading behavior receives further corrections both in inverse powers
of $N$ and $\lambda$.

\medskip

>From the {\sl t-channel} and {\sl u-channel} of the same correlators, we
extract the leading terms in the OPE of two $\O_\phi$'s and of two $\O_C$'s
as follows,
\bea
\label{OPEOO}
\O _\phi (x_1) \O_\phi (x_3) & = & S(x_1,x_3)
+ C_{\phi \phi} [\O _\phi \O_\phi]_\mu
+ C_{\phi c} [\O _C \O_C]_\mu + C_{\phi T} [TT]_\mu + \cdots
\nonumber \\
\O _C (x_1) \O_C (x_3) & = & S(x_1,x_3)
+ C_{c \phi} [\O _\phi \O_\phi]_\mu
+ C_{c c} [\O _C \O_C]_\mu + C_{c T} [TT]_\mu + \cdots
\eea
where the term $S(x_1,x_3)$ contains all the {\sl power singular terms} in
the expansion, and is given schematically by
\bea
S(x_1,x_3) = {I \over x_{13}^8} + {T (x_3) \over x_{13}^4} + {\p T (x_3)
\over x_{13}^3} + {\p \p T (x_3) \over x_{13}^2} + {\p \p \p T (x_3) \over
x_{13}}
\eea
The coefficient functions may be extracted from the logarithmic behavior as
before,
\bea
\label{OPEC}
C_{\phi \phi } = C_{cc} & = & 1 - {208 \over 21 N^2} \ln (x_{13}\mu)
\nonumber \\
C_{\phi c} = C_{c \phi} & = & 1 + {128 \over 21 N^2} \ln (x_{13}\mu)
\eea
Unfortunately, the coefficient functions $C_{\phi T}$ and $C_{cT}$ are not
known at this time as their evaluation would involve the highly complicated
calculation involving two external stress tensor insertions.

\medskip

To make progress, we make use of a continuous symmetry of supergravity,
namely $U(1)_Y$ {\sl hypercharge invariance}. Most important
for us here is that the operator
\bea
\O _B \equiv {1 \over \sqrt 2} \{ \O_\phi + i \O _C\}
\eea
has hypercharge $Y=2$, which is the unique highest values attained amongst
the canonical supergravity fields, as may be seen from the Table 7.  We
may now re-organize the OPE's of operators $\O_\phi$ and $\O_C$ in terms of
$\O_B$ and $\O_B ^*$. The OPE of $\O_B$ with $\O_B ^*$ contains the identity
operator, the stress tensor and its derivatives and powers, as well as the
$Y=0$ operator $[\O_B \O_B ^*]$,
\bea
\O_B (x_1) \O_B ^* (x_2) = S(x_1,x_2) + C_{BT} [TT]_\mu + C_{BB^*} [\O_B
\O_B ^*]_\mu + \cdots
\eea
while the $Y=4$ channel of the OPE is given by
\bea
\label{OBOB}
\O _B (x_1) \O_B (x_2) = (C_{\phi \phi} - C_{\phi c}) \Re [\O_B \O_B]_\mu
+ i A_{\phi c} \Im [\O_B \O_B]_\mu
\eea
Since the smallest dimensional operator of hypercharge $Y=4$ is the
composite $[\O_B \O_B]$, we see that the power singularity terms
$S(x_1,x_3)$ indeed had to be the same for both OPE's in (\ref{OPEOO}).
By the same token, the rhs of (\ref{OBOB}) must be proportional to $ [\O_B
\O_B]_\mu$, so we must have $C_{\phi \phi} - C_{\phi c} = A_{\phi c}$,
which is indeed borne out by the explicit calculational results of
(\ref{OPEA}) and (\ref{OPEC}). In summary, we have a single simple OPE
 \bea
\O _B (x_1) \O_B (x_2) = A_{\phi c} (x_{12}\mu) [\O_B \O_B]_\mu +\cdots
\eea
from which the anomalous dimension may be found to be, \cite{DMMR99},
\bea
 \gamma _{[\O_B \O_B]} = \gamma _{[\O_B^* \O_B^*]} = - 16/N^2
\eea
 There is
another operator occurring in this OPE channel of which we know the
anomalous dimension. Indeed, the double-trace operator $[\O _2 \O
_2]_{105}$ is 1/2 BPS, and thus has vanishing anomalous dimension. Its
maximal descendant $Q^4
\bar Q^4 [\O _2 \O _2]_{105}$ therefore has $Y=0$ and unrenormalized
dimension 8. For more on the role of the $U(1)_Y$ symmetry, see
\cite{Hoffmann:2001pa}. The study of the OPE via the 4-point function has also
revealed some surprising non-renormalization effects, not directly related to
the BPS nature of the intermediate operators. In the OPE of two half-BPS {\bf
20'} operators, for example, the {\bf 20'} intermediate state is {\sl not}
chiral. Yet, to lowest order at strong coupling, its dimension was found to be
protected; see \cite{Arutyunov:2001qw} and \cite{Arutyunov:2001mh}.

\subsection{Check of N-dependence}

The prediction for the anomalous dimension of the operator $[\O_B \O_B]$ to
order $1/N^2$ obtained from supergravity calculations holds for infinitely
large values of the `t~Hooft coupling $\lambda = g_{YM}^2 N$ on the SYM
side. As the regimes of couplings for possible direct calculations do not
overlap, we cannot directly compare this prediction with a calculation on
the SYM side. However, it is very illuminating to reproduce the $1/N^2$
dependence of the anomalous dimension from standard large $N$ counting
rules in SYM theory \cite{DMMR99}. We proceed by expansing $\N=4$ SYM in
$1/N$, while keeping the `t~Hooft coupling fixed (and perturbatively
small). The strategy will be to isolate the general structure of the
expansion  and then to seek the limit where $\lambda \to \infty$.

\medskip

To be concrete, we study the correlator $\< \O_\phi \O_c \O_\phi \O_c\>$,
though our results will apply generally.

\begin{fig}[htp]
\centering
\epsfxsize=4in
\epsfysize=3in
\epsffile{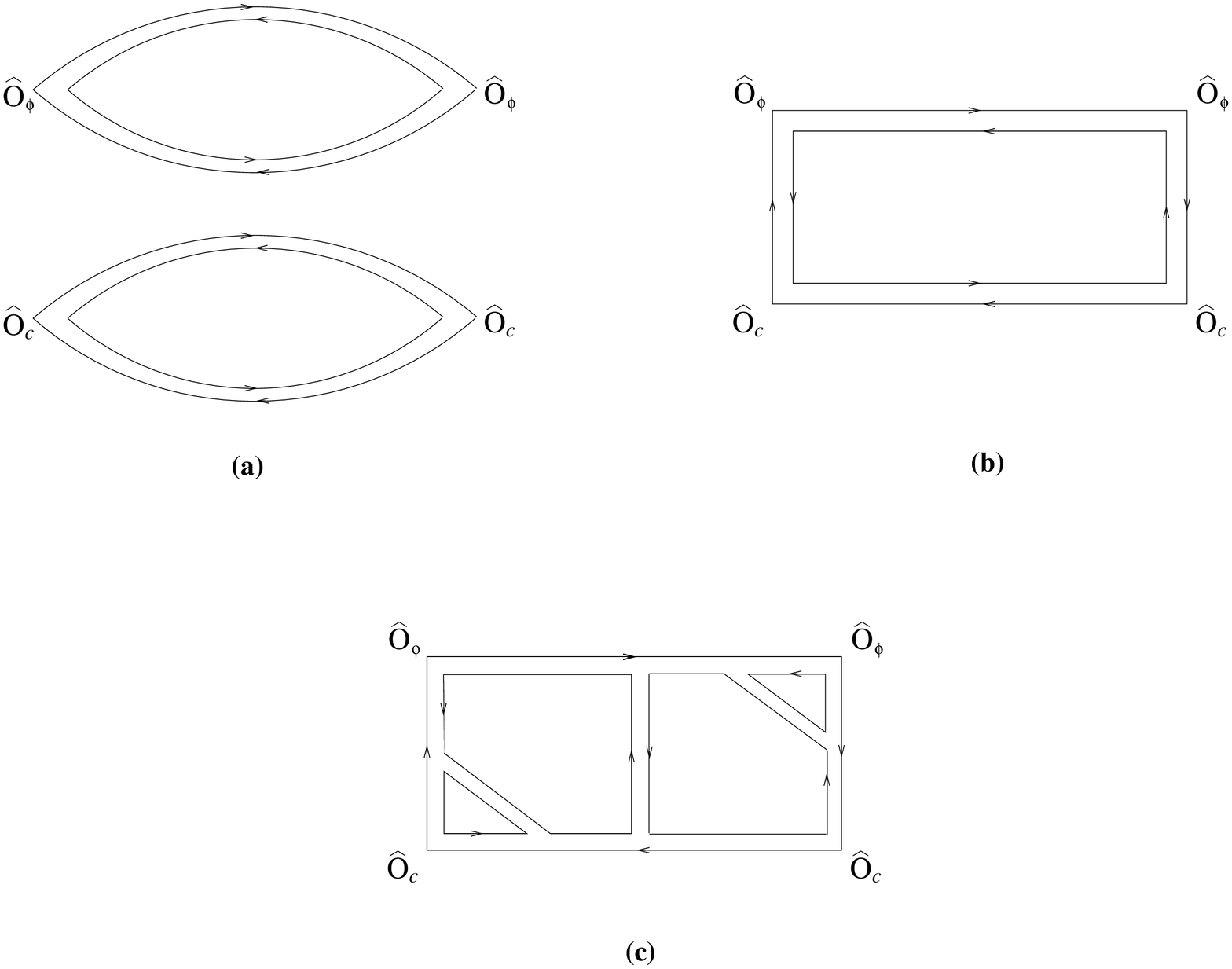}
\caption{Large $N$ counting for the 4-point function}
\label{fig:10}
\end{fig}

First, we normalize the
individual operators via their 2-point functions, which to leading order in
large $N$ requires
\bea
\O_c = {1 \over N} \tr F_{\mu \nu} \tilde  F^{\mu \nu} + \cdots
\qquad \quad
\O_\phi = {1 \over N} \tr F_{\mu \nu} F^{\mu \nu} + \cdots
\eea
In computing the 4-point function $\< \O_\phi \O_c \O_\phi \O_c\>$, there
will first be a {\sl disconnected contribution} of the form $\< \O_\phi
 \O_\phi \> \< \O_c \O_c\>$, which thus contributes precisely to order
$N^0$. The simplest connected contribution is a Born graph with a single
gluon loop; the operator normalizations contribute $N^{-4}$, while the two
color loops contribute $N^2$, thereby suppressing the connected
contribution by a factor of $N^{-2}$ compared to the disconnected one. This
Born graph has no logarithms because it is simply the product of 4
propagators. Perturbative corrections with internal interaction vertices
will however generate logarithmic corrections, and thus contributions to
anomalous dimensions. In the large $N$ limit, planar graphs will dominate
and the only corrections are due to non-trivial $\lambda$-dependence with
the same expansion order $N^{-2}$ and the connected contribution will take
the form
\bea
\< \O_\phi \O_c \O_\phi \O_c\>_{\rm conn} = {1\over N^2} f(\lambda) + O({1
\over N^4})
\eea
For the anomalous dimensions, a similar expansion will hold,
\bea
\gamma (N,\lambda) = {1 \over N^2} \bar \gamma  (\lambda) + O({1 \over N^4})
\eea
The above results were established perturbatively in the `t~Hooft coupling.
To compare with the supergravity results, $f$ and $\bar \gamma$ should
admit well-behaved $\lambda \to \infty$ limits. Our supergravity
calculation in fact established that $\bar \gamma _{[\O_B \O_B]} (\lambda
=\infty) = -16$, a result that could of course not have been gotten from
Feynman diagrams in SYM theory.

\medskip

The calculation of AdS four point functions in weak coupling perturbation
theory was carried out in \cite{Eden1} and \cite{Eden2}; string
corrections to 4-point functions  were considered early on in
\cite{Banks:1998nr} and \cite{Brodie:1998ke};  further 4-point
function calculations in the AdS setting may be found in
\cite{Lee:1999pj}, \cite{Herzog:2000di} and \cite{Chalmers:1998wu}. More
general correlators of 4-point functions and higher corresponding to the
insertion of currents and tensor forms may be found in
\cite{l'Yi:1998eu}, \cite{Osborn:1999az}, \cite{Kalkkinen:2000uk} and
\cite{Bertola:1999vq}. Finally, an approach to correlation functions
based on the existence of a higher spin field theory in Anti-de Sitter
space-time may be found in a series of papers \cite{vassiliev}; see also 
\cite{Metsaev:1999ui}. Finally,
effects of instantons on SYM and AdS/CFT correlators were explored recently in
\cite{Bianchi:1998nk}, \cite{Gopakumar:1999xh}, \cite{Dorey:1999pd},
\cite{Bonelli:1999it}, \cite{Bianchi:2000vh}, \cite{Dorey:2000dt}. Possible
constraints on correlators in AdS/CFT and $\N=4$ SYM from S-duality have been
investigated by \cite{Chalmers:2000vq}. Finally, very recently, correlators have 
been evaluated exactly for strings propagating in AdS$_3$ in 
\cite{Maldacena:2000hw}.

\vfill\eject


\section{How to Calculate CFT\protect\boldmath$_d$ Correlation Functions
from
  AdS\protect\boldmath$_{d+1}$Gravity}
\setcounter{equation}{0}

The main purpose of this chapter is to discuss the techniques
used to calculate correlation functions in $ \N \! =4,\,d=4$ SYM
field theory from Type IIB $ D=10$ supergravity. We will begin
with a quick summary of the basic ideas of the correspondence
between the two theories. These were discussed in more detail in
earlier sections, but we wish to make this chapter
self-contained. Other reviews we recommend to readers are the
broad treatment of \cite{magoo} and the 1999 TASI lectures of
Klebanov \cite{kleb} in which the AdS/CFT correspondence is
motivated from the viewpoints of $D$-brane and black hole
solutions, entropy and absorption cross-sections.

\medskip

The $\N \! =4$ SYM field theory is a 4-dimensional gauge theory
with gauge group $SU(N)$ and $R$-symmetry or global symmetry
group $SO(6) \sim SU(4)$. Elementary fields are all in the
adjoint representation of $SU(N)$ and are represented by
traceless Hermitean $N \times N$ matrices. There are 6 elementary
scalars $X^i(x)$, 4 fermions $\psi^a (x)$, and the gauge
potential $A_j(x)$. The theory contains a unique coupling
constant, the gauge coupling $g_{YM}$. It is known that the only
divergences of elementary Green's functions are those of wave
function renormalization which is unobservable and
gauge-dependent. The $\b$-function $\b(g_{YM})$ vanishes, so the
theory is conformal invariant. The bosonic symmetry group of the
theory is the direct product of the conformal group $SO(2,4) \sim
SU(2,2)$ and the R-symmetry $SU(4)$. These combine with 16
Poincar\'e and 16 conformal supercharges to give the superalgebra
$SU(2,2|4)$ which is the over-arching symmetry of the theory.

\medskip

Observables in a gauge theory must be gauge-invariant quantities,
such as:
\vspace{-2ex}
\begin{enumerate}
\item Correlation functions of gauge invariant local composite
  operators --- the subject on which we focus, \vspace{-2ex}
\item Wilson loops --- not to be discussed, See, for example,
  \cite{malda2, semenoff,grossdruk}\vspace{-1ex}
\end{enumerate}
\vspace{-1ex}
Our primary interest is in correlation functions of the chiral
primary operators\footnote{the normalization factor $N^{1- {k}
    \over {2}}$ is chosen so that all correlation functions of these
  operators are of order $N^2$ for large $N$.}
\be
\tr X^k \equiv N^{1- {k} \over {2}} \tr \biggl ( X^{\{
  i_1}X^{i_2}...X^{i_k\}}\biggr ) - \mbox{traces}
\ee
These operators transform as rank $k$ symmetric traceless $SO(6)$
tensors -- irreducible representations whose Dynkin designation
is $[0,k,0]$.  For $k=2,3,4$ the dimensions of these
representations are $20,50,105$, respectively.
\vspace{-1ex}
\begin{ex}
What is the dimension of the [0,5,0] representation?
\end{ex}
\vspace{-1ex}

The $\tr X^k$ are lowest weight states of {\bf short}
representations of $SU(2,2|4)$. The condition for a short
representation is the relation $\D_{\tr X^k} = k$ between scale
dimension $\D$ and $SO(6)$ rank. Since the latter must be an
integer, the former is quantized. The scale dimension of chiral
primary operators (and all descendents) is said to be
``protected'' It is given for all $g_{YM}$ by its free-field value
(i.e. the value at $g_{YM}=0$). This is to be contrasted with the many
composite operators which belong to {\bf long} representations of
$SU(2,2|4)$. For example, the Konishi operator $K(x) =
\tr [X^iX^i]$ is the primary of a long representation. In the weak
coupling limit, it is known \cite{agj} that $\D_K = 2 +
3g_{YM}^2N/4\pi^2 + \O(g_{YM}^4)$. The existence of a gauge invariant
operator with anomalous dimension is one sign that the field
theory is non-trivial, not a cleverly disguised free theory.

\medskip

In Sec. 3.4 it was discussed how $SU(2,2|4)$
representations are ``filled out'' with descendent states
obtained by applying SUSY generators with $\D = \frac{1}{2}$ to
the primaries. Descendents can be important.  For example, the
descendents of the lowest chiral primary $\tr X^2$ include the 15
SO(6) currents, the 4 supercurrents, and the stress tensor.

\medskip 

Some years ago `t~Hooft taught us (for a review, see \cite{coleman})
that it is useful to express amplitudes in an $SU(N)$ gauge
theory in terms of $N$ and the `t~Hooft coupling $\l=g_{YM}^2N$. Any
Feynman diagram can be redrawn as a sum of color-flow diagrams with
definite Euler character $\chi$ (in the sense of graph theory).
$n$-point functions of the operators $\tr X^k$ are of the form
\be \label{top}
N^{\chi} F(\l,x_i) = N^{\chi}[ f_0(x_i) + \l f_1(x_i) +
\cdots]
\ee
The right side shows the beginning of a weak coupling expansion.
One can see that planar diagrams (those with $\chi=2$) dominate
in the large $N$-limit.

\medskip

The extremely remarkable fact of the AdS/CFT correspondence 
is that the planar
contribution to $n$-point correlation functions of operators
$\tr X^k$ and descendents can be calculated (in the limit $N \to
\infty, ~ \l >>1$) from {\bf classical} supergravity, a strong
coupling limit of a QFT$_4$ without gravity from classical
calculations in a D=5 gravity theory. Results are interpreted as
the sum of the series in (\ref{top}).
Information about operators
in long representations can be obtained by including string scale
effects. It is known that their scale dimensions are of order
$\l^{{1}\over{4}}$ in the limit above. They decouple from
supergravity correlators.

This claim brings us to the supergravity side of the duality,
namely to type $II B,\, D=10$ supergravity which has the product
space-time \AdS\ as a classical ``vacuum solution''.
The first hint of some relation to $\N=4$ SYM theory is the
match of the isometry group $SO(2,4) \times SO(6)$ with the
conformal and $R$-symmetry groups of the field theory. The vacuum
solution is also invariant under $16+16$ supercharges and thus has the
same $SU(2, 2|4)$ superalgebra as the field theory.

\medskip

Type IIB supergravity is a complicated theory whose
structure was discussed in Secs. 4.4 and 4.5. Here we
describe only the essential points necessary to understand the
correspondence with $\N=4$ SYM theory. Since the supergravity
theory is the low energy limit of IIB string theory, the $10$D
gravitational coupling may be expressed in terms of the
dimensionless string coupling $g_s$ and the string scale $\ap$
(of dimension $l^2$). The relation is $\k_{10}^2 = 8 \pi G_{10}=
64 \pi^7 g_s^2 \ap^4$.
The length scale of the AdS$_5$ and $S_5$
factors of the vacuum space-time is $L$ with $L^4=4 \pi \ap^2
g_sN$. The integer $N$ is determined by the flux of the self-dual
$5-$form field strength on $S_5$.  The volume of $S_5$ is $\pi^3L^3$ so the
effective $5D$ gravitational constant is
\bea
\label{G5}
{\kappa _5 ^2 \over 8 \pi } = G_5 =
\frac{G_{10}}{{\rm Vol}(S_5)}=\frac{\pi L^3}{2N^2}
\eea

Among the bosonic fields of the theory, we single out the $10$D
metric $g_{MN}$ and $5$-form $F_{MNPQR}$, which participate in
the vacuum solution, and the dilaton $\phi$ and axion $C$. Other
fields consistently decouple from these and the subsystem is
governed by the truncated action (in Einstein frame)
\be
\begin{array}{r}
S_{\rm IIB}=\lf{1}{16 \pi G_{10}} \int d^{10} z \sqrt{g_{10}}  \{
R_{10} - \lf{1}{2\cdot 5!} F_{MNPQR} F^{MNPQR} - \frac{1}{2} \p_M
\phi \p^M \phi \\ - \lf{1}{2} e^{2 \phi} \p_M C \p^M C \}
\end{array}
\ee
Actually there is no covariant action which gives the self-dual
relation $F_5=\ast F_5$ as an Euler-Lagrange equation, and the
field equations from $S_{\rm IIB}$ must be supplemented by this extra
condition.

\medskip

Using $x^i,\, i=0,1,2,3$ as Cartesian coordinates of Minkowski
space with metric $\eta_{ij}=(-+++)$ and $y^a,\, a=1,2,3,4,5,6$
as coordinates of a flat transverse space, we write the following
ansatz for the set of fields above:
\be
\label{d3}
\begin{array}{l}
ds_{10}^2 = \frac{1}{\sqrt{H(y^a)}} \eta_{ij}dx^i dx^j + \sqrt{H(y^a)}
    \d_{ab} dy^a dy^b \\
F=dA + \ast dA \hspace{5em} A =\frac{1}{H(y^a)}dx^0\wedge dx^1\wedge
dx^2\wedge
dx^3
\\
\phi = C \equiv 0
\end{array}
\ee
Remarkably the configuration above is a solution of the equations
of motion provided that $H(y^a)$ is a harmonic function of $y^a$,
$\ie$
\be \label{multi}
\sum^6_{a=1} \frac{\p^2}{\p y^a \p y^a} H = 0
\ee
\begin{ex}
  Verify that the above is a solution. Compute the connection and
  curvature of the metric as an intermediate step. See the
  discussion of the Cartan structure equations in Section 9 for
  some guidance.
\end{ex}
\vspace{-1ex}

The appearance of harmonic functions is typical of $D$-brane
solutions to supergravity theories.  The solutions (\ref{d3}) are
$\half\!-\!BPS$ solutions which support $16$ conserved
supercharges. This fact may be derived by studying the
transformation rules of Type IIB supergravity to find the
Killing spinors.
A quite general harmonic function is given by
\be \label{harm}
H=1+ \sum^M_{I=1} \frac{L^4_I}{(y-y_I)^4} \hspace{5em} L_I^4 = 4
\pi \a^{\prime \,2}g_s N_I
\ee
This describes a collection of $M$ parallel stacks of
$D3$-branes, with $N_I$ branes located at position $y^a=y^a_I$ in
the transverse space. This ``multi-center'' solution of IIB
supergravity defines a $10$-dimensional manifold with $M$
infinitely long throats as $y \rightarrow y_I$ and which is
asymptotically flat as $y \rightarrow \infty$. The curvature
invariants are non-singular as $y \rightarrow y_I$, and these
loci are simply degenerate horizons. The solution has an
AdS/CFT interpretation as the dual of a Higgs branch vacuum
state of $\N = 4$ SYM theory, a vacuum in which conformal symmetry
is spontaneously broken.  However, we are jumping too far ahead.

\medskip

Let's consider the simplest case of a single stack of $N$ D3-branes at
$y_I=0.$  We replace the $y^a$ by a radial coordinate $r=\sqrt{y^ay^a}$
plus 5 angular coordinates $y^\a$ on an $S_5$. At the same time we take the
near-horizon limit. The physical and mathematical arguments for this
limit are rather complex and discussed in Sec 5.2 above, in
\cite{magoo} and elsewhere.
We simply state that it is the throat region of the geometry that determines
the physics of AdS/CFT. We therefore restrict to the throat simply by
dropping the 1 in the harmonic function $H(r)$. Thus we have the metric
\be
\label{adsxs}
ds_{10}^2 = {{r^2} \over {L^2}} \eta_{ij} dx^i dx^j +{{L^2
    dr^2}\over{r^2}} + L^2 d\Omega_5^2
\ee
where $ d\Omega_5^2$ is the $SO(6)$ invariant metric on the unit
$S_5$. The metric describes the product space \AdS. The coordinates $(x^i,
r)$ are collectively called $z_\m$ below. These coordinates give the
Poincar\'{e} patch of the induced metric on the hyperboloid embedded in
$6$-dimensions~\cite{magoo}.
\be
Y_0^2 + Y_5^2 - Y_1^2 - Y_2^2 - Y_3^2 - Y_4^2 = L^2
\ee
\begin{ex}
  Show that the curvature tensor in the $z_\m$ directions has the
  maximal symmetric form $R_{\m\n\r\l} = -{{1} \over {L^2}}
  (g_{\m\r} g_{\n\l} - g_{\m\l} g_{\n\r})$.
\end{ex}
\vspace{-1ex}

\medskip

The bulk theory may now be viewed as a supergravity theory in the
AdS$_5$ space-time with an infinite number of $5$D fields obtained
by Kaluza-Klein analysis on the internal space $S_5$. We will
discuss the KK decomposition process and the properties of the 5D
fields obtained from it. The main point is to emphasize the 1:1
correspondence between these bulk fields and the composite
operators of the $\N \! =4$ SYM theory discussed above.

The linearized field equations of fluctuations about the
background (\ref{adsxs}) were analyzed in \cite{krvn}. All fields
of the $D=10$ theory are expressed as series expansions in
appropriate spherical harmonics on $S_5$. Typically the
independent $5$D fields are mixtures of KK modes from different
$10$D fields. For example the scalar fields which correspond to
the chiral primary operators are superpositions of the trace
$h^\a_\a$ of metric fluctuations on $S_5$ with the $S_5$
components of the $4$-form potential $A_{\a\b\c\d}$. The
independent $5$D fields transform in representations of the
isometry group $SU(4) \sim SO(6)$ of $S_5$ which are determined
by the spherical harmonics.

\medskip

The analysis of \cite{krvn} leads to a graviton multiplet plus an
infinite set of KK excitations. We list the fields of the
graviton multiplet, together with the dimensionalities of the
corresponding $SO(6)$ representations: graviton $h_{\m\n},\, 1$,
gravitini $\psi_\m,\, {\bf 4}\oplus {\bf 4}^*$, 2-form potentials
$A_{\m\n},\, {\bf 6}_c$, gauge potentials $A_\m, \,{\bf 15}$, spinors $\l,\,
{\bf 4}\oplus {\bf 4}^* \oplus {\bf 20} \oplus {\bf 20}^* =48$, and finally
scalars $\phi,\, {\bf 20'} \oplus {\bf 10} \oplus {\bf 10}^* \oplus {\bf 1}_c = 
42.$ In
this notation ${\bf 10}^*$ denotes the conjugate of the complex
irrep ${\bf 10}$, while ${\bf 6}_c$ denotes a doubling of the real
6-dimensional (defining) representation of $SO(6)$.

Each of these fields is the base of a KK tower. For the scalar
primaries one effectively has the following expansion, after
mixing is implemented,
\be
\f(z,y) = \sum_{k=2}^\infty \f_{k}(z) Y^{k}(y)
\ee
Here $Y^{k}(y)$ denotes a spherical harmonic of rank $k$, so that
$\f_{k}(z)$ is a scalar field on AdS$_5$ which
transforms\footnote{indices for components of this irrep are
  omitted on both $\f_k$ and $Y^k$.} in the $[0,k,0]$ irrep of
$SO(6)$. In the same way that every scalar field on Minkowski
space contains an infinite number of momentum modes, each $\f_k$
contains an infinite number of modes classified in a unitary
irreducible representation of the AdS$_5$ isometry group
$SO(2,4)$. We will describe these irreps briefly. For more
information, see \cite{heid,brfr,Dobrev85,ffz}. The group has
maximal compact subgroup $SO(2) \times SO(4)$ and irreps are
denoted by $(\D,s,s^{\prime})$. The generator of the $SO(2)$
factor is identified with the energy in the physical setting, and
$\D$ is the lowest energy eigenvalue that occurs in the
representation. The quantum numbers $s,s^{\prime}$ designate the
irrep of $SO(4)$ in which the lowest energy components transform.
Unitarity requires the bounds
\bea
\D \ge 2 + s +s^{\prime}  \qquad    {\rm if}\, ss^{\prime}> 0 ~
\qquad \D \ge 1 + s      \qquad  {\rm if}\, s^{\prime}=0.
\eea
In general $\Delta$ need not be integer, but our KK scalars
$\f_k$ transform in the irrep $[0,\D=k,0]$ in which the energy
and internal symmetry eigenvalues are locked, a condition which
gives a short representation of $SU(2,2|4)$.

\medskip

Each $\f_k(z)$ satisfies an equation of motion of the form
\be
\label{eom}
(\Box_{AdS} -M^2)\f_k  = \mbox{~nonlinear interaction terms}
\ee
The symbol $\Box$ is the invariant Laplacian on AdS$_5$,
\begin{ex}
  Obtain its explicit form from the metric in (\ref{adsxs}).
\end{ex}
\vspace{-1ex}
Each KK mode has a definite mass $M^2 = m^2/L^2$ and the
dimensionless $m^2$ is essentially determined by $SO(6)$ group
theory\footnote{In the simplest case of the dilaton field, whose
  linearized 10D field equation is uncoupled, the masses in the
  KK decomposition are simply given by the eigenvalues of the
  Laplacian on $S_5$, namely $m^2=k(k+4)$. The mass formula which
  follows differs because of the mixing discussed above.} to be
$m^2=k(k-4)$. Formulas of this type are important in the
AdS/CFT correspondence, because the energy quantum number,
$\D=k$ in this case, is identified with the scale dimension of
the dual operator in the $\N \! =4$ SYM theory. Later we will see
how this occurs.

\medskip

Since the superalgebra $SU(2,2|4)$ operates in the dimensionally
reduced bulk theory all KK modes obtained in the decomposition
process can be classified in representations of $SU(2,2|4)$. It
turns out that one gets exactly the set of short representation
discussed above for the composite operators of the field theory.
There is thus a 1:1 correspondence between the KK fields of Type
IIB $D=10$ supergravity and the composite operators (in short
representations) of $\N \! =4$ SYM theory. The $\f_k$ we have been
discussing are dual to the chiral primary operators $\tr X^k$.
Within the lowest $k=2$ multiplet, the 15 bulk gauge fields
$A_\m$ are dual to the conserved currents $\cj_i$~of the $SO(6)$
R-symmetry group, and the AdS$_5$ metric fluctuation $h_{\m\n}$
is dual to the field theory stress tensor $T_{ij}$.

\medskip

Critics of the AdS/CFT correspondence legitimately ask whether
results are due to dynamics or simply to symmetries. It thus must
be admitted that the operator duality just discussed was
essentially ensured by symmetry. The superalgebra representations
which can occur in the KK reduction of a gravity theory whose
``highest spin'' field is the metric tensor $g_{MN}$ are strongly
constrained. In the present case of $SU(2,2|4)$ there was no
choice but to obtain the series of short representations which
were found. So what we have uncovered so far is just the working
of the same symmetry algebra in two different physical settings,
a field theory without gravity in $4$ dimensions and a gravity
theory in $5$ dimensions. The more dynamical aspects of the
correspondence involve the interactions of the dimensionally
reduced bulk theory, $\eg$ the nonlinear terms in (\ref{eom}). It
is notoriously difficult to find these terms,\footnote{Except in
  subsectors such as that of the $15\, A_\m$ where non-abelian
  gauge invariance in $5$ dimensions governs the situation.} but
fortunately enough information has been obtained to give highly
non-trivial tests of AdS/CFT, some of which are discussed
later.

\subsection{AdS$_{d+1}$Basics---Geometry and Isometries}
\label{adssec}

We now begin our discussion of how to obtain information on
correlation functions in conformal field theory from classical
gravity. For applications to the ``realistic'' case of $\N
=4$ SYM theory, we will need details of Type IIB
supergravity, but we can learn a lot from toy models of the bulk
dynamics. In most cases we will use Euclidean signature models in
order to simplify the discussions and calculations.

Consider the Euclidean signature gravitational action in $d+1$
dimensions
\be
\label{toyact}
 S = {{-1} \over {16 \pi G}} \int d^{d+1}z \sqrt{g}(R -\Lambda)
\ee
with $\Lambda = -d(d-1)/L^2$. The maximally symmetric solution is Euclidean
AdS$_{d+1}$~which should be more properly called the hyperbolic space
$H_{d+1}$. The metric can be presented in various coordinate systems, each
of which brings out different features. For now we will use the upper
half-space description
\be
\label{toymet}
\begin{array}{ll}
ds^2 &= \lf{L^2}{z_0^2} (dz_0^2 + \sum_{i=1}^d dz_i^2 )\\
&= \bar{g}_{\m\n} dz^\m dz^\n
\end{array}
\ee
\begin{ex}
  Calculate the curvatures $R_{\m\n} = {{-d} \over {L^2}}
  \bar{g}_{\m\n}, \,\, R= {{-d(d+1)} \over {L^2}}$.
\end{ex}
\vspace{-1ex}
The space is conformally flat and one may think of the
coordinates as a $(d+1)$-dimensional Cartesian vector which we
will variously denote as $z_\m = (z_0,z_i) =(z_0,\vec{z})$, with
$z_0 >0$. Scalar products $z \cdot w$ and invariant squares $z^2$
involve a sum over all $d+1$ components, $\eg$ $z\cdot w =
\d^{\m\n} z_\m w_\n$.

The plane $z_0=0$ is at infinite geodesic distance from any
interior point. Yet it is technically a boundary. Data must be
specified there to obtain unique solutions of wave equations on
the spacetime, as we will see later. We will usually set the
scale $L=1$. Equivalently, all dimensionful quantities are
measured in units set by $L$.

The continuous isometry group of Euclidean AdS$_{d+1}$~is
$SO(d+1,1)$. This consists of rotations and translations of the
$z_i$ with $\half d(d-1) \,+d$ parameters, scale transformations
$z_\m \rightarrow \l z_\m$ with 1 parameter, and special conformal
transformations whose infinitesimal form is $\d z_\m = 2 c\cdot z
z_{\mu} - z^2 c_\m$, with $c_\m = (0,c_i)$ and thus $d$ parameters.
The total number of parameters is $(d+2)(d+1)/2$ which is the
dimension of the group $SO(d+1,1)$.
\vspace{-2ex}
\begin{ex}
  Verify explicitly the Killing condition $D_\m K_\n +D_\n
  K_\m=0$ for all infinitesimal transformations. The covariant
  derivative $D_\m$ includes the Christoffel connection for the
  metric (\ref{toymet}).
\end{ex}
\vspace{-3ex}
\begin{ex}
  (Extra credit !) Since AdS$_{d+1}$is conformally flat, it has
  the same conformal group $SO(d+2,1)$ as flat
  $(d+1)$-dimensional Euclidean space. There are $d+2$ additional
  conformal Killing vectors $\bar{K}_\m$ which satisfy  $D_\m
  \bar{K}_\n +D_\n\bar{K}_\m - {{2} \over {d+1}} \bar{g}_{\m\n}
  D^\r \bar{K}_\r=0$. Find them!
\end{ex}
\vspace{-2ex}

The AdS$_{d+1}$space also has the important {\bf discrete}
isometry of {\bf inversion}. We will discuss this in some detail
because it has applications to the computation of AdS/CFT
correlation functions and in conformal field theory itself. Under
inversion the coordinates $z_\m$ transform to new coordinates
$z'_\m$ by $z_\m = z'_\m / z'^2$, and it is not hard to show that
the line element (\ref{toymet}) is invariant under this
transformation.
\vspace{-2ex}
\begin{ex}
Show this explicitly.
\end{ex}
\vspace{-2ex}
Inversion is also a discrete conformal isometry of flat Euclidean
space.

The Jacobian of the transformation is also useful,
\be
\label{eq:1.14}
\begin{array}{l}
\lf{\p z_\m}{\p z'_\n} = \lf{1}{z'^2} J_{\m\n}(z) \\
 J_{\m\n}(z) = J_{\m\n}(z^{\prime}) = \d_{\m\n} - \lf{2z_\m z_\n}{z^2}\\
\end{array}
\ee
The Jacobian tells us how a tangent vector of the manifold
transforms under inversion.
\vspace{-2ex}
\begin{ex}
  View $J_{\m\n}(z)$ as a matrix. Show that it satisfies
  $J_{\m\r}(z) J_{\r\n}(z) = \d_{\m\n}$ and has $d$ eigenvalues
  $+1$ and 1 eigenvalue $-1$.
\end{ex}
\vspace{-2ex}
$J_{\m\n}$ is thus a matrix of the group $O(d+1)$ which is not in
the proper subgroup $SO(d+1)$. As an isometry, inversion is an
improper reflection which cannot be continuously connected to the
identity in $SO(d+1,1)$.
\vspace{-1ex}
\begin{ex}
  (Important but tedious !) Let $z_\m,\,w_\m$ denote two vectors
  with $z'_\m,\,w'_\m$ their images under inversion. Show that
\end{ex}
\vspace{-1ex}
\be
{{1} \over {(z-w)^2}} = {{(z^{\prime})^2\,(w^{\prime})^2} \over
  {(z^{\prime}-w^{\prime})^2}}
\ee
\be
J_{\m\n}(z-w) = J_{\m\m^{\prime}}(z^{\prime})
J_{\m^{\prime}\n^{\prime}}(z^{\prime}-w^{\prime})
J_{\n^{\prime}\n}(w^{\prime})
\ee
\subsection{Inversion and CFT Correlation Functions}

Although we have derived the properties of inversion in the
context of AdS$_{d+1}$, the manipulations are essentially the
same for flat d-dimensional Euclidean space. We simply replace
$z_\m, \,w_\m$ by $d$-vectors $x_i, \,y_i$ and take $x_i =
\lf{x^{\prime}_i}{x^{\prime 2}}$, etc. Inversion is now a
  conformal isometry and in most cases \footnote{Inversion is an
    improper reflection similar to parity and is not always a
    symmetry of a field theory action containing fermions.} a
  symmetry of CFT$_d$.  Under the inversion $x_i \rightarrow
  x'_i$, a scalar operator of scale dimension $\D$ is transformed
  as $\O_{\D}(x) \rightarrow \O_{\D}^{\prime}(x) = x'^{2\D}
  \O_{\D}(x')$. Correlation functions then transform covariantly
  under inversion, viz.
\be
\label{eq:1.17}
\begin{array}{l}
\< \O_{\D_1}(x_1) \O_{\D_2}(x_2) \cdots \O_{\D_n}(x_n)\>   \\
\qquad =
(x'_1)^{2\D_1} (x'_2)^{2\D_2} \cdots (x'_n)^{2\D_n}
\< \O_{\D_1}(x'_1) \O_{\D_2}(x'_2) \cdots \O_{\D_n}(x'_n)\>
\end{array}
\ee

It is well known that the spacetime forms of $2$- and $3$-point
functions are unique in any CFT$_d$, a fact which can be
established using the transformation law under inversion. These
forms are
\bea
\label{eq:1.18}
\< \O_{\D}(x) \O_{\D^{\prime}}(y) \> = \frac{c \d_{\D \D'}}{(x-y)^{2\D}}\\
\label{eq:1.19}
\< \O_{\D_1}(x) \O_{\D_2}(y) \O_{\D_3}(3)\> =
\frac{\tilde{c}}{(x-y)^{\D_{12}}(y-z)^{\D_{23}}(z-x)^{\D_{31}}}\\
\label{eq:1.20}
\D_{12} =\D_1 +\D_2 - \D_3, \mbox{~and cyclic permutations}
\nonumber
\eea
It follows immediately from the exercise above that they do
transform correctly.

\medskip

Operators such as conserved currents $\cj_i$ and the conserved
traceless stress tensor $T_{ij}$ are important in a CFT$_d$.
Under inversion $\cj_i(x) \rightarrow J_{ij}(x') x'^{2(d-1)}
\cj_j(x')$ with an analogous rule for $T_{ij}$. The $2$-point
function of a conserved current takes the form
\be \label{curcorr} \begin{array}{ll}
\<\cj_i(x)\cj_j(y)\> & \approx (\p_i \p_j -\Box \d_{ij}) \lf{1}
{(x-y)^{(2d-4)}} \\  & \sim \lf{J_{ij}(x-y)}{(x-y)^{(2d-2)}}
\end{array} \ee
The exercise above can be used to show this tensor does transform
correctly.  Here are some new exercises.
\vspace{-1ex}
\begin{ex}
  show that the second line in (\ref{curcorr}) follows from the
  manifestly conserved first form and obtain the missing
  coefficient.
\end{ex}
\vspace{-1ex}
\begin{ex}
  Use the projection operator $\pi{ij}=\p_i \p_j -\Box \d_{ij}$
  to write the 2-point correlator of the stress tensor and then
  convert to a form with manifestly correct inversion properties,
\end{ex}
\vspace{-1ex}
\be
\label{tcorr}
\begin{array}{ll}
\<T_{ij}(x)T_{kl}(y)\> &= \left[2\pi_{ij}\pi_{kl}-3(\pi_{ik}\pi_{jl}
  +\pi_{il}\pi_{jk})\right] \lf{c}{(x-y)^{(2d-4)}} \\ & \sim
\lf{J_{ik}(x-y)J_{jl}(x-y)+  k \leftrightarrow l - {{2} \over
    {d}} \d_{ij}\d_{kl}}{(x-y)^{2d}}
\end{array}
\ee
This form is unique. For $d\ge4$ there are two independent tensor
structures
for a 3-point function of conserved currents and three structures
for the 3-point function of $T_{ij}$. For more information on the
tensor structure of conformal amplitudes, see the work of Osborn
and collaborators, for example \cite{stanev,osborn}.

It is useful to mention that any finite special conformal
transformation can be expressed as a product of
(inversion)(translation)(inversion).

\medskip

\vspace{-2ex}
\begin{ex}
  Show that the finite transformation is $x_i \rightarrow (x_i +
  x^2 a_i)/(1 +2a\cdot x + a^2x^2)$. Show that the flat Euclidean
  line element transforms with a conformal factor under this
  transformation. Show that the commutator of an infinitesimal
  special conformal transformation and a translation involves a
  rotation plus scale transformation.
\end{ex}
\vspace{-2ex}

The behavior of amplitudes under rotations and translations
is rather trivial to test. Special conformal
symmetry is more difficult, but it can be reduced to inversion.
Thus the behavior under inversion essentially establishes
covariance under the full conformal group.

We will soon put the inversion to good use in our study of the
AdS/CFT correspondence, but we first need to discuss how the
dynamics of the correspondence works.

\subsection{AdS/CFT Amplitudes in a Toy Model}
\label{amplitudes}

Let us consider a toy model of a scalar field
$\phi(z)$ in an AdS$_{d+1}$~Euclidean signature background. The action
is
\be
\label{scact}
S=\frac {1}{8\pi G}\int d^{d+1}z \sqrt{\bar{g}} \left(\half
  \p_\m \phi \p^\m \phi +\half m^2\phi^2 + \frac{1}{3}b \phi^3 +
  \cdots \right)\\
\ee
We will outline the general prescription for correlation functions due to
Witten~\cite{witten} and then give further details. The first step is to
solve the non-linear classical field equations
\be
\label{sceom}
\frac{\d S}{\d \phi} = (-\Box + m^2)\phi + b \phi^2 +\cdots = 0\\
\ee
with the boundary condition
\be
\label{deltam}
\begin{array}{c}
\phi(z_0,\vec{z}) \underset{z_0\rarrow 0}{\longrightarrow} z_0^{d-\D}
\bar{\phi}(\vec{z})\\
\D = \frac{d}{2} + \half \sqrt{d^2+4m^2}
\end{array}
\ee
This is a modified Dirichlet boundary value problem with boundary
data $\bar{\phi}(\vec{z})$. The scaling rate $z_0^{d-\D}$ is that
of the leading Frobenius solution of the linearized
version\footnote{\label{restrict}To simplify the discussion we
  restrict throughout to the range $m^2 >-\frac{d^2}{4}$ and
  consider $\D >\half d$. See \cite{klebwit} for an extension to the
  region $\half d \ge \D \ge \half (d-2)$ close to the unitarity
  bound; see also \cite{Minces:2001zy}.}
of~(\ref{sceom})

\medskip

Exact solutions of the non-linear equation~(\ref{sceom}) with general
boundary data are beyond present ability, so we work with the iterative
solution
\bea \label{linsol}
\phi_0(z) = \int d^d \vec{x} K_\D (z_0, \vec{z}-\vec{x})
\bar{\phi} (\vec{x})\\
\phi(z) = \phi_0(z) + b \int d^{d+1}w \sqrt{\bar{g}} G(z,w) \phi_0^2(w) +
    \cdots  
\eea
The linear solution $\phi_0$ involves the bulk-to-boundary propagator
\bea
\label{bbdy}
K_\D (z_0,\vec{z}) =C_\D \biggl (\frac{z_0}{z_0^2 +\vec{z}^2} \biggr )^\D
\hskip .7in
C_\D = \frac{\Gamma(\D)}{\pi^{\frac{d}{2}} \Gamma(\D-\frac{d}{2})},
\eea
which satisfies $(\Box + m^2)K_\D (z_0,\vec{z}) =0$.  Interaction
terms require the bulk-to-bulk propagator $G(z,w)$ which
satisfies $(-\Box_z + m^2)G(z,w) = \d(z,w)/\sqrt{\bar{g}}$ and is
given by the hypergeometric function
\bea
G_\Delta (u) & = &
  \tilde C _\Delta (2u^{-1})^\Delta
  F \left(\Delta , \Delta -d +\half; 2\Delta -d +1; -2u^{-1} \right) 
  \\
  \tilde C _\Delta & = & {\Gamma (\Delta) \Gamma (\Delta -{d \over 2} +\half)
                        \over  (4\pi)^{(d+1)/2} \Gamma (2 \Delta
                        -d+1)} 
                        \nonumber \\
 u  & = & {{(z-w)^2} \over {2z_0 w_0}}.
\nonumber
\eea
This differs from the form given in Sec. 6.3 by a quadratic
hypergeometric transformation, see \cite{DF99}.

For several purposes in dealing with the AdS/CFT correspondence
it is appropriate to insert a cutoff at $z_0=\e$ in the bulk
geometry and consider a true Dirichlet problem at this boundary.
This is the situation of 19th century boundary value problems
where Green's formula gives a well known relation between $G$ and
$K$. Essentially $K$ is the normal derivative at the boundary of
$G$. The cutoff region has less symmetry than full AdS. Exact
expressions for $G$ and $K$ in terms of Bessel functions in the
$\vec{p}$-space conjugate to $\vec{z}$ are straightforward to
obtain, but the Fourier transform back to $z_0,\vec{z}$ is
unknown. See Sec. 8.5 below.

\medskip

The next step is to substitute the solution $\phi(z)$ into the
action~(\ref{scact}) to obtain the on-shell action
$S[\bar{\phi}]$ which is a functional of the boundary data. The
key dynamical statement of the AdS/CFT correspondence is that
$S[\bar{\phi}]$ is the generating functional for correlation
functions of the dual operator $\O (\vec{x})$ in the boundary
field theory, so that
\be
\< \O (\vec{x_1}) \cdots \O (\vec{x_n}) \> = (-)^{n-1} { \d
  \over \d \bar{\phi}(\vec{x_1})} \cdots {\d \over \d
\bar{\phi}(\vec{x_n})} S[\bar{\phi}] \mbox{\LARGE{$\vert$}}_{\bar{\phi}=0}
\ee

Another way to state things is that the boundary data for bulk fields
play the role of sources for dual field theory operators. The
integrals in the on-shell action diverge at the boundary and must be cut off
either as discussed above or by a related method \cite{hs,deharo}.
However we will proceed formally here.

\begin{fig}[htp]
\centering
\epsfxsize=1.5in
\epsfysize=1.5in
\epsffile{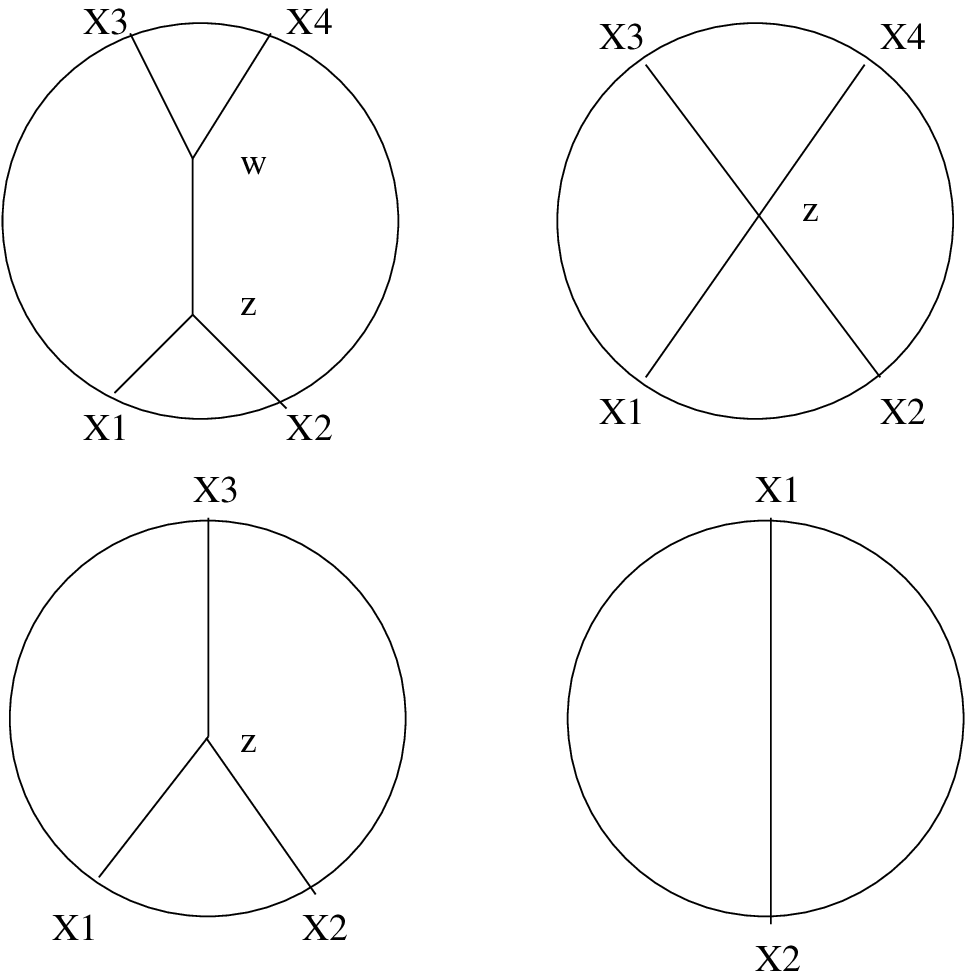}
\caption{Some Witten Diagrams}
\label{figwitten}
\label{fig:11}
\end{fig}

From the expansion of $S[\bar{\phi}]$ in powers of $\bar{\phi}$,
one obtains a diagrammatic algorithm (in terms of Witten
diagrams) for the correlation functions. Some examples are given
in Figure~\ref{figwitten}. In these diagrams the interior and
boundary of each disc denote the interior and boundary of the
AdS geometry. The rules for interpretation and computation
associated with the diagrams are as follows:

\vspace{-1ex}
\begin{alphal}
\item boundary points $\vec{x_i}$ are points of flat
  Euclidean$_d$ space where field theory operators are inserted.\vspace{-1ex}
\item bulk points $z,w \,\e\,AdS_{d+1}$ and are integrated as
  $\int d^{d+1}z \sqrt{\bar{g}(z)}$\vspace{-1ex}
\item Each bulk-to-boundary line carries a factor of $K_\D$ and
   each bulk-to-bulk line a factor of $G(z,w)$\vspace{-1ex}
\item An $n$-point vertex carries a coupling factor from the interaction
   terms of the bulk Lagrangian, e.g. $\L ={1 \over 3}b \phi^3 + {1 \over 4}
   c \phi^4 + \cdots$ with the same combinatoric weights as for Feynman-Wick
   diagrams. This is most clearly derived using the cutoff discussed
above.\vspace{-1ex}
\end{alphal}

Let us examine this construction more closely beginning with the linear
solution for bulk fields.
\vspace{-1ex}
\begin{ex}
Show that the linearized field equation can be written as
\be
\label{lineom}
(z_0^2 \partial_0^2 -(d-1)z_0\partial_0 + z_0^2 \nabla^2 -m^2)\phi =0
\ee
and that $K(z_0,\vec{z})$ given above is a solution. Plot $K(z_0,\vec{z})$
as a function of $|\vec{z}|$ for several fixed values of $z_0$. Note that
it becomes more and more like $\d(\vec{z})$ as $z_0 \rightarrow 0$.
\end{ex}
\vspace{-1ex}

The exercise shows that $\phi_0(z)$ in~(\ref{linsol}) is indeed a solution
of~(\ref{lineom}) and suggests that it satisfies the right boundary condition.
Let's verify that it has the correct normalization at the boundary.
Because of translation symmetry there is no loss of generality in taking
$\vec{z}=0$. We then have
\be
\begin{array}{ll}
\phi(z_0,0) & = C_\D \int d^d\vec{x} ({z_0 \over z_0^2
+\vec{x}^2})^\D \bar{\phi}(\vec{x})\\
&= C_\D z_0^{d-\D} \int d^d \vec{y} ({1 \over 1+\vec{y}^2})^\D
\bar{\phi}(z_0 \vec{y})\\
& \underset{z_0\rightarrow 0}{\longrightarrow} C_\D z_0^{d-\D} I_\D
\bar{\phi}(0)\\
I_\D & = \int {d^{d}\vec{y}\over (1+ \vec{y}^2)^{\D}}
\end{array}
\ee
Thus we do satisfy the boundary condition~(\ref{deltam})
provided that $C_\D = {1 \over I_\D}$ and the integral does indeed
give the value of $C_\D$ in~(\ref{bbdy}).

\subsection{How to calculate $3$-point correlation functions}
\label{3point}
Two-point correlations do not contain a bulk integral and turn
out to require a careful cutoff procedure which we discuss
later. For these reasons $3$-point functions are the prototype
case, and we now discuss them in some detail. The basic integral
to be done is:
\be \label{3pt}
\begin{array}{c}
A(\vec{x}, \vec{y}, \vec{z}) = \int \lf{d w_0 d^d
  \vec{w}}{w_0^{d+1}} \left(\lf{w_0}{(w-\vec{x})^2}
\right)^{\D_1} \left(\lf{w_0}{(w-\vec{y})^2} \right)^{\D_2}
\left(\lf{w_0}{(w-\vec{z})^2}\right)^{\D_3} \\ [2ex]
(w-\vec{x})^2 \equiv w_0^2 + (\vec{w} -
\vec{x})^2
\end{array}
\ee

Let us first illustrate the use of the method of inversion. We
change integration variable by $w_\m = w^{\prime}_\m / w^{\prime
  2}$ and at the same time refer boundary points to their
inverses, \ie $\vec{x} = \vec{x}^{~\prime}/(\vec{x}^{~\prime})^2$
and the same for $\vec{y}, \vec{z}$. The bulk-to-boundary
propagator transform very simply
\be
K_\D (w, \vec{x}) = |\vec{x}{\prime}|^{2\D} K_\D (w^{\prime},
\vec{x}^{~\prime})
\ee
with the prefactor associated with a field theory operator
$\O_\D(\vec{x})$ clearly in evidence. The AdS volume
element is invariant, \ie $d^{d+1}w/w_0^{d+1} =
d^{d+1}w^{\prime}/w_0^{\prime d+1}$ since inversion is an
isometry.
\vspace{-1ex}
\begin{ex}
  Use results of previous exercises to prove these important
  facts.
\end{ex}
\vspace{-1ex}
We then find that
\be
A(\vec{x}, \vec{y}, \vec{z}) = | \vec{x}^{~\prime} |^{2 \D_1} |
\vec{y}^{~\prime} |^{2 \D_2} | \vec{z}^{~\prime} |^{2 \D_3}
A(\vec{x}^{~\prime}, \vec{y}^{~\prime}, \vec{z}^{~\prime})
\ee
Thus the AdS/CFT procedure produces a $3$-point function which
transforms correctly under inversion. See~(\ref{eq:1.17}).

This is a very general property which holds for {\bf all}
AdS/CFT correlators. Suppose you wish to calculate $\< \cj_i^a
\cj_j^b \cj_k^c \>$. The Witten amplitude is the product (see
\cite{fmmr}) of $3$ vector bulk-to-boundary propagators, each
given by
\be
G_{\m i} (w, \vec{x}) = \frac{1}{2} c_d \frac{w_0^{d-1}}{(w -
  \vec{x})^{d-1}} J_{\m i} (w - \vec{x}),
\ee
in which the Jacobian ~(\ref{eq:1.14}) appears. The bulk indices
are contracted with a vertex rule from the Yang-Mills interaction
$f^{abc} A_{\m}^a A_{\nu}^b \p_\m A_{\nu}^c$. If you
try to do the change of variable in detail, you get a mess. But
the process is guaranteed to produce the correct inversion
factors for the conserved currents, namely $| \vec{x}^{~\prime}
|^{2(d-1)} J_{ii^{\prime}}(\vec{x}^{~\prime})$, etc, because
inversion is an isometry of AdS$_{d+1}$and all pieces of the
amplitude conspire to preserve this symmetry.
\vspace{-1ex}
\begin{ex}
Show that $G_{\m i}(w, \vec{x})$ satisfies the bulk Maxwell
equation
\be
\p_\m \sqrt{\bar{g}} \bar{g}^{\m \nu} \left( \p_\n G_{\r i}(w,
  \vec{x}) - \p_\r G_{\n i} (w, \vec{x}) \right)=0
\ee
where $\p_\m = \p/\p w_\m$. Express $G_{\m i}(w, \vec{x})$ in
terms of the inverted $G_{\m^{\prime} i^{\prime}}(w^{\prime},
\vec{x}^{~\prime})$.
\end{ex}

We can conclude that all AdS/CFT amplitudes are conformal
covariant! A transformation of the $SO(d+1,1)$ isometry group of
the bulk is dual to an $SO(d+1, 1)$ conformal transformation on
the boundary. Since there is a unique covariant form for scalar
$3$-point functions, given in~(\ref{eq:1.17}), the AdS/CFT integral
$A(\vec{x}, \vec{y}, \vec{z})$ is necessarily a constant multiple
of this form. Our exercise also shows conclusively that a scalar
field of AdS mass $m^2$~ is dual to an operator $\O_{\D}(\vec{x})$
of dimension $\D$~given by (\ref{deltam}).

\medskip

We still need to {\bf do} the bulk integral to obtain the
constant $\tilde{c}$. It is hard to do the integral in the
original form~(\ref{3pt}) because it contains $3$ denominators and
the restriction $w_0 > 0$. But we can simplify it by using
inversion in a somewhat different way. We use translation
symmetry to move the point $\vec{z} \longrightarrow 0$, \ie
$A(\vec{x}, \vec{y}, \vec{z}) = A(\vec{x}-\vec{z},
\vec{y}-\vec{z}, 0) \equiv A(\vec{u}, \vec{v}, 0)$. The integral
for $A(\vec{u}, \vec{v}, 0)$ is similar to~(\ref{3pt}) except that
the third propagator is simplified,
\be
\left(\lf{w_0}{(w- \vec{z})^2} \right)^{\D_3} \longrightarrow
\left(\lf{w_0}{w^2} \right)^{\D_3} = (w_0^{\prime})^{\D_3}.
\ee
There is no denominator in the inverted frame since $\vec{z}=0
\longrightarrow \vec{z}^{~\prime} = \infty$. After inversion the
integral is
\be
A(\vec{u}, \vec{v}, 0) =
\lf{1}{|\vec{u}|^{2\D_1}|\vec{v}|^{2\D_2}} \int \lf{d^{d+1}
  w^{\prime}}{(w_0^{\prime})^{d+1}}
  \left(\lf{w_0^{\prime}}{(w^{\prime}-\vec{u}^{~\prime})^2}\right)^{\D_1}
  \left(\lf{w_0^{\prime}}{(w^{\prime}-\vec{v}^{~\prime})^2}\right)^{\D_2}
(w_0^{\prime})^{\D_3}
\ee
The integral can now be done by conventional Feynman parameter
methods, which give
\be
\begin{array}{l}
A(\vec{u}, \vec{v}, 0) = \lf{1}{|\vec{u}|^{2\D_1}|\vec{v}|^{2\D_2}}
\lf{a}{|\vec{u}^{~\prime} - \vec{v}^{~\prime}|^{\D_1 + \D_2 -
    \D_3}} \\[2ex]
a = \lf{\pi^{d/2}}{2} \lf{\Gamma(\frac{1}{2}(\D_1 + \D_2 -
  \D_3)) \Gamma(\frac{1}{2}(\D_2 + \D_3 - \D_1))
  \Gamma(\frac{1}{2}(\D_3 + \D_1 - \D_2)) }{\Gamma(\D_1)
  \Gamma(\D_2) \Gamma(\D_3)} \Gamma[\frac{1}{2} (\D_1+\D_2+\D_3 -d)]
\end{array}
\ee
\vspace{-1ex}
\begin{ex}
Repristinate the original variables $\vec{x}, \vec{y}, \vec{z}$ to obtain
the form~(\ref{eq:1.19}) with $\tilde{c}=a$.
\end{ex}

The major application of this result was already discussed in
Sec. 6.7. ~A Princeton group~\cite{lmrs} obtained the cubic
couplings $b_{klm}$ of the Type IIB supergravity modes on \AdS\ which are
dual to the chiral primary operators
$\tr X ^k$, etc. of $\N \!=\!4$ SYM theory. They combined these
couplings with the Witten integral above and observed that the
AdS/CFT prediction
\be
\< \tr X^k (\vec{x}) \tr X^l (\vec{y}) \tr
X^m(\vec{z})\> = b_{klm} c_kc_lc_mA(\vec{x}, \vec{y}, \vec{z})
\ee
for the large $N$, large $\lambda$ supergravity limit agreed with
the {\bf free field} Feynman amplitude for these correlators. They
conjectured a broader non-renormalization property. It was
subsequently confirmed in weak coupling studies in the field
theory that order $g^2, g^4$ and non-perturbative instanton
contributions to these correlations vanished for all $N$ and all
gauge groups. General all orders arguments for
non-renormalization have also been developed. The
non-renormalization of $3$-point functions of chiral primaries
(and their descendents) was a surprise and the first major
new result about $\N=4$ SYM obtained from AdS/CFT. (See the
references cited in Sec 6.7.)

\subsection{$2$-point functions}
\label{2point}

This is an important case, but more delicate, since a cutoff
procedure is required to obtain a concrete result from the formal
integral expression. Since $3$-point functions do not require a
cutoff, one way to bypass this problem is to study the $3$-point
function $\<\cj_i(z) \O_\D(x) \O_\D^{\ast}(y)\>$ of a conserved
current and a scalar operator $\O_\D(x)$ assumed to carry one unit
of $U(1)$ charge.\footnote{When no ambiguity arises we will
denote boundary points by $x,y,z$ etc. rather than
$\vec{x},\vec{y}, \vec{z}.$}
The Ward identity relates $\<\cj_i \O_\D
\O_\D^{\ast}\>$ to $\<\O_\D \O_\D^{\ast}\>$. There is a unique
conformal tensor for $\<\cj_i \O_\D \O_\D^{\ast}\>$ in any CFT$_d$,
namely
\be
\label{eq:4.43}
\< J_i (z) \O_\D (x) \O_\D^{\ast} (y)\> = -i \x
\lf{1}{(x-y)^{2\D-d+2}} \lf{1}{(x-z)^{d-2}(y-z)^{d-2}} \left[
  \lf{(x-z)i}{(x-z)^2} - \lf{(y-z)i}{(y-z)^2} \right]
\ee
and the Ward identity is
\be
\label{eq:4.44}
\begin{array}{c}
\lf{\p}{\p z_i} \< J_i(z) \O_\D(x) \O_\D^{\ast}(y) \> = i
[\delta(x-z) - \delta(y-z)] \< \O_\D(x) \O_\D^{\ast}(y)\> \\
= i[\delta(x-z)-\delta(y-z)] \lf{2 \pi^{d/2}}{\Gamma (d/2)}
\x \lf{1}{(x-y)^{2 \D}}
\end{array}
\ee
\vspace{-1ex}
\begin{ex}
Derive~(\ref{eq:4.44}) from~(\ref{eq:4.43}).
\end{ex}
\vspace{-1ex}
To implement the gravity calculation of $\<\cj_i \O_\D
\O_\D^{\ast}\>$ we extend the bulk toy model~(\ref{scact}) to
include a $U(1)$ gauge coupling
\be
L = \frac{1}{4} F_{\m \nu}F^{\m \nu} + \bar{g}^{\m \nu} ( \p_\m
+ iA_\m) \phi^{\ast} (\p_{\nu} - iA_{\nu})\phi
\ee
In application to the duality between Type IIB sugra and $\N=4$ SYM, the
$U(1)$ would be interpreted as a subgroup of the $SO(6)$ R-symmetry group.
The cubic vertex leads to the AdS integral
\be
\< \cj_i (z) \O_\D (x) \O_\D^{\ast}(y) \> = -i \int
\frac{d^{d+1}w}{w_0^{d+1}} G_{\m i}(w, \vec{z}) w^2_0 K_\D (w,
\vec{x}) \stackrel{\longleftrightarrow}{\mbox{\small{$\frac{\p}{\p w_\m}$}}}
K_\D(w, \vec{y}).
\ee
\vspace{-1ex}
\begin{ex}
The integral can be done by the inversion technique, please do it.
\end{ex}
\vspace{-1ex}
The result is the tensor form~(\ref{eq:4.43})
with coefficient
\be
\x = \lf{(\D-d/2) \Gamma (\frac{d}{2}) \Gamma (\D)}{\pi^{d/2}
  \Gamma (\D-d/2)}
\ee
Using~(\ref{eq:4.44}) we thus obtain the $2$-point function
\be \label{fromward}
\< \O_\D(x) \O^{\ast}_\D(y)\> = \lf{(2 \D-d) \Gamma(\D)}{\pi^{d/2}
  \Gamma(\D - d/2)} \lf{1}{(x-y)^{2\D}}
\ee

We now discuss a more direct computation \cite{Gubs,fmmr} of $2$-point
correlators from a Dirichlet boundary value problem in the AdS
bulk geometry with cutoff at $z_0=\e$. This method illustrates the use of
a systematic cutoff, and it may be applied to (some) $2$-point
functions in holographic RG  flows for which the $3$-point
function $\< \cj_i \O_\D \O_\D^{\ast}\> $ cannot readily be
calculated.

The goal is to obtain a solution of the linear problem
\be
\label{eq:4.51}
\begin{array}{c}
(\Box - m^2) \phi(z_0, \vec{z})=0 \\[2ex]
\phi(\epsilon, \vec{z})= \bar{\phi}(\vec{z})
\end{array}
\ee
The result will be substituted in the bilinear part of the toy
model action to obtain the on-shell action. After partial
integration we obtain the boundary integral
\be
S[\bar{\phi}] = \frac{1}{2 \epsilon^{d-1}} \int d^d \vec{z}
\bar{\phi}(\vec{z}) \p_0 \phi(\epsilon, \vec{z})
\ee

Since the cutoff region $z_0 \geq \epsilon$ does not have the
full symmetry of AdS, an exact solution of the Dirichlet
problem is impossible in $x$-space, so we work in
$p$-space. Using the Fourier transform
\be
\phi(z_0, \vec{z}) = \int d^d \vec{p} e^{i \vec{p} \cdot \vec{z}}
\phi(z_0, \vec{p})
\ee
we find the boundary value problem
\be
\begin{array}{c}
[z_0^2 \p_0^2 - (d-1) z_0 \p_0 - (p^2 z_0^2 + m^2)] \phi
(z_0, \vec{p}) = 0 \\[2ex]
\phi(\epsilon, \vec{p}) = \bar{\phi}(\vec{p})
\end{array}
\ee
where $\bar{\phi}(\vec{p})$ is the transform of the boundary
data. The differential equation is essentially Bessel's equation,
and we choose the solution involving the function
$z_0^{d/2}K_\nu(p z_0)$, where $\nu =\D - d/2, p=|\vec{p}|$,
which is exponentially damped as $z_0 \rightarrow \infty$ and
behaves as $z_0^{d - \D}$ as $z_0 \rightarrow 0$. The second
solution $z_0^{d/2} I_{\nu} (p z_0)$ is rejected because it
increases exponentially in the deep interior. The normalized
solution of the boundary value problem is then
\be
\phi (z_0, \vec{p}) = \lf{z_0^{d/2} K_\nu (pz_0)}{\epsilon^{d/2}
  K_\nu(p \epsilon)} \bar{\phi}(\vec{p}),
\ee
The on-shell action in $p$-space is
\be
S[\bar{\phi}] = \lf{1}{2 \epsilon^{d-1}} \int d^d p d^d q (2
\pi)^d \delta(\vec{p} + \vec{q}) \phi(\epsilon, \vec{p}) \p_0
\phi(\epsilon, q)
\ee
which leads to the cutoff correlation function
\be
\label{2pt1}
\begin{array}{ll}
\< \O_\D (\vec{p}) \O_\D (\vec{q}) \>_{\epsilon} & = -
\lf{\delta^2 S}{\delta \bar{\phi} (\vec{p}) \delta \bar{\phi}(p)}
  \\[2ex]
& = -\lf{(2 \pi)^d \delta(\vec{p}+\vec{q})}{\epsilon^{d-1}}
\lf{d}{d \e} \ln (\e^{d/2} K_\nu(p \e))
\end{array}
\ee
To extract a physical result, we need the boundary asymptotics of
the Bessel function $K_\nu(p \e)$. The values of $\nu=\D-d/2$
which occur in most applications of AdS/CFT are integer. The
asymptotics were worked out for continuous $\nu$ in the Appendix
of~\cite{fmmr} with an analytic continuation to the final answer.
Here we assume integer $\nu$, although an analytic continuation
will be necessary to define Fourier transform to $x$-space. The
behavior of $K_{\nu}(u)$ near $u=0$ can be obtained from a
standard compendium on special functions such as \cite{grad}. For
integer $\nu$, the result can be written schematically as
\be
K_{\nu}(u)=u^{-\nu}(a_0 + a_1u^2 +a_2u^4 + \cdots) + u^{\nu} ln(u)\,(b_0 +
b_1u^2 + b_2u^4 + \cdots)
\ee
where the $a_i, b_i$ are functions of $\nu$ given in \cite{grad}.
This expansion may be used to compute the right side
of~(\ref{2pt1}) leading to
\be
\begin{array}{ll}
\< \O_\D(\vec{p}) \O_\D(\vec{q})\>_\e = \lf{(2 \pi)^d
  \delta(\vec{p}+\vec{q})}{\e^d} & [-\frac{d}{2} +
  \nu(1+c_2 \e^2 p^2+ c_4 \e^4 p^4 + \cdots) \\
& -\lf{2\nu b_0}{a_0} \e^{2\nu} p^{2\nu} ln(p \e)(1+ d_2 \e^2p^2 +
\cdots)]
\end{array}
\ee
where the new constants $c_i, d_i$ are simply related to $a_i,
b_i$. From \cite{grad} we obtain the ratio
\be
{2\n b_0 \over a_0} = {(-)^{(\n-1)} \over 2^{(2\n-2)}\Gamma(\n)^2}
\ee
which is the only information explicitly needed.
\vspace{2ex}

This formula is quite important for applications of AdS/CFT
ideas to both conformal field theories and RG  flows where
similar formulas appear. The physics is obtained in the limit as
$\e \rightarrow 0$, and we scale out the factor $\e^{2(\D-d)}$
which corresponds to the change from the true Dirichlet boundary
condition to the modified form~(\ref{deltam}) for the full AdS
space. We also drop the conventional momentum conservation factor
$(2 \pi)^d \delta(\vec{p}+ \vec{q})$ and study
\be \label{2pt2}
\< \O_\D(p) \O_\D(-p) \> = \lf{\b_0 + \b_1 \e^2 p^2 + \cdots +
  \b_{\nu}(\e p)^{2(\nu-1)}}{\e^{2\D-d}} - \lf{2\nu b_0}{a_0} p^{2
  \nu} \ln (p \e) + \O(\e^2)
\ee

The first part of this expression is a sum of non-negative
integer powers $p^{2m}$ with singular coefficients in $\e$. The
Fourier transform of $p^{2m}$ is $\Box^m
\delta(\vec{x}-\vec{y})$, a pure contact term in the
$\vec{x}$-space correlation. Such terms are usually physically
uninteresting and scheme dependent in quantum field
theory. Indeed it is easy to see that the singular powers
$\e^{2(m-\D)+d}$ carried by the terms corresponds to their
dependence on the ultraviolet cutoff $\Lambda^{2(\D - m) - d}$ in
a field theory calculation. This gives rise to the important
observation that the $\e-$cutoff in AdS space which cuts off
long distance effects in the bulk corresponds to an ultraviolet
cutoff in field theory.
Henceforth we drop the polynomial contact terms in (\ref{2pt2}).

The physical $p$-space correlator is then given by
\be
\< \O_\D(p) \O_\D(-p) \> = -\frac{2\nu b_0}{a_0} p^{2 \nu} \ln p.
\ee
This has an absorptive part which is determined by unitarity in
field theory. Its Fourier transform is proportional to
$1/(x-y)^{2 \D}$ which is the correct CFT behavior for $\<
\O_\D \O_\D\>$. The precise constant can be obtained using
differential regularization~\cite{diffreg} or by analytic
continuation in $\nu$ from the region where the Fourier transform
is defined.

The result agrees exactly with the $2$-point function calculated
from the Ward identity in~(\ref{fromward}).

\subsection{Key AdS/CFT results for $\N \!=\!4$ SYM
  and CFT$_d$ correlators.}

We can now summarize the important results discussed in this chapter and
earlier ones for CFT$_d$ correlation functions from the AdS/CFT
correspondence.
\vspace{-1ex}
\begin{ronums}
\setlength{\itemsep}{-1ex}
\item the non-renormalization of $\< \tr X^k \tr X^l \tr
  X^m\>$ in $\N \!=\!4$ SYM theory
\item $4$-point functions are less constrained than $2$- and $3$-
  point functions in any CFT. In general they contain arbitrary
  functions $F(\x,\eta)$ of two invariant variables, the cross
  ratios
\be
\x = \lf{x_{13}^2 x_{24}^2}{x_{12}^2 x_{34}^2} \hspace{3em}
\eta=\lf{x_{14}^2 x_{24}^2}{x_{12}^2 x_{34}^2}  \hspace{3em} x_{ij}
= x_i - x_j
\ee
One way to extract the physics of $4$-point functions is to use the
operator product expansion. This is written
\be
\O_\D (x) \O_{\D^{\prime}} (y) \underset{x \rarrow
  y}{\longrightarrow} \sum_p \lf{a_{\D \D^{\prime}
    \D_p}}{(x-y)^{\D+\D^{\prime}-\D_p}} \O_{\D_p}(y)
\ee
which is interpreted to mean that at short distance inside any
correlation function, the product of two operators acts as a sum
of other local operators with power coefficients. For simplicity
we have indicated only the contributions of primary operators.
Thus, in the limit where $|x_{12}|, |x_{34}| \ll |x_{13}|$, a
$4$-point function must factor as
\be
\hspace{-2ex}
\< \O_{\D_1} (x_1) \O_{\D_2} (x_2) \O_{\D_3} (x_3) \O_{\D_4}
(x_4) \> \approx \sum_p \lf{a_{12p}}{(x_{12})^{\D_1+\D_2-\D_p}}
\lf{c_p}{(x_{13})^{2 \D_p}}
\lf{a_{34p}}{(x_{34})^{\D_3+\D_d-\D_p}}
\ee
One must expect that AdS/CFT amplitudes satisfy this property
and indeed they do in a remarkably simple way. The amplitude of a
Witten diagram for exchange of the bulk field $\phi_p(z)$ dual to
$\O_{\D_p}(\vec{z})$ factors with the correct coefficients $c_p,
a_{12p}, a_{34p}$ determined from $2$- and $3$-point functions.
This holds for singular powers, \eg $\D_1 + \D_2 - \D_p > 0$.

\medskip

The
AdS/CFT amplitude also contains a $\ln (\x)$ term in its short
distance asymptotics. This is the level of the $OPE$ at which
$\O_p = : \O_{\D_1} (\vec{x}) \O_{\D_2} (\vec{y}):$
contributes. In $\N=4$ SYM theory the normal product is a double
trace operator, \eg $:\tr X^k (y) \tr X^l(y):$,
which has components in irreps of $SO(6)$ contained in the direct
product $(0, k, 0) \otimes (0, l, 0)$. The irreducible components
are generically primaries of long representations of $SU(2,
2|4)$. Their scale dimensions are not fixed, and have a large $N$
expansion of the form $\D_{kl} = k+l+\gamma_{kl}/N^2+\cdots$. The
contribution $\D \gamma_{kl}$ can be read from the $\ln (\x)$ term of
the $4$-point function. It is a strong coupling prediction of
AdS/CFT, which cannot yet be checked by field theoretic
methods.

\item Another surprising fact about $\N=4$ SYM correlators suggested by
the AdS/CFT correspondence is that {\bf extremal} $n$-point functions
are not renormalized. The extremal condition for $4$-point
functions is $\D_1 = \D_2 + \D_3 + \D_4$. The name extremal comes
from the fact that the correlator vanishes by $SO(6)$ symmetry
for any larger value of $\D_1$. As discussed in detail in
Secs. 6.8 and 6.9, the absence of radiative corrections was
suggested by the form of the supergravity couplings and Witten
integrals. This prediction was confirmed by weak coupling
calculation and general arguments in field theory. Field theory then
suggested that next-to-external correlators $(\D_1 = \D_2 + \D_3
+ \D_4 -2)$ were also not renormalized, and this was subsequently
verified by AdS/CFT methods.
\end{ronums}
\vspace{-1ex}

\medskip

It is clear that the AdS/CFT correspondence is a new principle
which stimulated an interplay of work involving both supergravity and
field theory methods. As a result we have much new information
about the $N\!=\!4$ SYM theory. It confirms that AdS/CFT has
quantitative predictive power, so we can go ahead and apply it in
other settings.

\vfill\eject

\section{Holographic Renormalization Group Flows}
\setcounter{equation}{0}

We have already seen that AdS/CFT has taught us a great deal of
useful information about $\N \! =4$~SYM theory as a CFT$_4$. But
years of elegant work in CFT$_2$ has taught us to consider both
the pure conformal theory and its deformation by relevant
operators. The deformed theory exhibits RG flows in the space of
coupling constants of the relevant deformations. For general dimension
$d$ we can also consider the CFT$_d$ perturbed by
relevant operators. For $\N \! =4$ SYM theory, the perturbed
Lagrangian would take the form
\be
\L = \L_{\N \! = 4} + \half m_{ij}^2 \tr X^i X^j +
\half M_{ab} \tr \psi^a \psi^b + b_{ijk} \tr X^i X^j X^k.
\ee
For $d>2$ there is the additional option of Coulomb and Higgs
phases in which gauge symmetry is spontaneously broken. The
Lagrangian is not changed, but certain operators acquire vacuum
expectation values, e.g. $\<X^i\> \ne 0$ in $\N \! =4$ SYM. In
all these
cases conformal symmetry is broken because a scale is introduced.
The resulting theories have the symmetry of the Poincar\'{e}
group in $d$ dimensions which is smaller than the conformal group
$SO(1,d+1)$. Our purpose in this chapter is to explore the
description of such theories using $D=d+1$ dimensional gravity.  We will
focus on relevant operator deformations.

\subsection{Basics of RG flows in a toy model}
\label{flowbasic}
The basic ideas for the holographic description of field theories
with RG flow were presented in \cite{gppz1,dz}.
We will discuss these ideas in a simple model in which Euclidean
$(d+1)$-dimensional gravity interacts with a single bulk scalar
field with action

\be
S = \lf{1}{4 \pi G} \int d^{d+1} \sqrt{g} \left[-{1 \over 4} R + \half
\p_{\m} \phi \p^{\m} \phi + V(\phi)\right]\\
\ee
We henceforth choose units in which $4\pi G =1$. In these units
$\phi$ is dimensionless and all terms in the Lagrangian have
dimension $2$. We envisage a potential $V(\phi)$ which has one or
more critical points, \ie $V^{\prime}(\phi_i) = 0$, at which
$V(\phi_i) < 0$. We consider both maxima and minima. See
Figure~\ref{figpotential}.

\begin{fig}[htp]
\centering
\epsfxsize=1.5in
\epsfysize=1.5in
\epsffile{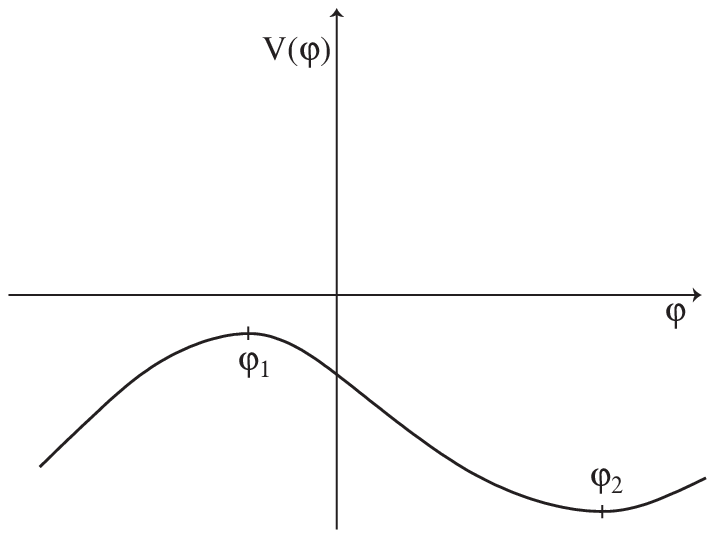}
\caption{Potential $V(\phi)$}
\label{figpotential}
\label{fig:12}
\end{fig}

The Euler-Lagrange equations of motion of our system are
\be
\label{eom1}
{1 \over \sqrt{g}} \p_{\m} (\sqrt{g} g^{\m\n} \p_{\nu} \phi)
- V^{\prime}(\phi) = 0
\ee
\be
\label{eom2}
R_{\m\n} - \half g_{\m\n} R = 2 \left[\p_\m \phi \p_\n \phi
  -g_{\m\n}\left(\half (\p \phi)^2 +V(\phi)\right)\right] =
2T_{\m\n}
\ee
For each critical point $\phi_i$ there is a trivial solution of
the scalar equation, namely $\phi(z) \equiv \phi_i$. The Einstein
equation then reduces to
\be
R_{\m\n} - \half g_{\m\n} R = -2 g_{\m\n} V(\phi_i).
\ee
This is equivalent to the Einstein equation of the action
(\ref{toyact}) if we identify $\Lambda_i =4V(\phi_i)= -d(d-1)/L_i^2$.
Thus constant scalar fields with AdS$_{d+1}$geometries of scale
$L_i$ are solutions of our model. We refer to them as critical solutions.

\medskip

However, more general solutions in which the scalar field is not
constant are needed to describe the gravity duals of RG flows in
field theory. Since the symmetries must match on both sides of
the duality, we look for solutions of the
$D=d+1$-dimensional bulk equations with $d$-dimensional
Poincar\'{e} symmetry. The most general such configuration is
\be
\label{domwall}
\begin{array}{ll}
ds^2 & = e^{2A(r)} \d_{ij} d x^i d x^j + d r^2\\
\phi & = \phi(r)
\end{array}
\ee
This is known as the domain wall ansatz. The coordinates separate
into a radial coordinate $r$ plus $d$ transverse coordinates
$x^i$ with manifest Poincar\'{e} symmetry. Several equivalent
forms which differ only by change of radial coordinate also
appear in the literature.

\medskip

Domain wall metrics have several modern applications, and it is
worth outlining a method to compute the connection and curvature.
Symbolic manipulation programs are very useful for
this purpose, but analytic methods can also be useful, and we
discuss a method which uses the Cartan structure equations.
A similar method works quite well for brane metrics such as
(\ref{d3}). One
proceeds as follows using the notation of differential forms:
\vspace{-1ex}
\begin{enumerate}
\setlength{\itemsep}{-1ex}
\item The first step is to choose a basis of frame 1-forms $e^a =
  e^a_{\m} dx^{\m}$ such that the metric is given by the inner
  product $ds^2 = e^a \d_{ab} e^b$.
\item The torsion-free connection $1$-form is then defined by
  $de^a + \omega^{ab} \wedge e^b = 0$ with the condition
  $\omega^{ab} = -\omega^{ba}$. The connection is valued in the Lie
  algebra of $SO(d+1)$.
\item The curvature 2-form is
\be
R^{ab} = d \omega^{ab} + \omega^{ac} \wedge \omega^{cb} = \half
R^{ab}_{cd} ~e^c \wedge e^d.
\ee
\end{enumerate}
\vspace{-1ex}
The general formulas for $\omega^{ab}_{\m}$ and $R^{ab}_{\m\n}$
which appear in textbooks can be deduced from these definitions.
However, for a reasonably simple metric ansatz and suitable
choice of frame, it is frequently more convenient to use the
definitions and compute directly. It takes some experience to
learn to use the $d$ and $\wedge$ operations efficiently. One
must also remember to convert from frame to coordinate components
of the curvature as needed.

\medskip

For the domain wall metric a convenient frame is given by the
transverse forms $e^{\hat{i}} = e^{A(r)} d x^i, \,\, i = 1 \cdots
d$, and the radial form $e^D = dr$.
\vspace{-1ex}
\begin{ex}
  Use the Cartan structure equations with the frame $1$-forms above
  to obtain the domain wall connection forms:
\be
\omega^{\hat{i} \hat{j}} = 0 \,\,\,\, \omega^{D\hat{i}} =
A^{\prime}(r) e^{\hat{i}}
\ee
Find next the curvature $2$-forms:
\be
\begin{array}{ll}
R^{\hat{i}\hat{j}} & = -A^{\prime 2} e^{\hat{i}} \wedge
e^{\hat{j}} \\
R^{\hat{i}D} & = - (A^{\prime \prime} + A^{\prime 2}) e^{\hat{i}} \wedge e^D
\end{array}
\ee
Next obtain the curvature tensor (with coordinate indices)
\be
\begin{array}{ll}
R^{ij}_{kl} & = -A^{\prime 2}\left(\d^i_k \d^j_l - \d^i_l \d^j_k\right)\\[2ex]
R^{iD}_{jD} & = - (A^{\prime \prime} + A^{\prime 2}) \d^i_j \\[2ex]
R^{ij}_{kD} & = 0
\end{array}
\ee
The final task is to find the Ricci tensor components
\be
\label{ricci}
\begin{array}{ll}
R_{ij} & = -e^{2A}(A^{\prime \prime} + d A^{\prime 2}) \d_{ij} \\[2ex]
R_{DD} & = -d (A^{\prime \prime} +  A^{\prime})^2 \\[2ex]
R_{iD} & = 0
\end{array}
\ee
\end{ex}
\vspace{-1ex}
\begin{ex}
If you still have some energy compute the non-vanishing
components of the Christoffel connection, namely
\be \G^D_{ij} = -e^{2A} A^{\prime} \d_{ij} \qquad
\G^i_{jD}=A^{\prime}\d^i_j
\ee
\end{ex}
\vspace{-1ex}
The fact that certain connection and curvature components vanish
could have been seen in advance, since there are no possible
Poincar\'{e} invariant tensors with the appropriate symmetries.
We can introduce a new radial coordinate $z$, defined by
$\frac{dz}{dr}=e^{-A(r)}$. This brings the domain wall metric to
conformally flat form. It's Weyl tensor thus vanishes.

\medskip

We now ask readers to manipulate the Einstein equation $G^\m_\n
\equiv R^\m_\n -\half \d^\m_\n R = 2 T^\m_\n$ for the domain wall
and deduce a simple condition on the scale factor $A(r)$.
\vspace{-1ex}
\begin{ex}
Deduce that
\be
\begin{array}{l}
G^D_D = \lf{d(d-1)}{2} A^{\prime 2} = 2 T^D_D \\[2ex]
G^i_j = \d^i_j (d-1)\left(A^{\prime \prime} + \frac{1}{2}
  d A^{\prime 2}\right) = 2 T^i_j
\end{array}
\ee
Compute $G^i_i - G^D_D$ for any fixed diagonal component (no sum on $i$)and
deduce that
\be
\label{adoub}
A^{\prime \prime} ={2 \over d-1}\left(T^i_i - T^D_D\right) = - {2
  \over d-1} \phi^{\prime 2}
\ee
\end{ex}

Thus we certainly have $A^{\prime \prime} < 0$ in the dynamics of
the toy model. However there is a much more general result,
namely $T^i_i - T^D_D < 0$ for any Poincar\'{e} invariant matter
configuration in all conventional models for the bulk dynamics,
for example, several scalars with non-linear $\s$-model kinetic
term. In Lorentzian signature, the condition above is one of the
standard energy conditions of general relativity. Later we will
see the significance of the fact that $A^{\prime \prime}(r) < 0.$
\begin{ex}
  Complete the analysis of the Einstein and scalar equations of
  motion for the domain wall and obtain the equations
\be
\label{domwalleq}
\begin{array}{ll}
A^{\prime 2} & = \lf{2}{d(d-1)}[\phi^{\prime 2} - 2V(\phi)] \\[2ex]
\phi^{\prime \prime} + d A^{\prime} \phi^{\prime} & = {dV(\phi)
  \over d\phi}
\end{array}
\ee
\end{ex}
\vspace{-1ex}
It is frequently the case that the set of equations obtained from
a given ansatz for a gravity-matter system is not independent
because of the Bianchi identity. Indeed in our system the
derivative of the $A^{\prime 2}$ equation combines simply with
the the scalar equation to give (\ref{adoub}). We can thus view
the system (\ref{domwalleq}) as independent.

\medskip

It is easy to see how the previously discussed critical solutions
fit into the domain wall framework. At each critical point
$\phi_i$ of the potential, the scalar equation is satisfied by
$\phi(r) \equiv \phi_i$. The $A^{\prime 2}$ equation then gives
$A(r) = \pm {r \over L_i} +a_0$. The integration constant $a_0$
has no significance since it can be eliminated by scaling the
coordinates $x^i$ in (\ref{domwall}). The sign
above is a matter of convention and we choose the positive sign.
The metric (\ref{domwall}) is then equivalent to our previous
description of AdS$_{d+1}$with the change of radial coordinate $z_0 =
L_i e^{-{r \over L_i}}$. With this sign convention we find that $r
\rightarrow + \infty$ is the boundary region and $r \rightarrow -
\infty$ is the deep interior.

\medskip

Our main goal now is to discuss more general solutions of the
system (\ref{domwalleq}) in a potential of the type shown in
Figure~\ref{figpotential}. We are interested in solutions which
interpolate between two critical points, producing a domain wall
geometry which approaches the boundary region of an AdS space
with scale $L_1$ as $r \rarrow + \infty$ and the deep interior of
another AdS with scale $L_2$ as $r \rarrow - \infty$. Such
geometries are dual to field theories with RG  flow.

\medskip

To develop this interpretation let's first look at the quadratic
approximation to the potential near a critical point,
\be \label{vquad}
V(\phi) \approx V(\phi_i) +\half {m_i^2 \over L_i^2}h^2,
\ee
where we use the fluctuation $h = \phi -\phi_i$ and the scaled
mass $m_i^2 = L_i^2 V^{\prime \prime}(\phi_i)$ with $V(\phi_i) =
-d(d-1)/4L_i^2$. Let's recall the basic AdS/CFT idea that the
boundary data for a bulk scalar field is the source for an
operator in quantum field theory. We apply this to the
fluctuation $h(r,\vec{x})$ which will be interpreted as the bulk
dual of an operator $\O_\D(\vec{x})$ whose scale dimension is
related to the mass $m_i^2$ by (\ref{deltam}). Given the discussion of
Sec. \ref{amplitudes} it is reasonable to suppose that a general solution
of the non-linear scalar equation of motion (\ref{eom1}) will
approach the critical point with the following boundary
asymptotics for the fluctuation,
\be
\begin{array}{ll}
h(r,\vec{x}) & \underset{r\rightarrow \infty}{\longrightarrow} e^{(\D-d)r}
\tilde{h}(\vec{x})\\
& = e^{(\D-d)r}(\bar{\phi} + \bar{h}(\vec{x})).
\end{array}
\ee
in which $\tilde{h}(\vec{x})$ contains $\bar{\phi}$, describing
the boundary behavior of the domain wall profile plus a
remainder $\bar{h}(\vec{x})$. We can form the on-shell action
$S[\bar{\phi} +\bar{h}]$ which is a functional of this boundary
data.\footnote{A
  complete discussion should include the bulk metric which is
  coupled to $\phi(r,\vec{x})$. We have omitted this for
  simplicity. See \cite{flow,holoren} for a recent general
  treatment.}

A neat way to package the statement that the bulk on-shell action
generates correlation functions in the boundary field theory is
through the generating functional relation
\be
\< e^{-[S_{\rm CFT} + \int d^d \vec{x} \O_\D(\vec{x}) (\bar{\phi} +
  \bar{h} (\vec{x}))]} \> = e^{-S[\bar{\phi} + \bar{h}]}
\ee
in which $\<\cdots\>$ on the left side indicates a path integral in
the field theory. This is a simple generalization of a formula
which we have implicitly used in Sec. \ref{amplitudes} for CFT
correlators, and $S_{\rm CFT}$ must still appear. The natural
procedure in the present case is to define correlation functions
by
\be
\frac{(-)^{n-1}\d^n}{\d \bar{h}(\vec{x_1}) \cdots \d \bar{h}(\vec{x_n}))}
S[\bar{\phi} + \bar{h}] \mbox{\LARGE{$\vert$}}_{\bar{h}=0}.
\ee
The term $\D S \equiv \, \int d^d \vec{x} \O_\D (\vec{x})
\bar{\phi}$ then remains in the QFT Lagrangian and describes an
operator deformation of the CFT with coupling constant
$\bar{\phi}$. If $0 \, > \, m^2 \, > \, -\frac{d^2}{4} < 0$, that
is if the critical point $\phi_i$ is a local maximum which is not
too steep, then $d >\D>\half d$, and  we are describing a
{\bf relevant deformation} of a CFT$_{UV}$,
one which will give a new long distance realization of
the field theory. It is worth remarking that the lower bound
agrees exactly with the stability criterion \cite{bf,mt} for
field theory in Lorentzian AdS$_{d+1}$. It is the lower mass
limit for which the energy of normalized scalar field
configurations is conserved and positive.

If the critical point is a local minimum, then $m_i^2 > 0$, and the
dual operator has dimension $\D > d$. We thus have the
deformation of the CFT by an {\bf irrelevant} operator, exactly
as describes the approach of an RG  flow to a CFT$_{IR}$ at long
distance. We thus see the beginnings of a gravitational
description of RG  flows in quantum field theory!

\subsection{Interpolating Flows, $I$}
\label{intflow1}

Interpolating flows are solutions of the domain wall
equations~(\ref{domwalleq}) in which the scalar field $\phi(r)$ approaches
the maximum $\phi_1$ of $V(\phi)$ in Fig. \ref{figpotential} as $r \rarrow
+\infty$ and the minimum $\phi_2$, as $r \rarrow - \infty$. The
associated metric approaches an AdS geometry in these limits as
discussed in the previous section. Exact solutions of the second
order non-linear system~(\ref{domwalleq}) are difficult (although we
discuss an interesting method in the next section). However, we
can learn a lot by linearizing about each critical point.

\medskip

We thus set $\phi(r)=\phi_i+h(r)$ and $A^{\prime}=\frac{1}{L_i} +
a^{\prime}(r)$ and work with the quadratic approximate potential
in~(\ref{vquad}). (See Footnote \ref{restrict}.) The linearized
scalar equation of motion and its general solution are
\be
h^{\prime \prime} + \frac{d}{L_i} h^{\prime} -
\frac{m_i^2}{L_i^2} h = 0
\ee
\be
h(r)=B e^{(\D_i -d)r/L_i} +Ce^{-\D_ir/L}
\ee
\be
\D_i=\left(d + \sqrt{d^2 + 4m^2_i}\right)/2
\ee
One may then linearize the scale factor equation in
~(\ref{domwalleq}) to find $a^{\prime}=\O(h^2)$ so
that the scale factor $A(r)$ is not modified to linear order
\vspace{-2ex}
\begin{ex}
Verify the statements above.
\end{ex}
\vspace{-1ex}

The basic idea of linearization theory is that there is an exact
solution of the nonlinear equations of motion that is well
approximated by a linear solution near a critical point. Thus as
$r \rarrow +\infty$, we assume that the exact solution behaves as
\be
\phi(r) \underset{r \gg 0}{\approx} \phi_1 + B_1 e^{(\D_1
  -d)r/L_1} + C_1 e^{-\D_1 r/L_1}.
\ee
The fluctuation must disappear as $r \rarrow +\infty$. For a
generic situation in which the dominant $B$ term is present,
this requires $\frac{d}{2} < \D_1 < d$ or $m_1^2 < 0$. Hence the
critical point associated with the boundary region of the domain
wall must be a local maximum, and everything is consistent with
an interpretation as the dual of a QFT$_d$ which is a relevant
deformation of an ultraviolet CFT$_d$.

Near the critical point $\phi_2$, which is a minimum, we have
$m_2^2 > 0$ so $\D_2 > d$. This critical point must be approached
at large negative $r$, where the exact solution is approximated
by
\be
\phi(r) \underset{r \ll 0}{\approx} \phi_2 + B_2 e^{(\D_2-d)r/L_2}
+ C_2e^{-\D_xr/L_2}.
\ee
The second term diverges, so we must choose the solution with
$C_2=0$. Thus the domain wall approaches the deep interior region
with the scaling rate of an irrelevant operator of scale
dimension $\D_2 > d$ exactly as required for infrared fixed
points by RG  ideas on field theory.

\medskip

The non-linear equation of motion for $\phi(r)$ has two
integration constants. We must fix one of them to ensure $C=0$ as
$r \rarrow -\infty$. The remaining freedom is just the shift $r
\rarrow r + r_0$ and has no effect on the physical picture. A
generic solution with $C=0$ in the IR would be expected to
approach the UV critical point at the dominant rate
$Be^{(\D_1-d)r/L_1}$ which we have seen to be dual to a relevant
operator deformation of the CFT$_{UV}$. It is possible (but
exceptional) that the $C=0$ solution in the IR would have
vanishing $B$ term in the UV and approach the boundary as $C_1
e^{- \D_1 r/L_1}$. In this case the physical interpretation is
that of the deformation of the CFT$_{UV}$ by a vacuum
expectation value, $\<\O_{\D_1}\> \sim C_1 \neq 0$.
See~\cite{bkl,bklt,klt,klebwit}.

\medskip

The domain wall flow ``sees'' the AdS$_{IR}$ geometry only in the
deep interior limit. To discuss the CFT$_{IR}$ and its operator
perturbations in themselves, we must think of extending this
interior region out to a complete AdS$_{d+1}$geometry with scale
$L_{IR}=L_2$.

\medskip

The interpolating solution we are discussing is plotted in
Figure~\ref{figprofile}. The scale factor $A(r)$ is concave
downward since $A^{\prime \prime}(r)<0$ from~(\ref{adoub}). This
means that slopes of the linear regions in the deep interior and near
boundary are related by $1/L_{IR} > 1/ L_{UV}$ (where we
have set $L_{UV}=L_1$). Hence,
\be
V_{IR} = \lf{-d(d-1)}{4L_{IR}^2} < V_{UV} =
\lf{-d(d-1)}{4L_{UV}^2}.
\ee
Thus the flow from the boundary to the interior necessarily goes
to a deeper critical point of $V(\phi)$. Recall that the
condition $A^{\prime \prime}(r)<0$ is very general and holds in
any physically reasonable bulk theory, \eg a system of many
scalars $\phi^{I}$ and potential $V(\phi^{I})$. Thus any
Poincar\'e invariant domain wall interpolating between AdS
geometries is {\bf irreversible}.

\begin{fig}[htp]
\centering
\epsfxsize=2in
\epsfysize=2in
\epsffile{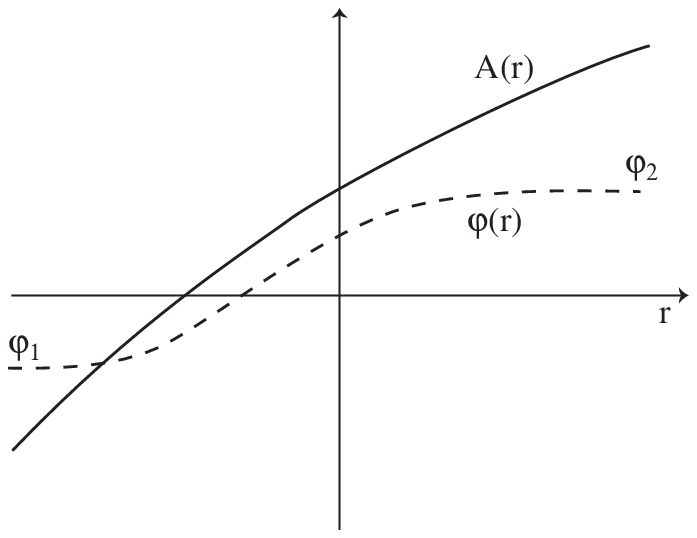}
\caption{Profile of the scale factor $A(r)$}
\label{figprofile}
\label{fig:13}
\end{fig}

The philosophy of the AdS/CFT correspondence suggests that any
conspicuous feature of the bulk dynamics should be dual to a
conspicuous property of quantum field theory. The irreversibility
property reminds us of Zamolodchikov's $c$-theorem \cite{zam} which implies
that RG  flow in QFT$_2$ is irreversible. We will discuss
the $c$-theorem and its holographic counterpart later.
Our immediate goals are to present a very interesting
technique for exact solutions of the non-linear flow
equations~(\ref{domwalleq}) and to discuss a ``realistic'' application of
supergravity domain walls to deformations of $\N \!=\!4$ SYM theory.

\subsection{Interpolating Flows, $II$}
\label{intflow2}

The domain wall equations
\be
\label{dom1}
\phi^{\prime \prime} + d A^{\prime} \phi^{\prime} = {dV(\phi)
  \over d\phi}
\ee
\be
\label{dom2}
A^{\prime 2} = \frac{2}{d(d-1)} \left(\phi^{\prime 2} -
  2V(\phi)\right)
\ee
constitute a non-linear second order system with no apparent
method of analytic solution. Nevertheless, a very interesting
procedure which does give exact solutions in a number of
examples has emerged from the literature
\cite{tow84,fgpw1,skentow,dfgk}.

Given the potential $V(\phi)$, suppose we could solve the
following differential equation in field space and obtain an
auxiliary quantity, the superpotential $W(\phi)$:
\be
\label{hj}
\frac{1}{2} \left(\frac{dW}{d \phi} \right)^2 - \frac{d}{d-1} W^2
= V(\phi)
\ee
We then consider the following set of first order equations
\be
\label{flow1}
\frac{d \phi}{d r} = \frac{d W}{d \phi}
\ee
\be
\label{flow2}
\frac{d A}{d r} = - \frac{2}{d-1} W \left(\phi (r)\right)
\ee
These decoupled equations have a trivial structure and can be
solved sequentially, the first by separation of variables, and
the second by direct integration. (We assume that the two
required integrals are tractable.) It is then easy to show that
{\bf any solution of the first order
  system~(\ref{hj}, \ref{flow1}, \ref{flow2}) is also a solution of
  the original second order system~(\ref{dom1}, \ref{dom2})}.
\vspace{-2ex}
\begin{ex}
Prove this!
\end{ex}
\vspace{-2ex}
It is also elementary to see that any critical point of $W(\phi)$
is also a critical point of $V(\phi)$ but not conversely.
\vspace{-2ex}
\begin{ex}
  Suppose that $W(\phi)$ takes the form $W \approx - \frac{1}{L_i}(
  \lambda + \frac{1}{2} \m h^2)$ near a critical point. Show that
  $\l, \m$ are related to the parameters of the approximate
  potential in~(\ref{vquad}) by $\l
  = \half(d-1)$ and $m^2= \m(\m -d)$. Show that the solution to
  the flow equation~(\ref{flow1}) approaches the critical point
  at the rate $h \sim e^{-\m r/L}$ with $\m = \D$, the vev rate,
  or $\m = d - \D$ the operator deformation rate.
\end{ex}
\vspace{-1ex}

This apparently miraculous structure generalizes to bulk
theories with several scalars $\phi^{I}$ and Lagrangian
\be
\label{lag3}
L = - \frac{1}{4}R + \half \p_\m \phi^{I} \p^{\m} \phi^{I} +
V(\phi^{I})
\ee
The superpotential $W(\phi^{I})$ is defined to satisfy the
partial differential equation
\be
\label{hj1}
\half \sum_{I} \left(\frac{\p W}{\p \phi^{I}}\right)^2 -
\frac{d}{d-1}W^2=V
\ee
The first order flow equations
\be
\label{flow3}
\frac{d \phi^{I}}{dr} = \frac{\p W}{\p \phi^{I}}
\ee
\be
\label{flow4}
\frac{d A}{d r} = - \frac{2}{d-1} W
\ee
automatically give a solution of the second order Euler-Lagrange
equations of~(\ref{lag3}) for Poincar\'e invariant domain walls.
\vspace{-2ex}
\begin{ex}
Prove this and derive first order flow equations with the same
property for the non-linear $\s$-model (in which the kinetic term
of~(\ref{lag3}) is replaced by $\half G_{IJ}(\f^K)\p_\m \f^I
\p^\m \f^J$.
\end{ex}
\vspace{-1ex}

\medskip

The equations~(\ref{flow3}) are conventional gradient flow
equations. The solutions are paths of steepest descent for
$W(\f^I)$, everywhere perpendicular to the contours
$W(\f^I)=\mbox{~const}$.  In applications to RG  flows, the
  $\f^I(r)$ represent scale dependent couplings of relevant
  operators in a QFT Lagrangian, so we are talking about
gradient flow in the space of couplings---an idea which is
frequently discussed in the RG  literature!

\medskip

There are two interesting reasons why there are first order flow
equations which reproduce the dynamics of the second order system
(\ref{dom1}, \ref{dom2}).
\vspace{-1ex}
\begin{enumerate}
\setlength{\itemsep}{-1ex}
\item They emerge as $BPS$ conditions for supersymmetric domain
  walls in supergravity theories. For a review, see~\cite{cvet}.
  The superpotential $W(\f^I)$ emerges by algebraic analysis of
  the quantum transformation rule. Bulk solutions have Killing
  spinors, and bulk supersymmetry is matched in the boundary
  field theory which describes a supersymmetric deformation of an
  SCFT.
\item They are the Hamilton-Jacobi equations for the dynamical
  system of gravity and scalars~\cite{dbvv}. The
  superpotential $W(\f^I)$ is the classical Hamilton-Jacobi
  function, and one must solve~(\ref{hj}) or~(\ref{hj1}) to
  obtain it from the potential $V(\f^I)$. This is very
  interesting theoretically but rather impractical because it is
  rare that one can actually use the Hamilton-Jacobi formulation
  to solve a dynamical system explicitly. Numerical and
  approximate studies have been instructive~\cite{dfgk,martelli}.
  However, most applications involve superpotentials from $BPS$
  conditions in gauged supergravity. One may also employ a toy
  model viewpoint in which $W(\f^I)$ is postulated with potential
  $V(\f^I)$ defined through~(\ref{hj}) or~(\ref{hj1}).
\end{enumerate}

\subsection{Domain Walls in $D=5, \, \N=8$, Gauged Supergravity}
\label{gaugedsg}

The framework of toy models is useful to illustrate the
correspondence between domain walls in $(d+1)$-dimensional
gravity and RG  flows in QFT$_d$. However it is highly desirable
to have ``realistic examples'' which describe deformations of
$\N=4$ SYM in the strong coupling limit of the AdS/CFT
correspondence. There are two reasons to think first about
supersymmetric deformations. As just discussed, the bulk dynamics
is then governed by a superpotential $W(\f^I)$ with first order
flow equations. Further the methods of Seiberg dynamics can give
control of the long distance non-perturbative behavior of the
field theory, so that features of the supergravity description
can be checked.

\medskip

We can only give a brief discussion here. We begin by discussing
the relation between $D=10$ Type IIB sugra dimensionally reduced
on \AdS\ of~\cite{krvn} and the $D=5,\, \N=8$
supergravity theory with gauge group $SO(6)$ first completely
constructed in~\cite{grw}. As discussed in Section~\ref{intflow2}
above, the spectrum of the first theory consists of the graviton
multiplet, whose fields are dual due to all relevant and
marginal operators of $\N=4$ SYM plus Kaluza-Klein towers of
fields dual to operators of increasing $\D$. On the other hand
gauged $\N \!=\!8$ supergravity is a theory formulated in $5$
dimensions with only the fields of the graviton
multiplet above, namely
\be
\begin{array}{cccccc}
g_{\m \nu} & \Psi_\m^a & A_\m^{A} & B_{\m \nu} & X^{abc} & \f^I
\\
1 & 8 & 15 & 12 & 48 & 42
\end{array}
\ee
It is a complicated theory in which the scalar dynamics is that
of a nonlinear $\s$-model on the coset $E(6,6)/USp(8)$ with a
complicated potential $V(\f^I)$.

\medskip

Gauged $\N=8$ supergravity has a maximally symmetric ground state
in which the metric is that of AdS$_5$. The global symmetry is $SO(6)$
with $32$ supercharges, so that the superalgebra is $SU(2, 2|4)$.
Symmetries then match the vacuum configuration of Type IIB sugra on
\AdS. Indeed $D=5, \, \N=8 $ sugra is believed to be
the {\bf consistent truncation} of $D=10$ Type IIB sugra to the fields of
its graviton multiplet. This means that {\bf every} classical solution
of $D=5, \, \N=8$ sugra can be ``lifted'' to a solution of
$D=10$ Type IIB sugra. For example the $SO(6)$ invariant AdS$_5$ ground
state solution lifts to the \AdS\ geometry
of~(\ref{adsxs}) (with other fields
either vanishing or maximally symmetric). There is not yet a
general proof of consistent truncation, but explicit lifts of
nontrivial domain wall solutions have been given
\cite{pw1,pw2,kw,johnson}. Consistent truncation has been established
in other similar theories ~\cite{dwn,vann}.

\medskip

In the search for classical solutions with field theory duals it
is more elegant, more geometric, and more ``braney'' to work at the
level of $D=10$ Type IIB sugra. There are indeed very interesting
examples of Polchinski and Strassler~\cite{ps} and Klebanov and
Strassler~\cite{ks}. Another example is the multi-center
$D3$-brane solution of~(\ref{d3},\ref{harm}) which is dual to a Higgs
deformation of $\N=4$ SYM in which the $SU(N)$ gauge symmetry is
broken spontaneously to $SU(N_1) \otimes SU(N_2) \otimes \cdots
\otimes SU(N_M)$. In these examples the connection with field
theory is somewhat different from the emphasis in the present
notes. For this reason we confine our discussion to domain wall
solutions of $D=5~\N=8$ sugra.  This is a realistic framework since
the $D=5$ theory contains all relevant deformations of $\N=4$ SYM,
and experience indicates that $5D$ domain wall solutions can be
lifted to solutions of $D=10$ Type IIB sugra.

\medskip

For domain walls, we can restrict to the metric and scalars of
the theory which are governed by the action
\be
S = \int d^5 z \sqrt{g} \left[- \frac{1}{4}R + \half G_{IJ}(\f^K)
  \p_\m \f^I \p^{\m} \f^J + V(\f^k) \right]
\ee
The $42$ scalars sit in the $27$-bein matrix $V_{AB}^{ab}(\f^K)$
of $E(6,6)/USp(8)$. The indices $a,b$ and $AB$ have $8$
values. They are anti-symmetrized (with symplectic trace removed)
in most expressions we write. The coset metric $G_{IJ}$, the
potential $V$ and other quantities in the theory are constructed
from $V^{ab}_{AB}$. Symmetries govern the construction, but the
nested structure of symmetries makes things very complicated.

A simpler question than domain walls is that of critical points
of $V(\f^k)$. The AdS/CFT correspondence requires that every
stable critical point with $V < 0$ corresponds to a
CFT$_4$. Stability means simply that mass eigenvalues of
fluctuations satisfy $m^2 > -4$ so that bulk fields transform in
unitary, positive energy representations of $SO(2,4)$ (for
Lorentz signature).

\medskip

Even the task of extremizing $V(\phi^k)$ is essentially
impossible in a space of $40$ variables, ($V$ does not depend on
the dilaton and axion fields), so one uses the following simple
but practically important trick, \cite{warner}:
\vspace{-1ex}
\begin{alphal}
\setlength{\itemsep}{-1ex}
\item Select a subgroup $H$ of the invariance
  group  $SO(6)$ of $V(\f^K)$.
\item The $42 \f^K$ may be grouped into fields $\f$ which are
  singlets of $H$ and others $\xi$ which transform in non-trivial
  representations of $H$.
\item It follows from naive group theory that the expansion of
  $V$ takes the form $V(\f, \xi)=V_0(\f)+ V_2(\f) \xi^2 +
  \O(\xi^3)$ with no linear term.
\item Thus, if $\widehat{\f}$ is a stationary point of $V_0(\f)$,
  then $\widehat{\f}, \xi=0$ is a stationary point of $V(\f,
  \xi)$. The problem is then reduced to minimization in a much
  smaller space.
\end{alphal}
\vspace{-1ex}
The same method applies to all solutions of the equations of
motion $\lf{\p S}{\p \f^K}=0$, and to the Killing spinor
problem since that gives a solution to the equations of
motion. The general principle is that if $S$ is invariant under
$G$, in this case $G=SO(6)$, and $H \subset G$ is a subgroup, then a
consistent $H$-invariant solution to the dynamics can be obtained
by restricting, ab initio, to singlets of $H$.

\medskip

All critical points with preserved symmetry $H \supseteq SU(2)$ are
known \cite{kpw}. There are $5$ critical points of which $3$
are non-supersymmetric and unstable \cite{grw,KP}. There are two
$SUSY$ critical points of concern to us. The first with $H=SO(6)$ and
full $\N\!=\!8~SUSY$ is the maximally symmetric state discussed above,
and the second has $H=SU(2)\otimes U(1)$ and $\N\!=\!2\:SUSY$. The
associated critical bulk solutions are dual to the undeformed
$\N\!=\!4$ SYM and the critical IR limit of a particular deformation
of $\N\!=\!4$ SYM.

\medskip

The search for supersymmetric domain walls in $\N \! =8$ gauged
supergravity begins with the fermionic transformation
rules\footnote{Conventions for spinors and $\g$-matrices are
  those of \cite{fgpw1} with spacetime signature $+----$.} which
have the form:
\be
\label{kspin1}
\delta\psi_{\mu}^{a}= D_{\mu}\epsilon^{a}-{1 \over
  3}W_{b}^{a} \gamma_{\mu}\epsilon^{b}
\ee
\be
\label{kspin2}
\delta\chi^{A}=\left(\gamma^{\mu}P^{A}_{aI}\p \phi^I-Q^{A}_{a}
  (\varphi)\right) \epsilon^{a}
\ee
where $A$ is an index for the 48 spinor fields $\chi^{abc}$. The
$\e^a$ are 4-component symplectic Majorana spinors \cite{fgpw1}
(with spinor indices suppressed and $a=1,\cdots,8$). The matrices
$W^{a}_{b}$, $P^A_{aI}$ and $Q^{A}_{a}$ are functions of the
scalars $\varphi^{I}$ which are part of the specification of the
classical supergravity theory.  Killing spinors
$\epsilon^{a}(\vec{x},r)$ are spinor configurations which satisfy
$\delta\psi_{\mu}^{a}$=0 and $\delta\chi^{A}$=0. The process of
solving these equations leads both to the
$\epsilon^{a}(\vec{x},r)$ and to conditions which determine the
domain wall geometry which supports them. These conditions, in
this case the first order field equations (\ref{flow3},
\ref{flow4}), imply that the bosonic equations of motion of the
theory are satisfied.
\vspace{-1ex}
\begin{ex}
For a generic SUSY or sugra theory, show that if there are
Killing spinors for a given configuration of bosonic fields, then
that configuration satisfies the equations of motion. Hint:
\be
\d S = \int \left[{\d S \over \d B} \d B +{\d S \over \d \psi}\d
  \psi\right] \equiv 0,
\ee
where $B$ and $\psi$ denote the boson and fermion fields of the
theory and $\d B$ and $\d \psi$ their transformation rules.
\end{ex}

We now discuss the Killing spinor analysis to outline how the
first order flow equations arise.
\vspace{-1ex}
\begin{ex}
Using the spin connection of Ex: 20, show that the condition
$\delta\psi_{j}^{a}$=0 can be written out in detail as
\be
\delta\psi_{j}^{a}=\partial_{j}\epsilon-\frac{1}{2}A'(r)\gamma_{j}\gamma_{5}
\epsilon^{a}-\frac{1}{3}W^{a}_{b}\gamma_{j}\epsilon^{b}=0
\ee
\end{ex}
\vspace{-1ex}
We can drop the first term because the Killing spinor must be
translation invariant. What remains is a purely algebraic
condition, and we can see that the flow equation (\ref{flow2})
for the scale factor directly emerges with superpotential
$W(\phi)$ identified as one of the eigenvalues of the tensor
$W^a_b$. In detail one actually has a symplectic eigenvalue
problem, with $4$ generically distinct $W$'s as solutions. Each of
these is a candidate superpotential. One must then examine the $48$
conditions
\be
\d \chi^A = (\g^5 P^A_{aI} \partial_r \phi^I - Q^A_a)\e^a =0
\ee
to see if $SUSY$ is supported on any of the eigenspaces.  One can
see how the gradient flow equation (\ref{flow3}) can emerge.
Success is not guaranteed, but when it occurs, it generically
occurs on one of the four (symplectic) eigenspaces. The $5D$ Killing
spinor solution satisfies a $\g^5$ condition effectively yielding
a $4d$ Weyl spinor,
giving ${\cal N}=1 ~SUSY$ in the dual field
theory. Extended ${\cal N}>1~ SUSY$ requires further degeneracy
of the eigenvalues. (The  $\d \Psi^a_r =0$ condition which has not
yet been mentioned gives a differential equation for the
$r$-dependence of $\e^a(r)$)
\vspace{-1ex}
\begin{ex}
It is a useful exercise to consider a simplified version of the
Killing spinor problem involving one complex (Dirac) spinor with
superpotential $W(\phi)$ with one scalar field. The equations are
\be
\begin{array}{ll}
(D_\m - \frac{1}{3}iW \g_\mu)\e &=0\\
\\
(-i\g^\m\partial_{\m}\phi -\frac{dW}{d\phi})\e &=0
\end{array}
\ee
Show that the solution of this problem yields the flow equations
(\ref{flow1},\ref{flow2}) and
\be
\e = e^{\frac{A}{2}}\eta
\ee
where $\eta$ is a constant eigenspinor of $\gamma^5$. Show that
at a critical point of W, there is a second Killing spinor (which
depends on the transverse coordinates $x^i$). See \cite{Shuster:1999zf}.
This appears because of the
the doubling of supercharges in superconformal $SUSY$.
\end{ex}
\vspace{-1ex}

\medskip

Needless to say the analysis is impossible on the full space of
$42$ scalars. Nor do we expect a solution in general, since many
domain walls are dual to non-supersymmetric deformations and
cannot have Killing spinors. In~\cite{fgpw1} a symmetry reduction
to singlets of an $SU(2)$ subgroup of $SO(6)$ was used. After
further simplification it was found that $\N=1~SUSY$ with $SU(2)
\times U(1)$ global symmetry was supported for flows involving
two scalar fields, $\f_2$ a field with $\D=2$ in the
$20^{\prime}$ of $SO(6)$ in the full theory, and $\f_3$ a field
with $\D=3$ in the $10+ \Bar{10}$ representation.
(In~\cite{fgpw1}, these fields were called $\f_3, \f_1$
respectively.) The fields $\f_2, \f_3$ have canonical kinetic
terms as in~(\ref{lag3}). Using $\rho=e^{{\f_2 \over \sqrt{6}}}$,
the superpotential is
\be \label{supot}
W(\f_2, \f_3) = \frac{1}{4L \r^2} \left[{\rm cosh} (2 \f_3)(\r^6 - 2)
  - 3 \r^6 -2 \right]
\ee
%
\vspace{-1ex}
\begin{ex}
Show that $W(\f_2, \f_3)$ has the following critical points:
\vspace{-1ex}
\begin{ronums}
\setlength{\itemsep}{-1ex}
\item a maximum at $\f_2=0, \f_3=0$, at which $W=-\lf{3}{2L}$
\item a saddle point at $\f_2=\frac{1}{\sqrt{6}} \ln 2, \f_3= \pm
  \frac{1}{2} \ln 3$ at which $W=-{2^{2/3} \over L}$. (The two solutions are
  related by a $Z_2$ symmetry and are equivalent).
\end{ronums}
\vspace{-1ex}
\end{ex}
\vspace{-1ex}
Thus there is a possible domain wall flow interpolating between
these two critical points. The flow equation~(\ref{flow3}) cannot
yet be solved analytically for $W$ of (\ref{supot}), but a
numerical solution and its asymptotic properties were discussed
in~\cite{fgpw1}. See Figure~\ref{figcontours}.

\begin{fig}[htp]
\centering
\epsfxsize=2in
\epsfysize=2in
\epsffile{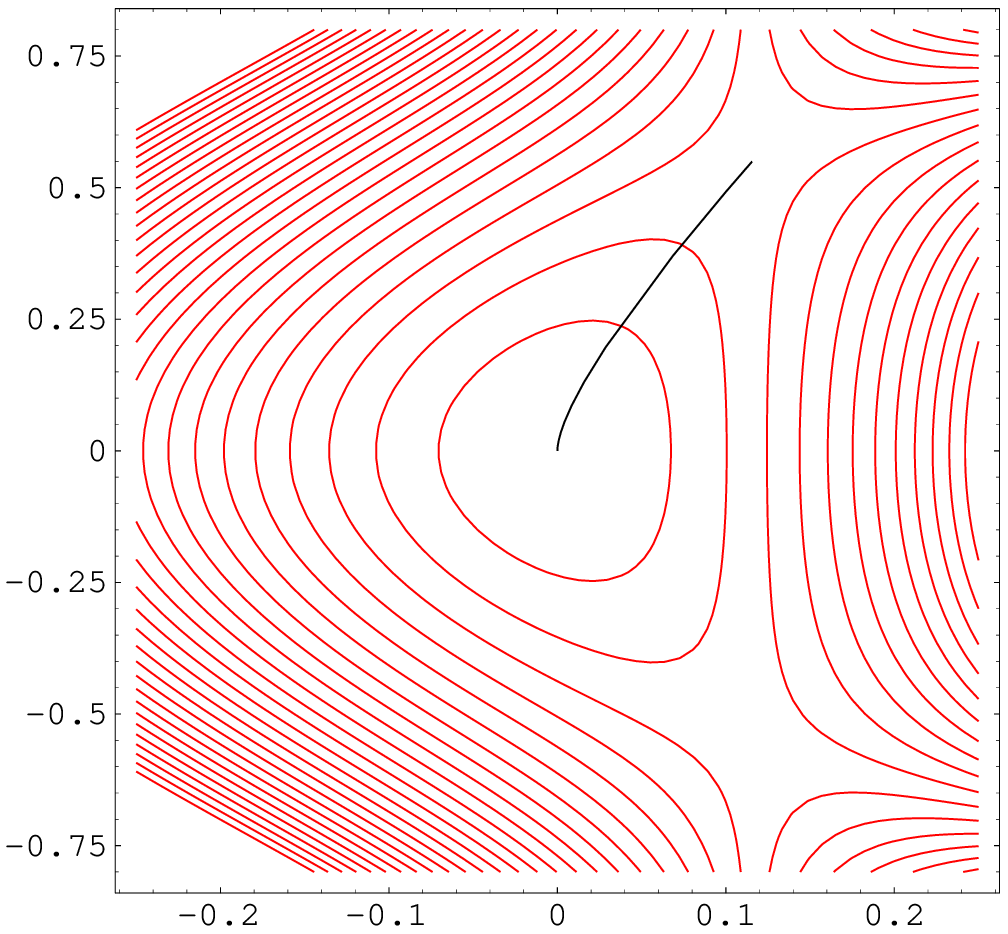}
\caption{Contour plot of $W(\phi_2,\phi_3)$}
\label{figcontours}
\label{fig:14}
\end{fig}

In accord with the general discussion of Section~\ref{intflow1},
the solution should be dual to a relevant deformation of $\N\!=\!4 ~
SYM$ theory which breaks $SUSY$ to $\N=1$ and flows to an SCFT$_4$
at long distance. In the next section we discuss this field
theory and the evidence that the supergravity description is
correct.

\medskip

In the space of the two bulk fields $\f_2, \f_3$ there is a
continuously infinite set of gradient flow trajectories emerging
from the $\N\!=\!4$ critical point. One must tune the initial
direction to find the one which terminates at the $\N\!=\!1$
point. All other trajectories approach infinite values in
field space, and the associated geometries, obtained from the flow
equation~(\ref{flow4}) for $A(r)$ have curvature
singularities. There are analytic domain wall flows with $\f_3
\equiv 0$ \cite{fgpw2,bs} and in other sectors \cite{gppz2} of
the space of scalars of $\N\!=\!8$ sugra, and a number of $2$-point
correlation functions have been
computed~\cite{theisen,Anatomy,bs2,mueck,flow,holoren}.
Nevertheless the curvature singularities
are at least a conceptual problem for the AdS/CFT
correspondence. In any case we do not discuss singular flows
here.

\subsection{SUSY Deformations of $\N\!=\!4$ SYM Theory}
\label{susydef}

It is useful for several purposes to describe the $\N\!=\!4$ SYM
theory in terms of $\N\!=\!1$ superfields. The $4$ spinor fields
$\l^{\a}$ are regrouped, and $\l^{4}$ is paired with gauge
potential $A_j$ in a gauge vector superfield $V$. The remaining
$\l^{1,2,3}$ may be renamed $\psi^{1,2,3}$ and paired with
complex scalars $z^1 = X^1+iX^4, z^2=X^2+iX^5,
z^3=X^3+iX^6$ to form $3$ chiral superfields $\Phi^i$. In the
notation of Section~2.5, the
Lagrangian consists of a gauge kinetic term plus matter terms
\be
\L = \int d^4 \q \tr \!\left(\bar{\F}^i e^{gV} \F^i \right) + \int
d^2 \q g \tr \F^3 [\F^1, \F^2] + h.c.
\ee
The manifest supersymmetry is $\N=1$ with $R$-symmetry $SU(3)
\otimes U(1)$. Full symmetry is regained after re-expression in
components because the Yukawa coupling $g$ is locked to the
$SU(N)$ gauge coupling. This formulation is commonly used to
explore perturbative issues since the $\N=1$ supergraph formalism
(first reference in \cite{west}) is quite efficient. 
This formulation suits our main
purpose which is to discuss $SUSY$ deformations of the theory.

\medskip

A general relevant $\N=1$ perturbation of $\N=4$ SYM is obtained
by considering the modified superpotential
\be
U = g \tr \F^3 [\F^1, \F^2] + \half M_{\a \b} \tr \F^{\a} \F^{\b}
\ee
This framework is called the $\N=1^{\ast}$ theory. The moduli
space of vacua, $\lf{\p U}{\p \F^{\a}}=0$, has been studied
\cite{vafa,donwit,ps} and describes a rich panoply of dynamical
realizations of gauge theories, confinement and Higgs-Coulomb
phases, and as we shall see, a superconformal phase.

\medskip

We discuss here the particular deformation with a mass term for
one chiral superfield only
\be
\label{uls}
U=g \tr \F^3 [\F^1, \F^2] + \half m \tr (\F^3)^2
\ee
The $R$-symmetry is now the direct product of $SU(2)$ acting on
$\F^{1,2}$ and $U(1)_R$ with charges $(\half, \half, 1)$ for
$\F^{1,2,3}$. The massive field $\F^3$ drops out of the long
distance dynamics, leaving the massless fields $\F^{1,2}$. We
thus find symmetries which match those of the supergravity flow
of the last section. However to establish the duality, it needs
to be shown that long distance dynamics is conformal. We will
briefly discuss the pretty arguments of Leigh and
Strassler~\cite{ls} that show this is the case.

\medskip

The key condition for conformal symmetry is the vanishing of
$\b$-functions for the various couplings in the Lagrangian. In a
general $\N=1$ theory with gauge group $G$ and chiral superfields
$\F_\a$ in representations $R_{\a}$ of $G$, the exact $NSVZ$
gauge $\b$-function is
\be
\b (g) = -\frac{g^3}{8 \pi^2}\: \lf{3T(G)- \sum_{\a} T(R_{\a})(1 -2
  \c_{\a})}{1- \lf{g^2T(G)}{8 \pi^2}}
\ee
where $\c_{\a}$ is the anomalous dimension of $\F^{\a}$ and
$T(R_{\a})$ is the Dynkin index of the representation. (If $T^a$
are the generators in the representation $R_{\a}$, then $\tr T^a
T^b \equiv T(R_{\a}) \d^{ab})$. In the present case, in which
$G=SU(N)$ and all fields are in the adjoint, we have
$T(R_{\a})=T(G)=N$, and
\be
\b(g) \sim 2N (\c_1 + \c_2 + \c_3)
\ee
In addition we need the $\b$-function for various invariant field
monomials $\Tr(\F^1)^{n_1}(\F^2)^{n_2}(\F^3)^{n_3}$,
\be
\b_{n_1, n_2, n_3} = 3 - \sum^3_{\a =1} n_{\a} - \sum^3_{\a=1}
n_{\a} \c_{\a}
\ee
This form is a consequence of the non-renormalization theorem
for superpotentials in $\N=1~SUSY$. The first two terms are fixed
by classical dimensions and the last is due to wave function
renormalization. For the two couplings in the
superpotential~(\ref{uls}) we have
\be
\b_{1,1,1}= \c_1 + \c_2 + \c_3
\ee
\be
\b_{0,0,2} = 1 - 2 \c_3.
\ee
One should view the $\c_{\a}(g,m)$ as functions of the two
couplings. The conditions for the vanishing of the $3$
$\b$-functions have the unique $SU(2)$ invariant solution
\be
\label{anomdim}
\c_1 = \c_2 = - \half \c_3 = - \frac{1}{4}
\ee
which imposes one relation between $ g, m$, suggesting that the theory
has a fixed line of couplings. The $\b=0$ conditions are
necessary conditions for a superconformal realization in the
infrared, and Leigh and Strassler give additional arguments that
the conformal phase is realized.

$\N=1$ superconformal symmetry in $4$ dimensions is governed by
the superalgebra $SU(2,2, | 1)$. This superalgebra has several
types of short representations. (See Appendix of \cite{fgpw1}).
For example, chiral superfields,
either elementary or composite, are short multiplets in which
scale dimensions and $U(1)_R$ charge are related by $\D =
\frac{3}{2} r$. For elementary fields $\D_{\a} = 1 + \c_{\a}$,
and one can see that the $\c_{\a}$ values in~(\ref{anomdim}) are
correctly related to the $U(1)_R$ charges of
the $\F^{\a}$.

\medskip

The observables in the SCFT$_{IR}$ are the correlation functions
of gauge invariant composites of the light
superfields\footnote{the index of the field strength superfield
  $W_a$ is that of a Lorentz group spinor, while that of
  $\F^{\a}$ is that of $SU(2)$ flavor.} $W_{\a}, \F^1, \F^2$. We
list several short multiplets together with the scale dimensions
of their primary components
\be
\begin{array}{cccccc}
\O & \tr \F^{\a} \F^{\b} & \tr W_{\a} \F^{\b} & \tr W_{\a}
W^{\a} & \tr \F^{+}T^A \F & \tr  (W_{\a} W_{\dot{\b}} + \cdots) \\
\D & \frac{3}{2} & \frac{9}{4} & 3 & 2 & 3
\end{array}
\ee
The first $3$ operators are chiral, the next is the multiplet containing
the $SU(2)$ current, and the last is the multiplet containing
$U(1)_R$ current, supercurrent, and stress tensor. Each
multiplet has several components.

\subsection{AdS/CFT Duality for the Leigh-Strassler Deformation}

We now discuss the evidence that the domain wall of $\N=8$ gauged
supergravity of Section~\ref{gaugedsg} is the dual of the mass
deformation of $\N=4$ SYM of Section~\ref{susydef}. There are
two types of evidence, the match of dimensions of operators,
discussed here, and the match of conformal anomalies discussed in
the next chapter. Critics may argue that much of the detailed
evidence is a consequence of symmetries rather than dynamics. But
it is dynamically significant that the potential $V(\F^k)$
contains an IR critical point with the correct symmetries and
the correct ratio $V_{IR}/V_{UV}$ to describe the IR fixed
point of the Leigh-Strassler theory. The
AdS/CFT correspondence would be incomplete if $D=5~\N=8$~sugra
did not contain this SCFT$_4$.

\medskip

Whether due to symmetries or dynamics, much of the initial
enthusiasm for AdS/CFT came from the $1:1$ map between bulk
fields of Type IIB sugra and composite operators of $\N=4$ SYM. The map
was established using the relationship between the AdS masses
of fluctuations about the \AdS\ solution and scale
dimensions of operators. The same idea may be applied to
fluctuations about the IR critical point of the flow of
Section~\ref{gaugedsg}. One can check the holographic description
of the dynamics by computing the mass eigenvalues of all fields
in the theory, namely all fields of the graviton multiplet
listed in Section~\ref{gaugedsg}. This task is complicated
because the Higgs mechanism acts in several sectors. Scale
dimensions are then assigned using the formula in ~(\ref{deltam}) for
scalars and its generalizations to other spins. The next step is
to assemble component fields into multiplets of the $SU(2, 2|1)$
superalgebra. One finds exactly the $5$ short multiplets listed
at the end of Section~\ref{susydef} together with $4$ long
representations. The detailed match of short multiplets confirms
the supergravity description, while the scale dimensions of
operators in long representations are non-perturbative
predictions of the supergravity description.

It would be highly desirable to study correlation functions of
operators in the Leigh-Strassler flow, but this requires an
analytic solution for the domain wall, which is so far
unavailable.

\subsection{Scale Dimension and AdS Mass}

For completeness we now list the relation between $\Delta$
and the mass for the various bulk fields which occur in a
supergravity theory. For $d=4$ some results were given in
\cite{ffz}. For the general case of for ${\rm AdS}_{d+1}$, the
relations are given below with references. There are exceptional
cases in which the lower root of $\pm$ is appropriate.

\vspace{-1ex}
\begin{enumerate}
\setlength{\itemsep}{-1ex}
\item
scalars \cite{witten}:\quad $\Delta_{\pm} = \half(d \pm \sqrt{d^2
+4m^2})$,
\item spinors \cite{henning}:\quad  $\Delta = \half (d + 2|m|)$,
\item  vectors\quad
$ \Delta_{\pm} = {1 \over 2} (d \pm \sqrt{(d-2)^2 + 4m^2})$,
\item  $p$-forms \cite{l'Yi:1998eu}:\quad
$ \Delta = {1 \over 2} (d \pm \sqrt{(d-2p)^2 + 4m^2})$,
\item first-order $(d/2)$-forms ($d$ even):
$\Delta={1\over 2}(d + 2|m|)$,
\item  spin-3/2  \cite{AVol,KosRyt}:\quad
$\Delta = \half(d + 2|m|)$,
\item  massless spin-2 \cite{poli}:\quad
$\Delta = \half (d + \sqrt{d^2 + 4 m^2} )$.

\end{enumerate}
\vspace{-1ex}

\vfill\eject

\section{The $c$-theorem and Conformal Anomalies}
\setcounter{equation}{0}

In this chapter we develop a theme introduced in Sec.
\ref{intflow1}, the irreversibility of domain walls in
supergravity and the suggested connection with the $c$-theorem
for RG  flows in field theory. The $c$-theorem is related to the
conformal anomaly. We discuss this anomaly for 4d field theory
and the elegant way it is treated in the AdS/CFT
correspondence. This suggests a simple form for a holographic
$c$-function, and monotonicity follows from the equation
$A^{\prime \prime}(r) < 0$. It follows that any RG  flow which can
be described by the AdS/CFT correspondence satisfies the
$c$-theorem. The holographic computation of anomalies agrees with
field theory for both the undeformed $\N \! =4$ SYM theory and
the $\N=1$ Leigh-Strassler deformation.

\subsection{The $c$-theorem in Field Theory}

We briefly summarize the essential content of Zamolodchikov's
$c$-theorem \cite{zam} which proves that RG  flows in QFT$_2$ are
irreversible. We consider the correlator
$\<T_{zz}(z, \bar{z})T_{zz}(0)\>$ in a flow from a CFT$_{UV}$ to a
CFT$_{IR}$. It has the form
\be \label{tt}
\< T_{zz}(z,\bar{z})T_{zz}(0)\> = \frac {c(M^2 z\bar{z})}{z^4}
\ee
where $M^2$ is a scale that is present since conformal symmetry
is broken. The function $c(M^2 z\bar{z})$ has the properties:
\vspace{-1ex}
\begin{enumerate}
\setlength{\itemsep}{-1ex}
\item $c(M^2 z\bar{z}) \rightarrow c_{UV}$ as $|z|\rightarrow 0$
  and $c(M^2 z\bar{z}) \rightarrow c_{IR}$ as $|z|\rightarrow
  \infty$ where $c_{UV}$ and $c_{IR}$ are central charges of the
  critical theories CFT$_{UV}$ and CFT$_{IR}$.
\item $c(M^2 z\bar{z})$ is not necessarily monotonic, but there
  are other (non-unique) $c$-functions which decrease
  monotonically toward the infrared and agree with $c(M^2
  z\bar{z})$ at fixed points. Hence $c_{UV} > c_{IR}$
  which proves irreversibility of the flow!
\item the central charges are also measured by the curved space
  Weyl anomaly in which the field theory is coupled to a fixed
  external metric $g_{ij}$ and one has
\be
\< \theta \> = - \frac{c}{12} R
\ee
for both CFT$_{UV}$ and CFT$_{IR}$.
\end{enumerate}
\vspace{-1ex}

The intuition for the $c$-theorem comes from the ideas of
Wilsonian renormalization and the decoupling of heavy particles
at low energy. Since $T_{ij}$ couples to all the degrees of
freedom of a theory, the $c$-function measures the effective
number of degrees of freedom at scale $x=\sqrt{z\bar{z}}$. This
number decreases monotonically as we proceed toward longer
distance and more and more heavy particles decouple from the low
energy dynamics. These are fundamental ideas and we should see if
and how they are realized in QFT$_4$ and AdS$_5$/CFT$_4$.

\medskip

First we define two projection operators constructed from the
basic $\pi_{ij}=\partial_i \partial_j -\d_{ij} \Box$,
\be
\begin{array}{ll}
\P^{(0)}_{ijkl} & = \pi_{ij} \pi_{kl} \\
\\
\p^{(2)}_{ijkl} & = 2 \pi_{ij} \pi_{kl}
- 3(\pi_{ik} \Pi_{jl} + \pi_{il} \pi_{jk})
\end{array}
\ee
In any QFT$_4$ the  $\< TT\>$ correlator then takes the form
\be
\label{ttcorr}
\< T_{ij}(x)T_{kl}(0)\> = \frac{-1}{48\pi^4} P^{(2)}_{ijkl}
\frac{c(m^2x^2)}{x^4} + P^{(0)}_{ijkl} \frac{f(M^2x^2)}{x^4}
\ee

In a flow between two CFT's, the central function
\cite{anselmi} $c(m^2x^2)$ approaches central charges
$c_{UV},c_{IR}$ in the appropriate limits, but $f(M^2 x^2)
\rightarrow 0$ in the UV and IR since effects of the trace
$T^i_i$ must vanish in conformal limits.

The correlators of $T_{ij}$ can be obtained from a generating
functional formally constructed by coupling the flat space theory
covariantly to a non-dynamical background metric $g_{ij}(x)$. For
example, in a gauge theory one would take
\be
S[g_{ij},A_k] \equiv \frac{1}{4}\int d^4x \sqrt{g}
g^{ik}g^{jl}F_{ij}F_{kl}
\ee
The effective action is then defined as the path integral over
elementary fields, \eg
\be \label{seff}
e^{-S_{eff}[g]} \equiv \int [dA_i] e^{-S[g,A]}
\ee
Correlation functions are obtained by functional
differentiation, viz.
\be
\< T_{i_1j_1}(x_1) \cdots T_{i_nj_n}(x_n)\>
=\frac{(-)^{n-1}2^n}{\sqrt{g(x_1)} \cdots \sqrt{g(x_n)}}
\frac{\d^n}{\d g^{i_1j_1}(x_1) \cdots g^{i_nj_n}(x_n)} S_{eff}[g]
\ee
with $g_{ij}\rightarrow \d_{ij}$.

\medskip

Consider two background metrics related by a Weyl transformation
$g^{\prime}_{ij}(x) = e^{2\sigma(x)} g_{ij}(x)$. Since the trace of
$T_{ij}$ vanishes in a CFT and $\< T^i_i \> = -\d S/\d\sigma$,
one might expect that $S_{eff}[g] = S_{eff}[g\prime]$. However,
$S_{eff}[g]$ is divergent and must be regulated. This must be
done even for a free theory (such as the pure $U(1)$ Maxwell
theory). In a free theory the correlators of composite operators
such as $T_{ij}$ are well defined for separated points but must be
regulated since they are too singular at short distance to have a
well defined Fourier transform. Regularization introduces a scale
and leads to the Weyl anomaly, which is expressed as
\be
\label{weyl}
\< T^i_i \> =\frac{c}{16\pi^2} W_{ijkl}^2
-\frac{a}{16\pi^2}\tilde{R}_{ijkl}^2 + \a \Box R + \b R^2
\ee
where the Weyl tensor and Euler densities are
\be
\begin{array}{c}
W_{ijkl}^2 = R_{ijkl}^2 -2 R_{ij}^2 +\frac{1}{3}R^2\\
\\
(\half \e_{ij}{}^{mn}R_{mnkl})^2 = R_{ijkl}^2 -4 R_{ij}^2 +R^2
\end{array}
\ee
The anomaly must be local since it comes from ultraviolet
divergences, and we have written all possible local terms of
dimension 4 above. One can show that $\b R^2$ violates the
Wess-Zumino consistency condition while $\Box R$ is the variation of
the local term $\int d^4x \sqrt{g} R^2$ in $S_{eff}[g]$. Finite
local counter terms in an effective action depend on the
regularization scheme and are usually considered not to carry
dynamical information.  (But see \cite{anselmi2} for a proposed
$c$-theorem based on this term. See \cite{duff2} for a more
extensive discussion of the Weyl anomaly.)

\medskip 

For the reasons above attention is usually restricted to the
first two terms in (\ref{weyl}). The scheme-independent
coefficients $c,a$ are central charges which characterize a CFT$_4$. One
can show by a difficult argument \cite{cfl,afgj} that $c$
for the critical theories CFT$_{UV}$, CFT$_{IR}$ agrees with the
fixed point limits $c_{UV},c_{IR}$ of $c(M^2x^2)$ in
(\ref{ttcorr}). The central charge $a$ is not measured in
$\<TT\>$ but agrees with constants $a_{UV},\,a_{IR}$  obtained in
short and long distance limits
of the 3-point function $\< TTT\>$, see
\cite{osborn,cappelli}

\medskip 

What can be said about monotonicity? One might expect
$c_{UV}\,>\, c_{IR}$, since the Weyl central charge is related to
$\< TT\>$ and thus closer to the notion of unitarity which was
important in Zamolodchikov's proof. However this inequality fails
in some field theory models. Cardy \cite{cardy} conjectured that
the inequality $a_{UV}\,>\,a_{IR}$ is the expression of the
c-theorem in QFT$_4$. This is plausible since $a$ is related to
the topological Euler invariant in common with $c$ for QFT$_2$,
and Cardy showed that the inequality is satisfied in several
models.  Despite much effort (see \cite{forte} and references
therein), there is no generally accepted proof of the $c$-theorem
in QFT$_4$.

\medskip

The values of $c,a$ for free fields have been known for years.
They were initially calculated by heat kernel methods, as
described in \cite{duff2}. The free field values agree with
$c_{UV},\,a_{UV}$ in any asymptotically free gauge theory, since
the interactions vanish at short distance. For a theory of $N_0$
real scalars, $N_{\half}$ Dirac fermions, and $N_1$ gauge bosons,
the results are
\be
\begin{array}{ll}
c_{UV} &= \frac{1}{120}[N_0 + 5N_{\half} +12N_1]\\
\\
a_{UV} &= \frac{1}{360}[N_0 +11N_{\half} +62N_1]
\end{array}
\ee
In a $SUSY$ gauge theory, component fields assemble into chiral
multiplets (2 real scalars plus 1 Majorana (or Weyl) spinor) and
vector multiplets (1 gauge boson plus 1 Majorana spinor). For a
theory with $N_{\chi}$ chiral and $N_V$ vector multiplets, the
numbers above give
\be
\begin{array}{ll}
c_{UV} &= \frac{1}{24}[N_{\chi} + 3N_{V}]\\
\\
a_{UV} &= \frac{1}{48}[N_{\chi} + 9N_{V}].
\end{array}
\ee

It is worthwhile to present some simple ways to calculate these
central charges which are directly accessible to field theorists.
Because of the relation to $\<T_{ij}T_{kl}\>$ detailed above, the
values of $c_{UV}$ can be easily read from a suitably organized
calculation of the free field 1-loop contributions of the various
spins. For gauge bosons one must include the contribution of
Faddeev-Popov ghosts.
\vspace{-1ex}
\begin{ex}
  Do this. Work directly in $x$-space at separated points. No
  integrals and no regularization is required. Organize the
  result in the form of the first term of (\ref{ttcorr}).
\end{ex}
\vspace{-1ex}

In $SUSY$ gauge theories the stress tensor has a supersymmetric
partner, the $U(1)_R$ current $R_i$. There are anomalies
when the theory is coupled to $g_{ij}$ and/or an external vector
$V_i(x)$ with field strength $V_{ij}$. Including both sources one
can write the combined anomalies as \cite{afgj}
\be
\begin{array}{ll}
\label{extanom}
\< T^i_i \> &=\frac{c}{16\pi^2} W_{ijkl}^2
-\frac{a}{16\pi^2}\tilde{R}_{ijkl}^2 +\frac{c}{6\pi^2} V_{ij}^2\\
\\
\<\partial_i \sqrt{g}R^i\> &= \frac{c-a}{24\pi^2}
R_{ijkl}\tilde{R}^{ijkl} + \frac{5a-3c}{9\pi^2} V_{ij}\tilde{V}^{ij}
\end{array}
\ee
Anomalies in the coupling of a gauge theory to external sources
may be called external anomalies. There are also internal or
gauge anomalies for both $\< T^i_i \>$ and $R_i$.   The gauge
anomaly of $R_i$ is described by an additional term
in (\ref{extanom}) proportional to $\b(g)F_{ij}\tilde{F}^{ij}$,
but this term vanishes in a CFT.

\medskip

The formula (\ref{extanom}) can be used to obtain $c,a$ from
1-loop fermion triangle graphs for both the UV and IR
critical theories.  The triangle graph for $\<R^iT_{jk}T_{lm}\>$
is linear in the $U(1)_R$ charges $r_{\hat{\a}}$ of the fermions
in the theory, while the graph for $\<R^iR_jR_k\>$ is cubic. We
consider a general $\N \! =1$ theory with gauge group $G$ and
chiral multiplets in representations $R_\a$ of $G$. Comparing
standard results for the anomalous divergences of triangle graphs
with (\ref{extanom}), one finds (see \cite{afgj,aefj}),
\bea
\label{anom1}
c-a & = -\frac{1}{16} ({\rm dim}G + \sum_{\a} {\rm dim}R_\a(r_\a-1))\\
\\
\label{anom2}
5a-3c & = \frac{9}{16}({\rm dim}G + \sum_{\a} {\rm dim}R_\a(r_\a-1)^3
\eea
We incorporate the facts that the $U(1)_R$ charge of the gaugino
is $r_\l =1$ while the charge of a fermion $\psi_\a$ in a chiral
multiplet is related to the charge of the chiral superfield
$\Phi^\a$ by $r_{\hat{\a}}\,=\,r_\a -1$.

\medskip

If asymptotic freedom holds, then the CFT$_{UV}$ is free, and one
obtains its central charges $c_{UV},\,a_{UV}$ using the free
field $U(1)_R$ charges, $r_\l =1$ for the gaugino and $r_\a
=\frac{2}{3}$ for chiral multiplets. The situation is more
complex for the CFT$_{IR}$ since the central charges are
corrected by interactions. Seiberg and others following his
techniques have found a large set of $SUSY$ gauge theories which
do flow to critical points in the IR \cite{seiberg88}.
The $\N\!=\!1$ superconformal
algebra $SU(2,2|1)$ contains a $U(1)_R$ current $S_i$ which is in
the same composite multiplet as the stress tensor. In many models
this current is uniquely determined as a combination of the free
current $R_i$ plus terms which cancel the internal (gauge)
anomalies of the former. Of course, the  current $S_i$ must also
be conserved classically.  Thus the $S$-charges of each $\Phi_\a$
arrange so that all terms in the superpotential $U(\Phi_\a)$ have
charge $2$.It is the $S$-current which is used
to  show that anomalies
match between Seiberg duals. These anomalies can be calculated from
1-loop graphs because the external anomalies are 1-loop exact for
currents with no gauge anomaly. This is just the standard
procedure of 't Hooft anomaly matching. The charge assigned by
the $S_i$ current is $r_\l =1$ for gauginos and uniquely
determined values $r_\a$ for chiral multiplets. It can be
shown~\cite{kogan,afgj} that $c_{IR}, \, a_{IR}$ are
obtained by inserting these values in (\ref{anom1}).

\medskip

Given this theoretical background it is a matter of simple algebra to
obtain the UV and IR central charges and subtract to deduce
the following formulas for their change in an RG  flow:
\bea
\label{cflow}
c_{UV} - c_{IR} & = \frac{1}{384} \sum_{\a}
dim R_a(2-3 r_{\a})[(7-6r_\a)^2-17]\\
\\
a_{UV} -a_{IR} &= \frac{1}{96}
\sum_{\a}dimR_\a(3r_\a-2)^2(5-3r_\a)
\eea
These formulas were applied \cite{aefj} to test the proposed $c$-theorem in
the very many Seiberg models of $SUSY$ gauge theories with IR
fixed points. Results indicated that the sign of $c_{UV} -c_{IR}$
is model-dependent, but $a_{UV} -a_{IR} >0$ in {\bf all} models.
Thus there is a wealth of evidence that the Euler central charge
satisfies a $c$-theorem, even though a fundamental proof is
lacking.
\vspace{-2ex}
\begin{ex}
Serious readers are urged to verify as many statements about the
anomalies as they can. For minimal credit on this exercise please
obtain the flow formulas (\ref{cflow}) from (\ref{anom1}).
\end{ex}
\vspace{-1ex}

Let us now apply some of these results to the field theories of
most concern to us, namely the undeformed $\N \! =4$ theory and its
$\N \! =1$ mass deformation. We can view the undeformed theory as
the UV limit of the flow of its $\N \! =1$ deformation.  The free
$R$-current assigns the charges
$(1,-\frac{1}{3},-\frac{1}{3},-\frac{1}{3})$ to the gaugino and
chiral matter fermions of the $\N=1$ description, while the
$S$-current of the mass deformed theory with superpotential in
(\ref{uls}) assigns $(1,-\frac{1}{2},-\frac{1}{2},0)$. In both
cases these are elements of the Cartan subalgebra of $SU(4)_R$ and
have vanishing trace. It is easy to see that the formula
(\ref{anom1}) for $c-a$ is proportional to this trace and
vanishes. The same observation establishes that both currents
have no gauge anomaly. The formula (\ref{anom2}) then becomes
\be
a=c=\frac{9}{32}(N^2-1)(1 +\sum{(r_\a-1)^{3}}).
\ee
Applied to the free current and then the $S$-current, this gives
\be \label{n=4anoms}
\begin{array}{ll}
a_{UV} = c_{UV} & = \frac{1}{4}(N^2-1) \\
\\
a_{IR} = c_{IR} & = \frac{27}{32} \frac{1}{4}(N^2-1).
\end{array}
\ee

The relation $\D = \frac{3}{2}r$ between scale dimension and
$U(1)_R$ charge also leads to the assignment of charges we have
used. In the UV limit we have the $\N \! =4$ theory with chiral
superfields $W_\a, \Phi^\b$ with dimensions $\frac{3}{2},1$. In
the {IR} limit we must consider the SU(2) invariant split
$W_\a,\Phi^{1,2},\Phi^3$, and the Leigh-Strassler argument for a
conformal fixed point which requires $\D
=\frac{3}{2},\frac{3}{4},\frac{3}{4},\frac{3}{2}.$ These values
give the fermion charges used above. It is no accident that $r=0$
for the fermion $\psi^3$. The $\Phi^3$ multiplet drops out at
long distance and thus cannot contribute to IR anomalies.

\subsection{Anomalies and the $c$-theorem from AdS/CFT}

One of the early triumphs of the AdS/CFT was the calculation of
the central charge $c$ for $\N \! =4$ SYM from the $\<TT\>$
correlator whose absorptive part was obtained from the
calculation of the cross-section for absorption of a graviton
wave by
the D3-brane geometry, \cite{Gubs}. This was reviewed in
\cite{kleb} and we will take a different viewpoint here.

We will describe in some detail the general approach of
Henningson and Skenderis \cite{hs} to the holographic Weyl
anomaly. This leads to the correct values of the central charges
and suggests a simple monotonic $c$-function.

We focus on the gravity part of the toy model action of Sec
\ref{flowbasic}
\be
\label{bulks}
S = \frac{-1}{16 \pi G} [\int d^{5}z \sqrt{g}(R +
\frac{12}{L^2}) + \int d^{4}z \sqrt{\gamma} 2K]
\ee
in which we have added the Gibbons-Hawking surface term which we
will explain further below. Lower spin bulk fields can be added
and do not change the gravitational part of the conformal anomaly.

\medskip

One solution of the Einstein equation is the AdS$_{d+1}$ geometry
which we previously wrote as
\be
ds^2 = e^{\frac{2r}{L}} \d_{ij} dx^i dx^j + dr^2
\ee
We introduce the new radial coordinate $\r = e^{-\frac{2r}{L}}$ in
order to follow the treatment of \cite{hs}. The boundary is now
at $\r=0$.
\vspace{-1ex}
\begin{ex}
Show that the transformed metric is
\be
ds^2 = L^2[\frac{d\r^2}{4\r^2} +\frac{1}{\r} \d_{ij} dx^i dx^j]
\ee
\end{ex}
\vspace{-1ex}
This is just AdS$_5$ in new coordinates. We now consider more
general solutions of the form
\be
\label{gensol}
ds^2 = L^2[\frac{d\r^2}{4\r^2} +\frac{1}{\r}g_{ij}(x,\r)dx^idx^j]
\ee
with non-trivial boundary data on the transverse metric, viz.
\be
g_{ij}(x,\r) \underset{\r \rarrow 0}{\longrightarrow} \bar{g}_{ij}(x)
\ee
The reason for this generalization may be seen by thinking of the
form $\bar{g}_{ij}(x) = \d_{ij} +h_{ij}(x)$. The first term
describes the flat boundary on which the CFT$_4$ lives, while
$h_{ij}(x)$ is the source of the stress tensor $T_{ij}$. We can use
the formalism to compute $\<T_{ij}\>,\,\<T_{ij}T_{kl}\>$, etc.

\vspace{-1ex}
\begin{ex}
Consider the special case of (\ref{gensol}) in which
$g_{ij}(x,\r)=g_{ij}(x)$ depends only on the transverse
$x^i$. Let $R_{ijkl}, R_{ij}$ and $R$ denote Riemann, Ricci and
scalar curvatures of the $4d$ metric $g_{ij}(x)$. Show that the
$5D$ metric thus defined satisfies the EOM $R_{\m\n} =
-4g_{\m\n}$ if $R_{ij}=0$. Show that the $5D$ curvature invariant
is
\be
R_{\m\n\r\s}R^{\m\n\r\s} = {\r^2 \over L^4} R_{ijkl}R^{ijkl} -{4\r
  \over L^2}R + {40 \over L^4}
\ee
Thus, as observed in \cite{cvpope}, if $R_{ij}=0$, we have a
reasonably generic solution of the $5D$ EOM's with a curvature
singularity on the horizon, $\r \rarrow \infty$.
\end{ex}
\vspace{-1ex}

As we will see shortly we will need to introduce a cutoff at
$\r=\e$ and restrict the integration in (\ref{bulks}) to the
region $\r \ge \e$. The induced metric at the cutoff is
$\gamma_{ij}={g_{ij} \over \r}$. The measure $\sqrt{\gamma}$
appears in the surface term in (\ref{bulks}) as does the trace of
the second fundamental form
\be
K = \gamma^{ij}K_{ij} = -g^{ij} \r \frac{\partial}{\partial \r}
\left({g_{ij}(x,\r) \over \e}\right) \lvert_{\r=\e}
\ee

We now consider a particular type of infinitesimal 5D
diffeomorphism first considered in this context in
\cite{isty}:
\be \label{diffeo}
\begin{array}{ll}
\r  & = \r^\prime (1 -2 \s(x^\prime)) \\
x^i & = x^{\prime i} + a^i(x^{\prime},\r^{\prime})\\
\end{array}
\ee
with
\be
a^i(x,\r) =\frac{L^2}{2}\int_{0}^\r d\hat{\r}
g^{ij}(x,\hat{\r})\partial_j \s(x)
\ee
\vspace{-1ex}
\begin{ex}
Show that $g_{55}^\prime =g_{55}$ and $g_{5i}^\prime =g_{5i}=0$
under this diffeomorphism, but that
\be
g_{ij} \rightarrow g{\prime}_{ij} =g_{ij} +2\s(1-\r{\partial \over
  \partial_\r})g_{ij} +\nabla_i a_j + \nabla_j a_i
\ee
\end{ex}
\vspace{-1ex}
In the boundary limit, $a_i \rarrow 0$ and $\r{\partial \over
  \partial_\r}g_{ij} \rarrow 0$, so that
\be
\bar{g}_{ij}(x) \rarrow \bar{g}\prime_{ij}(x)
=(1+2\s(x))\bar{g}_{ij}(x)
\ee
Hence the effect of the $5$D diffeomorphism is a Weyl
transformation of the boundary metric!

\medskip

This raises a puzzle. Consider the on-shell action
$S[\bar{g}_{ij}]$ obtained by substituting the solution
(\ref{gensol}) into (\ref{bulks}). Since the bulk action and the
field equations are invariant under diffeomorphisms, we would
expect $S[\bar{g}_{ij}] =S[\bar{g}\prime_{ij}]$. But AdS/CFT
requires that $S[\bar{g}_{ij}] = S_{eff}[\bar{g}_{ij}]$, and we
know that, due to the Weyl anomaly, $S_{eff}[\bar{g}_{ij}] \neq
S_{eff}[\bar{g}\prime_{ij}]$.

\medskip

The resolution of the puzzle is that $S[\bar{g}_{ij}]$ as we
defined it is meaningless since it diverges. This isn't the
somewhat fuzzy-wuzzy divergence usually blamed on the functional
integral for $S_{eff}[\bar{g}_{ij}]$ in quantum field theory. It
is very concrete; when you insert a solution of Einstein's
equation with the boundary behavior above into (\ref{bulks}), the
radial integral diverges near the boundary.

\medskip

Therefore we define a cutoff action $S_\e[\bar{g}_{ij}]$ as the
on-shell value of (\ref{bulks}) with radial integration restricted
to $\r \ge \e$. One can study its dependence on the cutoff to
obtain and subtract a counterterm action $S_\e[\bar{g}_{ij}]_{ct}$
to cancel singular terms as $\e \rarrow 0\,\,\,$.
$S_\e[\bar{g}_{ij}]_{ct}$ is an integral over the hypersurface $\r
=\e$ of a local function of the induced metric $\gamma_{ij}$ and
its curvatures, and it is not Weyl invariant. The renormalized
action is defined as
\be
\label{sren}
S_{ren}[\bar{g}] \equiv =\lim_{\e \rarrow 0} (S_\e[\bar{g}]
-S_e[\bar{g}]_{ct})
\ee

We now outline how the calculation of correlation functions and
the conformal anomaly proceeds in this formalism and then discuss
further necessary details. The variation of $S_{ren}$ is
\be
\label{vevt}
\d S_{ren}[\bar{g}] \equiv \half \int d4x
\sqrt{\bar{g}}\<T_{ij}\> \d \bar{g}^{ij}.
\ee
The variation defines the quantity $\<T_{ij}(x)\>$ which, in the
light of (\ref{seff}), is interpreted as the expectation value of
the field theory stress tensor in the presence of the source
$\bar{g}_{ij}$, and it depends non-locally on the source.
Correlation functions in the CFT are then obtained by further
differentiation, \eg
\be
\label{2pttt}
\<T_{ij}(x)T_{kl}(y)\> = -{2 \over \sqrt{\bar{g}(y)}} \frac {\d}{\d
  \bar{g}^{kl}(y)} \<T_{ij}(x)\> \lvert_{\bar{g}_{ij}=\d_{ij}}
\ee
The contributions to $\<T_{ij}(x)\>$ come from the surface term in
the radial integral in $S_\e[\bar{g}]$ and from
$S_\e[\bar{g}]_{ct}$. Possible contributions involving bulk
integrals vanish by the equations of motion.

The variation $\d \bar{g}^{ij}$ is arbitrary; let's choose it to
correspond to a Weyl transformation, \ie $\d \bar{g}^{ij}
=-2\bar{g}^{ij} \d\s$. Then (\ref{vevt}) gives
\be
\<T^i_i\> = \bar{g}^{ij}\<T_{ij}\>=- \frac{\d
  S_{ren}[\bar{g}]}{\d\s}
\ee
which is a standard result in quantum field theory in curved
space. The quantity $\<T^i_i\>$ is to be identified with the
conformal anomaly of the CFT and must therefore be local. It is
local, and the holographic computation gives (as we derive below)
\be
\label{holoanom}
\<T^i_i\> =\frac{L^3}{8\pi G} (\frac{1}{8} R^{ij}R_{ij}
-\frac{1}{24}R^2)
\ee
(The 2-point function (\ref{2pttt}) must be non-local, and it is.
See \cite{flow,holoren} for recent studies in the present formalism,
and \cite{mueck} for a closely related treatment.)

The holographic result may be compared with the field theory
$\<T^i_i\>$ in (\ref{weyl}). The absence of the
invariant $R_{ijkl}^2$ in (\ref{holoanom}) requires $c=a$. Thus
we deduce that any CFT$_4$ which has a holographic dual in this
framework must have central charges which satisfy $c=a$ (at least
as $N \rarrow \infty$ when the classical supergravity
approximation is valid.) This is satisfied by $\N\!=\!4$ SYM but not
by the conformal invariant $\N \! =2$ theory with an $SU(N)$ gauge
multiplet and $2N$ fundamental hypermultiplets.

\vspace{-1ex}
\begin{ex}
Show that when $c=a$ the QFT trace anomaly of (\ref{weyl})
reduces to
\be
\<T^i_i\> =\frac{c}{8\pi^2} (R^{ij}R_{ij}-\frac{1}{3}R^2)
\ee
\end{ex}
\vspace{-1ex}

Thus agreement with the holographic result (\ref{holoanom})
requires $c =\frac{\pi L^3}{8G}$. To check this recall that $G$
is the 5D Newton constant, so that $G = \frac{G_{10}}{Vol S_5} =
\frac{\pi L^3}{2N^2}$, where the last equality incorporates the
requirement that \AdS\ with 5-form flux $N$ is a
solution of the field equations of $D=10$ Type IIB sugra. This gives
the anomaly of undeformed $\N\! =4$ SYM theory on the nose!

The Henningson-Skenderis method is very elegant and has useful
generalizations \cite{deharo}. It is worth discussing in
more detail. The treatment starts with the mathematical result
\cite{feff} that the general solution of the Einstein equations
can be brought to the form (\ref{gensol}), and that the
transverse metric can be expanded in $\r$ near the boundary as
\be
\label{feffexp}
g_{ij}(x,\r) = \bar{g}_{ij} +\r g_{(2)ij} +\r^2g_{(4)ij} +\r^2 \ln \r
h_{(4)ij}+\cdots
\ee
The tensor coefficients are functions of the transverse
coordinates $x^i$. The tensors $g_{(2)ij},\, h_{(4)ij}$ can be
determined as local functions of the curvature $\bar{R}_{ijkl}$
of the boundary metric $\bar{g}_{ij}$. One just needs to
substitute the expansion (\ref{feffexp}) in the 5D field
equations $R_{\m\n}=-4g_{\m\n}$ and grind out a term-by-term
solution.
\vspace{-1ex}
\begin{ex}
Do this and derive $g_{(2)ij} =\half (\bar{R}_{ij}-\frac{1}{6}
\bar{R}\bar{g}_{ij})$. Very serious readers are encouraged to
obtain the more complicated result for $h_{(4)ij}$ given in (A.6) of
\cite{deharo}.
\end{ex}
\vspace{-1ex}

\medskip

The tensor $g_{(4)ij}$ is only partially determined by this
process of near-boundary analysis. Specifically its divergence
and trace are local in the curvature $\bar{R}_{ijkl}$, but
transverse traceless components are left undetermined. This is
sensible since the $EOM$'s are second order, and the single
Dirichlet boundary condition does not uniquely fix the solution.
At the linearized level the extra condition of regularity at
large $\r$ (the deep interior) is imposed. The transverse
traceless part of $g_{(4)ij}$ then depends non-locally on
$\bar{g}_{ij}$ and eventually contributes to $n$-point
correlators of $T_{ij}$ in the dual field theory.

\medskip

The local tensors in (\ref{feffexp}) are sufficient to determine
the divergent part of $S_\e[\bar{g}]$. It is tedious, delicate
(but straightforward!) to substitute the expansion in
(\ref{bulks}), integrate near the boundary and identify the
counterterms which cancel divergences. The result is
\be \label{sct}
S_\e[\bar{g}]_{ct} =\frac{1}{4\pi G} \int d^4x
\sqrt{\gamma}(\frac{3}{2 L^2}
-\frac{\hat{R}}{8}-\frac{L^2\ln\e}{32}(\hat{R}^{ij}\hat{R}_{ij}
-\frac{1}{3}\hat{R}^2))
\ee
where $\hat{R}_{ij}$ and $\hat{R}$ are the Ricci and scalar curvatures
of the induced metric $\gamma_{ij} = {g_{ij}(x,\e) \over \e}$. The first
two terms in (\ref{sct}) thus have power singularities as $\e \rarrow 0$.
Recall the discussion of cutoff dependence in Section~\ref{2point}.
In the $\r=z_0^2$ coordinate, the bulk cutoff $\e$ should be
identified with $1/\Lambda^2$~where $\Lambda$ is the UV cutoff in QFT.
Thus we find the quartic, quadratic, and logarithmic divergences
expected in QFT$_4$! ~(See Appendix B of \cite{deharo} for details of
the computation of (\ref{sct}).)

The next step is to calculate
\be
\<T^i_i(x)\>= - {\rm lim}_{\e\rarrow 0} \frac{\d}{\d\s (x)}
(S_\e[\bar{g}] -S_\e[\bar{g}]_{ct}).
\ee
However, one must vary the boundary data $\d\bar{g}_{ij} =2\d\s
\bar{g}_{ij}$ while maintaining the fact that the interior
solution corresponds to that variation. Thus one is really
carrying out the diffeomorphism of (\ref{diffeo}) so that $\d \e =
2\e\d\s(x)$. All terms of $S_\e[\bar{g}]$ are invariant under
the combined change of coordinates and change of shape of the
cutoff hypersurface. The first two terms in $S_\e[\bar{g}]_{ct}$
are also invariant. There is the explicit variation $\d\ln\e =
-2\d\s(x)$ in the logarithmic counterterm, and this is the only
variation since the boundary integral is the difference of the
Weyl$^2$ and Euler densities and is invariant.  Thus we find the result
(\ref{holoanom}) stated earlier in a strikingly simple way!

The method just described may be applied to the calculation of
holographic conformal anomalies in any {\bf even} dimension
\cite{hs,deharo}. However for {\bf odd} dimension the structure
of the near-boundary expansion (\ref{feffexp}) changes. There is
no $\ln\r$ term and no logarithmic counterterm either. Hence no
conformal anomaly in agreement with QFT in odd dimension.

\subsection{The Holographic $c$-theorem}

The method just discussed can be extended to apply to the Weyl
anomalies of the critical theories at end-points of holographic
RG  flows. In general we can consider a domain wall
interpolating between the region of an AdS$_{UV}$ with scale
$L_{UV}$ and the deep interior of an AdS$_{IR}$ with scale
$L_{IR}$. The holographic anomalies are
\be
c_{UV}=\frac{\pi}{8G}L_{UV}^3\quad c_{IR}=\frac{\pi}{8G}L_{IR}^3
\ee
The first result can be derived by including relevant scalar
fields in the previous method, and latter by applying the method
to an entire AdS geometry with scale $L_{IR}$.

For any bulk domain wall one can consider the following
scale-dependent function (and its radial derivative):
\be
\begin{array}{ll}
C(r) &= \frac{\pi}{8G} \frac{1}{A^{\prime 3}}\\
\\
C^{\prime}(r) & =  \frac{\pi}{8G}
\frac{-3A^{\prime\prime}}{A^{\prime 4}}
\end{array}
\ee
We have $C^{\prime}(r) \ge 0$ as a consequence of the
condition $A^{\prime\prime} \le 0$ derived from the domain wall
$EOM$'s in Section~\ref{intflow1}. Thus $C(r)$ is an essentially perfect
holographic $c$-function:
\vspace{-1ex}
\begin{enumerate}
\setlength{\itemsep}{-1ex}
\item It decreases monotonically along the flow from $UV \rarrow IR$.
\item It interpolates between the central charges $c_{UV}$ and
  $c_{IR}$.
\item If perfect, it would be stationary only if conformal
  symmetry holds. This is true if the domain wall is the solution
  of the first order flow equations discussed in Sec
  \ref{intflow2} and thus true for SUSY flows.
\end{enumerate}
\vspace{-1ex}

The moral of the story is that the $c$-theorem for RG  flows,
which has resisted proof by field theory methods, is trivial when
the theory has a gravity dual since $A^{\prime\prime} \le 0$. See
\cite{gppz1,fgpw1}.

Finally, we note that for the mass deformed $\N\!=\!4$ theory the
ratio $(\frac{L_{IR}}{L_{UV}})^3 = (\frac{W_{UV}}{W_{IR}})^3 =
\frac{27}{32}$. Thus the
holographic prediction of $c_{IR}=a_{IR}$ agrees with the field
theory result in (\ref{n=4anoms})! See \cite{klm}

There is much more to be said about the active subject of
holographic RG  flows and many interesting papers that deserve
study by interested theorists. We hope that the introduction to
the basic ideas contained in these lecture notes will stimulate
that study.

\bigskip

\section{Acknowledgements}

We are grateful to Massimo Bianchi, Umut Gursoy, Krzysztof Pilch
and Kostas Skenderis for many
helpful comments on the manuscript. We would also like to thank
the organizers of TASI 2001, Steven Gubser, Joe~Lykken and the general director  
K.T.~Mahanthappa
for their invitation to this stimulating Summer School. Finally, these
lectures were based in part on material presented by E.D. at Saclay,
S\` ete and Ecole Normale Superieure, and by D.Z.F. during the Semestre
Supercordes at the Institut Henri Poincar\' e and at the Les Houches summer
school. We wish to thank Costas Bachas, Eug\` ene Cremmer, Michael Douglas, 
Ivan
Kostov, Andr\' e Neveu, Kelley Stelle, Aliocha Zamolodchikov and Jean-Bernard
Zuber for invitations to present those lectures.

\vfill\eject

\end{document}